\newif\ifcern
\newif\ifijmp

\cernfalse
\ijmptrue

\cerntrue
\ijmpfalse

\ifcern
  \documentclass[ALICE,manyauthors]{cernphprep}
  \def\refcite{\cite}
  \def\tbl{\caption}
  \def\toprule{\hline}
  \def\colrule{\hline}
  \def\botrule{\hline}
\fi

\ifijmp
  \documentclass{ws-ijmpa}
  \def\bib{B\kern-.05em{I}\kern-.025em{B}\kern-.08em}
  \def\btex{B\kern-.05em{I}\kern-.025em{B}\kern-.08em\TeX}
  \def\Irefn#1{\textsuperscript{\ref{#1}}}
  \def\And{, }
  \def\Aref#1{$^{\mathrm{#1})}$} 
  \newenvironment{Authlist}{\center}{\endcenter}
  \gdef\Adef#1{\label{LA#1}}
  \gdef\Aref#1{\textsuperscript{,\ref{LA#1}}}
  \gdef\Idef#1{\label{LI#1}\gdef\INTlatex{#1}}
  \gdef\Irefn#1{\textsuperscript{\ref{LI#1}}}

  \gdef\And{, }
  \gdef\NNospacing{\labelsep=0pt\itemsep=0pt\topsep=0pt\partopsep=0pt\parskip=0pt\parsep=0pt}
  \renewenvironment{Authlist}{\vspace*{-1.5ex}\enumerate\NNospacing}
                              {\endlist\vspace*{-1.5ex}}
\fi

\usepackage{color}
\usepackage{lineno}
\usepackage{amssymb}
\usepackage{xspace}
\usepackage[colorlinks,linkcolor=red,urlcolor=blue]{hyperref}

\newcommand{\pp}         {pp\xspace}
\newcommand{\ee}         {{e$^+$e$^-$}\xspace}
\newcommand{\ppb}        {\mbox{p--Pb}\xspace}
\newcommand{\pbp}        {\mbox{Pb--p}\xspace}
\newcommand{\pbpb}       {\mbox{Pb--Pb}\xspace}

\newcommand{\sqrts}      {\ensuremath{\sqrt{s}}\xspace}
\newcommand{\sqrtsnn}    {\ensuremath{\sqrt{s_\mathrm{NN}}}\xspace}
\newcommand{\gevc}       {~\ensuremath{\mathrm{GeV}\!/c}\xspace}
\newcommand{\gevcc}      {~\ensuremath{\mathrm{GeV}\!/c^2}\xspace}
\newcommand{\mevc}       {~\ensuremath{\mathrm{MeV}\!/c}\xspace}
\newcommand{\mevcc}      {~\ensuremath{\mathrm{MeV}\!/c^2}\xspace}
\newcommand{\lumi}       {\ensuremath{\mathcal{L}}\xspace}
\newcommand{\mum }       {~\ensuremath{\mu\mathrm{m}}\xspace}

\newcommand{\pim}        {\ensuremath{\pi^-}\xspace}
\newcommand{\pip}        {\ensuremath{\pi^+}\xspace}
\newcommand{\pizero}     {\ensuremath{\pi^0}\xspace}
\newcommand{\kam}        {K\ensuremath{^-}\xspace}
\newcommand{\kap}        {K\ensuremath{^+}\xspace}
\newcommand{\kzerol}     {K\ensuremath{^0_L}\xspace}
\newcommand{\kzeros}     {K\ensuremath{^0_\mathrm{S}}\xspace}
\newcommand{\jpsi}       {J/\ensuremath{\psi}\xspace}

\newcommand{\pt}         {\ensuremath{p_\mathrm{T}}\xspace}

\newcommand{\Et}         {\ensuremath{E_\mathrm{T}}\xspace}

\newcommand{\nch}        {\ensuremath{N_\mathrm{ch}}\xspace}
\newcommand{\dndy}       {\ensuremath{\mathrm{d}\nch/\mathrm{d}y}\xspace}
\newcommand{\dndeta}     {\ensuremath{\mathrm{d}\nch/\mathrm{d}\eta}\xspace}

\newcommand{\ezdc}       {\ensuremath{E_\mathrm{ZDC}}\xspace}
\newcommand{\degr}       {\ensuremath{^\circ}\xspace}

\newcommand{\npart}      {\ensuremath{N_\mathrm{part}}\xspace}
\newcommand{\ncol}       {\ensuremath{N_\mathrm{coll}}\xspace}
\newcommand{\dedx}       {\ensuremath{\mathrm{d}E/\mathrm{d}x}\xspace}
\newcommand{\vd}         {\ensuremath{v_\mathrm{drift}}\xspace}
\newcommand{\tz}         {\ensuremath{t_0}\xspace}

\newcommand{\0}          {\hphantom{0}} 
\newcommand{\trueepsilon}{\text{\usefont{OML}{cmr}{m}{n}\symbol{15}}}

\newcommand{\ptjet}{\ensuremath{\pt^{\rm jet}}\xspace}

\newcommand{\Eclust}{\ensuremath{E_{\rm{clust}}}\xspace}
\newcommand{\sump}{\ensuremath{\Sigma_{\rm{p}}}\xspace}
\newcommand{\fsub}{\ensuremath{f_{\rm{sub}}}\xspace}

\newcommand{\Ecorr}{\ensuremath{E_{\rm{corr}}}\xspace}

\newcommand{\DEcorr}{\ensuremath{\Delta\Ecorr}\xspace}

\newcommand{\Rcorr}{\ensuremath{R_{\rm{corr}}}\xspace}

\newcommand{\invub}{\mu\mathrm{b}^{-1}}      
\newcommand{\invubps}{\invub\mathrm{s}^{-1}} 

\newcommand{\invnb}{\mathrm{nb}^{-1}}        

\newcommand{\invpb}{\mathrm{pb}^{-1}}        

\begin{document}

\ifijmp
  \markboth{The ALICE Collaboration} {Performance of the ALICE Experiment at the CERN LHC}
  \catchline{}{}{}{}{}
\fi

\ifcern
  \begin{titlepage}
  \PHnumber{031}         
  \PHyear{2014}          
  \PHdate{18 February}   
\fi

\title{Performance of the ALICE Experiment at the CERN LHC}

\ifcern
  \ShortTitle{Performance of the ALICE Experiment}   
  \Collaboration{ALICE Collaboration\thanks{See Appendix~\ref{app:collab} for the list of collaboration members}}
  \ShortAuthor{ALICE Collaboration}      
\fi

\ifijmp
  \author{The ALICE Collaboration\footnote{\ref{app:collab}}}
  \maketitle
\fi


\begin{abstract}
ALICE is the heavy-ion experiment at the CERN Large Hadron Collider. 
The experiment continuously took data during the first physics campaign 
of the machine from fall 2009 until early 2013, using proton and lead-ion beams. 
In this paper we describe the running environment and the data handling 
procedures, and discuss the performance of the ALICE detectors and analysis 
methods for various physics observables. 
\end{abstract}
\ifcern
  \end{titlepage}
  \setcounter{page}{2}
\fi

\tableofcontents

\clearpage\newpage\section{ALICE apparatus}
\label{sect:intro}

ALICE~\cite{Carminati:2004fp,Alessandro:2006yt,Aamodt:2008zz} 
(A Large Ion Collider Experiment) is a major experiment at the Large 
Hadron Collider (LHC), Geneva, which is optimized for 
the study of QCD matter created in high-energy collisions 
between lead nuclei. Analysis based on QCD (quantum chromodynamics) 
lead to a prediction of the existence of a state of deconfined quarks 
and gluons at energy densities above 1~GeV/fm$^3$. 
The transition to this state is accompanied by chiral symmetry 
restoration, in which the quarks assume their current masses. 
This state of matter occurred in the early universe after the 
electroweak phase transition, i.e. at the age of \mbox{10$^{-12}$--10$^{-5}$ s} 
(for a recent review see Ref.~\refcite{BraunMunzinger:2009zz}.)
High-energy nuclear collisions allow such energy densities to be reached, albeit 
in a small volume and for a limited duration. Assessing the properties of 
the created matter requires a sound understanding of the underlying 
collision dynamics. For this, the heavy-ion (AA) collision studies in the 
new energy regime accessible at the LHC have to be complemented by 
proton-proton (\pp) and proton-nucleus (pA) collision experiments. 
These control measurements, besides being interesting in themselves, 
are needed to separate the genuine QCD-matter signals from the cold-matter 
initial- and final-state effects. The physics goals of ALICE are described 
in detail in Refs.~\refcite{Carminati:2004fp,Alessandro:2006yt}; 
the results obtained to date are accessible at Ref.~\refcite{alicepub}. 

\begin{figure}[b]
\centering\rotatebox{90}{\scalebox{0.9}{\includegraphics{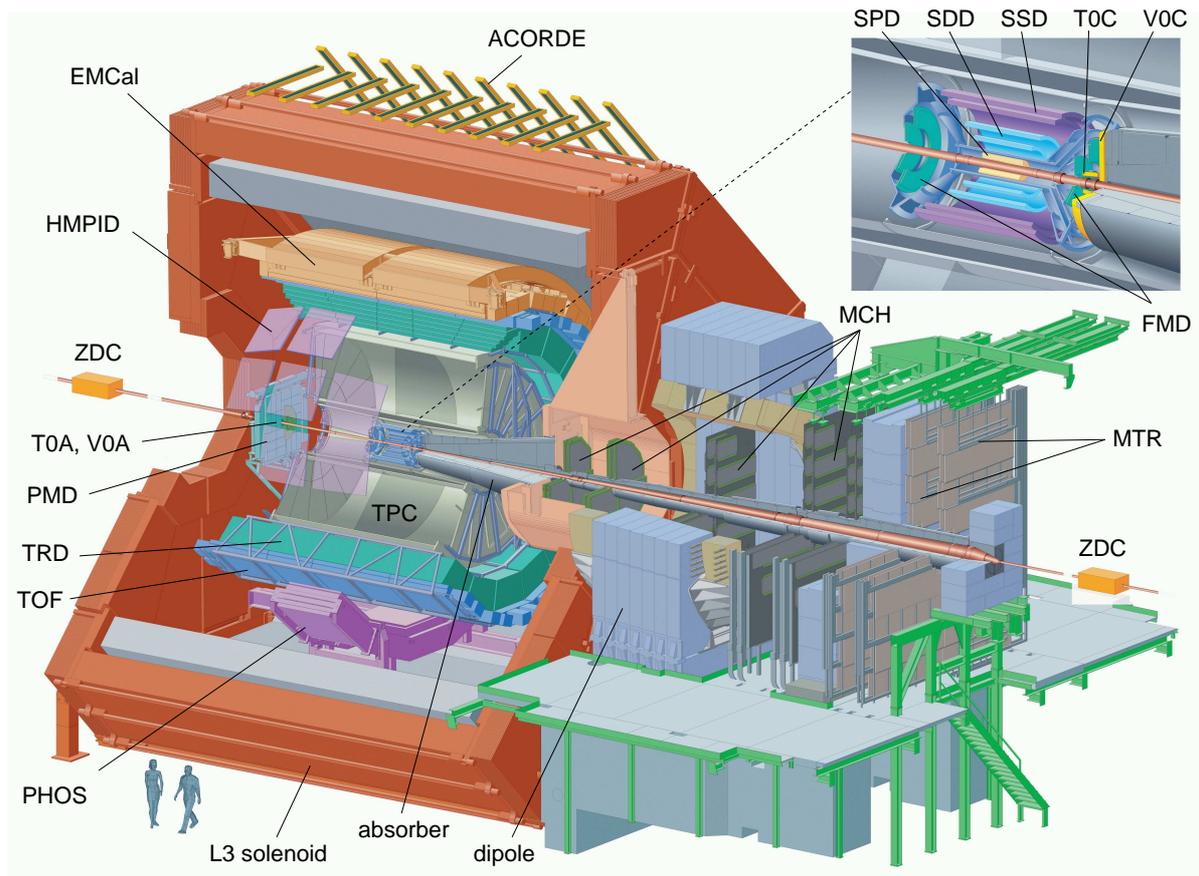}}}
\caption{The ALICE experiment at the CERN LHC. The central-barrel detectors 
(ITS, TPC, TRD, TOF, PHOS, EMCal, and HMPID) are embedded in a solenoid with magnetic field 
$B=0.5$~T and address particle production at midrapidity. 
The cosmic-ray trigger detector ACORDE is positioned on top of the magnet. 
Forward detectors (PMD, FMD, V0, T0, and ZDC) are used for triggering, event 
characterization, and multiplicity studies. 
The MUON spectrometer 
covers $-4.0<\eta<-2.5$, $\eta=-\ln\tan(\theta\!/2)$.  
}
\label{fig:setup}
\end{figure}

The ALICE apparatus (Fig.~\ref{fig:setup}) 
has overall dimensions of 16$\times$16$\times$26~m$^3$ and a total weight of $\sim$10~000~t. 
It was designed to cope with the particle densities expected in central \pbpb collisions 
at the LHC. 
The experiment has a high detector granularity, a low transverse momentum threshold 
$\pt^\mathrm{min} \approx0.15$\gevc, and good particle identification capabilities 
up to 20\gevc.
The seventeen ALICE detector systems, listed in Table~\ref{tab:dets}, fall into three categories: 
central-barrel detectors, forward detectors, and the MUON spectrometer. 
In this section, a brief outline of their features is given. Specifications and a 
more detailed description can be found in Ref.~\refcite{Aamodt:2008zz}. 

\begin{table}[b]
\ifijmp
  \renewcommand{\arraystretch}{1.15}
\fi
\centering
{\begin{tabular}{@{\extracolsep{-4.5mm}}lccccl@{}} \toprule
  Detector &          \multicolumn{2}{c}{Acceptance}      & Position               & Technology         & Main purpose         \\
           & Polar            & Azimuthal &               &                        &                                           \\
\colrule
SPD*       &  $|\eta|<2.0\0$  &   full                    & $r=\03.9$~cm           & Si pixel           & tracking, vertex     \\
           &  $|\eta|<1.4\0$  &   full                    & $r=\07.6$~cm           & Si pixel           & tracking, vertex     \\
SDD        &  $|\eta|<0.9\0$  &   full                    & $r=15.0$~cm            & Si drift           & tracking, PID        \\
           &  $|\eta|<0.9\0$  &   full                    & $r=23.9$~cm            & Si drift           & tracking, PID        \\
SSD        &  $|\eta|<1.0\0$  &   full                    & $r=38$~cm              & Si strip           & tracking, PID        \\
           &  $|\eta|<1.0\0$  &   full                    & $r=43$~cm              & Si strip           & tracking, PID        \\
TPC        &  $|\eta|<0.9\0$  &   full                    & $\085<r/{\rm cm}<247$  & Ne drift+MWPC      & tracking, PID        \\
TRD*       &  $|\eta|<0.8\0$  &   full                    & $290<r/{\rm cm}<368$   & TR+Xe drift+MWPC\0 & tracking, $e^\pm$ id \\
TOF*       &  $|\eta|<0.9\0$  &   full                    & $370<r/{\rm cm}<399$   & MRPC               & PID                  \\
PHOS*      &  $|\eta|<0.12$   &$220\degr<\phi<320\degr$   & $460<r/{\rm cm}<478$   & PbWO$_4$           & photons              \\
EMCal*     &  $|\eta|<0.7\0$  &$\080\degr<\phi<187\degr$  & $430<r/{\rm cm}<455$   & Pb+scint.          & photons and jets     \\
HMPID      &  $|\eta|<0.6\0$  &$\0\01\degr<\phi<\059\degr$& $r=490$~cm      & \!\! C$_6$F$_{14}$ RICH+MWPC\0 & PID              \\
ACORDE*    &  $|\eta|<1.3\0$  &$\030\degr<\phi<150\degr$  & $r=850$~cm             & scint.             & cosmics              \\
PMD        &  $2.3<\eta<3.9$  &   full                    & $z=367$~cm             & Pb+PC              & photons              \\
FMD        &  $3.6<\eta<5.0$  &   full                    & $z=320$~cm             & Si strip           & charged particles    \\
           &  $1.7<\eta<3.7$  &   full                    & $z=\080$~cm            & Si strip           & charged particles    \\
           & $-3.4<\eta<-1.7$ &   full                    & $z=-70$~cm             & Si strip           & charged particles    \\
V0*        &  $2.8<\eta<5.1$  &   full                    & $z=329$~cm             & scint.             & charged particles    \\
           & $-3.7<\eta<-1.7$ &   full                    & $z=-88$~cm             & scint.             & charged particles    \\
T0*        &  $4.6<\eta<4.9$  &   full                    & $z=370$~cm             & quartz             & time, vertex         \\
           & $-3.3<\eta<-3.0$ &   full                    & $z=-70$~cm             & quartz             & time, vertex         \\
ZDC*       &  $|\eta|>8.8$    &   full                    & $z=\pm113$~m           & W+quartz           & forward neutrons     \\
           & $6.5<|\eta|<7.5$ & $|\phi|<10\degr$          & $z=\pm113$~m           &\0\0brass+quartz\0\0& forward protons      \\
           &  $4.8<\eta<5.7$  & $|2\phi|<32\degr$         & $z=7.3$~m              & Pb+quartz          & photons              \\
MCH        & $-4.0<\eta<-2.5$ &   full                    &$-14.2<z/{\rm m}<-5.4$  & MWPC               & muon tracking        \\
MTR*       & $-4.0<\eta<-2.5$ &   full                    &$-17.1<z/{\rm m}<-16.1$ & RPC                & muon trigger         \\
\botrule
\end{tabular} 
\tbl{The ALICE detectors. 
The transverse (for HMPID, radial) and longitudinal coordinates $r$, $z$ are measured 
with respect to the ALICE interaction point (IP2). The $z$ axis points along the anticlockwise 
LHC beam. The detectors marked with an asterisk (*) are used for triggering. 
As of 2013, 13/18 of the TRD modules and 3/5 of the PHOS modules have been installed.
The ZDCs were moved from $|z|\approx114.0$~m to $|z|\approx112.5$~m during the winter 
shutdown 2011/2012. 
The $\eta$ and $\phi$ ranges specified for the proton ZDC are purely geometrical and do 
not take into account how charged particles are transported through the magnetic elements 
of the beam line.
}
\label{tab:dets}}
\end{table}

The central-barrel detectors -- Inner Tracking System (ITS), Time Projection 
Chamber (TPC), Transition Radiation Detector (TRD), Time Of Flight (TOF), 
Photon Spectrometer (PHOS), Electromagnetic Calorimeter (EMCal), and 
High Momentum Particle Identification Detector (HMPID) -- 
are embedded in the L3 solenoid magnet which has $B$=0.5~T. 
The first four cover the full azimuth, 
with a segmentation of 20\degr, at midrapidity ($|\eta|\lesssim0.9$). 
The ITS and the TPC are the main charged-particle tracking detectors of ALICE. 
The ITS is composed of six tracking layers, two Silicon Pixel Detectors (SPD), two 
Silicon Drift Detectors (SDD), and two Silicon Strip Detectors (SSD). 
The TPC has a 90~m$^3$ drift volume filled with Ne--CO$_2$ and is divided into two 
parts by the central cathode, which is kept at -100~kV. The end plates 
are equipped with multiwire proportional chambers (MWPC). 
In addition to tracking, SDD and TPC provide charged-particle identification 
via measurement of the specific ionization energy loss \dedx. 
The TRD detector consists of six layers of Xe--CO$_2$-filled MWPCs, with a fiber/foam 
radiator in front of each chamber. It is used for charged-particle tracking and for 
electron identification via transition radiation and \dedx. 
The TOF detector, which is based on Multigap Resistive Plate Chamber (MRPC) technology, 
is used for particle identification at intermediate momenta. 
Finally, the cylindrical volume outside TOF is shared by two electromagnetic 
calorimeters with thicknesses of $\sim\!\!20 \ X_0$ (radiation lengths) 
and $\sim\!\!1 \ \lambda_{\rm int}$ (nuclear interaction length), 
the high-resolution PHOS and the large-acceptance EMCal, along with the 
ring-imaging Cherenkov detector HMPID, which has a liquid C$_6$F$_{14}$ 
radiator and a CsI photo-cathode for charged-hadron identification at intermediate momenta. 

The central barrel detectors have an 18-fold segmentation in azimuth. The ITS, TPC, 
and TOF cover the entire azimuthal range, which is of significant advantage for 
measurements of angular distributions and correlations. Modules of TRD, PHOS, 
and EMCal were successively added during the first years of running. The 
installation history of these detectors is given in Table~\ref{tab:installation}. 
\begin{table}[h]
\ifijmp
  \renewcommand{\arraystretch}{1.15}
\fi
\centering
{\begin{tabular}{@{\hspace{7mm}}@{\extracolsep{7mm}}cccc@{\hspace{7mm}}} \toprule
        &       TRD      &     PHOS      &      EMCal     \\
        & $|\eta|<0.8\0$ & $|\eta|<0.12$ & $|\eta|<0.7\0$ \\
\colrule
2008    &        4       &       1       &        0       \\
2009    &        7       &       3       &        2       \\
2010    &        7       &       3       &        2       \\
2011    &       10       &       3       &        5       \\
2012    &       13       &       3       & 5$\,^1$\!/\scriptsize{3} \\
2013    &       13       &       3       & 5$\,^1$\!/\scriptsize{3} \\
Goal    &       18       &       5       & 5$\,^1$\!/\scriptsize{3} \\
\botrule
\end{tabular} 
\tbl{Number of sectors (20\degr in azimuth each) of the central barrel covered 
by TRD, PHOS, and EMCal in the first years of ALICE running.}
\label{tab:installation}}
\end{table}

The ALICE forward detectors include the preshower/gas-counter Photon Multiplicity 
Detector (PMD) and the silicon Forward Multiplicity Detector (FMD), which are 
dedicated to the measurement of photons and charged particles around $|\eta|\approx3$, 
respectively. 
The quartz Cherenkov detector T0 delivers the time and the longitudinal position 
of the interaction. 
The plastic scintillator detector V0\footnote{In ALICE physics papers 
an alternative notation, VZERO, is used to avoid conflict with V$^0$, the neutral 
particle decaying into two charged tracks (see Section~\ref{sect:secondary}). 
In this article we follow the original notation 
from~Refs.~\refcite{Carminati:2004fp,Alessandro:2006yt,Aamodt:2008zz}.}
measures charged particles at $-3.7<\eta<-1.7$ 
and $2.8<\eta<5.1$, and is mainly used for triggering and for the determination of 
centrality and event plane angle in \pbpb collisions~\cite{Abbas:2013taa}. 
The centrality can also be measured with the Zero Degree Calorimeter (ZDC). The ZDC  
consists of two tungsten-quartz neutron (ZN) and two brass-quartz proton (ZP) calorimeters, 
placed symmetrically on both sides of the Interaction Point and used to count spectator nucleons. 
The ambiguity between the most central (few spectator nucleons) and the most peripheral 
(spectator nucleons bound in nuclear fragments) collisions is resolved by using 
an electromagnetic calorimeter (ZEM), which consists of two modules placed symmetrically 
on both sides of the beam pipe at $4.8<\eta<5.7$. 

The MUON spectrometer, with a hadron absorber of $\sim\!\!10 \; \lambda_{\rm int}$, 
a dipole magnet of 3~Tm, and five tracking stations with two pad chambers each 
(Muon Chambers, MCH), is used to measure quarkonium and light vector meson 
production in a region of $-4.0<y<-2.5$. 
The measurement of high-\pt muons, which predominantly come from the decay of charm 
and beauty, also falls within the scope of the spectrometer.
Single-muon and muon-pair triggers with an adjustable transverse-momentum threshold 
are provided by two further stations (Muon Trigger, MTR) placed behind an additional 
$7\lambda_{\rm int}$ absorber. 

The physics goals and a detailed description of the detectors and their expected 
performance can be found in Refs.~\refcite{Carminati:2004fp,Alessandro:2006yt,Aamodt:2008zz}. 
In this paper we report the actual performance achieved in the LHC data taking campaign 
2009-2013 (LHC Run 1). 
The collision systems and energies inspected by ALICE are summarized in 
Table~\ref{tab:statistics-of-blocks} in Section~\ref{sect:datataking}. In the following, 
we start from a description of the running conditions, data taking and calibration, 
and then review the performance of the experiment in terms of various physics observables. 

The ALICE Coordinate System, used in Table~\ref{tab:dets} 
and throughout the paper, is a right-handed orthogonal Cartesian system 
defined as follows~\cite{refsystem}. 
The origin is at the LHC Interaction Point 2 (IP2).  
The $z$ axis is parallel to the mean beam direction at IP2 and points along the LHC Beam~2 
(i.e. LHC anticlockwise). 
The $x$ axis is horizontal and points approximately towards the center of the LHC. 
The $y$ axis, consequently, is approximately vertical and points upwards. 

\clearpage\newpage\section{Beam conditions}
\label{sect:conditions}
 
\subsection{Beam parameters}

ALICE is situated at the interaction point IP2 of the LHC, close to the Beam~1 Transfer 
Line TI~2 injection region. The ALICE design, optimized for nuclear 
collisions~\cite{Alessandro:2006yt}, requires a reduced luminosity in \pp 
interactions at IP2.  After three years of operation at the LHC, experience 
has shown that the maximum \pp interaction rate at which all ALICE detectors 
can be safely operated is around 700~kHz (including the contribution of both 
beam--beam and beam--gas collisions). Typical target luminosity values for the 
\hbox{ALICE} \pp data taking range from \lumi~$\simeq$~10$^{29}$~s$^{-1}$cm$^{-2}$ 
(during minimum bias data taking) to \lumi~$\simeq$~10$^{31}$~s$^{-1}$cm$^{-2}$ 
(when accumulating rare triggers). The average number of interactions per 
bunch crossing ($\mu$) varies from ~0.05 to ~0.3. 

During LHC Run 1, the instantaneous luminosity delivered to ALICE in \pp 
collisions was adjusted by the machine to the required level by optimizing 
the following parameters: number of interacting bunches; value of the amplitude function 
at the interaction point\footnote{In accelerator physics, the amplitude function 
$\beta(z)$ describes the single-particle motion and determines the variation of 
the beam envelope as a function of the coordinate along the beam orbit, $z$ 
(see e.g. Ref.~\refcite{accPhys}). The parameter $\beta^*$ denotes the value 
of $\beta(z)$ at the interaction point.} $\beta^*$ and crossing angles; 
and separation of colliding beams (in the plane orthogonal to the crossing plane). 
Typically, the beams had to be separated at IP2 by 1.5--3.5 times the RMS 
of the transverse beam profile, depending on the values of $\beta^*$, bunch 
intensity, and emittance. In 2012, the machine was operated at the highest 
beam intensities so far (up to $\simeq$2$\times$10$^{14}$~protons/beam). 
In order to ensure the necessary levelling 
of \lumi and $\mu$ at IP2, a ``main--satellite'' bunch collision scheme was 
adopted: ALICE took data by triggering on the encounters of the main bunches of 
one beam with the satellite bunches of the other beam, sitting 10 RF buckets (25~ns) 
away from the nearest main bunch. The intensity of the satellite bunches is 
typically 0.1\% of that of the main bunches ($\sim$~1.6$\times$10$^{11}$ p), 
therefore the luminosity per colliding bunch pair was reduced by the same 
factor. The very low $\mu$ was balanced by the large ($>$~2000) number of 
main--satellite encounters per LHC orbit, thus allowing the required \lumi to 
be achieved with collisions quite uniformly distributed along the LHC orbit, with low pileup. 

The rate of \pbpb collisions in 2010 and 2011 was well below the ALICE 
limits and ALICE was able to take data at the highest achievable luminosity, on the 
order of 10$^{25}$~s$^{-1}$cm$^{-2}$ in 2010 and 10$^{26}$~s$^{-1}$cm$^{-2}$ in 2011, 
with the corresponding hadronic 
$\mu$ being on the order of 10$^{-5}$--10$^{-4}$ and  10$^{-4}$--10$^{-3}$, 
respectively. The maximum manageable interaction rate for \ppb collisions 
was 200~kHz, roughly corresponding to a luminosity of 
1$\times$10$^{29}$~s$^{-1}$cm$^{-2}$, only slightly below the LHC peak 
luminosity in 2013. The hadronic interaction probability in such conditions 
is about~0.06. 

The $\beta^*$ parameter at IP2 was 3.5~m for most of 2010, including the 
\pbpb run. In 2011 it was 10~m for the \pp runs and 1~m for the \pbpb run. 
Finally, a value of 3~m was used in 2012, and it was reduced to 0.8~m for 
the \ppb run at the beginning of 2013. The corresponding beam RMS widths for 
typical emittance values range from 15 to 150~$\mu$m. The longitudinal size 
of the luminous region depends mainly on the bunch length. Its typical RMS 
value is about 6~cm. The size of the luminous region was determined from 
ALICE data, via the distribution of interaction vertices (see 
Section~\ref{sect:tracking}) and was monitored online. 

Due to the muon spectrometer dipole magnet and its respective compensator 
magnet, there is an intrinsic (internal) vertical crossing angle at IP2, 
which varies with the energy per nucleon ($E$), charge ($Z$), and mass number 
($A$) of the beam particles as 
\begin{equation}
\alpha_\mathrm{int}=\frac{Z}{A} \; \frac{E_0}{E} \; \alpha_0 \, ,
\end{equation} 
with $E_0$~=~3.5~TeV/nucleon and $\alpha_0=280 \>\, \mu\mathrm{rad}$. 
In addition, an external vertical crossing angle $\alpha_{\rm{ext}}$ can be 
applied by means of a suitable magnet current setup dependent 
on $E$ and $\beta^*$ in order to control long range beam--beam effects and 
to prevent parasitic collisions in the vicinity of the IP. During \pbpb runs 
the external crossing angle is combined with the internal crossing 
angle in a way that minimizes the net crossing angle, in order to 
prevent acceptance losses in the ZDCs due to shadowing of the spectator 
neutron spot by the LHC tertiary collimators~\cite{zdcTCTV}. 
 
The main beam parameters at IP2 during Run 1 are summarized in 
Table~\ref{tab:parameters}.
 
\begin{table}[hbt]
\renewcommand{\arraystretch}{1.15}
\centering
 {\begin{tabular}{@{}lcccccc@{}} \toprule
  Year  & Mode  &  \sqrtsnn (TeV) &  $\beta^*$ (m) & $\alpha_\mathrm{int}$ ($\mu$rad)  & $\alpha_\mathrm{ext}$ ($\mu$rad) & Colliding bunches\\  \hline
  2009 & \pp &  0.9 & 10 & 2180 & 0 &  $\leq$~~2\\
  2009 & \pp &  2.36 & 10 & 830 & 0 &  $\leq$~~2\\
  2010 & \pp &  7 & 2; 3.5 & 280 & 0; 220 & $\leq$~~16 \\
  2010 & \pbpb &  2.76 & 3.5 & 280 & -280 & $\leq$~130 \\
  2011 & \pp & 2.76 & 10 & 710 & 0 & $\leq$~~64 \\
  2011 & \pp & 7  & 10 & 280 & 160 & $\leq$~~39 \\
  2011 & \pbpb & 2.76 & 1 & 280 & -160 & $\leq$~336 \\
  2012 & \pp & 8 & 3 & 245(-245) & -180(+290) & 0 (main--main) ; $\leq$~2500 (main--sat.) \\ 
  2012 & \ppb & 5.02 & 10 & -245 & -290 & $\leq$~8 \\
  2013 & \ppb & 5.02 & 0.8 & -245 & 125 & $\leq$~338 \\
  2013 & \pp & 2.76  & 10 & 710 & 170 & $\leq$~36 \\
\botrule
\end{tabular} 
 \tbl{Summary of beam parameters for ALICE during the first four years of 
LHC operation.}
\label{tab:parameters}}
\end{table}
 
\subsection{Machine induced background}
 
\subsubsection{Background sources}

The operation and performance of detectors at the LHC can be affected 
by machine-induced background (MIB), a particle flux originating from 
the beams interacting with matter in the machine. This background 
scales with beam intensity and depends mainly on the residual gas 
pressure in the beam pipe and on the cleaning efficiency of collimator 
systems. 
The most relevant component of beam background at IP2 is produced 
close to the experimental region by 
inelastic beam--gas (BG) interactions in the first 40~m of the so-called 
Long Straight Section 2 (LSS2), 270~m on either side of IP2. 

Given the requirement of a reduced luminosity, in \pp running the 
background rate  in ALICE can be of the same order of magnitude as 
the interaction rate. Since ALICE has been designed to perform tracking 
for up to 1000 times the \pp multiplicity, the tracking performance is 
not affected by such a background level. However, MIB affects the 
operation of gaseous detectors, leading to HV trips due to large 
charge deposits. Such trips were observed during the highest-intensity 
\pp running periods in 2011 and 2012 and concerned mainly the TPC and MCH 
detectors. Furthermore,  MIB  can cause cumulative radiation damage from 
high integral doses and neutron fluence~\cite{Morsch}, thus accelerating 
the ageing of detectors. For these reasons, in the high beam intensity 
\pp running ALICE was switching on its detectors only after the background 
interaction rate dropped to an acceptable level (up to several hours after 
the beginning of the fill). 

Large background from BG interactions was observed in 2011 and 
2012 during the \pp runs, increasing faster than linearly with the 
number of circulating bunches and bunch intensity. 
Vacuum deterioration inside the beam pipe can be caused by synchrotron 
radiation-induced desorption, beam-induced RF heating, and electron 
cloud formation in various sections of the 
accelerator~\cite{Baglin:2011zz,Neupert:2011za,Salvant:1569436}. 
In particular, a large pressure increase was observed with circulating 
beams inside the TDI (beam absorber for injection protection) and the large 
recombination chamber located in 
LSS2~\cite{Lanza:1470601,Iadarola:1572988,Salvant:1470298}. 

\begin{figure}[h]
\centering
\includegraphics[width=0.95\textwidth]{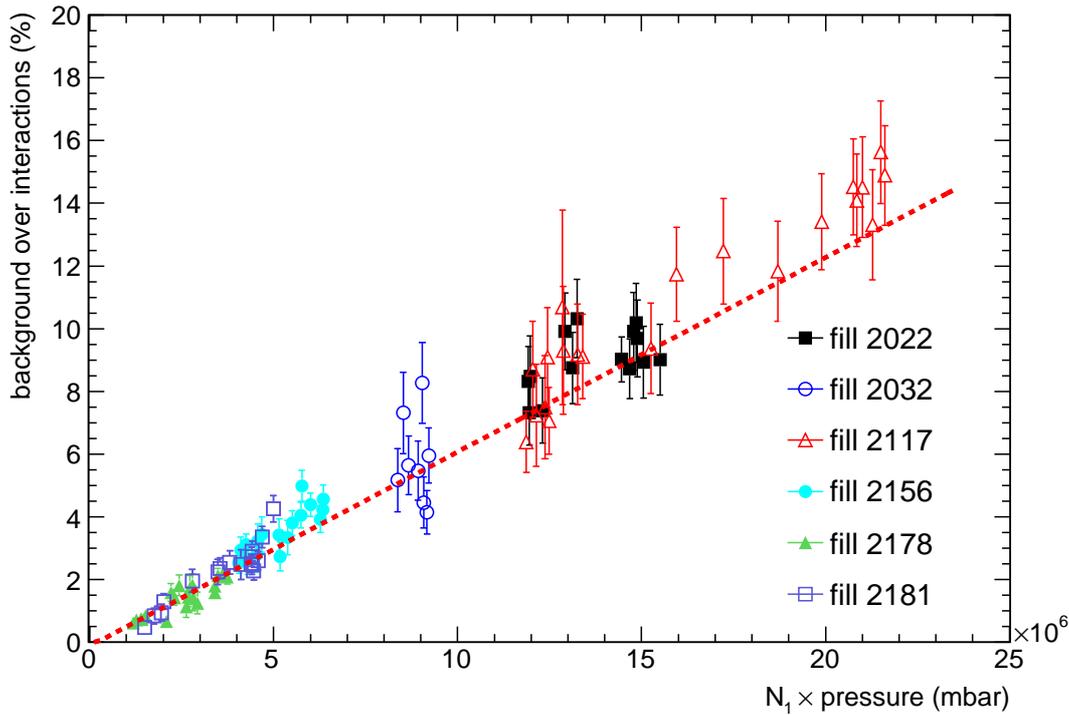}
\caption{Background rate observed during several fills as a function of 
the product of the intensity of Beam~1, N$_1$, and the sum of the measured 
pressures from three vacuum gauges on the left LSS2.}
\label{fig:bkgVSvacuum}
\end{figure}

\begin{figure}[p]
\centering
\includegraphics[width=0.75\textwidth]{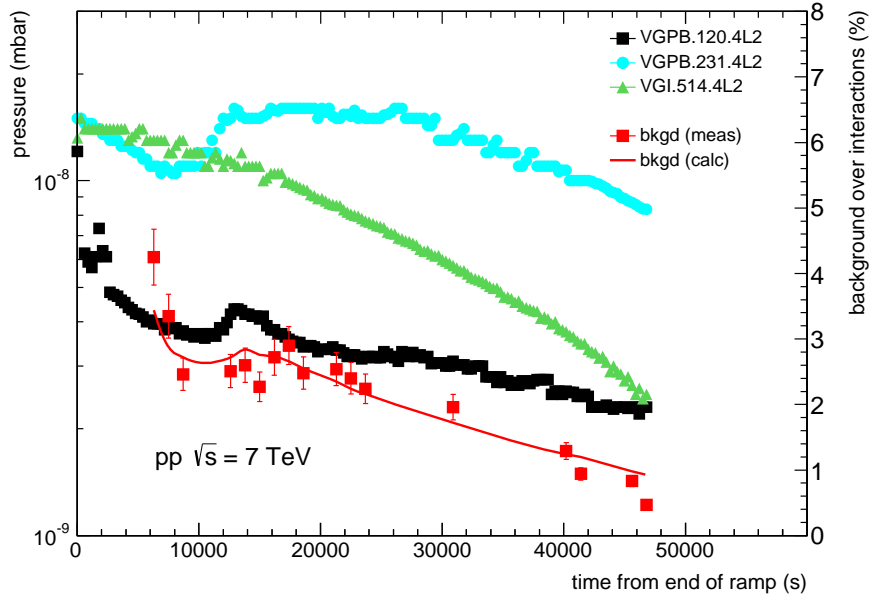}
\caption{Beam pipe pressure and background rate in 
fill 2181. The expected background rate has been estimated using the linear 
parameterization shown in Fig.~\ref{fig:bkgVSvacuum}. VGPB.120.4L2, 
VGPB.231.4L2, and VGI.514.4L2 are the pressure gauges located in front of 
the Inner triplet (at 69.7 m from IP2), on the TDI beam stopper (at 80~m 
from IP2), and on the large recombination chamber (at 109 m from IP2), 
respectively.}
\label{fig:bkgFill2181}
\end{figure}

A detailed study has been performed to characterize the dependence of the 
observed background rate\footnote{The background from BG interactions 
is measured via the V0 detector timing information, as will be described 
in Section~\ref{subsec:BGrej}.} on vacuum conditions and beam charge. A linear 
correlation was found between the background rate and the product of the 
beam charge and the sum of the pressures measured by the 
vacuum gauges along the LSS2, on both sides of IP2 (Fig.~\ref{fig:bkgVSvacuum}).   
Figure~\ref{fig:bkgFill2181} shows a comparison between the measured background 
rate for a given LHC fill\footnote{A fill is a period during which beams 
are circulating in the LHC: it starts with the injection and ends with the beam 
dump.} and that estimated using the linear dependence described 
in Fig.~\ref{fig:bkgVSvacuum}, confirming the validity of the model. 

\begin{figure}[p]
\centering
\includegraphics[width=0.8\textwidth]{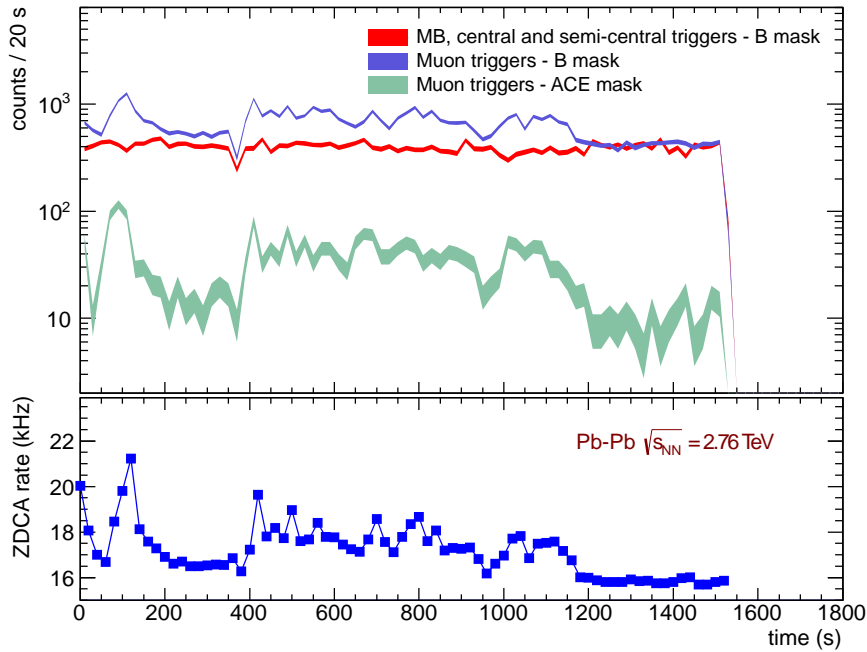}
\caption{Top: minimum bias, centrality, and muon triggers as a function of 
time during \pbpb data taking (run 169721). The B mask selects the LHC bunch 
slots where collisions between bunches of Beam~1 and Beam~2 are expected at 
IP2, while the ACE mask selects slots where no beam--beam collision is 
expected. Bottom: ZDC-A trigger rate as a function of time in the same run.}
\label{fig:bkgPbPb}
\end{figure}

The residual gas pressure is always nominal in the \pbpb physics mode, 
since the total beam charge is about two 
orders of magnitude smaller than in \pp. Thus, all processes which degrade 
the vacuum in the proton physics mode, in particular TDI heating and 
electron cloud formation,  are suppressed.
Minimum bias and centrality triggers are not affected by any beam background; 
however, some of the trigger inputs, such as the ZDC  and muon triggers, 
showed large rate fluctuations (Fig.~\ref{fig:bkgPbPb}). 
A detailed analysis of all fills has shown that the observed fluctuations are 
always correlated with Beam~1 losses on the tertiary collimator (TCTH) located 
a few meters upstream of one of the ZDCs (ZDC-A). A clear correlation was observed between the 
ZDC-A trigger rate (which is sensitive to both beam--beam and beam--gas collisions) 
and the losses recorded by Beam~1 BLM (Beam Loss Monitor) located on the TCTH. 
Generally, an increase towards the end of the fill has been observed, which 
could be explained by a degradation of the beam quality and interactions with 
the collimation system.
\clearpage
 
\subsubsection{Background rejection in ALICE}\label{subsec:BGrej}

\begin{figure}[hbt]
\centering
\includegraphics[width=0.8\textwidth]{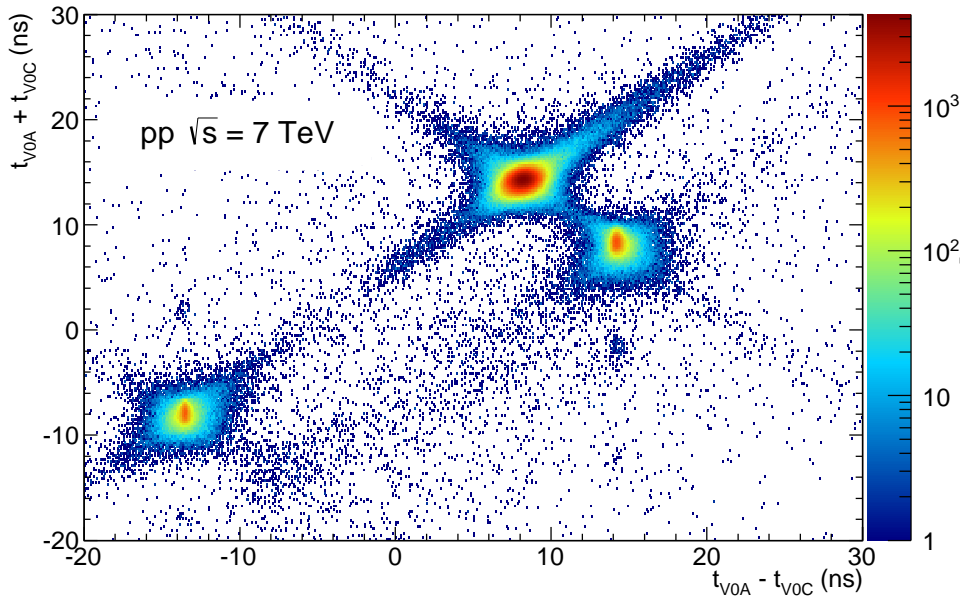}
\caption{Correlation between the sum and difference of signal times in V0A and 
V0C. Three classes of events -- collisions at  (8.3 ns, 14.3 ns), background 
from Beam~1 at (-14.3 ns, -8.3 ns), and background from Beam~2 at (14.3 ns, 8.3 ns) 
-- can be clearly distinguished.}
\label{fig:v0sumVSdiff}
\end{figure}

Background estimation for \pp running is performed with the V0 detector, a small-angle 
detector consisting of two circular arrays of 32 scintillator counters each, called 
V0A and V0C, which are installed on either side of the ALICE interaction 
point~\cite{Abbas:2013taa}. 
As described in Section~\ref{sect:intro}, the V0A detector is located 329 cm 
from IP2 on the side opposite to the muon spectrometer, whereas V0C is 
fixed to the front face of the hadronic absorber, 88 cm from IP2. 
The signal arrival time in the two V0 modules is exploited in order to discriminate 
collision events from background events related to the passage of LHC Beam~1 
or Beam~2.  The background caused by one of the beams is produced upstream of 
the V0 on the side from which the beam arrives. It thus produces an ``early'' 
signal when compared with the time corresponding to a collision in the nominal 
interaction point. The difference between the expected beam and background 
signals is about 22.6~ns in the A side and 6~ns in the C side.
As shown in Fig.~\ref{fig:v0sumVSdiff}, 
background events accumulate mainly in two peaks in the time sum--difference 
plane, well separated from the main (collision) peak. 
With the experience gained during the first years of 
data taking, in 2012 the V0 time gates used to set the trigger conditions 
on collision or background events have been refined and the MIB contamination 
has been reduced to $\sim$~10\%, depending on vacuum conditions and 
luminosity. 
 
\begin{figure}[t]
\centering
\includegraphics[width=0.9\textwidth]{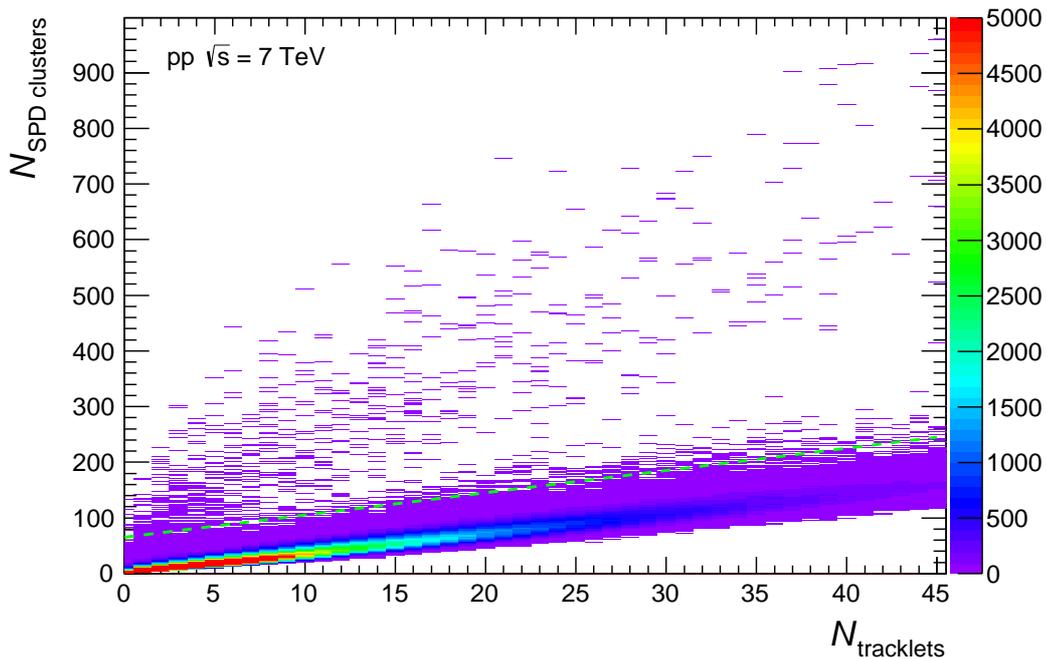}
\caption{Correlation between reconstructed SPD clusters and tracklets. 
  Two bands corresponding to the collisions and MIB are visible. 
  The dashed cyan line represents the cut used in the offline selection: events 
  lying in the region above the line are tagged as BG and rejected.}
\label{fig:cluster_vs_tracklets}
\end{figure}

The collected events are further selected offline to validate the online 
trigger condition and to remove any residual contamination from MIB and 
satellite collisions. As a first step, the online trigger logic is validated 
using offline quantities.  The V0 arrival time is computed using a weighted 
average of all detector elements. Then, 
MIB events are rejected using the timing information measured in the V0 
complemented, in \pp physics mode, by a cut on the correlation between 
clusters and tracklets reconstructed in the SPD. Background particles usually 
cross the pixel layers in a direction parallel to the beam axis. Therefore, 
only random combinations of BG hits can build a reconstructed track 
pointing to the vertex. Hence, one needs a large number of clusters  to have a 
significant probability for this to happen (Fig.~\ref{fig:cluster_vs_tracklets}). 
This cut requires a large multiplicity in order to be effective and rejects 
a negligible number of events beyond those already rejected by the V0.
Only a very small fraction of background events survive the above-mentioned 
cuts in \pbpb collisions. The overall contamination can be determined by an 
analysis of control data taken with only one of the beams crossing the 
ALICE interaction point and is found to be smaller than 0.02\%. In \pp collisions, 
the amount of background surviving the cuts is strongly dependent on the running 
conditions and on the specific trigger configuration under study. While the 
fraction of background events in the physics-selected minimum bias triggers 
amounts to about 0.3\% in the data taken during the 2010 run, it can reach 
values above 10\% at the beginning of a fill in the 2011 and 2012 runs. 
Whenever relevant for the normalization of the results, the residual background 
was subtracted in the physics analyses, based on the information obtained 
from the control triggers.
 
\begin{figure}[h]
\centering 
\includegraphics[width=0.8\textwidth]{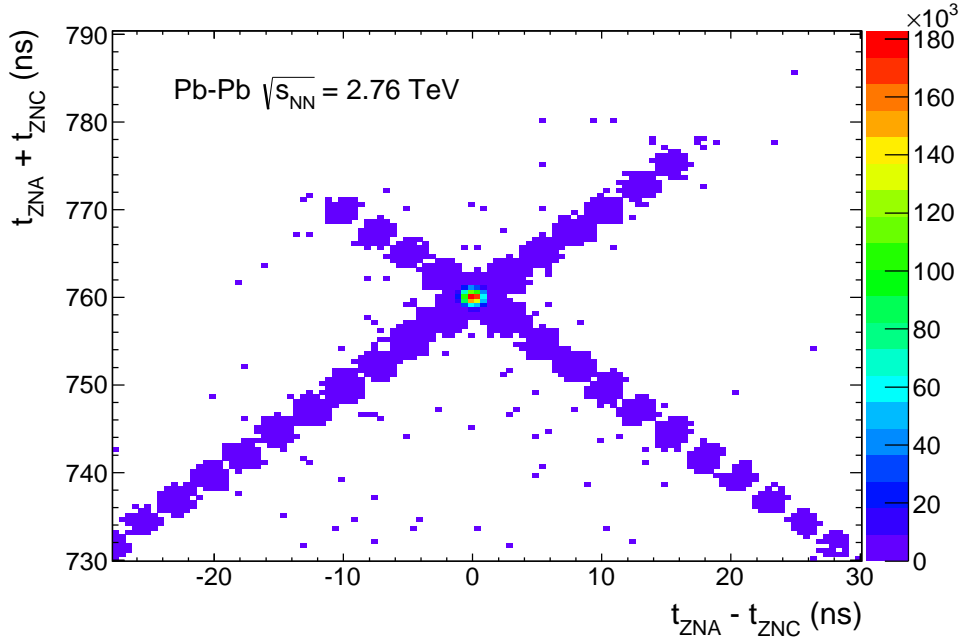}    
\caption{Correlation between the sum and the difference of times recorded 
  by the neutron ZDCs on either side (ZNA and ZNC) in \pbpb collisions. The large cluster in the middle 
  corresponds to collisions between ions in the nominal RF bucket on both 
  sides, while the small clusters along the diagonals (spaced by 2.5~ns in 
  the time difference) correspond to collisions in which one of the 
  ions is displaced by one or more RF buckets.}
\label{fig:zdc_time}
\end{figure}

The parasitic collision of main bunches with satellite bunches located a few 
RF buckets away from a main bunch are also a source of background in the 
standard analyses. The background from 
main--satellite collisions is non-negligible in the \pbpb running mode where 
the satellite population is larger than in \pp. Main--satellite collisions 
occur at positions displaced by multiples of 
$2.5~\mathrm{ns}/2\cdot c = 37.5~\mathrm{cm}$, with respect to the nominal 
interaction point. This is well outside the standard fiducial vertex region 
$\left|V_{z}\right|\lesssim 10~\mathrm{cm}$.  Satellite events are rejected 
using the correlation between the sum and the difference of times measured 
in the ZDC, as shown in Fig.~\ref{fig:zdc_time}. 
 
\subsection{Luminosity determination}
\label{sect:lumi}

\subsubsection{Introduction}

Cross section measurements in \pp collisions are essential for the ALICE 
physics program because particle production in nucleus--nucleus (A--A) collisions is often 
compared with the extrapolation from elementary \pp collisions via binary 
nucleon--nucleon collision scaling (nuclear modification factor, $R_{AA}$). 
The precision of $R_{AA}$ measurements needed to quantify the importance of nuclear 
effects is typically $\simeq$10\%. Thus, a precision on the order of 5\% or 
better on the \pp cross section (including luminosity normalization) is 
desired. 

Although it is not crucial for $R_{AA}$\footnote{As is shown in Section~\ref{Section:Centrality}, 
a centrality-dependent normalization factor can be obtained via the Glauber model.}, 
the determination of the absolute luminosity in \pbpb collisions is needed for cross 
section studies in electromagnetic and ultraperipheral interactions. 

\subsubsection{van der Meer scanning technique}

The measurement of the cross section $\sigma_{\rm{R}}$ for a chosen reference 
process is a prerequisite for luminosity normalization. 
Reference (or visible) cross sections can be measured in van der Meer (vdM) 
scans~\cite{vdM}, where the two beams are moved across each other in the 
transverse direction. Measurement of the rate $R$ of a given process as a 
function of the beam separation $\Delta x$, $\Delta y$ allows one to 
determine the head-on luminosity \lumi for a pair of colliding bunches 
with particle intensities $N_1$ and $N_2$ as:
\begin{equation}
\lumi = \frac{N_1 \> N_2 \> f_{\rm{rev}}}{h_x h_y} \ ,
\end{equation}
where $f_{\rm{rev}}$ is the accelerator 
revolution frequency and $h_x$ and $h_y$ are the effective beam widths in 
the $x$ and $y$ directions: they are measured as the area below the 
\hbox{$R(\Delta x,0)$} and \hbox{$R(0,\Delta y)$} curve, respectively, 
when divided by the head-on rate \hbox{$R(0,0)$}. Under the assumption 
that the beam profiles are Gaussian, 
the effective width can simply be obtained as the Gaussian standard deviation 
parameter (obtained from a fit to the curve) multiplied by $\sqrt{2\pi}$. 
However, the Gaussian assumption is not necessary for the validity of the 
method; thus, other functional forms can be used, as well as numerical 
integration of the curve. The cross section $\sigma_{\rm{R}}$ for the chosen 
reference process can be obtained as \hbox{$\sigma_{\rm{R}} = R(0,0)/\lumi$}. 

\subsubsection{van der Meer scan analysis and results} \label{sect:conditions-vdm-pbpb}

In this section, results from five scans carried out at the LHC 
are summarized. Two scans were performed in 2010 for \pp collisions at 
\mbox{\sqrts = 7 TeV}. Another \pp scan was done in 2011 at \mbox{\sqrts = 2.76 TeV}. 
Furthermore, two \pbpb scans were performed at \mbox{\sqrtsnn = 2.76 TeV} in 2010 and 2011. 
More details on these measurements can be found in Ref.~\refcite{lumiDaysProc}.

\begin{figure}[t]
\centering
\includegraphics[width=0.77\textwidth]{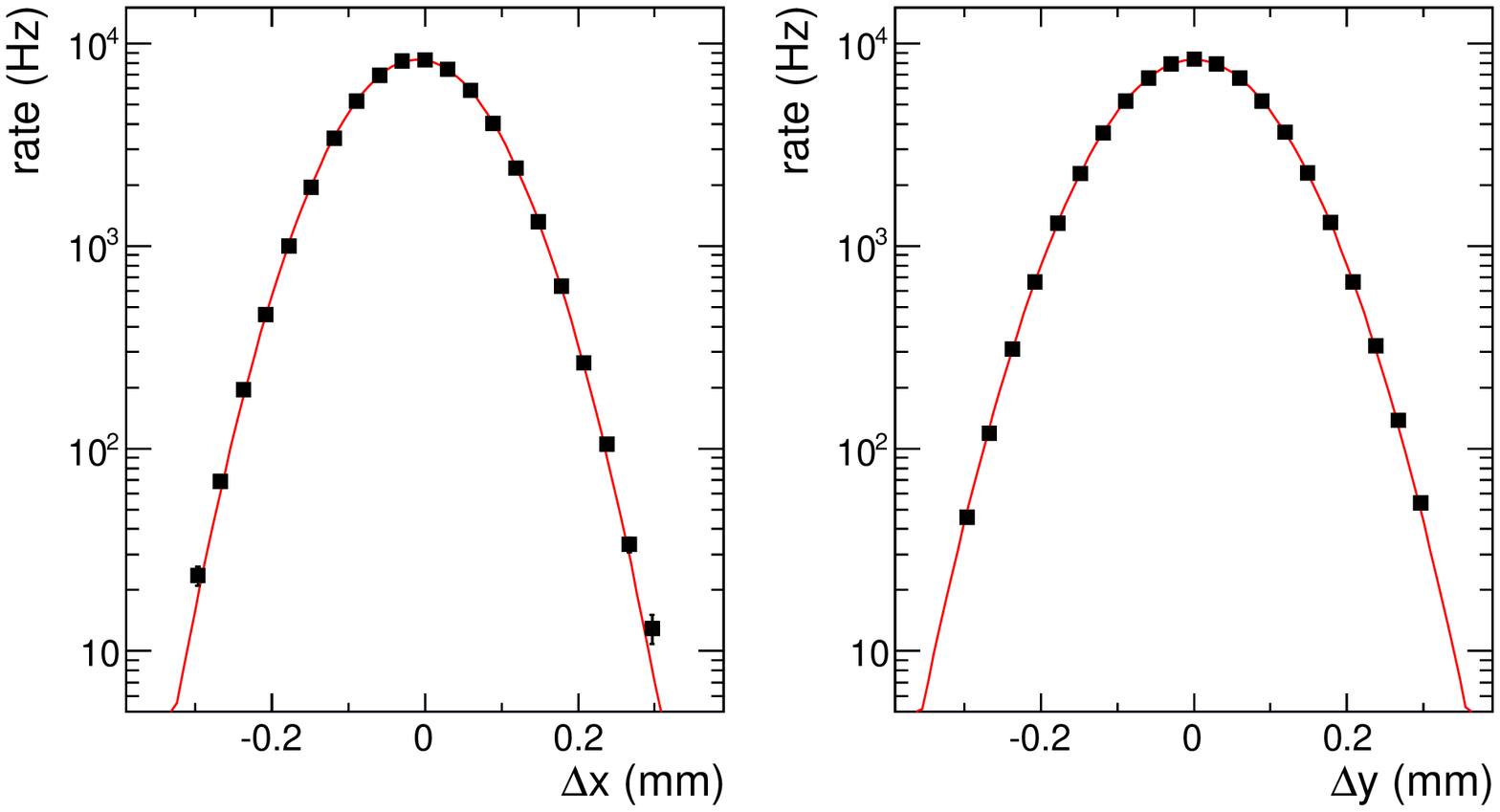}
\includegraphics[width=0.8\textwidth]{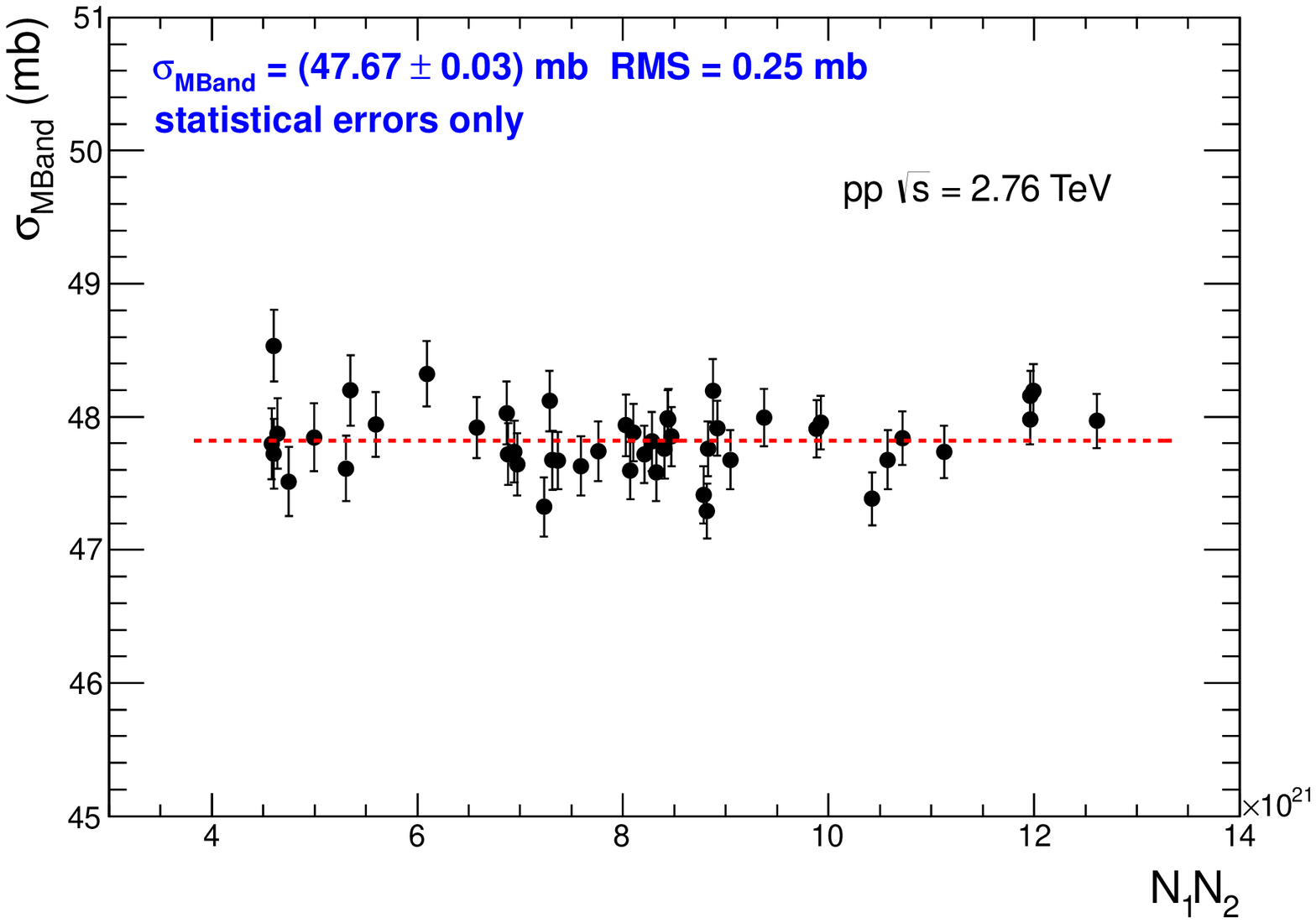}
\caption{Top: MBand trigger rate vs. beam separation in $x$ and $y$ obtained 
  during the May 2010 van der Meer scan. Double Gaussian fits to 
  the data are shown as lines. Bottom: Measured MBand cross section for 48 
  colliding bunch pairs in the March 2011 scan, as a function of the product 
  of colliding bunch intensities N$_1$N$_2$.}
\label{fig:scan}      
\end{figure}

The conditions, results, and systematic uncertainties of the three \pp scans 
are specified in Table~\ref{tab:scans}. 
The chosen reference process (MBand) for all of these scans is the coincidence of hits 
in the V0 detectors on the A and C sides. 
The MBand rate was measured as a function of the beam separation 
(upper panels of Fig.~\ref{fig:scan}). The scan areas were obtained via numerical 
integration. In the March 2011 scan, the cross section was measured 
separately for the 48 colliding bunch pairs (as shown in the bottom panel of 
Fig.~\ref{fig:scan}) and then averaged. The resulting spread among different 
bunches is less than 0.5\% (RMS).  
A set of corrections must be applied throughout the data analysis procedure, 
namely: pileup correction
(up to 40\%); length scale calibration, needed for a precise determination 
of the beam separation and performed by displacing the beams in the same 
direction and measuring the primary vertex displacement with the pixel 
detector (SPD); satellite (displaced) collisions of protons captured in 
non-nominal RF slots, detected via the arrival time difference in the two 
V0 arrays~\cite{bcnwg2}; background from beam--gas interactions; and variation of 
the luminosity during the scan due to intensity losses and emittance growth. 
In October 2010, two scans were performed in the same fill, in order to check 
the reproducibility of the measurement. The two results agree within 0.4\%: 
they have been averaged and the difference included in the systematic 
uncertainties. 
The beam intensity is measured separately for each circulating bunch by 
the LHC beam current transformers, and provided to the experiments after 
detailed analysis~\cite{bcnwg2,bcnwg1,bcnwg3,bcnwg4,bcnwg5}. 
In the March 2011 scan, the uncertainty on the bunch intensity was much 
lower compared with the 2010 scans~\cite{bcnwg3,bcnwg4}, so certain
additional sources of uncertainty were also investigated. These were: 
coupling between horizontal and vertical displacements; variation of 
$\beta^*$ during the scan resulting from beam--beam effects; and afterpulses
in the V0 photomultipliers arising from ionization of the residual gas 
inside the photomultiplier tube.  For the 2010 scans, these 
additional sources are negligible when compared with the 
uncertainty on the beam current.
\begin{table}[h]
  \renewcommand\arraystretch{1.1}
\centering
{\begin{tabular}{@{}lccc@{}} \toprule
 Scan  & May 2010  & October 2010 &  March 2011 \\  \hline
 \sqrts (TeV) & 7 & 7 &  2.76 \\ 
 $\beta^*$ (m) & 2& 3.5& 10\\
 Net crossing angle ($\mu$rad) & 280 & 500 & 710 \\
 Colliding bunch pairs in ALICE & 1 & 1  & 48 \\ \hline
  $\sigma_{\rm{MBand}}$ (mb) & 54.2$\pm$2.9 & 54.3$\pm$1.9 & 47.7$\pm$0.9 \\ \hline
  Uncertainties  &   &  &   \\  
 Bunch intensity & 4.4\% & 3.2\% & 0.6\%   \\ 
 Length scale & 2.8\% & 1.4\% &  1.4\% \\
 Luminosity decay   & 1\% & negligible & 0.5\% \\
 V0 afterpulses &negligible & negligible & 0.2\% \\
 Background  subtraction & negligible & negligible & 0.3\%\\
 Same fill reproducibility & negligible & 0.4\% & 0.4\% \\
 $x$--$y$ displacement coupling & negligible & negligible & 0.6\% \\
 $\beta^*$ variation during the scan & negligible & negligible & 0.4\%\\
 Total & 5.4\% & 3.5\% & 1.9\%  \\
 \botrule
 \end{tabular}
\tbl{Details of the colliding systems and measured MBand cross sections and 
uncertainties for the three \pp vdM scans performed at the LHC IP2.}
 \label{tab:scans}}
\end{table}

The ALICE luminosity determination in \pp collisions has been compared with the other LHC experiments 
via the cross section for a candle process, defined as a \pp interaction with 
at least one charged particle produced with $\pt>$~0.5\gevc and $|\eta|<$~0.8. 
This was determined as \hbox{$\sigma_{\rm{candle}}$~=~f$_{\rm{candle}}$ $\sigma_{\rm{MBand}}$}, 
where the scaling factor $f_{\rm{candle}}$~=~(0.817$\pm$0.004) was determined from data with 
a small ($\simeq$3\%) Monte Carlo efficiency correction. 
The obtained result (from the May 2010 scan) is \hbox{$\sigma_{\rm{candle}} = 44.3 \pm 2.1$~mb}, 
in good agreement with the ATLAS ($42.3 \pm 2.1$~mb) and CMS ($44.0 \pm 2.0$~mb) 
results~\cite{heinemann}. 
The quoted uncertainties represent the statistical and systematic errors
combined in quadrature; part of the uncertainty of the beam intensity determination,
that is common to all experiments~\cite{bcnwg1}, is not included.

The main parameters for the two \pbpb scans are reported in Table~\ref{tab:PbScans}. 
Given the low hadronic interaction rate in 2010, the scan was 
based on the detection of neutrons from electromagnetic \pbpb interactions 
by the ZDC~\cite{ALICE:2012aa}. The chosen reference process is the logical 
OR of hits in either of the two neutron calorimeters (ZNor). 
The scanned process in 2011 was a semicentral (SC) trigger based on the 
coincidence of V0A and V0C, with signal amplitude thresholds chosen in such 
a way that the trigger efficiency is 100\% for events belonging to the 
0--50\% centrality percentile, and drops rapidly for more peripheral events.

The analysis technique is the same as described for the \pp scans. Since the 
bunch-by-bunch measurement of the reference process rate was not available 
in 2010, the analysis of the November 2010 scan was performed for the 
``inclusive" rate, i.e. the sum of all bunch rates, thus measuring an 
``average" beam profile. The bias arising from this limitation was estimated 
in two ways: 
by simulation with realistic bunch intensities and emittances, and by 
computing the difference between the two methods for the 2011 scan. The second 
approach resulted in a larger discrepancy (2\%), which was added to the 
systematic uncertainties.

The result and uncertainties for the \pbpb scans are reported in Table~\ref{tab:PbScans}. The 
main source of uncertainty is the fraction of ghost charge in the measured 
beam current, consisting of ions circulating along the LHC rings outside of 
nominally filled bunch slots, which do not contribute to the 
luminosity~\cite{bcnwg5}.

The analysis of the 2012 (\pp) and 2013 (\ppb) vdM scans is ongoing. For these scans, 
along with the MBand trigger, another luminosity signal is available, based on the T0 
detector. The T0 provides a vertex trigger defined as the coincidence between T0A and T0C, 
with the additional requirement that the difference in their signal times corresponds to 
an interaction happening within 30~cm from IP2. 
The latter condition provides excellent rejection of beam-gas and satellite background. 
Indeed, a background contamination below 0.1\% was obtained in \ppb collisions at a 
luminosity of 10$^{29}$~s$^{-1}$cm$^{-2}$. 

\begin{table}[t]
\renewcommand{\arraystretch}{1.1}
\centering
{\begin{tabular}{@{}lcc@{}} \toprule
 Scan  & November 2010 & December 2011 \\  \hline
 \sqrtsnn (TeV) & 2.76  & 2.76 \\ 
 $\beta^*$ (m) & 3.5 & 1 \\
 Crossing angle ($\mu$rad) & $\simeq$0 & 120 \\
 Colliding bunch pairs in ALICE & 114  & 324\\ \hline
  $\sigma_{\rm{ZNor}}$ (b) & 371 $_{-19}^{+24}$ & - \\
  $\sigma_{\rm{SC}}$ (b) & - & 4.10 $_{-0.13}^{+0.22}$ \\ \hline
 Uncertainties  &  &  \\  
 Bunch intensity &  -3.0\% +4.7\%  & -1.6\% +4.4\%\\ 
 Length scale & 2.8\%  & 1.4\%\\
 Luminosity decay   & 2\% & 2\%\\
 Unknown bunch-by-bunch profile & 2\% & - \\  
 Background  subtraction & 1\% & 1\%\\
 Scan-to-scan reproducibility & 1\% & 1\% \\
 Total &  -5.2\% +6.4\% & -3.1\% +5.3\%\\
 \botrule
 \end{tabular} 
\tbl{Details of the colliding systems and measured cross sections and 
uncertainties for two \pbpb vdM scans performed at the LHC IP2.}
\label{tab:PbScans}}
\end{table}

\subsubsection{Application of the vdM scan results in luminosity and cross
section measurements\label{vdMapplications}}

The van der Meer scan results in \pp collisions at \mbox{\sqrts = 2.76} and 7~TeV were used 
to measure the inelastic cross sections at the two energies~\cite{sigmaInel}. A Monte Carlo 
simulation, tuned so as to reproduce the fractions of diffractive events observed in data, 
was used to determine the efficiency of the MBand trigger for inelastic \pp interactions. 
The MBand cross sections were then corrected for this efficiency, giving the result 
$\sigma_{\rm{INEL}}$~=~62.8$\pm$1.2~(vdM)~$_{-4.0}^{+2.4}$~(MC)~mb at
\mbox{\sqrts = 2.76 TeV} and
$\sigma_{\rm{INEL}}$~=~73.2$\pm$2.6~(vdM)~$_{-4.6}^{+2.0}$~(MC)~mb at
\mbox{\sqrts = 7 TeV}. 

In all the other ALICE analyses involving cross section measurements\footnote{With the
exception of Ref.~\refcite{Abelev:2012ba}, where a theoretical reference cross section was used 
instead.}, the reference cross sections (MBand, ZNor, SC) measured in the van der 
Meer scans (Tables~\ref{tab:scans} and \ref{tab:PbScans}) 
were used for indirect determination of the integrated luminosity.
In cases where the trigger condition used for the physics analysis coincided
with the reference trigger 
(as was the case in Ref.~\refcite{ALICE:2012aa}), the luminosity was simply measured as the 
number of analyzed events divided by the trigger cross section. 
In all other cases, the number of triggered events was converted into an equivalent number of 
reference triggers via a scaling factor, computed either from data (as for example in 
Refs.~\refcite{Aamodt:2011gj} and \refcite{Abelev:2012kr}) or via the ratio of the trigger rates, 
measured with scalers (as in Ref.~\refcite{upcMid}). Depending on the analysis, this scaling
procedure resulted in additional systematic uncertainties of up to 3\%.

\clearpage\newpage\section{Data taking}
\label{sect:datataking}

\subsection{Running periods}

ALICE took data for all the collision systems and energies offered by the LHC.
The data taking started in fall 2009 with \pp collisions at the LHC injection 
energy, \mbox{\sqrts = 0.9 TeV}. 
In 2010, the proton beam energy was brought up to half of its nominal value,  
3.5~TeV, and the luminosity was gradually increased.
In this period the interaction rate was low (between a few kHz and a few tens of kHz) 
and ALICE mostly triggered on minimum bias (MBor~\cite{sigmaInel,lumiDaysProc}) 
interactions using V0 and SPD, single muon trigger (MSL), and high-multiplicity trigger 
(HM) (see Section~\ref{sect:datataking:trigger} for a description of the ALICE triggers).
In the subsequent high-intensity \pp and \ppb running in 2011--2013, 
ALICE usually split its data-taking into minimum-bias (MB) and rare-trigger 
blocks, for which the interaction rate was reduced to $O$(10)~kHz and $O$(100)~kHz, 
respectively. 
Methods for reducing the luminosity are described in Section~\ref{sect:conditions}. 
The two limits correspond to the saturation of the readout with minimum-bias 
triggered events and to the maximum flux tolerated by the detectors, respectively. 
The two modes of operation are briefly discussed below. 

For minimum bias runs, the \pp and \ppb interaction rates were on the level 
of 10~kHz, enough to reach 95\% of the maximum detector readout rate while keeping 
the mean number of interactions per bunch crossing ($\mu$) low, nominally below 
0.05, in order to avoid significant same-bunch pileup. 

In the rare-trigger running mode, the luminosity in \pp and \ppb was increased to 
4--10~$\invubps$ and 0.1~$\invubps$, corresponding to inelastic interaction rates of 
200--500~kHz and 200~kHz, respectively. 
The luminosity limits were determined by the stability of the TPC and muon 
chambers under the load caused by interactions at IP2 and by background particles.  
During \pp and \ppb rare-trigger runs, the TPC event size increased by an 
order of magnitude due to pileup tracks within the drift time window of $\sim$100~$\mu$s.
The trigger dead time was kept at a level of 20--40\% in order to inspect as much 
of the luminosity delivered by the LHC as possible.

The luminosity reduction in the \pp running in 2012 was performed by colliding main 
bunches with satellite bunches (see Section~\ref{sect:conditions}). 
This resulted in a typical luminosity of $\sim\!7\;\invubps$ (up to a maximum of 
$20\;\invubps$) at the beginning of the fill and a rapid decay within the fill. 
Owing to this and to a background-interaction rate of the same order as the \pp rate 
(see Section~\ref{sect:conditions}), ALICE only took data in the second part of each 
fill, starting in the rare-trigger mode with a subset of detectors, and only switching 
to minimum-bias mode when the luminosity dropped to about 1--4~$\invubps$, a level 
tolerable for the V0 and the TPC. 
The downscaling factors for the MBand~\cite{sigmaInel,lumiDaysProc}, TJE, and SPI 
triggers were dynamically determined at the beginning of each run so as to keep the 
overall trigger live time at a level of 70--80\% over the duration of the fill.

During the 2011 \pbpb running period, the interaction rate provided by 
the LHC reached 3-4~kHz.
ALICE ran with the minimum bias, centrality, and rare triggers activated at the 
same time. With the multi-event buffering and with the minimum bias and centrality 
triggers downscaled, the effective trigger dead time was low (dead-time factor of 33\%). 
The situation will be similar in the LHC Run 2 (2015--2017), for which the 
expected collision rate is $O$(10)~kHz, still low enough to avoid pileup.

Table~\ref{tab:statistics-of-blocks} summarizes data taking with beams by 
ALICE together with the luminosity provided by the LHC, the obtained trigger 
statistics, and the recorded data volume.
\begin{table}[p]
\ifijmp
  \renewcommand\arraystretch{1.15}
\fi

\centering
{
\ifcern
  \small
\fi
\ifijmp
  \scriptsize
\fi
\begin{tabular}{@{\extracolsep{-0.8mm}}ccccccrlc@{}}
\toprule
Year & System,      & Running & Peak \lumi           & Duration     & Delivered       & \multicolumn{2}{@{\extracolsep{-4mm}}c}{ Recorded }   & Data read  \\
     & \sqrtsnn     & mode    &                      & beam [run]   & \lumi           & \multicolumn{2}{@{\extracolsep{-4mm}}c}{ statistics } & [recorded] \\
     & (TeV)        &         & ($\invubps$)         & (h)          &                 & \multicolumn{2}{@{\extracolsep{-4mm}}c}{ ($10^6$events) }   & (TB)  \\
\colrule
2009 & \pp          & MB      & $5.2\times 10^{-4}$  & n.a.         & 19.6 $\invub$   & MBor:    & 0.5                  & 0.41  \\
     & 0.9          &         &                      & [26.8]       &                 &          &                      & [0.43]\\
\colrule
     & \pp          & MB      & $1.0\times 10^{-4}$  & n.a.         & 0.87 $\invub$   & MBor:    & 0.04                 & 0.01  \\
     & 2.36         &         &                      & [3.1]        &                 &          &                      & [0.01]\\
\colrule
2010 & \pp          & MB      & $1.5\times 10^{-2}$  &  15.7        & 0.31 $\invnb$   & MBor:    & 8.5                  & 5.74  \\
     & 0.9          &         &                      & [13.0]       &                 &          &                      & [5.97]\\
\colrule
     & \pp          & MB+rare & 1.7*                 &  847         & 0.5 $\invpb$    & MB:      & 825                  &  755  \\
     & 7.0          & (mixed) &                      & [613]        &                 & HM:      & 26                   & [773] \\
     &              &         &                      &              &                 & MSL:     & 132                  &       \\
\colrule
     & \pbpb        & MB      & $2.8\times 10^{-5}$  &  223         & 9 $\invub$      & MB:      & 56                   & 810   \\
     & 2.76         &         &                      & [182]        &                 &          &                      & [811] \\
\colrule
2011 & \pp          & rare    & $4.4\times 10^{-1}$  &  35          & 46 $\invnb$     & MBor:    & 74                    &  100  \\
     & 2.76         &         &                      & [32]         &                 & HM:      & 0.0015               & [101] \\
     &              &         &                      &              &                 & E0:      & 0.78                 &       \\
     &              &         &                      &              &                 & MSL:     & 9.4                  &       \\
\colrule
     & \pp          & rare    & 9                    & 1332         & 4.9 $\invpb$    & MBor:    & 608                  &  1981 \\
     & 7.0          &         & (450 kHz)            & [841]        &                 & MBand:   & 163                  & [1572]\\
     &              &         &                      &              &                 & EJE:     & 27                   &       \\
     &              &         &                      &              &                 & EGA:     & 8                    &       \\
     &              &         &                      &              &                 & MUL:     & 7.6                  &       \\ 
\colrule
     & \pbpb        & rare    & $4.6\times 10^{-4}$  & 203          & 146 $\invub$    & MBZ:     & 9                    & 3151  \\
     & 2.76         &         &                      & [159]        &                 & CENT:    & 29                   & [908] \\
     &              &         &                      &              &                 & SEMI:    & 34                   &       \\
     &              &         &                      &              &                 & MSH:     & 23                   &       \\
     &              &         &                      &              &                 & EJE:     & 11                   &       \\
     &              &         &                      &              &                 & CUP:     & 7.9                  &       \\
     &              &         &                      &              &                 & MUP:     & 3.4                  &       \\
\colrule
2012 & \pp          & MB      & 0.2*                 & 1824         & 9.7 $\invpb$    & MBor:    & 38                   &  3211 \\
     & 8            &         & (10 kHz)             & [1073]       & (altogether)    & MBand:   & 270                  & [1286]\\
     &              &         &                      &              &                 & SPI:     & 63                   &       \\
%
%
     &              & rare    & 20                   &              &                 & MSH:     &  86                  &       \\
     &              &         & (1 MHz)              &              &                 & MUL:     &  12                  &       \\
     &              &         &                      &              &                 & EGA:     &  3.1                 &       \\
     &              &         &                      &              &                 & TJE:     &  21                  &       \\
\colrule
     & \ppb         & MB      & $9\times 10^{-5}$    &  7.6         & 1.5 $\invub$    & MBand:   & 2.43                 & 5.0   \\
     & 5.02         & (pilot) & (180 Hz)             & [6.6]        &                 &          &                      & [3.4] \\ 
\colrule
2013 & \ppb         & MB      & $5\times 10^{-3}$*   &  50.2        & 0.891 $\invnb$  & MBand:   & 134                  &  406  \\
     & 5.02         &         & (10 kHz)             & [46.8]       &                 & ZED:     & 1.1                  & [91]  \\
     &              & rare    & $1\times 10^{-1}$    &  70.1        & 14.0 $\invnb$   & MSH:     & 10                   &  472  \\
     &              &         & (200 kHz)            & [50.0]       &                 & MUL:     & 9.5                  & [97]  \\
     &              &         &                      &              &                 & EGA:     & 1.3                  &       \\
     &              &         &                      &              &                 & TJE:     & 0.59                 &       \\
     &              &         &                      &              &                 & MUP:     & 0.76                 &       \\
\colrule
%
     & \pbp         & rare    & $1\times 10^{-1}$    &  77.1        & 17.1 $\invnb$   & MSH:     & 18                   &  731  \\
     & 5.02         &         & (200 kHz)            & [61.8]       &                 & MUL:     & 24                   & [151] \\
     &              &         &                      &              &                 & EGA:     & 1.9                  &       \\
     &              &         &                      &              &                 & TJE:     & 1.0                  &       \\
     &              &         &                      &              &                 & MUP:     & 3.3                  &       \\
\colrule
     & \pp          & rare    & 2.2*                 &  27.4        & 129~nb$^{-1}$   & MBand:   & 20                   &  71   \\
     & 2.76         &         & (105 kHz)            & [24.9]       &                 & MSH:     &  0.89                & [16]  \\
     &              &         &                      &              &                 & MUL:     &  0.53                &  \\
     &              &         &                      &              &                 & EGA:     &  0.43                &  \\
     &              &         &                      &              &                 & TJE:     &  0.036               &  \\
\botrule
\end{tabular} 
\tbl{ALICE data taking in Run~1 (2009--2013). See text for details.}
\label{tab:statistics-of-blocks}}
\end{table}

Whenever the luminosity was reduced for ALICE, its final value is quoted and 
marked with an asterisk. 
The beam duration and run duration are the integrated time with stable beams 
and the time during which ALICE was recording collision data, respectively. 
The difference between the two represents the time spent on starting/stopping 
of runs, the recovery time after detector trips, and, for \pp runs in 2011 and 
2012, the time spent waiting for the particle flux to drop to a level acceptable 
for the detectors. 
The run duration is not corrected for the trigger/acquisition dead time. 
The delivered luminosity is the luminosity integrated over the beam duration. 
The abbreviations denoting various triggers are explained in Section~\ref{sect:datataking:trigger}.
The recorded data volume slightly exceeds the read one because of the header data. 
The large differences between these two numbers, starting from \pp in 2011, arise 
from the online compression discussed in Section~\ref{sect:datataking:compression}. 

In the context of Table~\ref{tab:statistics-of-blocks} one should note that 
many of the top ALICE physics goals involve measurements at low transverse momenta, 
where triggering cannot be used. This applies in particular to all measurements in 
the ALICE central barrel, where the vast majority of published papers are 
from minimum-bias data. Consequently, for the performance of ALICE the recorded 
statistics of minimum-bias events, where the data acquisition system runs with a 
significant dead time, is the main figure of merit. The evolution of the ALICE 
experiment towards Run 3 (see Section~\ref{sect:conclusion}) is, consequently, 
focused on continuous read-out of 50 kHz minimum-bias \pbpb collisions. 

In addition to the running blocks summarized in Table~\ref{tab:statistics-of-blocks}, ALICE
took data with cosmic ray triggers defined using ACORDE, TOF, and TRD for cosmic-ray studies and 
detector calibration purposes~\cite{Alessandro:2010zz}. 
The cosmic runs were usually performed in the absence of beams. In 2012, ALICE took
$\sim 4\times 10^{6}$ cosmic ray events in parallel with the collision data taking, 
using a high-multiplicity muon trigger (signal on at least 4 scintillator paddles) 
provided by ACORDE.

\subsection{Trigger}
\label{sect:datataking:trigger}

The trigger decision is generated by the Central Trigger Processor (CTP) of 
ALICE~\cite{ALICECTPDAQHLTTDR,ALICECTPWEB} based on detector signals and 
information about the LHC bunch filling scheme. 
The detectors that provide input to the trigger decision are listed in 
Table~\ref{tab:triggerdetectors}. 
The CTP evaluates trigger inputs from the trigger detectors every machine clock 
cycle ($\sim$25~ns).
The Level~0 trigger decision (L0) is made $\sim$0.9~$\mu$s after the collision 
using V0, T0, EMCal, PHOS, and MTR.
The events accepted at L0 are further evaluated by the Level~1 (L1) trigger 
algorithm in the CTP.
The L1 trigger decision is made 260 LHC clock cycles ($\sim$6.5~$\mu$s) after L0.
The latency is caused by the computation time (TRD and EMCal) and propagation 
times (ZDC, 113~m from IP2).
The L0 and L1 decisions, delivered to the detectors with a latency of about 300~ns, 
trigger the buffering of the event data in the detector front-end electronics.
The Level~2 (L2) decision, taken after about 100~$\mu$s corresponding to the 
drift time of the TPC, triggers the sending of the event data to DAQ and, in parallel, 
to the High Level Trigger system (HLT).
During Run 1, all events with L1 were accepted by L2. 
In the future, in some running scenarios (e.g. when taking downscaled minimum bias 
events in parallel with rare triggers) L2 may be used to reject events with multiple 
collisions from different bunch crossings piled-up in the TPC (past--future protection). 
The events with L2 will subsequently be filtered in the HLT.
\begin{table}[b]
\centering
{\begin{tabular}{@{}llc@{}} \toprule
Detector & Function                                                                      & Level \\
\colrule
SPD      &  hit-multiplicity based trigger and hit-topology based trigger                & L0    \\
TRD      &  electron trigger, high-\pt particle trigger, charged-jet trigger             & L1    \\
TOF      &  multiplicity trigger, topological (back-to-back) trigger, cosmic-ray trigger & L0    \\
PHOS     &  photon trigger                                                               & L0    \\
EMCal    &  photon trigger, neutral-jet trigger                                          & L0/L1 \\
ACORDE   &  cosmic-ray trigger (single and multiple hits)                                & L0    \\
V0       &  coincidence based minimum-bias interaction trigger, centrality trigger       & L0    \\
T0       &  event-vertex selection trigger, interaction trigger                          & L0    \\
ZDC      &  minimum-bias interaction and electromagnetic-dissociation triggers in \pbpb  & L1    \\
MTR      &  single-muon trigger, dimuon trigger                                          & L0    \\
\botrule
\end{tabular}
\tbl{Trigger capabilities of the ALICE detectors.}
 \label{tab:triggerdetectors}}
\end{table}

Information about the LHC bunch filling scheme was used by CTP to suppress the background. 
The bunch crossing mask (BCMask) provides the information as to whether there are 
bunches coming from both A-side and C-side, or one of them, or neither, at a 
resolution of 25 ns.
The beam--gas interaction background was studied by triggering on bunches 
without a collision partner, and subtracted from the physics data taken with 
the requirement of the presence of both bunches. 

Table~\ref{tab:triggers} summarizes the most important trigger configurations 
used by ALICE. 
The minimum bias triggers (MBand and MBor) were used for all \pp data taking, 
as well as in \pbpb in 2010.
The high-efficiency MBor trigger was used at low luminosity. 
Once the luminosity and the background level increased, the high-purity MBand 
trigger became more advantageous. 
In the high luminosity \pbpb runs in 2011, the V0-based trigger was 
complemented by a requirement of signals in both ZDCs (MBZ) in order to suppress 
the electromagnetic interactions between the lead ions. 
The biased ``power-interaction'' trigger (SPI) required a certain number 
of hits (usually around 10) in the SPD. 
With thresholds on the summed-up signals, V0 was also used to generate central 
0--10\% (CENT) and
semicentral 0--50\% (SEMI) \pbpb triggers. 
The thresholds were applied separately to the sums of the output charges of 
V0A and V0C, then the coincidence of the two sides was required.
\begin{table}[b]
\vspace{5mm}
\centering
{\begin{tabular}{@{\extracolsep{-0mm}}llll@{}} \toprule
Trigger    & Description            & Condition                                                      \\
\colrule   
\multicolumn{3}{@{\extracolsep{-0mm}}l}{\it MB-type triggers}                                        \\
MBor       & minimum bias           & signals in V0 and SPD                                          \\
MBand      & minimum bias           & signals in V0A and V0C                                         \\
MBZ        & minimum bias           & MB and signals in both ZDC's                                   \\
SPI        & multiplicity trigger   & $n$ hits in SPD                                                \\
\colrule                             
\multicolumn{3}{@{\extracolsep{-0mm}}l}{\it Centrality triggers}                                     \\
CENT       & central                & V0 based centrality trigger for \pbpb (0--10\%)                \\
SEMI       & semicentral            & V0 based semicentral trigger for \pbpb (0--50\%)               \\
\colrule                             
\multicolumn{3}{@{\extracolsep{-0mm}}l}{\it EMCal rare triggers}                                     \\
E0         & EMCal L0               & EMCal L0 shower trigger in coincidence with MB                 \\
EJE        & neutral jet            & EMCal L1 jet algorithm following EMCal L0                      \\
EJE2       & neutral jet            & like EJE but with a lower threshold than EJE                   \\
EGA        & photon/electron        & EMCal L1 photon algorithm following EMCal L0                   \\
EGA2       & photon/electron        & like EGA but with a lower threshold than EGA                   \\
\colrule                             
\multicolumn{3}{@{\extracolsep{-0mm}}l}{\it TRD rare triggers}                                       \\
TJE        & charged jet            & $n$ charged particles in TRD chamber in coincidence with MB    \\
TQU        & electron for quarkonia & electron with $\pt>2$\gevc in TRD in coincidence with MB       \\
TSE        & electron for open beauty & electron with $\pt>3$\gevc in TRD in coincidence with MB     \\
\colrule                             
\multicolumn{3}{@{\extracolsep{-0mm}}l}{\it MUON rare triggers}                                      \\
MSL        & single muon low        & single muon in MTR in coincidence with MB                      \\
MSH        & single muon high       & like MSL but with a higher threshold                           \\
MUL        & dimuon unlike sign     & two muons above low threshold, unlike sign, in coinc. with MB  \\ 
MLL        & dimuon like sign       & two muons above low threshold, same sign, in coinc. with MB    \\ 
\colrule                             
\multicolumn{3}{@{\extracolsep{-0mm}}l}{\it Miscellaneous triggers}                                  \\
HM         & high multiplicity      & high multiplicity in SPD in coincidence with MB                \\
PH         & photon by PHOS         & PHOS energy deposit in coincidence with MB                     \\ 
EE         & single electron        & electron signal in TRD (sector 6--8) and EMCal                 \\
DG        & diffractive            & charged particle in SPD and no signal in V0                     \\
CUP        & barrel ultraperipheral& charged particle in SPD and no signal in V0, for \pbpb and \ppb \\
MUP        & muon ultraperipheral  & (di-)muon in MTR and no signal in V0A, for \pbpb and \ppb       \\
ZED  & electromagnetic dissociation & signal in any of the neutron ZDCs                              \\
COS        & cosmic trigger         & signal in ACORDE                                               \\
\botrule
\end{tabular} 
\tbl{Major ALICE triggers.}
\label{tab:triggers}}
\vspace{5mm}
\end{table}

The rest of the triggers in Table~\ref{tab:triggers} are rare triggers.
The high-multiplicity trigger (HM) was based on the hit multiplicity in the outer
layer of the SPD.
The multiplicity threshold was typically set to 80--100 hits, corresponding 
to 60--80 SPD tracklets (pairs of matching clusters in the two layers of SPD). 
This value was chosen in order to maximize the inspected luminosity without 
contaminating the sample with multiple-interaction events.
The PHOS and EMCal L0 triggers (PH and E0, respectively) required 
a certain energy deposit within a window of 4$\times$4 calorimeter cells. 
At L1, EMCal provided triggers on photons/electrons (EGA) and on jets (EJE). 
The EGA trigger has a higher threshold than E0 and a better handling of 
supermodule boundaries. The EJE trigger uses a window of 32$\times$32 cells 
and is primarily sensitive to neutral energy but also includes contributions 
from charged particles (see Section~\ref{sect:chapaendeinem}). 
The TRD trigger was introduced in the 2012 \pp runs. 
A fraction (limited to 10 to 25~kHz) of the minimum bias triggers at L0 were 
subject to a TRD L1 decision.
At L1, four algorithms were implemented: jet trigger (TJE), 
single electron trigger (TSE), quarkonium electron trigger (TQU), and TRD+EMCal 
electron trigger (EE). 
The jet trigger requires at least 3 charged particle tracks with $\pt>3\gevc$
to be detected in one TRD stack. A TRD stack consists of 6 layers of chambers 
in radial direction and covers $\Delta\eta\approx\Delta\phi\approx 0.1$.
13 TRD supermodules, five stacks each, were installed and operational in the 
2012 and 2013 runs.
The electron trigger required an electron PID based on a threshold 
for the electron likelihood calculated from the integrated signal of each layer. 
The quarkonia electron trigger required a lower \pt threshold of 2\gevc with a 
tighter electron likelihood cut.
This enables the detection of low momentum electrons from \jpsi and $\psi^\prime$ 
decays.
In contrast to the TJE, TQU, and TSE triggers, the high-purity electron trigger 
EE was inspecting all events with EMCal Level~0 trigger (E0).
The TRD trigger condition for EE was the same as for the single electron trigger; 
however, the acceptance was limited to the TRD sectors (supermodules 6, 7, and 8) that overlap 
with the acceptance of EMCal. 
A signal in the innermost TRD layer was required for all TRD electron triggers 
in order to suppress the background caused by late photon conversions. 

All of the muon triggers were implemented at Level~0.
There were two single-muon triggers (MSL and MSH) and two dimuon triggers (MUL and 
MLL), all in coincidence with MB.
A low \pt threshold was used for MSL, MUL, and MLL, and a high one 
for the MSH trigger.
The low-threshold single-muon trigger MSL was downscaled when used in parallel 
with MSH. 
The unlike-sign muon-pair trigger MUL, used for measuring mesons, was complemented 
by the like-sign (MLL) one for the combinatorial background estimation.
The low and high \pt thresholds were 0.5--1.0\gevc and 1.7--4.2\gevc, respectively, 
adjusted according to the run type. 

Several additional triggers were implemented in order to enhance events related 
to diffractive physics in \pp and ultraperipheral nuclear collisions, and to measure 
cosmic rays. 
The DG (double gap) trigger in \pp required a particle at midrapidity and no 
particles within the intermediate pseudorapidity ranges covered by the V0 detector. 
The CUP (central-rapidity ultraperipheral) trigger performed a similar selection 
in collision systems involving ions. 
An analogous condition, but with a forward muon rather than a midrapidity particle, 
was named the MUP (muon ultraperipheral) trigger. 
Finally, a cosmic trigger defined by ACORDE (COS) was active during most of 
2012 to collect high muon multiplicity cosmic events.

The rare triggers implemented in TRD, EMCal, and MUON are further discussed in 
Sections~\ref{sect:electrons}, \ref{sect:jets}, and \ref{sect:muons}. 
Physics results based on analyses of E0-, MSL-, and MUP-triggered events were 
published in Refs.~\refcite{Abelev:2013fn}, \refcite{Aamodt:2011gj}, and \refcite{Abelev:2012ba}, 
respectively.

\begin{figure}[p]
\centering
\includegraphics[width=1.0\textwidth]{{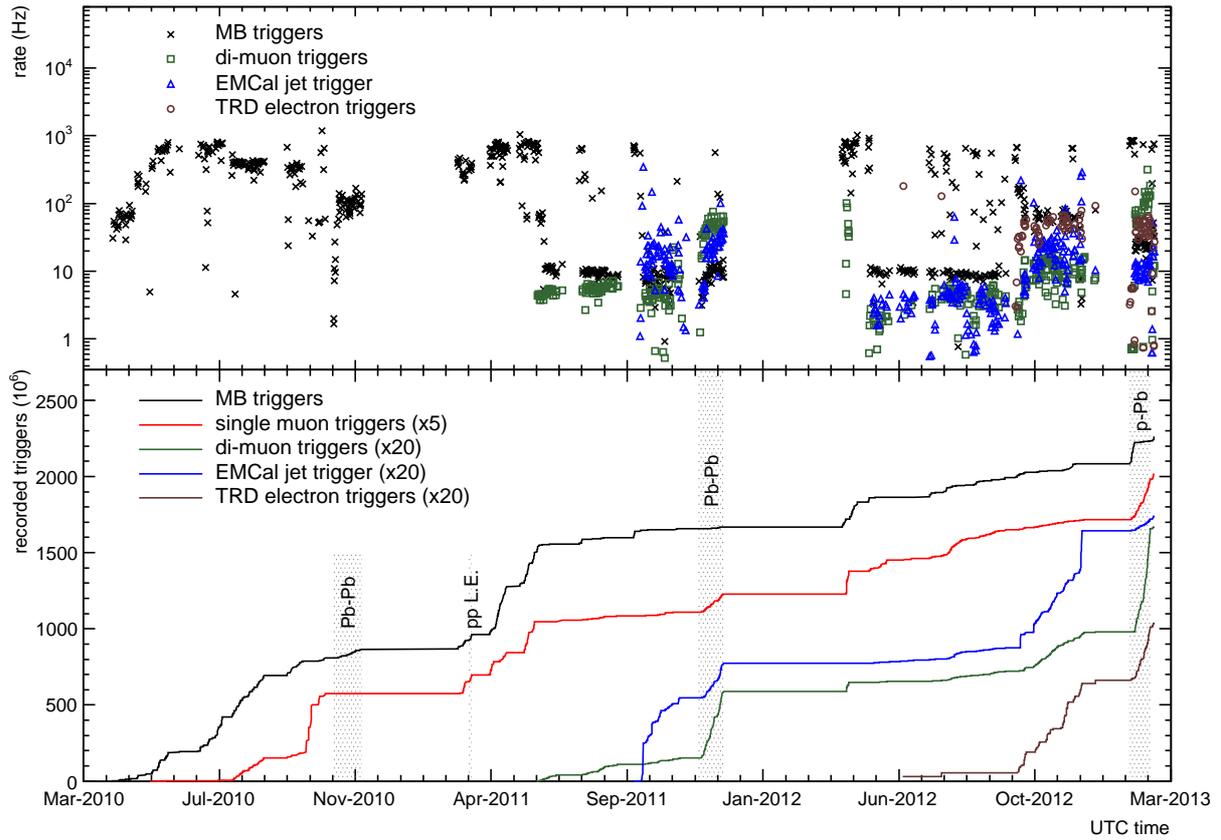}}
\caption{
Instantaneous rate (top) and number of collected events (bottom) for selected 
triggers in the running periods from 2010 to 2013.
Special running periods (\pbpb, \ppb, low energy \pp) are indicated by shaded 
areas; the rest represents \pp runs at the highest available energy. 
}
\label{fig:integrated_triggers}
\end{figure}

\begin{figure}[p]
\centering
\includegraphics[width=0.8\textwidth]{{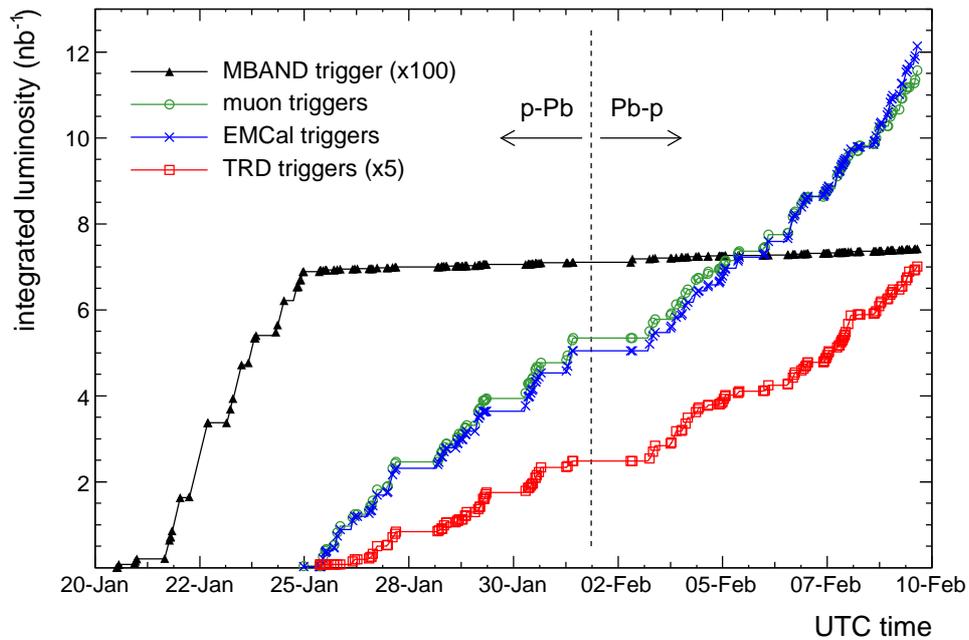}}
\caption{
Integrated luminosity in the 2013 \ppb run, collected in the minimum bias 
and the rare-trigger mode (before and after January 25, respectively).
}
\label{fig:pA_integrated_lumi}
\end{figure}

The instantaneous rate and the total number of collected events in Run 1 are 
shown for selected triggers in Fig.~\ref{fig:integrated_triggers}.
The minimum bias and rare-trigger running modes are illustrated in detail for 
the \ppb data taking in 2013 in Fig.~\ref{fig:pA_integrated_lumi}. 

The total number of recorded events and the inspected luminosity are shown in 
Table~\ref{tab:statistics-of-blocks} for selected minimum-bias and rare triggers, 
respectively. The values are based on raw trigger counts. 
The luminosities were determined for reference triggers as described in Section~\ref{sect:lumi} .
For rare triggers, for which no direct measurement of cross section was performed, 
the integrated luminosity was estimated by comparing their rates to that of a reference trigger.
The resulting uncertainty is typically about 10\%.
Another uncertainty of up to 20\% comes from the fact that this simple method 
does not account for the trigger purity. The actual statistics useful for physics analysis may 
thus fall significantly below the numbers given in the table.

\begin{table}[t!]
\centering
{\begin{tabular}{@{\extracolsep{3mm}}lrrrrrr@{}} \toprule
Detector   &\multicolumn{2}{c}{\pp}&\multicolumn{2}{c}{\pbpb}&\multicolumn{2}{c}{\ppb}\\
           & Busy time & Data size & Busy time & Data size & Busy time & Data size \\
           & ($\mu$s)  &  (kB)     & ($\mu$s)  &  (kB)     & ($\mu$s)  &  (kB)     \\
\colrule
SPD        &    0      & 7         & 0         & 26        & 0 or  370 & 7         \\
SDD        & 1024      & 22        & 1024      & 143       & 1024      & 16        \\
SSD        &  265      & 46        & 265       & 180       & 265       & 42        \\
TPC        &  500      & 6676      & 500       & 25740     & 350       & 15360     \\
TRD        &  300      & 181       & 450       & 3753      & 270       & 350       \\
TOF        &    0      & 23        & 0         & 63        & 0         & 23        \\
PHOS       &  850      & 25        & 850       & 72        & 850       & 35        \\
EMCal      &  270      & 22        & 300       & 53        & 270       & 25        \\
HMPID      &  220      & 15        & 300       & 22        & 220       & 18        \\
ACORDE     &  116      & 0.1       & 116       & 0.1       & 116       & 0.1       \\
PMD        &  170      & 10        & 220       & 50        & 170       & 8         \\
FMD        &  190      & 14        & 350       & 55        & 190       & 13        \\
V0         &    0      & 6         & 0         & 6         & 0         & 6         \\
T0         &    0      & 0.4       & 0         & 0.7       & 0         & 0.6       \\
ZDC        &  122      & -         & 122       & 0.8       & 122       & 0.7       \\
MCH        &  300      & 35        & 300       & 61        & 250       & 18        \\
MTR        &  160      &  7        & 160       & 7         & 160       & 7         \\
\botrule 
\end{tabular} 
\tbl{Average busy times and event sizes of the ALICE detectors observed 
     in typical rare-trigger \pp runs in 2012, \pbpb runs in 2011, and \ppb runs in 2013. 
     ZDC was not active in the 2012 \pp running therefore no value is given for the data size.
     In \ppb runs, SPD busy time was either 0 or 370 $\mu$s depending on the running mode. 
}
\label{tab:readout}}
\end{table}

\subsection{Readout}

The ALICE detectors are equipped with standardized optical fiber based 
data transmission devices working at a bandwidth of 200~MB/s.
Some of the detectors have multiple data transmission connections. 
Event data are sent to DAQ and HLT where event building and data 
compression are performed.
Trigger detectors provide low-voltage differential signals (LVDS) to the CTP inputs.
The CTP makes the global ALICE trigger decision as described in Section~\ref{sect:datataking:trigger}.
In conjunction with the LHC clock and bunch filling scheme, 
this decision is propagated to all detectors, to DAQ, and to HLT via the TTC 
(Timing, Trigger, and Control)~\cite{CERNTTC} passive optical transmission network system. 
The LHC clock is used to synchronize the data of all detectors with the bunch 
crossing.

The busy time of the data taking is mainly defined by the CTP waiting for the 
completion of the readout of all detectors.
In addition, L1-rejected events contribute to the busy time because of the 
latency of the L1 decision. 
The detector busy time due to readout, in general, depends on the event size 
and thus on the collision system and background conditions.  
The ability to buffer events, possessed by some of the detectors, reduces 
their respective average busy times by a rate-dependent factor. 
The typical readout performance of the ALICE detectors in recent \pp, \pbpb, and \ppb runs 
is summarized in Table~\ref{tab:readout}. 
By virtue of event buffering, SPD, TOF, T0, and V0 do not cause a ``detector busy'' state. 
TPC and TRD have multi-event buffers which efficiently reduce their busy times 
in rare-trigger \pp and \pbpb runs at event rates of 200--300~Hz.  
The TPC busy duration is identical in these two collision systems although the 
event sizes are very different.  
The TPC busy time includes a protection period of approximately 300~$\mu$s covering 
the electron drift and the ion collection times.

The ALICE data volume is dominated by the event size of the TPC. 
The latter scales with the charged-particle multiplicity, 
including pileup tracks from other interactions within the TPC drift time 
window of $\sim$100~$\mu$s. 
The maximum TPC event size, observed in central \pbpb collisions, was 70~MB. 

\subsection{Online data compression}
\label{sect:datataking:compression}

Over the course of preparations for the \pbpb run in 2011 it was estimated that 
the data rate would exceed the maximum bandwidth of the connection to mass storage. 
The data volume was then reduced by storing TPC cluster information instead of raw data, 
using online processing by HLT~\cite{Kollegger:2012cja,Richter:2012ne}.
The reduced data are further compressed by HLT using lossless compression with 
Huffman encoding~\cite{HUFFMAN}.
The procedure was tested during the \pp runs in 2011, and successfully used in the 
lead-ion run and all subsequent data taking. 
For integrity checks, 1\% of the events were recorded without compression. 
This way, a data compression by a factor of 5 was achieved for the TPC data.
As the TPC is the dominant contributor to the event size, the compression 
factor for the total data volume in 2012 \ppb running was about 4.
The effect of the compression can be seen from the difference between 
``data read'' and ``data recorded'' in Table~\ref{tab:statistics-of-blocks}.

\clearpage\newpage\section{Calibration strategy}
\label{sect:calibration}

The momentum resolution and the particle identification performance 
critically depend on the quality of the calibration. 
The actual positions of the detectors (alignment), maps of dead or 
noisy elements, and time and amplitude calibrations are used in the 
reconstruction. 
For the drift detectors (SDD, TPC, TRD), the gain and the time response 
are calibrated differentially in space (single readout pads for TPC 
and TRD) and time (units of 15 minutes for TPC). 
Finally, the geometry of the luminous region and (for \pbpb collisions) 
calibrated centrality and event plane are important for physics analysis. 

In this section we briefly describe the main sources of the various 
calibration parameters. 
Once determined, the calibration parameters are stored in the Offline 
Conditions Database (OCDB) and thus become accessible for reconstruction 
jobs running on the distributed computing Grid. 
The list of the calibration parameters, organized according to the source, 
is given in Table~\ref{tab:cali}. 

\begin{table}[p]
\ifijmp
  \renewcommand\arraystretch{1.2}
\fi
\centering
{
\ifcern
  \small
  \hspace*{-1cm}
\fi
\ifijmp
  \scriptsize
\fi
\begin{tabular}{@{\extracolsep{-2mm}}lllllll@{}} \toprule
System     & Condition data   & Special runs     & Physics runs online      & After cpass0        & After cpass1 & After full pass\\
\colrule
SPD        & trigger chip map &                  & half-stave status        &                     &              & pixel status \\
           & and thresholds   &                  & pixel noise              &                     &              & (only for MC)\\
\colrule
SDD        &                  & anode ped (peds) &                          &                     & anode gain$(r\phi)$ &       \\
           &                  & anode gain, status (puls) &                 &                     & anode $\vd(r\phi)$  &       \\
           &                  & anode \vd (inject) &                        &                     & anode \tz           &       \\
\colrule
SSD        &                  & strip ped, noise,&                          &                     &              & module-side  \\
           &                  & status (peds)    &                          &                     &              & gain         \\
\colrule
TPC        & $P, T(x,y,z)$    & pad gain (Kr)    & \vd (laser)              &$\vd(x,y)$,          & as in cpass0 &              \\
           & pad status       & pad noise (peds) &                          & $\tz(x,y)$,         & but higher   &              \\
           & trigger \tz      & \vd (laser)      &                          & alignment           & granularity; &              \\
           &                  & pad status (puls)&                          & gain, attachment    & Bethe-Bloch par. &              \\
\colrule
TRD        &                  & pad gain (Kr)    &                          & cham status,        &              & det PID      \\
           &                  & pad noise, status&                          & gain, \tz, \vd,     &              &              \\
           &                  & (peds)           &                          & $E\times B$         &              &              \\
\colrule
TOF        & channel status   & pad noise (peds) & pad noise (calib)        & \tz                 &              & pad \tz      \\
           & TDC \tz          &                  & TDC status               &                     &              & pad slewing  \\
           & BPTX \tz         &                  &                          &                     &              &              \\
\colrule
PHOS       &                  & cell gain (LED)  &                          &                     &              & cell gain, \tz\\
           &                  &                  &                          &                     &              & cell status  \\
\colrule
EMCal      & supermodule $T$  &                  & cell gain (LED)          &                     & cell status  & cell gain, \tz\\
\colrule
HMPID      & chamber $P$, HV  & pad ped, noise,  & pad noise, status        &                     &              &            \\
           & radiator $T$ and & status (peds)    &                          &                     &              &            \\
           & transparency     &                  &                          &                     &              &            \\
\colrule
ACORDE     &                  &                  &                          &                     &              &            \\
\colrule
PMD        &                  & cell ped         & cell gain                &                     &              & cell gain, status \\
\colrule
FMD        &                  & strip ped, noise (peds)&                    &                     &              & strip gain \\
           &                  & strip gain (puls)&                          &                     &              &            \\
\colrule
V0         & PMT HV, trigger  & PMT slewing      & PMT ped, noise,          &                     &              &            \\
           & thresh. and \tz  & (once)           & gain                     &                     &              &            \\
\colrule
T0         &                  & PMT slewing      & PMT \tz                  & \tz                 &              &            \\
\colrule
ZDC        &                  & ped (peds)       & PMT \tz                  &                     &              & PMT gain   \\
\colrule
MCH        & chamber HV       & pad ped, noise, status & noisy pads         &                     &              & alignment  \\
           &                  & (peds), pad gain (puls)  &                  &                     &              &            \\
\colrule
MTR        &                  & strip, board     & strip, board status      &                     &              &            \\
           &                  & status (puls)    & (special events)         &                     &              &            \\
\colrule
Luminous   &                  &                  & $x,y,z,\sigma_z$ from SPD& $x,y,z,\sigma_x,\sigma_y,\sigma_z$&&            \\
region     &                  &                  &                          &                     &              &            \\
\colrule
Centrality &                  &                  &                          & centrality from     &              & centrality \\
           &                  &                  &                          & SPD, V0             &              & from TPC   \\
\botrule
\end{tabular} 
\tbl{Calibration parameters used in the reconstruction, ordered according to 
their source. The following abbreviations are used: 
``\vd'' \ -- drift velocity, ``\tz'' \ -- time offset, ``$T$'' -- temperature, 
``$P$'' -- pressure, ``ped'' -- pedestals, ``peds'' -- pedestal runs, 
``inject'' -- injector runs, ``puls'' -- pulser runs, ``laser'' -- laser runs/events, 
``LED'' -- LED runs/events, ``Kr'' -- runs with radioactive krypton in the gas.}
\label{tab:cali}}
\end{table}

\subsection{Condition data and online calibration}

Condition data are monitored continuously and archived by the Detector 
Control System (DCS). 
Some of these data (e.g. temperatures and pressures) affect the detector 
response and thus are relevant to event reconstruction. 

Those calibration parameters that can be derived from raw data are extracted 
online, i.e. during data taking, from interaction events and/or dedicated 
calibration events. The latter can be collected in dedicated calibration runs 
or in parallel with the physics data taking. 
The data processing takes place on the computers of the Data Acquisition (DAQ) 
system~\cite{Chapeland:2010zz}. 

At the end of each run the condition data and the online calibration parameters 
are collected by the Shuttle system~\cite{calib:shuttlenote} and transported to 
the OCDB. A successful Shuttle termination triggers the first reconstruction 
pass of the run. 

\subsection{Offline calibration}

The first two reconstruction passes are performed on a sample of events from 
each run and serve for calibration and monitoring purposes. 
The first pass (cpass0) provides input for the calibration of TPC, TRD, TOF, T0, 
luminous region, and centrality. 
The second pass (cpass1) applies the calibration, and the reconstructed events 
are used as input for data quality assurance and for improved calibration of 
SDD, TPC, and EMCal. 
Once a data taking period (typically 4--6 weeks) is completed, a manual 
calibration spanning many runs is performed. 
The complete calibration is then verified by a validation pass (vpass) 
performed on a sample of events from all runs in the period. 
The subsequent physics reconstruction pass (ppass) is, in general, performed 
on all events and provides the input for physics analysis. 

The complete calibration reconstruction sequence is thus: cpass0, calibration, 
cpass1, quality assurance and calibration, manual multi-run calibration, 
validation pass, quality assurance, physics reconstruction pass, quality 
assurance. 

\subsection{Detector alignment}
The objective of the data-driven alignment of detectors is to account for 
deviations of the actual positions of sensitive volumes and material blocks 
from the nominal ones in the reconstruction and simulation software. 
In order to achieve this, first for those detectors for which standalone 
reconstruction is possible (ITS, TPC, TRD, MUON) an internal alignment (e.g. 
positions of ITS sensors with respect to the sensor staves and of staves 
with respect to the ITS center; relative positions of TRD chambers within 
a stack; etc.) was performed. 
This was done by iterative minimization of the residuals between the cluster 
positions (measured under the current assumption of alignment parameters) and 
the tracks to which these clusters were attached by the reconstruction procedure. 
Given the large number of degrees of freedom in the ITS and MUON detectors
(14622 and 1488, respectively) their alignment was performed using a modified 
version of the Millepede algorithm~\cite{Blobel:2002ax}.
The alignment of ITS~\cite{Aamodt:2010aa}, TPC, and TRD was initially performed 
using the cosmic muons data, and then it was refined using tracks reconstructed 
in the collision events collected in physics runs as well as in dedicated runs 
without magnetic field. For the alignment of the 
MUON detector, muon tracks from runs with and without magnetic field were used 
together with the information from the optical geometric monitoring 
system~\cite{Grigorian:2005sia,Grossiord:2005jia}. 
The precision of the internal alignment in the ITS is estimated 
to be on the level of $\sim$10(70), 25(20), and 15(500)\mum in the bending 
(non-bending) direction for SPD, SDD, and SSD layers, respectively. For MUON, 
the alignment precision is estimated to be better than \mbox{50--100(100--150)\mum} 
in the bending (non-bending) direction, depending on muon station.
The precision of the inter-sector alignment in the TPC is estimated to be 
$\sim$0.1~mm.

After the internal alignment, the ITS and TPC were aligned to each other to 
a precision of $\sim$30\mum and $\sim$0.1~mrad by applying a Kalman-filter based 
procedure of minimizing the residuals between the tracks reconstructed in 
each detector. 
The global alignment of other central-barrel detectors was performed by 
minimizing the residuals between their clusters and the extrapolation 
of the ITS--TPC tracks. 
The residual misalignment in the $r\phi$ and $z$ directions is estimated
to be smaller than $\sim$0.6~mm for the TRD, $\sim$5~mm for the TOF, 5--10~mm 
depending on chamber for HMPID, $\sim$6~mm for the PHOS, and $\sim$2~mm for 
the EMCal.
The global alignment of MUON is performed by requiring the 
convergence of the muon tracks to the interaction vertex.

The alignment is checked and, if necessary, redone after shutdowns and/or 
interventions that may affect the detector positions. 
In order to minimize the influence of the residual misalignment on the 
reconstructed data, the physics measurements in ALICE are routinely 
performed with both magnetic field polarities.

\clearpage\newpage\section{Event characterization}
\label{sect:characterization}

For spherical nuclei, the geometry of heavy-ion collisions is characterized by 
the impact parameter vector ${\mathbf b}$ 
connecting the centers of the two colliding nuclei in the plane transverse to the beams. 
In the experiment, the centrality (related to $b \, :=|{\mathbf b}|$) and 
the reaction-plane angle (azimuthal angle of ${\mathbf b}$) are 
estimated using the particle multiplicities and/or the zero-degree energy, 
and the anisotropies of particle emission, respectively. 
Below we sketch the methods and quote the resolution achieved in these variables. 
A more detailed discussion of the centrality determination in ALICE can be found 
in Ref.~\refcite{Abelev:2013qoq}. 

\subsection{Centrality} \label{Section:Centrality} 

It is customary to express the centrality of nuclear collisions not in terms of 
the impact parameter $b$ but via a percentage of the total hadronic interaction 
cross section $\sigma_{AA}$. 
The centrality percentile $c$ of an $AA$ collision with impact parameter 
$b$ is defined as 
\begin{equation}
c(b)=\frac{\int_0^{b}{\frac{\mathrm{d}\sigma}{\mathrm{d}b^{'}}} \, \mathrm{d}b^{'}}
  {\int_0^{\infty}{\frac{\mathrm{d}\sigma}{\mathrm{d}b^{'}}} \, \mathrm{d}b^{'}} 
= \frac{1}{\sigma_{AA}} \, \int_0^{b}{\frac{\mathrm{d}\sigma}{\mathrm{d}b^{'}}} \, \mathrm{d}b^{'} .
\label{cent-the}
\end{equation}
Experimentally, the centrality is defined as the fraction of cross section with the largest 
detected charged-particle multiplicity \nch or the smallest zero-degree energy \ezdc:
\begin{equation}
c \approx  \frac{1}{\sigma_{AA}} \, \int_{\nch}^{\infty}{\frac{\mathrm{d}\sigma}{\mathrm{d}\nch^{'}}} \, \mathrm{d}\nch^{'}
\approx \frac{1}{\sigma_{AA}} \, \int_{0}^{\ezdc}{\frac{\mathrm{d}\sigma}{\mathrm{d}\ezdc^{'}}} \, \mathrm{d}\ezdc^{'}. 
\label{cent-exp1}
\end{equation}
The cross section may be replaced with the number of observed events $n$ 
(corrected for the trigger efficiency and for the non-hadronic interaction 
background): 
\begin{equation}
c \approx \frac{1}{N_\mathrm{ev}} \, \int_{\nch}^{\infty}{\frac{\mathrm{d}n}{\mathrm{d}\nch^{'}}} \, \mathrm{d}\nch^{'}
  \approx \frac{1}{N_\mathrm{ev}} \, \int_{0}^{\ezdc} {\frac{\mathrm{d}n}{\mathrm{d}\ezdc^{'}}} \, \mathrm{d}\ezdc^{'}. 
\label{cent-exp2}
\end{equation}
Eqs.~(\ref{cent-exp1}) and (\ref{cent-exp2}) are based on the assumption that, 
on average, the particle multiplicity at midrapidity (the zero-degree energy) 
increases (decreases) monotonically with the overlap volume, i.e. with centrality. 
For the zero-degree energy measurement (\ref{cent-exp2}), this assumption holds only 
for central collisions $c \lesssim 50$\%, because nuclear fragments emitted 
in peripheral collisions may be deflected out of the acceptance of the 
zero-degree calorimeter, leading to low signals indistinguishable from 
those seen in central collisions. 

\begin{figure}[tp]
\centering
\scalebox{0.58}{\includegraphics{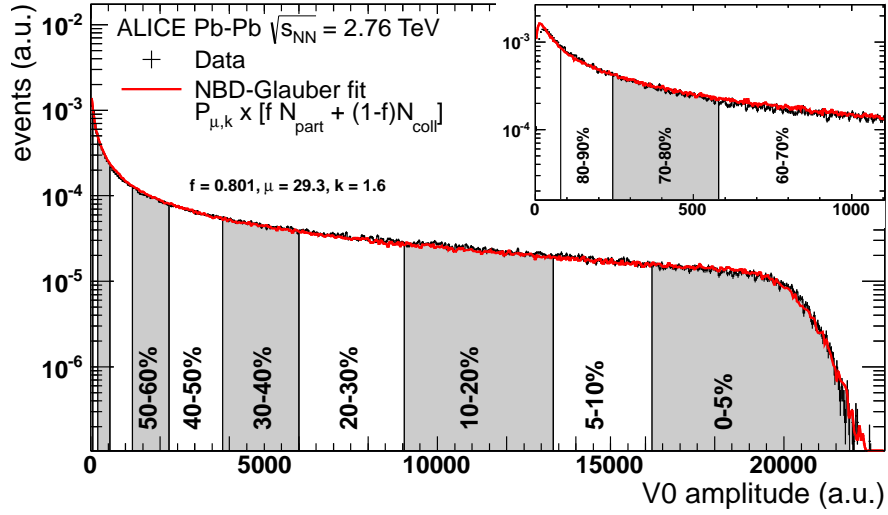}}
\caption{Distribution of the V0 amplitude (sum of V0A and V0C). 
The centrality bins are defined by integrating from right to left following 
Eq.~(\ref{cent-exp2}). 
The absolute scale is determined by fitting to a model (red line), see below 
for details. 
The inset shows a magnified version of the most peripheral region. 
\label{fig:vzeroglau}}
\end{figure}

The centrality determination via the particle multiplicity in V0 is 
illustrated in Fig.~\ref{fig:vzeroglau}. The V0 multiplicity (sum of 
V0A and V0C amplitudes) distribution was recorded in \pbpb collisions at 
\mbox{\sqrtsnn = 2.76 TeV}, requiring a coincidence of V0 and SPD, and using 
ZDC to reduce the electromagnetic dissociation background. 
Machine-induced background and parasitic collisions are
removed using the timing information from V0 and ZDC. 
The analysis is restricted to events with a vertex position within 
$\left|z_{\rm vtx}\right|\lesssim 10~\mathrm{cm}$.
The centrality bins are defined by integrating the charged-particle 
multiplicity distribution following Eq.~(\ref{cent-exp2}), and 
the absolute scale is determined by fitting to a model as described below. 

\begin{figure}[hbtp]
\centering
\scalebox{0.6}{\includegraphics{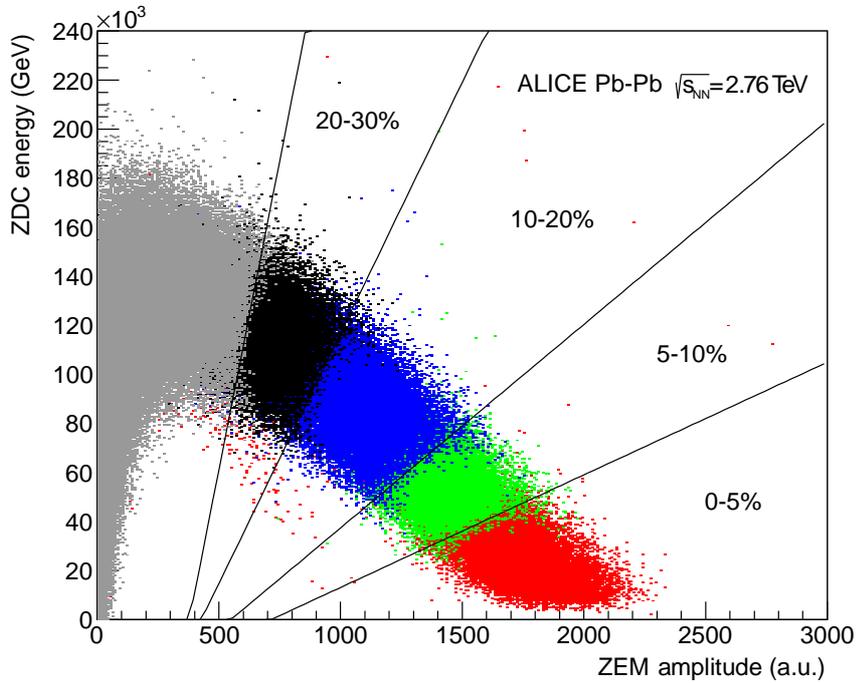}}
\caption{Correlation between the total energy deposited in the zero-degree 
calorimeters and the ZEM amplitude. 
The centrality bins defined based on this distribution (lines) 
are compared to the centrality from V0 (colored dots). 
\label{fig:zdczem}}
\end{figure}

The distribution of the energy deposited in the zero-degree calorimeter is 
shown in Fig.~\ref{fig:zdczem}. The ambiguity between central 
and peripheral collisions with undetected nuclear fragments is resolved 
by correlating the zero-degree signal with the amplitude of the 
electromagnetic calorimeter at $4.8<\eta<5.7$ (ZEM). 

\begin{figure}[h]
\centering
\scalebox{0.4}{\includegraphics{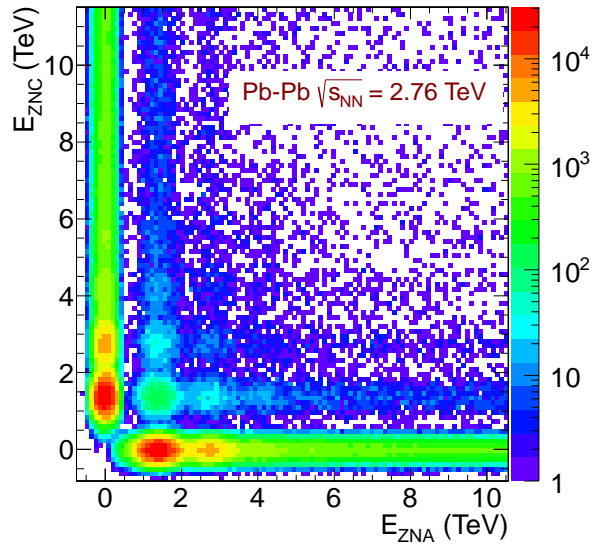}}
\caption{Correlation between signals in the two neutron zero-degree calorimeters. 
Single electromagnetic dissociation events produce a signal in only one of the 
calorimeters. 
Mutual dissociation and hadronic interactions populate the interior of the plot 
and can be distinguished from each other by the signal in ZEM. 
\label{fig:zncvszna}}
\end{figure}

An absolute determination of centrality according to Eqs.~(\ref{cent-exp1}) or 
(\ref{cent-exp2}) requires knowledge of the total hadronic cross section 
$\sigma_{AA}$ or the total number of events $N_\mathrm{ev}$, respectively. 
The total hadronic cross section $\sigma_{AA}$ for \pbpb at \mbox{\sqrtsnn = 2.76 TeV} was 
measured in ALICE in a special run triggering on signals in the neutron zero-degree 
calorimeters (ZNs) with a threshold well below the signal of a 1.38~TeV neutron~\cite{ALICE:2012aa}. 
The recorded event sample is dominated by the electromagnetic dissociation (EMD) 
of one or both nuclei. 
The single EMD events can be clearly identified in the correlation plot 
between the two ZNs (Fig.~\ref{fig:zncvszna}). 
An additional requirement of a signal in ZEM (see Section~\ref{sect:intro}) allows one to 
distinguish between the mutual EMD and the hadronic interaction events. 
With the absolute normalization determined by means of a van der Meer scan as 
described in Section~\ref{sect:conditions-vdm-pbpb}, a hadronic cross section of 
$\sigma_{\rm PbPb}=\left(7.7 \pm 0.1 {\rm (stat)}^{+0.6}_{-0.5} {\rm (syst)}\right)$~b 
was obtained. The centrality may then be derived from the calorimeter signals using 
Eq.~(\ref{cent-exp1}). 

A higher accuracy of the centrality calibration can be achieved by
normalizing the measured event yield to the total number of events
$N_\mathrm{ev}$ that would be registered in an ideal case,
i.e. without background interactions and with a perfect trigger
efficiency (Eq.~(\ref{cent-exp2})). This was the method of choice 
in ALICE. The high-multiplicity part of the multiplicity
distribution was fitted by the Glauber model (red line in
Fig.~\ref{fig:vzeroglau}), and the extrapolation of the model was used
to determine the unbiased number of events at low multiplicities.  The
Glauber model describes the collision geometry using the nuclear
density profile, assuming that nucleons follow straight line
trajectories and encounter binary nucleon-nucleon collisions according
to an inelastic nucleon-nucleon cross section $\sigma_{\rm NN}$. 
For the latter, 64~mb was assumed in the calculation; this value is consistent 
with the subsequent ALICE \pp measurement reported in Section~\ref{vdMapplications}.  
The number of binary NN collisions \ncol
and the number of participants \npart (nucleons which underwent a NN
collision) are determined for a given impact parameter.  The
multiplicity distribution was modeled assuming $f\npart+(1-f)\ncol$
particle sources, with each source producing particles following a
negative binomial distribution (NBD) with fit parameters $\mu$ 
and $k$. The parameter $f$ represents the contribution of soft processes 
to the particle production. The fit provides the integrated
number of events $N_\mathrm{ev}$ needed for the absolute centrality
scale and relates the number of participants and binary NN collisions 
to the centrality.  The latter relation is presented in detail 
in Ref.~\refcite{Abelev:2013qoq}.

\begin{figure}[hbt]
\centering
\includegraphics[width=0.78\textwidth]{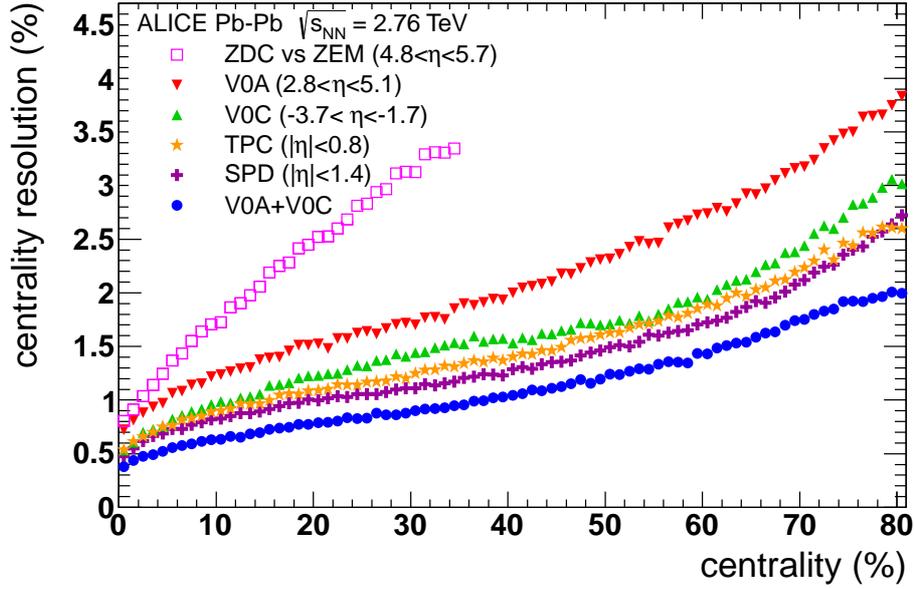}
\caption{Resolution of various centrality estimators. \label{fig:centers}}
\end{figure}

The centrality for each event can be independently calculated from the 
multiplicities seen in V0A, V0C, ZDC, SPD, and TPC. The resolution of 
each of these centrality estimators, defined as their r.m.s. for a sample 
of events with a fixed $b$, was determined by studying correlations 
between them and is shown in Fig.~\ref{fig:centers}. 
The resolution ranges from 0.5\% to 4\% depending on centrality and on 
the detector used. 
As expected, the resolution of each detector depends on its rapidity 
coverage, scaling with $\sim1/\sqrt{\nch}$. 

\subsection{Event plane} 
\label{Section:EventPlane}

The orientation of the reaction plane or, in case of flow fluctuations, 
the $n^{\rm th}$-harmonic collision symmetry plane is estimated with the 
$n^{\rm th}$-harmonic event-plane angle, $\Psi_n^{\rm EP}$~\cite{Voloshin:2008dg}.
For a given harmonic $n$, one constructs the two-dimensional event-plane vector 
${\bf Q}_n$ from the measured azimuthal distribution of particles produced in 
the event as follows: 
\begin{eqnarray}
{\bf Q}_n =  \left(Q_{n,x}, Q_{n,y}\right)=  \left(\sum_{i} \> w_i \> \cos n \phi_i, \; \sum_{i} \> w_i \> \sin n \phi_i\right).
\label{Eq:Q-vector}
\end{eqnarray}
Here the sum runs over all reconstructed tracks in the case of the TPC, 
or segments of the detectors with azimuthal segmentation like V0, FMD, ZDC, or PMD. 
The angle $\phi_i$ is the azimuthal emission angle of the particle $i$ 
or the azimuthal coordinate of the detector element $i$, respectively. 
For TPC tracks the weight $w_i$ can be unity or a specific function of
\pt~\cite{Voloshin:2008dg}. 
For segmented detectors, $w_i$ is the signal observed in the detector element $i$. 
Using the components of the ${\bf Q}$-vector one can calculate 
the $\Psi_n^{\rm EP}$~\cite{Voloshin:2008dg}:
\begin{eqnarray}
\Psi_n^{\rm EP} = \frac{1}{n}\arctan2(Q_{n,y},Q_{n,x}) . 
\label{Eq:Psi}
\end{eqnarray}

The correction for the finite event-plane angle resolution can be calculated
using the two- or three- \mbox{(sub-)detector} correlation technique.
The resolution correction factor, in the following for brevity called ``resolution'', is close 
to zero (unity) for poor (perfect) reconstruction of the collision symmetry plane.
In case of two \mbox{(sub-)detectors} $A$ and $B$ the subevent resolution is defined as 
\begin{eqnarray}
R_{n}^{\rm sub} = \sqrt{\langle \cos n (\Psi_{n}^{A} - \Psi_{n}^{B}) \rangle}
\label{Eq:2subevent-correlation}
\end{eqnarray}
where $\Psi_{n}^{A}$ and $\Psi_{n}^{B}$ are the event-plane angles 
of the two subevents, and the  
angle brackets denote the average over an ensemble of the events.
Typically, the same harmonic is used in the flow measurement and for 
the event-plane determination. In this case, the full event-plane 
resolution, i.e. the correlation between the event-plane angle
for the combined subevents and the reaction-plane angle, 
can be calculated from~\cite{Voloshin:2008dg} 
\begin{equation}
R_n(\chi) = \sqrt{\pi}/2 \ \chi \exp(-\chi^2/2) \ (I_0(\chi^2/2) + I_1(\chi^2/2)) \ ,
\label{Eq:chi2}
\end{equation}
\begin{equation}
R_n^{\rm full} = R_n (\sqrt{2}\chi_{\rm sub}).
\label{Eq:full-resolution}
\end{equation}
The variable $\chi$ represents the magnitude of flow normalized to the 
precision with which it can be measured, 
and $I_0$, $I_1$  are the modified Bessel functions. 
In case of (sub-)detectors 
with different kinematic coverages, 
such as V0A and V0C, a three-detector subevent technique can be used.
In this case, the resolution for a given detector
can be defined from the correlation between each detector pair 
\begin{equation}
R_n^{A} = \sqrt{\frac{\langle \cos n (\Psi_{n}^{A} - \Psi_{n}^{C}) \rangle \langle \cos n (\Psi_{n}^{A} - \Psi_{n}^{B}) \rangle}{\langle \cos n (\Psi_{n}^{B} - \Psi_{n}^{C}) \rangle}} 
\label{Eq:3-subevents}
\end{equation}
where $\Psi_{n}^{A}$ is the event-plane angle for which the resolution is calculated,
and $B$ and $C$ are any other two (sub-)detectors.
One can get the resolution for each of the three detectors by permutation of the
event-plane angles for all three detectors.
Note that non-flow correlations and the effects of flow fluctuations can
result in different resolutions being 
extracted for the same detector from two- or three- (sub-)detector correlations.

\subsubsection{Event plane from elliptic flow}
\label{Section:v2_resolution}
\begin{figure}[b]
\centering
\includegraphics[width=0.98\linewidth]{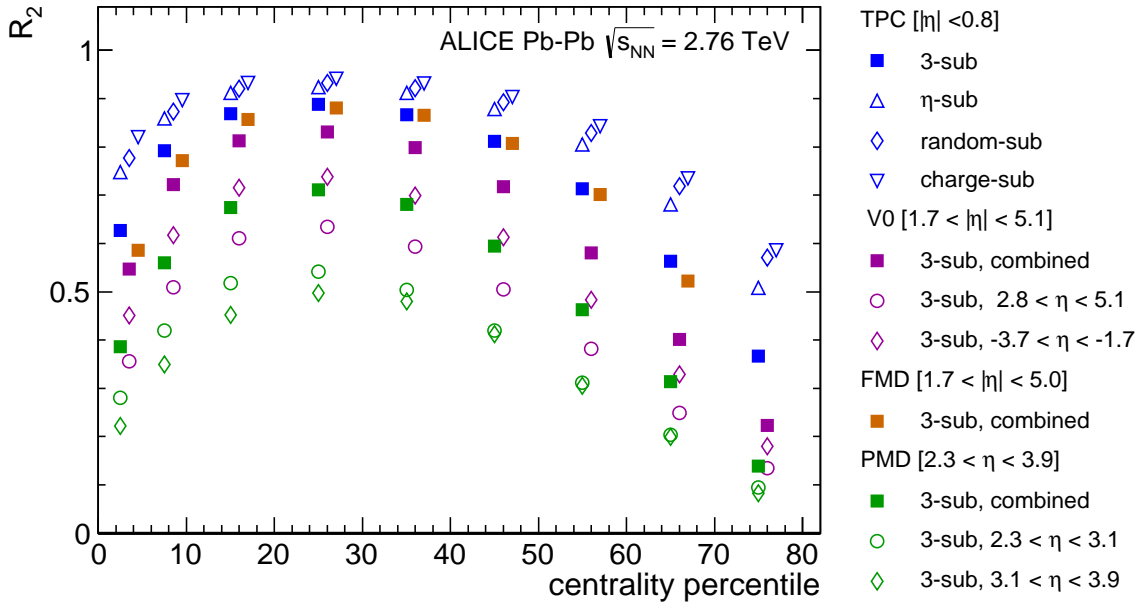}
\caption{Resolution of the second-order event-plane angle, $\Psi_{2}^{\rm EP}$, 
extracted from two- and three-detector subevent correlations for TPC, V0, FMD, and PMD.
}
\label{fig:v2_resolution}
\end{figure}

The dominant component of the anisotropic flow in mid-central collisions at 
LHC energies is the elliptic flow. 
Consequently, the resolution of the second-order event plane is the best.
Figure~\ref{fig:v2_resolution} 
shows the resolution $R_2$ of the second-order event-plane angle 
$\Psi_{2}^{\rm EP}$, extracted from two- and three-detector subevent 
correlations for TPC, V0, FMD, and PMD, vs. the collision centrality.
Effects from the azimuthal non-uniformity of the detectors, 
which may result in non-physical correlations, 
were corrected at the time of the event-plane angle calculations.
The resolution $R_2$ for charged particles measured in the
TPC detector was calculated using four different methods:
by randomly dividing particles into two subevents (denoted as `random-sub'),
by constructing subevents from particles with opposite charges (`charge-sub') or
particles separated by a rapidity gap of at least 0.4 units (`$\eta$-sub'),
and from three-detector subevent correlations
in combination with the V0A and V0C detectors (`3-sub').
Variations in the event-plane resolution calculated with different
methods indicate differences in their sensitivity to the correlations
unrelated to the reaction plane (non-flow) and/or flow fluctuations. 

\subsubsection{Event plane from higher harmonics}
\begin{figure}[hbt]
\centering
\includegraphics[width=0.98\linewidth]{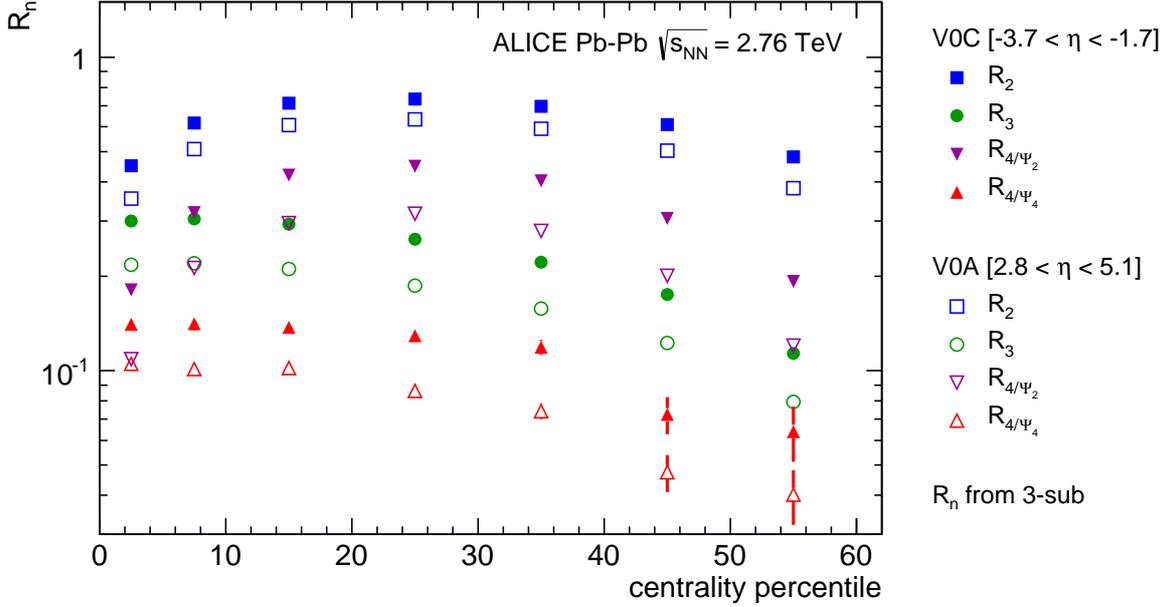}
\caption{Event-plane angle, $\Psi_{n}^{\rm EP}$, resolution for $n=2,3$, and $4$, calculated 
with a three-detector subevent technique separately for V0A and V0C detectors. 
}
\label{fig:vn_resolution}
\end{figure}

Figure~\ref{fig:vn_resolution} shows the resolution of the event-plane angle, $\Psi_{n}^{\rm EP}$,
for the $n=2,3$, and $4$ harmonics calculated with a three-detector subevent technique
separately for the V0A and V0C detectors. The TPC was used as a third, reference, detector.
The ordering of the resolutions for mid-central collisions in Fig.~\ref{fig:vn_resolution} 
reflects the fact that higher harmonics of the anisotropic flow are gradually suppressed.
At the same time, even with small signals we still can statistically resolve higher-harmonic 
event-plane angles with resolutions of the order of a few percent.

\subsubsection{Event plane from spectator deflection}
\label{Section:v1_resolution}
\begin{figure}[ht]
\centering
\includegraphics[width=0.7\linewidth]{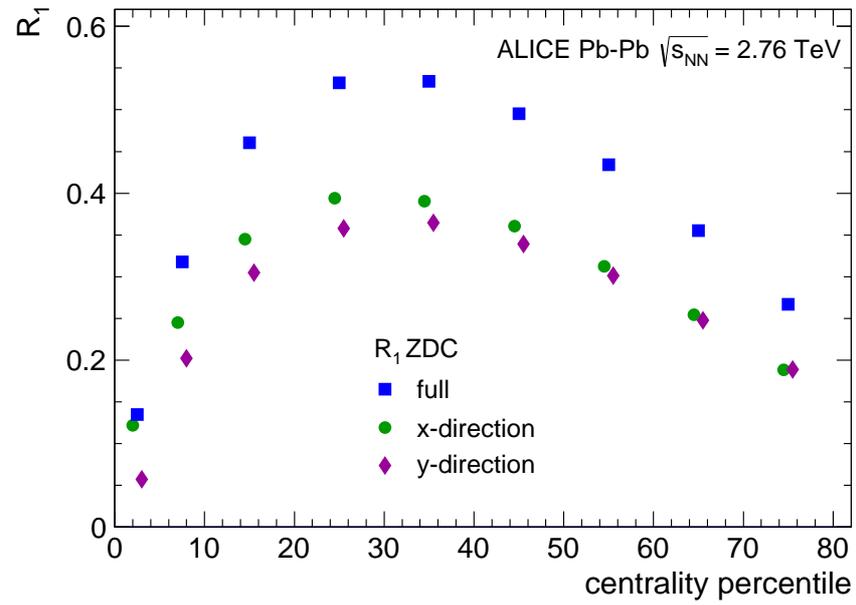}
\caption{Resolution of the first-harmonic event plane
estimated from spectator deflection, as measured by the two ZDCs.
}
\label{fig:v1_resolution}
\end{figure}

In non-central nuclear collisions at relativistic energies, the 
spectator nucleons are assumed to be deflected in the reaction plane 
away from the center of the system. 
The first-order event-plane angle, 
which provides an experimental estimate of the orientation of the
impact parameter vector ${\bf b}$, can be reconstructed using the
neutron ZDCs~\cite{Carminati:2004fp}.
Located about a hundred meters from the interaction point, these detectors 
are sensitive to neutron spectators at beam rapidity. 
Each ZDC, A-side ($\eta>0$) and C-side ($\eta<0$), has a $2\times 2$ tower geometry.
Event-by-event spectator deflection is estimated from
the ZDC centroid shifts ${\bf Q_1}$: 
\begin{equation}
{\bf Q_1} = {\sum\limits_{i=1}^{4}
{\bf r}_i E_i}/{\sum\limits_{i=1}^{4} E_i} \ ,
\label{equation:1}
\end{equation}
where ${\bf r}_i=(x_i,y_i)$ and $E_i$ are the coordinates and the recorded signal 
of the $i^{\rm th}$ ZDC tower, respectively. 
To correct for the time-dependent variation of the beam crossing position
and event-by-event spread of the collision vertex with respect to the center of the TPC
we perform the recentering procedure: 
\begin{equation}
{\bf Q_1'}= {\bf Q_1}-\langle {\bf Q_1}\rangle \ .
\label{equation:1cor}
\end{equation}
Recentering (subtracting the average centroid position
$\langle {\bf Q_1}\rangle$) is performed as a function of time, 
collision centrality, and transverse position of the collision vertex.
After recentering we observe an anticorrelation of the spectator deflections 
on the A and C sides. This demonstrates the capability to measure directed 
flow using the ZDCs. 
Figure~\ref{fig:v1_resolution} shows, as a function of centrality, the 
first-order event-plane resolution obtained from two different transverse 
directions $x$ and $y$ in the detector laboratory frame together with the combined 
resolution from both ZDCs.

\clearpage\newpage\section{Central barrel tracking}
\label{sect:tracking}

\begin{figure}[htbp]
\hspace*{10mm}\includegraphics[scale=0.9, clip, trim = -0cm -5mm -3cm -0cm]{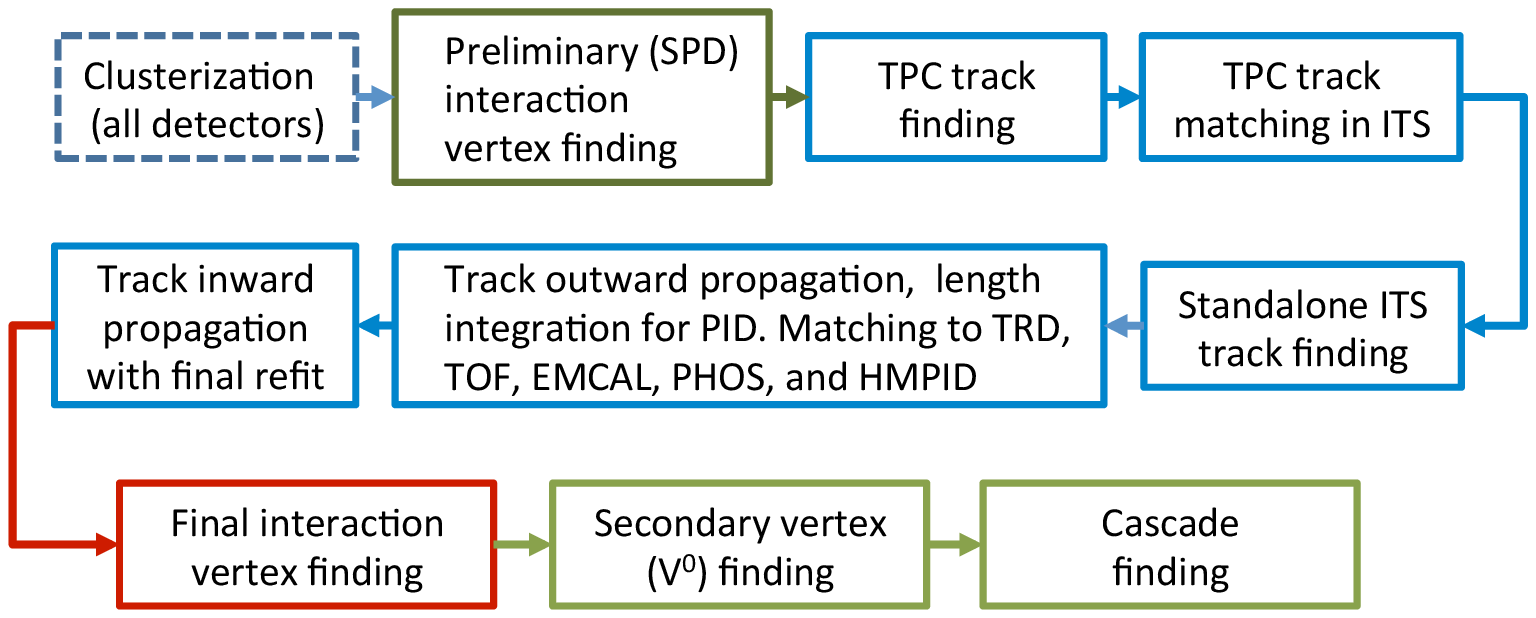}
\caption{\label{fig:recoscheme} 
Event reconstruction flow.}
\end{figure}

This section describes track finding in the central barrel. 
The procedure, shown schematically in Fig.~\ref{fig:recoscheme}, 
starts with the clusterization step, in which the detector data 
are converted into ``clusters'' characterized by positions, signal 
amplitudes, signal times, etc., and their associated errors. 
The clusterization is performed separately for each detector. 
The next step is to determine the preliminary interaction vertex
using clusters in the first two ITS layers (SPD). Subsequently, track 
finding and fitting is performed in TPC and ITS using the Kalman filter 
technique~\cite{Fruhwirth:1987fm}. The found tracks are matched 
to the other central-barrel detectors and fitted. The final interaction 
vertex is determined using the reconstructed 
tracks. A search for photon conversions and decays of strange hadrons 
\kzeros/$\Lambda$ (denoted as V$^0$), $\Xi^{\pm}$, and $\Omega^{\pm}$  
concludes the central-barrel tracking procedure. The steps 
are described in further detail in this section.

\subsection{Preliminary determination of the interaction vertex}
\label{sect:tracking-spd-vertex}

Tracking in the central barrel starts with the 
determination of the interaction vertex using the two innermost layers (SPD)
of the ITS. It is found as a space point to which a maximum number of 
tracklets (lines defined by pairs of clusters, one cluster in each SPD layer) 
converge. In \pp collisions, where interaction pileup is expected, the
algorithm is repeated several times, discarding at each iteration those clusters 
which contributed to already-found vertices. 
By construction, the first vertex found has the
largest number of contributing tracklets and is assumed to be the
primary one. When a single convergence point is not found 
(particularly in low-multiplicity events) the algorithm performs a one-dimensional
search of the maximum in the $z$-distribution of the points of closest
approach (PCA) of tracklets to the nominal beam axis.

\subsection{Track reconstruction}

Track finding and fitting is performed in three stages, following an 
inward--outward--inward scheme~\cite{Ivanov200670,Belikov:2003yr}. 

\begin{figure}[h]
\centering
\includegraphics[width=0.7\textwidth]{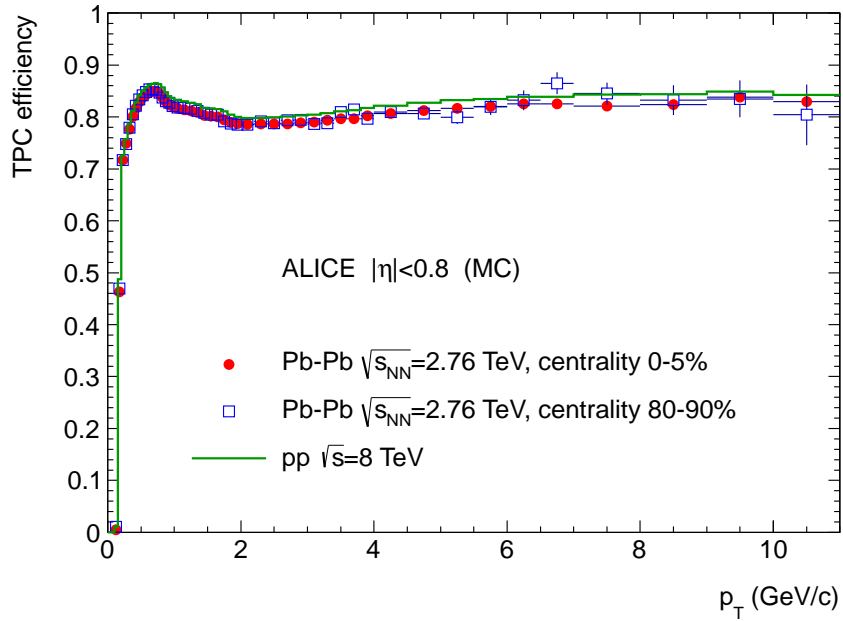}
\caption{\label{fig:tpcabseff} 
TPC track finding efficiency for primary particles in \pp and \pbpb collisions 
(simulation). The efficiency does not depend on the detector occupancy. }
\end{figure}

The first inward stage starts with finding tracks in the TPC. 
The TPC readout chambers have 159 tangential pad rows and thus a track 
can, ideally, produce 159 clusters within the TPC volume. 
The track search in the TPC starts at a large radius.
Track seeds are built first with two TPC clusters and the vertex point, 
then with three clusters and without the vertex constraint. 
The seeds are propagated inward and, at each step, updated with the nearest cluster
provided that it fulfils a proximity cut. 
Since the clusters can be reused by different seeds, the same physical 
track can be reconstructed multiple times. In order to avoid this, a
special algorithm is used to search for pairs of tracks with a fraction of common clusters 
exceeding a certain limit (between 25\% and 50\%). The worse of the two is rejected 
according to a quality parameter based on the cluster density, number of clusters, and momentum. 
Only those tracks that have at least 20 clusters (out of maximum 159 possible) and 
that miss no more than 50\% of the clusters expected for a given track position 
are accepted. 
These are then propagated inwards to the inner TPC radius. A preliminary particle 
identification is done based on the specific energy loss in the TPC gas 
(see Section~\ref{sect:hadrons}). 
The most-probable-mass assignment is used in the ionization energy loss 
correction calculations in the consecutive tracking steps. (Due to the 
ambiguity of electron identification, the minimum mass assigned is that of a pion).
Figure~\ref{fig:tpcabseff} shows the tracking efficiency, defined as the ratio between 
the reconstructed tracks and generated primary particles in the simulation, as a 
function of transverse momentum. While the drop below a transverse momentum of
$\sim$0.5$\gevc$ is caused by energy loss in the detector material, 
the characteristic shape at larger $\pt$ is determined by
the loss of clusters in the $\pt$-dependent fraction of the track 
trajectory projected on the dead zone between readout sectors.
The efficiency is almost independent of the occupancy in the detector. 
Even in the most central \pbpb collisions the contamination by tracks with more 
than 10\% wrongly associated clusters does not exceed 3\%.

\begin{figure}[h]
\includegraphics[width=0.495\textwidth]{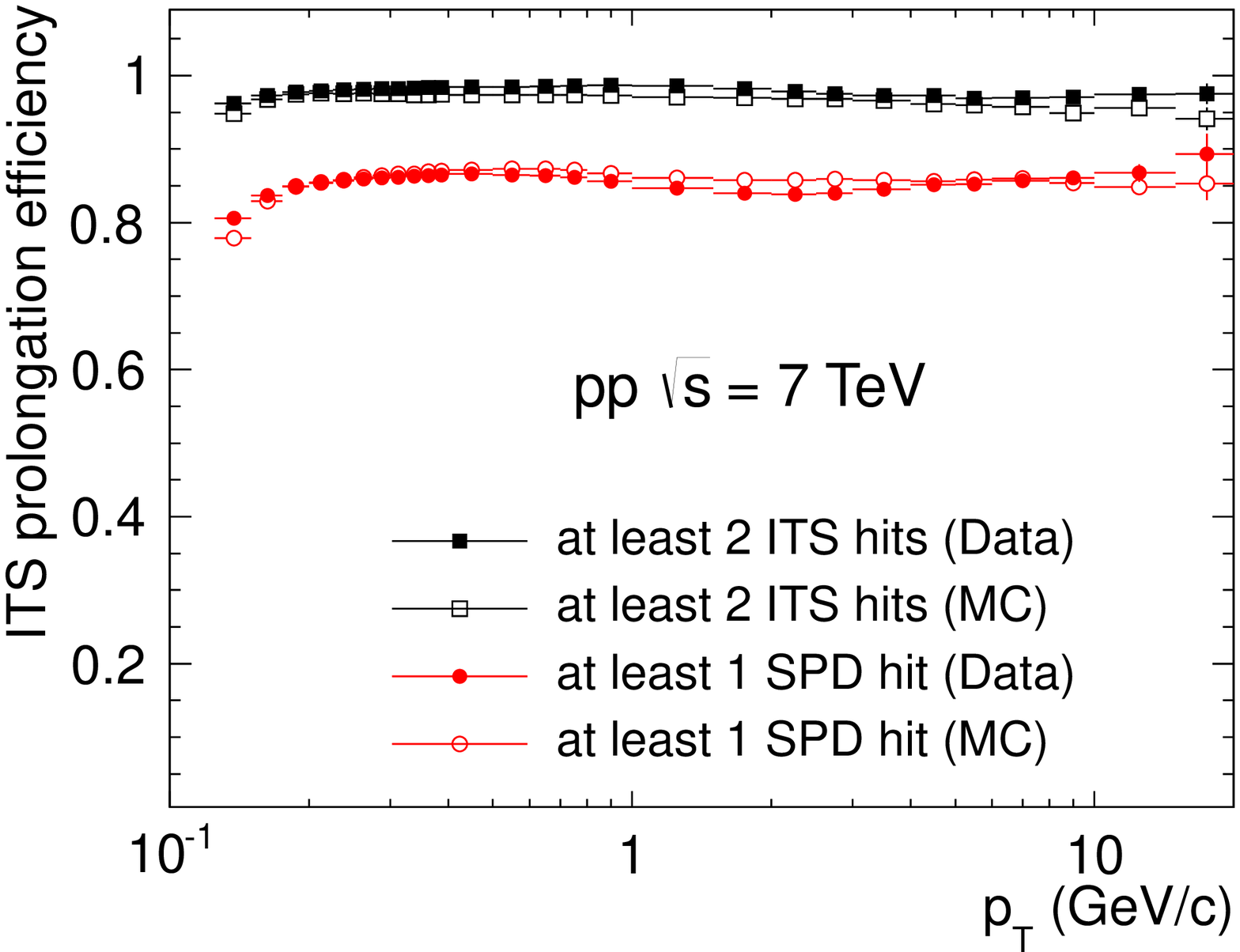}
\includegraphics[width=0.495\textwidth]{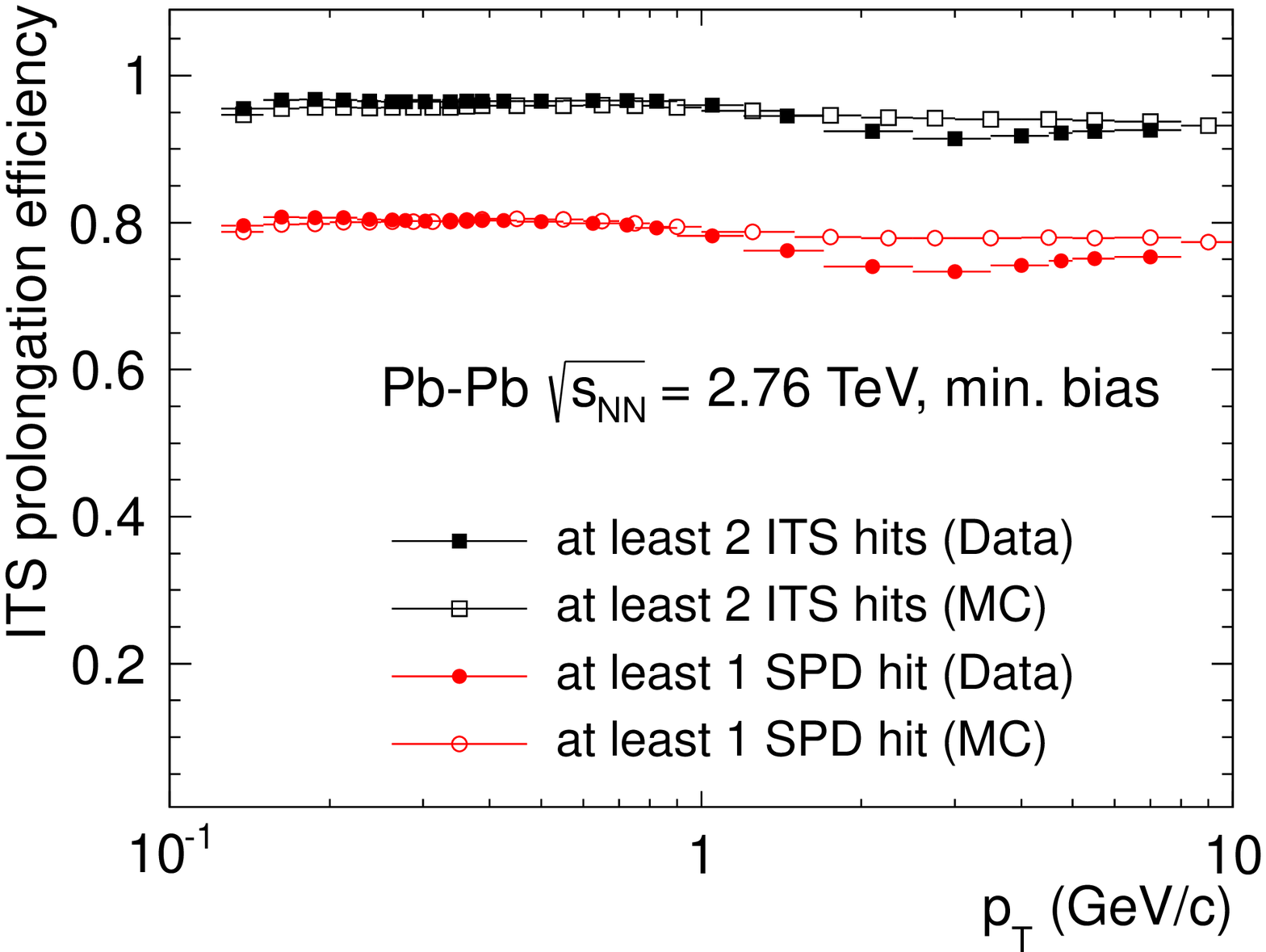}
\caption{\label{fig:its_tpc_matching} ITS--TPC matching efficiency
  vs. \pt for data and Monte Carlo for \pp (left) and \pbpb (right)
collisions.}
\end{figure}

The reconstructed TPC tracks are then propagated to the outermost ITS layer and become 
the seeds for track finding in the ITS. 
The seeds are propagated inward and are updated at each ITS layer by all 
clusters within a proximity cut, which takes into account positions and errors. 
The result of each update is saved as a new seed. 
In order to account for the detection inefficiency, seeds
without an update at a given layer are also used for further track finding. 
The $\chi^{2}$ of such seeds is increased by a penalty factor for a missing 
cluster (unless the seed extrapolation happened to be in the dead zone of the 
layer, in which case no cluster should be expected). 
Thus, each TPC track produces a tree of track hypotheses in the ITS. 
As is the case in the TPC, this seeding procedure is
performed in two passes, with and without vertex constraint. 
Once the complete tree of prolongation candidates
for the TPC track is built, the candidates are sorted according
to the reduced $\chi^{2}$. 
The candidates with the highest quality from each tree are checked for 
cluster sharing among each other. If shared clusters are found, 
an attempt is made to find alternative candidates 
in the involved trees. In the case of a failure to completely resolve the
conflict between two tracks, the worse of the two acquires a
special flag for containing potentially incorrectly matched (``fake'') 
clusters. Finally, the
highest quality candidate from each hypothesis tree is added to the reconstructed event. 
Figure~\ref{fig:its_tpc_matching} 
shows the TPC track prolongation efficiency to ITS in \pp and \pbpb collisions 
as a function of track transverse momentum, with different requirements of ITS 
layer contributions. The data and Monte Carlo (MC) efficiencies are shown by solid 
and open symbols, respectively. The fraction of tracks with at least one fake 
cluster in the ITS in the most central \pbpb collisions reaches $\sim$30$\%$ at 
$\pt<~0.2$\gevc, decreases to $\sim7\%$ at 1\gevc, and drops below 2\% at 10\gevc.

As one can see in Fig.~\ref{fig:tpcabseff}, the reconstruction efficiency in the TPC 
sharply drops at low transverse momentum. The cutoff is around 200\mevc for pions 
and 400\mevc for protons, and is caused by energy loss and multiple scattering in the 
detector material. 
For this reason, a standalone ITS 
reconstruction is performed with those clusters that were not used 
in the ITS--TPC tracks. The helical seeds are built using two 
clusters from the three innermost ITS layers and the
primary vertex point. Each such seed is propagated to the other
layers and updated with clusters within a proximity cut. 
Each matching cluster increments the number of
seed-completion hypotheses. For the final step of seed processing, all of the
hypotheses are refitted by a Kalman filter and the track with the best fit
$\chi^{2}$ is accepted, with its clusters being removed from 
further searches. In order
to increase the efficiency of tracking, the whole procedure is
repeated a few times, gradually opening the 
seed completion road widths. This algorithm enables 
the tracking of particles with transverse momenta down to about 80\mevc.

Once the reconstruction in the ITS is complete, all tracks are extrapolated
to their point of closest approach to the preliminary interaction vertex,
and the outward propagation starts. 
The tracks are
refitted by the Kalman filter in the outward direction using the
clusters found at the previous stage. 
At each outward step, the track length integral, as well as the time of flight 
expected for various particle species (e, $\mu$, $\pi$, K, p), are updated for
subsequent particle identification with TOF (see Section~\ref{sect:hadrons}).
Once the track reaches the TRD ($R=290$~cm),
an attempt is made to match it with a TRD tracklet (track segment within 
a TRD layer) in each of the six TRD layers.
Similarly, the tracks reaching the TOF detector are matched to 
TOF clusters. The track length integration and time-of-flight
calculation are stopped at this stage. The tracks are then 
propagated further for matching with signals in EMCal, PHOS, and
HMPID (see Sections~\ref{sect:hadrons} and \ref{sect:electrons} 
for the performance of matching to external detectors).
The detectors at a radius larger than that of the TPC are currently not used to
update the measured track kinematics, but their information is stored 
in the track object for the purposes of particle identification.

At the final stage of the track reconstruction, all tracks are
propagated inwards starting from the outer radius of the TPC. In each
detector (TPC and ITS), the tracks are refitted with the previously
found clusters. The track's position, direction, inverse curvature, and its
associated covariance matrix are determined. 

\begin{figure}[t]
\centering
\includegraphics[width=0.7\textwidth]{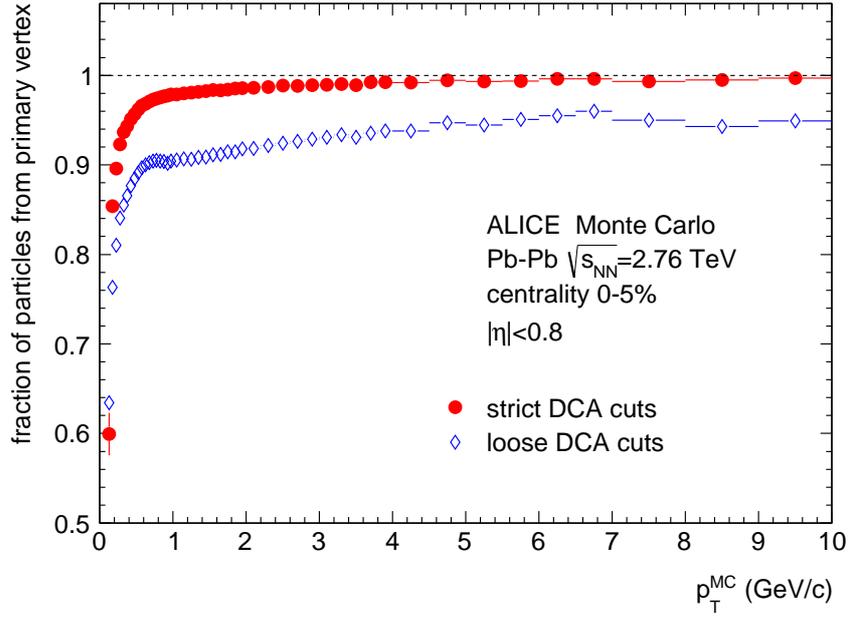}
\caption{
Fraction of reconstructed tracks coming from the primary interaction vertex. 
Two sets of cuts on the track distance of closest approach ($d_{0}$) to the primary vertex 
are shown: ``loose'' with $|{\rm d}_{0,z}|<3$~cm, ${\rm d}_{0,xy}<3$~cm and ``strict'' 
with $|{\rm d}_{0,z}|<2$~cm, ${\rm d}_{0,xy}<(0.0182+0.0350\gevc \>  \> \pt^{-1}$) cm.} 
\label{fig:contamination}
\end{figure}

The majority of tracks reconstructed with the described procedure come from 
the primary interaction vertex (Fig.~\ref{fig:contamination}). Secondary 
tracks, representing the products of decays and of secondary interactions in 
the detector material, can be further suppressed by cuts on the longitudinal 
and transverse distances of closest approach ($d_{0}$) to the primary vertex. 
The dedicated reconstruction of secondary tracks is the subject of 
Section~\ref{sect:secondary}.

\begin{figure}[b!]
\includegraphics[width=0.49\textwidth]{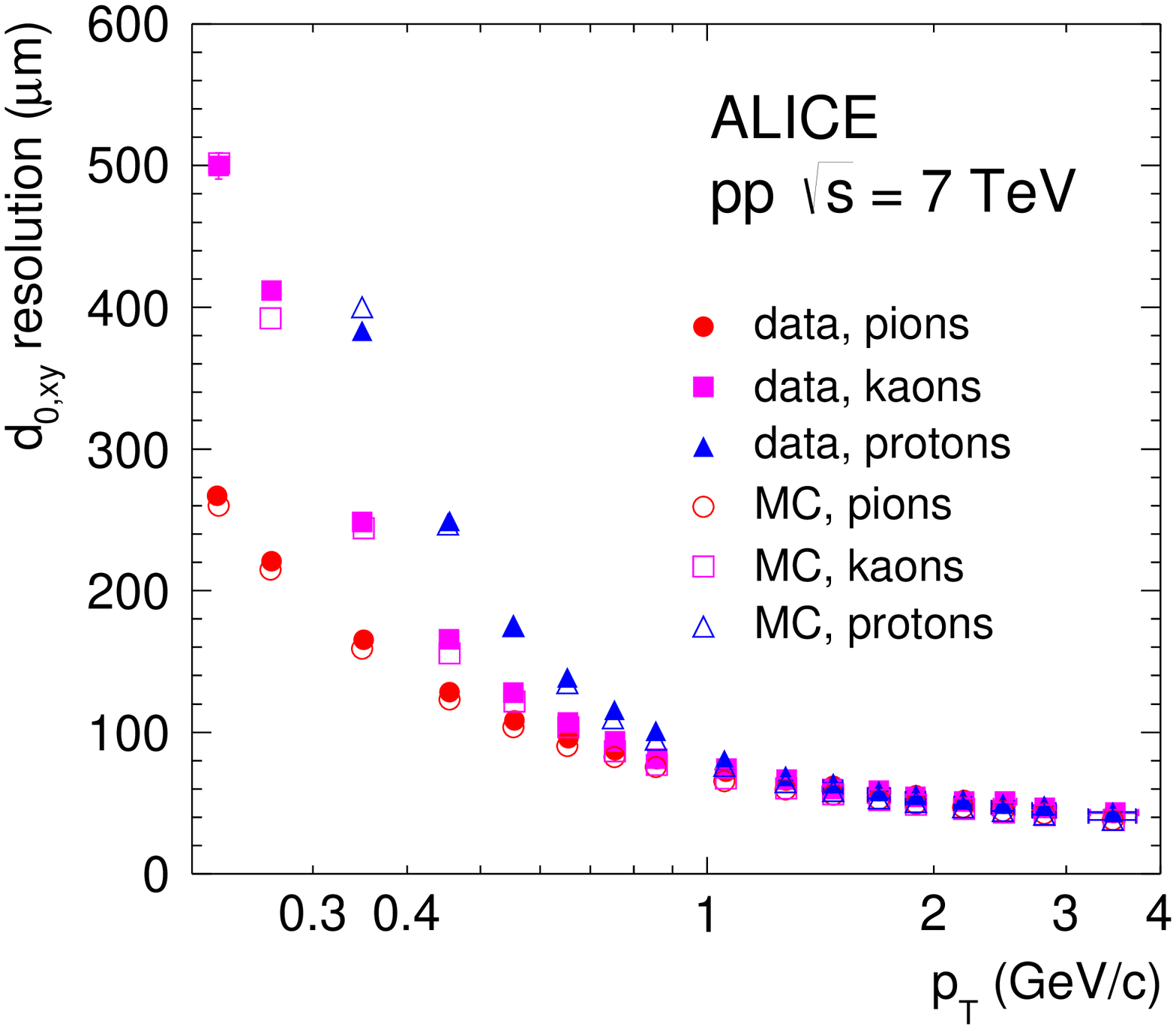}
\includegraphics[width=0.49\textwidth]{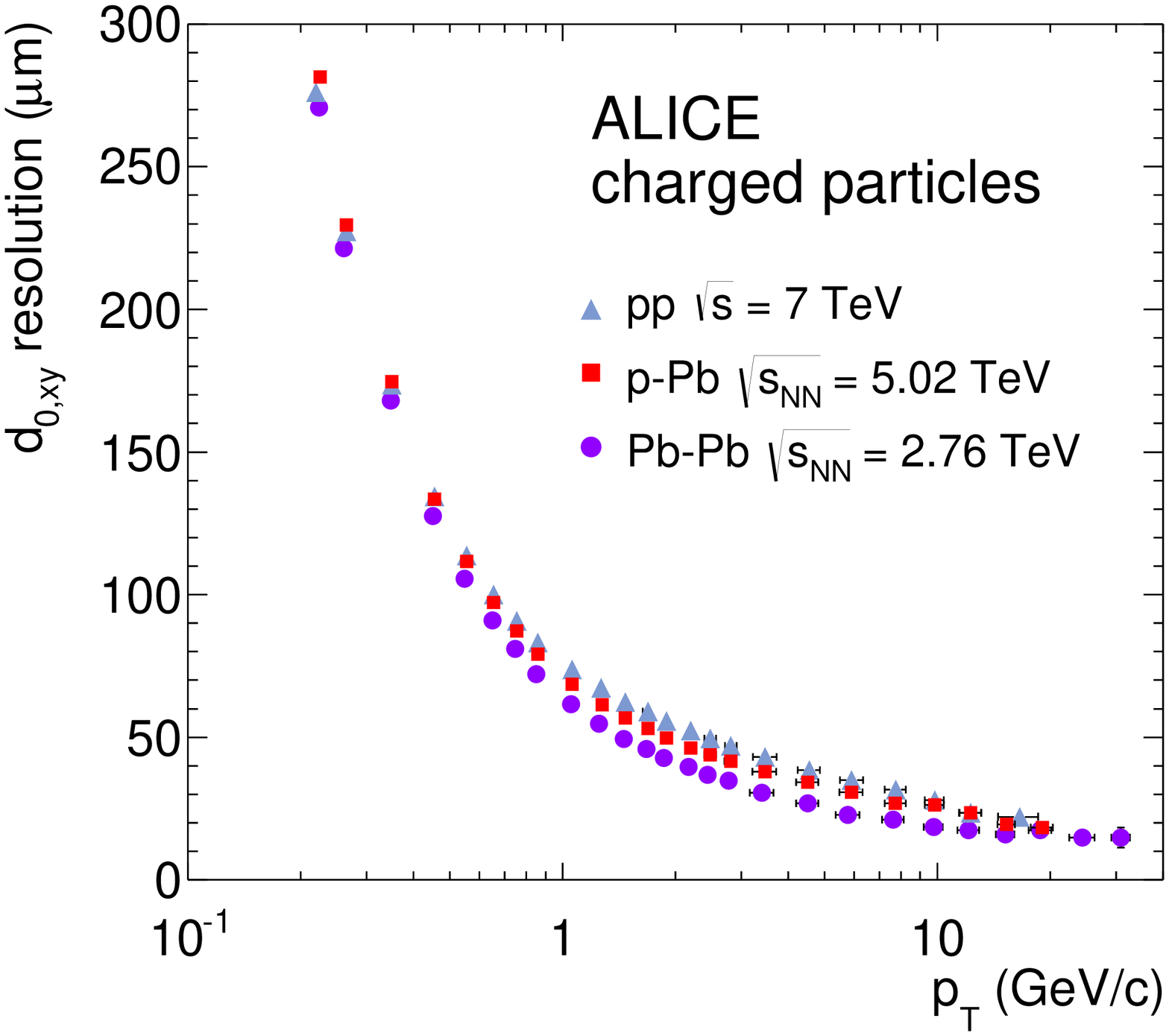}
\caption{\label{fig:d0res} 
Resolution of the transverse distance to the primary vertex for identified
particle global ITS--TPC tracks (left) and for all charged ITS--TPC tracks (right).
The contribution from the vertex resolution is not subtracted.}
\end{figure}

The left panel of Fig.~\ref{fig:d0res} shows the resolution
of the transverse distance to the primary vertex for identified ITS--TPC tracks in
\pp collisions, compared with simulation. The contribution from the vertex resolution
is not subtracted. The right panel of Fig.~\ref{fig:d0res} shows the same quantity
for all charged particle tracks for three colliding systems and with a higher \pt reach.
One can notice an improvement of the resolution in heavier systems thanks to the more precisely 
determined vertex for higher multiplicities. 

\begin{figure}[h]
\centering
\includegraphics[width=0.70\textwidth]{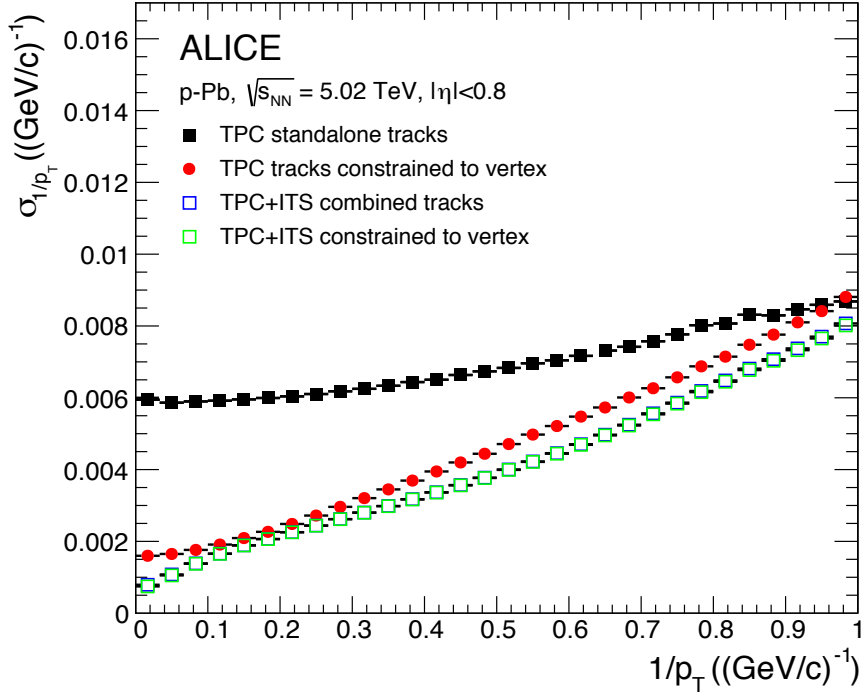}
\caption{\label{fig:ptres} 
The \pt resolution for standalone TPC and ITS--TPC matched tracks with and without 
constraint to the vertex. The vertex constrain significantly improves the resolution of TPC 
standalone tracks. For ITS--TPC tracks, it has no effect (green and blue squares overlap).
}
\end{figure}

The transverse momentum resolution for TPC standalone tracks and ITS--TPC combined tracks, 
extracted from the track covariance matrix, 
is shown in Fig.~\ref{fig:ptres}. 
The effect of constraining the tracks to the primary vertex is shown as well.
The inverse-\pt resolution, plotted in this figure, is connected to the relative 
transverse momentum resolution via 
\begin{equation}
  \frac{\sigma_{\pt}}{\pt} = \pt \ \sigma_{1/\pt} \ .
\end{equation}
The plot represents the most advanced reconstruction scheme that was applied to the 
data taken in the recent \ppb run. In central \pbpb collisions, the \pt resolution is 
expected to deteriorate by $\sim$10--15\% at high \pt due to the loss (or reduction) 
of clusters sitting on ion tails, cluster overlap, and fake clusters attached to the tracks. 

\begin{figure}[h]
\includegraphics[width=0.49\textwidth]{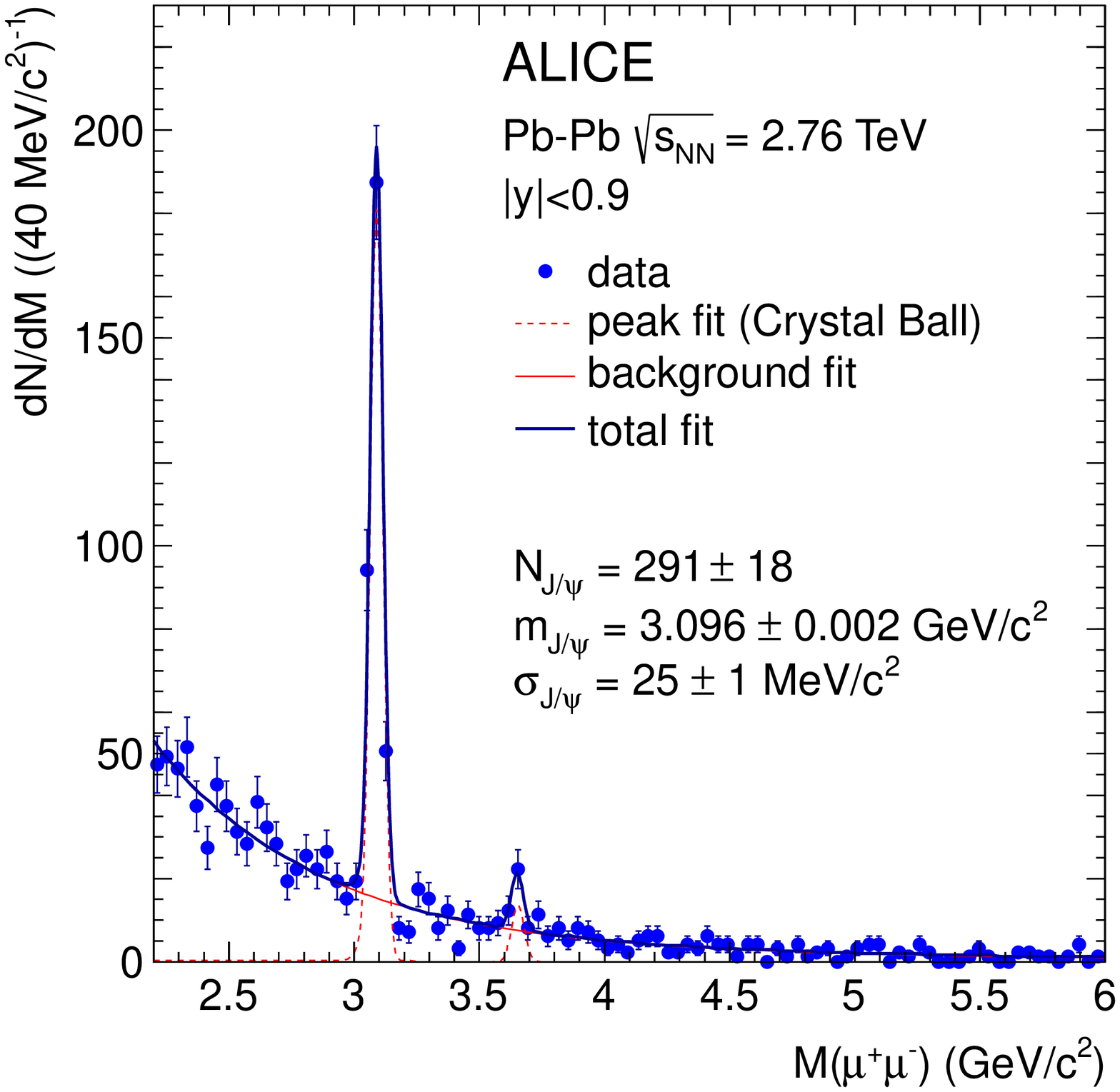}
\includegraphics[width=0.49\textwidth]{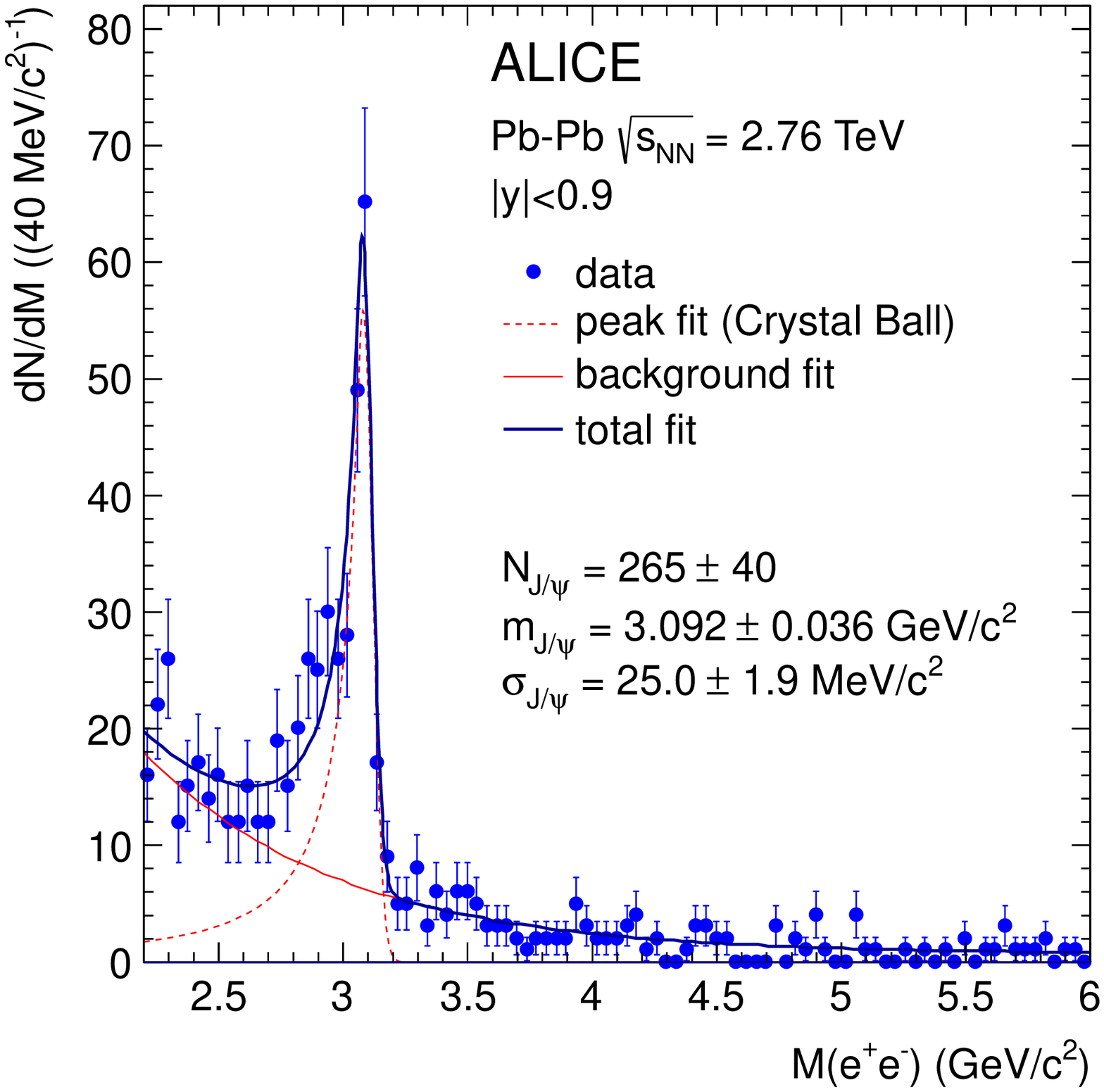}
\caption{
\label{fig:psimass} 
Invariant mass spectra of $\mu^+\mu^-$ (left) and \ee (right) pairs in
ultraperipheral \pbpb collisions. 
The solid and dotted lines represent the background (exponential) and peak 
(Crystal Ball\protect~\cite{Gaiser:1982yw}) fit components, respectively. 
The bremsstrahlung tail in the \ee spectrum is reproduced in simulation. 
The mass resolution is better than 1\%. }
\end{figure}

To demonstrate the mass resolution achievable with ITS--TPC global
tracks we show in Fig.~\ref{fig:psimass} the invariant mass spectra of 
$\mu^+\mu^-$ (left) and \ee (right) pairs measured in ultraperipheral 
\pbpb collisions at \mbox{\sqrtsnn = 2.76 TeV}. The mass resolution at the 
\jpsi peak is better than 1\%. 

Although it provides the best estimate of track parameters, the global ITS--TPC track 
reconstruction suffers from gaps in the ITS acceptance. 
In particular, in the innermost two SPD layers, up to 20\% and 30\% of the modules 
were inactive in the years 2010 and 2011, respectively. The inefficiency was 
reduced to $\sim$5\% in 2012 after solving problems with detector 
cooling. For those analyses that require a uniform
detector response, the parameters of the tracks fitted only in the TPC 
and constrained to the primary vertex can be used. 
The transverse momentum resolution of these tracks is comparable 
to that of the global tracks up to $\pt\approx 10$\gevc and significantly 
worse for higher momenta (red filled circles in Fig.~\ref{fig:ptres}). 

The ability to reconstruct pairs of close tracks is important for particle-correlation 
measurements. 
The track-separation dependent efficiency has to be either corrected for or, 
when dealing with ratios, close pairs\footnote{Two tracks that are so 
close to each other that the presence of one track affects the reconstruction 
efficiency of the other.} have to be removed in the 
numerator and denominator of the correlation function. 
In the first pion femtoscopy analysis in \pbpb collisions~\cite{Aamodt:2011mr}, 
those pairs of tracks that were separated by less than 10~mrad in $\theta$ 
and by less than 2.4~cm in $r\phi$ at a cylindrical radius of $r=1.2$~m were removed. 
This was sufficient to determine precisely the shape of the two-particle correlation 
function. 

\subsection{Final determination of the interaction vertex}

Global tracks, reconstructed in TPC and ITS, are used to find the
interaction vertex with a higher precision than with SPD tracklets alone. 
By extrapolating the tracks to the point of closest approach to the
nominal beam line and removing far outliers, the 
approximate point of closest approach of validated tracks is
determined. Then the precise vertex fit is performed using track weighting to
suppress the contribution of any remaining outliers~\cite{karimaki}.
In order to improve the transverse vertex position precision
in low-multiplicity events, the nominal beam position
is added in the fit as an independent measurement with errors 
corresponding to the transverse size of the luminous region.

\begin{figure}[t]
\includegraphics[width=0.49\textwidth]{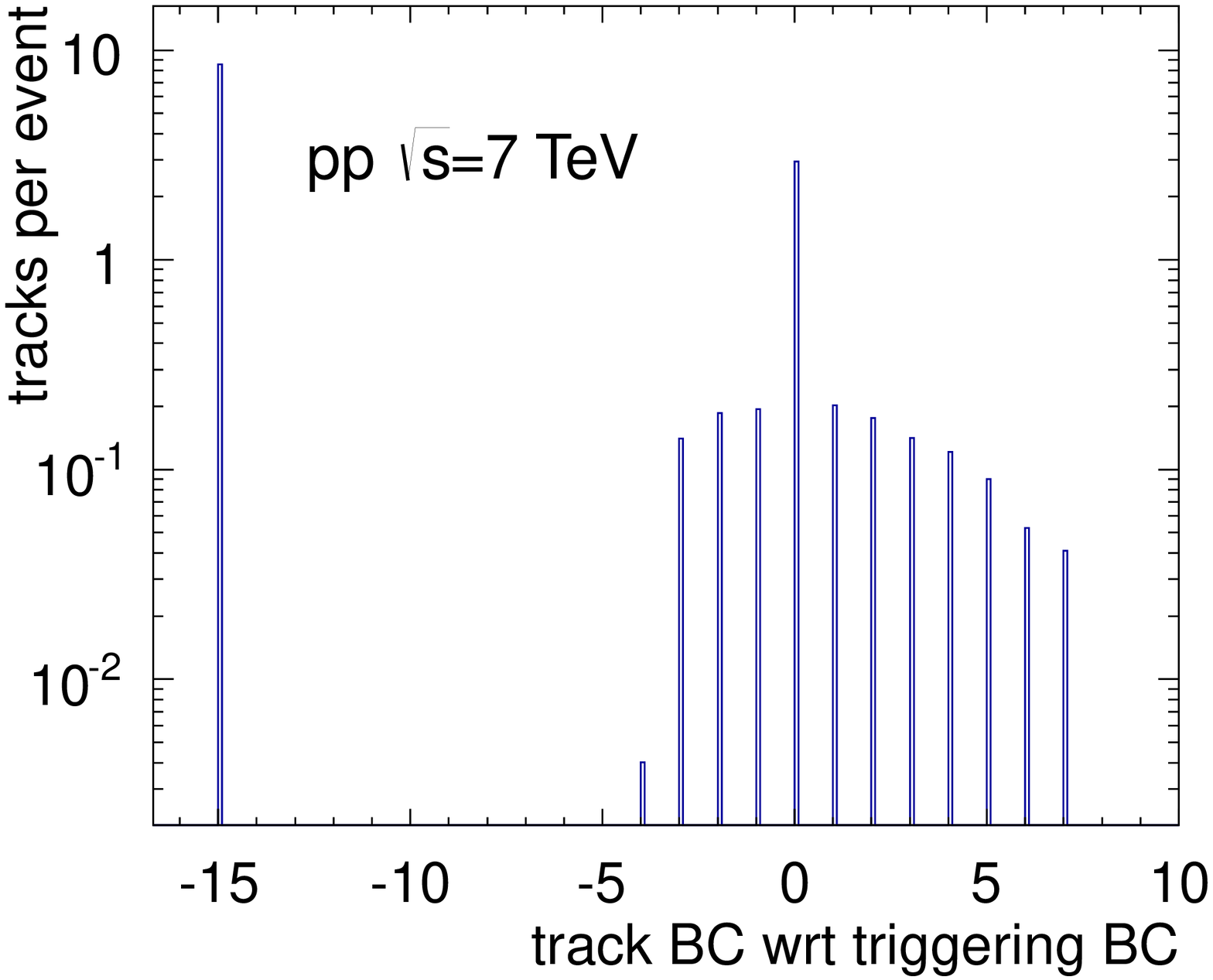}
\includegraphics[width=0.49\textwidth]{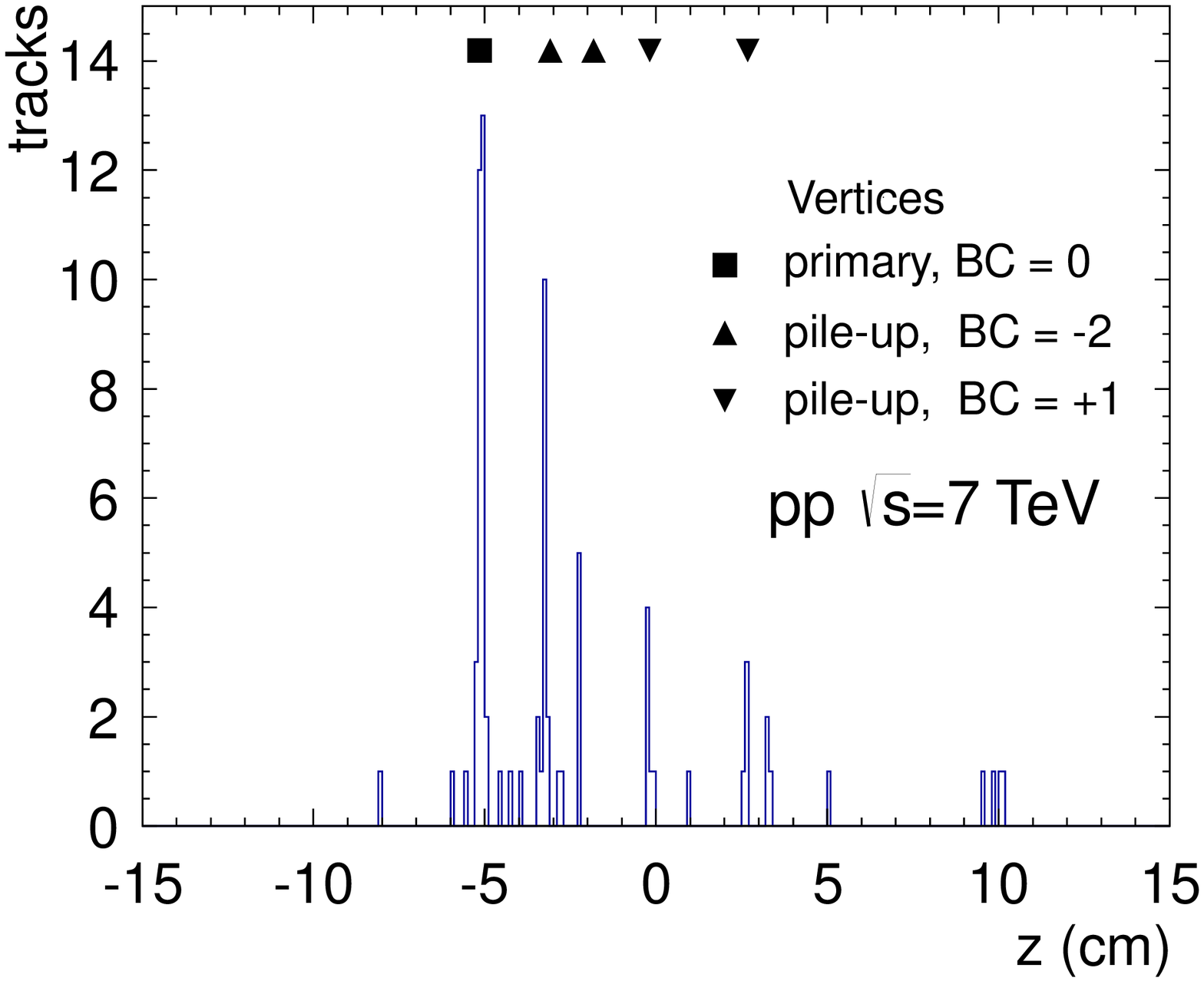}
\caption{\label{fig:plptof} Left: Bunch crossing (BC) ID of tracks 
obtained from the comparison of time of flight measured in the TOF 
detector and expected from the track kinematics. The ID is defined 
with respect to the BC in which the triggering interaction took place. 
The peak at -15 corresponds to tracks not matched in TOF (mostly from
the pileup in the TPC, outside of the TOF readout window of 500~ns). 
Right: $z$ coordinates of tracks' PCA to the beam axis in a single event 
with pileup; the positions of reconstructed vertices with attributed 
bunch crossings are shown by markers.}
\end{figure}

For data-taking conditions where a high pileup rate is expected,
a more robust version of vertex finding inspired by the algorithm 
from Ref.~\refcite{Agakishiev:1996sq} is used. It is based on
iterative vertex finding and fitting using Tukey bisquare weights~\cite{Tukey-ref} 
to suppress outliers. A scaling factor is applied to the errors on the tracks
extrapolated to the nominal beam axis and inflated until at least two
tracks with non-zero weights are found for an initial vertex
position. The fit, similar to Ref.~\refcite{karimaki} but accounting for 
these weights, is performed, and as the fitted vertex moves towards its 
true position, the scaling factor is decreased. The iterations stop when the
distance between successively fitted vertices is below 10~$\mu$m. 
If the scaling factor at this stage is still significantly larger than unity 
or the maximum number of iterations is reached, the vertex candidate is
abandoned and the search is repeated with a different seeding
position. Otherwise the final fit of the weighted tracks is done, the
vertex is validated, the tracks with non-zero weights are removed from
the pool, and the search for the next vertex in the same event is performed. 
The algorithm stops when no more vertices are found in the scan along the
beam direction. In order to reduce the probability of including tracks 
from different bunch crossings in the same vertex,
only tracks with the same or undefined bunch crossing
are allowed to contribute to the same vertex. 
Tracks are associated with bunch crossings using the time information
measured by the TOF detector.
The left plot of Fig.~\ref{fig:plptof} shows bunch-crossings assigned 
to ITS--TPC tracks in a typical high intensity \pp run. On the right an
example of a single event with identified pileup is shown. The
histogram shows the $z$ coordinate of tracks' closest
approach to the beam axis, while the positions of reconstructed vertices
with attributed bunch crossings are shown by markers.

\begin{figure}[b]
\includegraphics[width=0.48\textwidth]{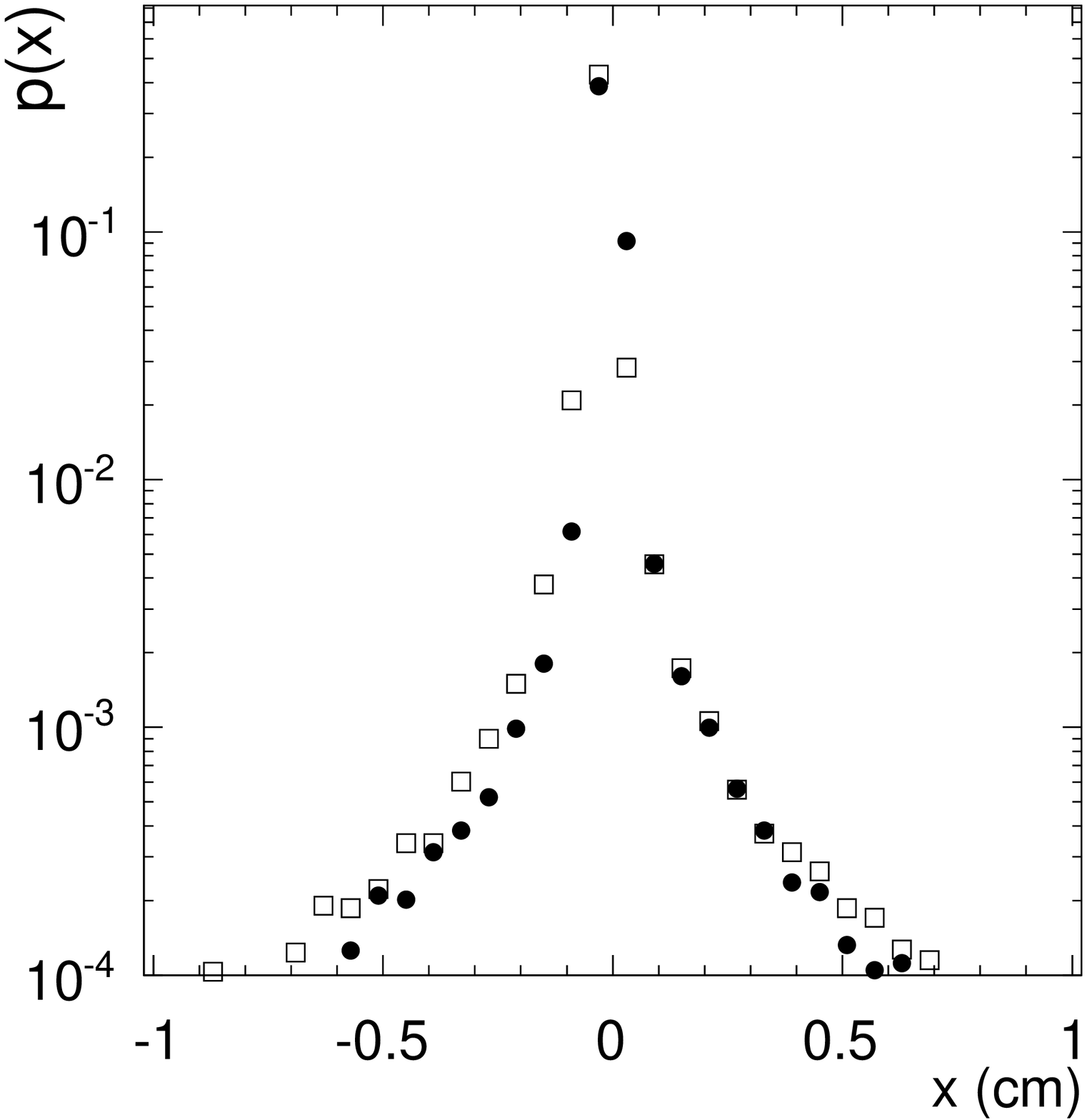}
\includegraphics[width=0.48\textwidth]{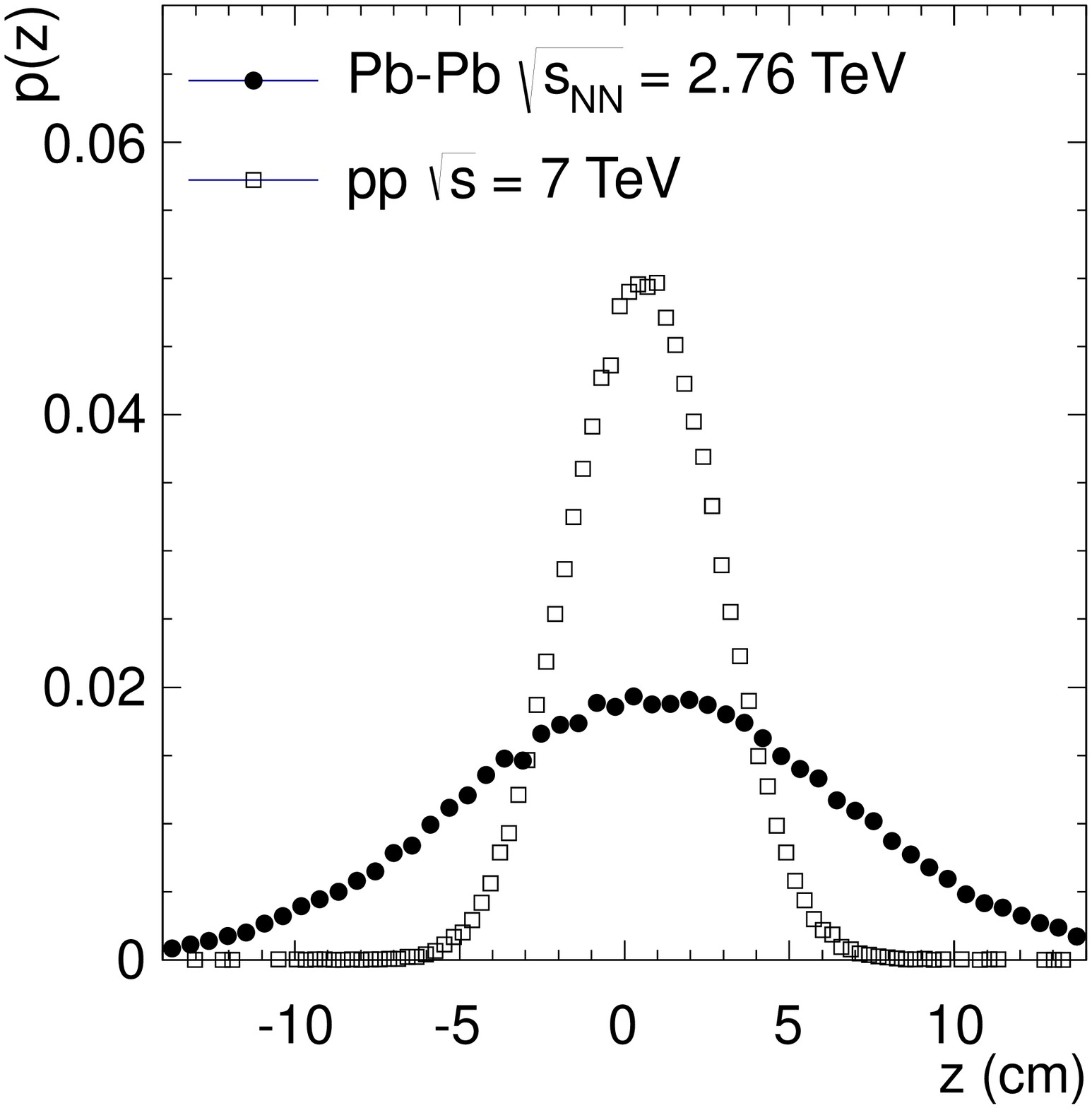}
\caption{\label{fig:vtxdist} The $x$ (left) and $z$ (right)
  projections of the luminous region obtained 
  from reconstructed vertices in \pp and \pbpb collisions (folded with vertex
  resolution).
}
\end{figure}

\begin{figure}[t]
\centering
\includegraphics[width=0.7\textwidth]{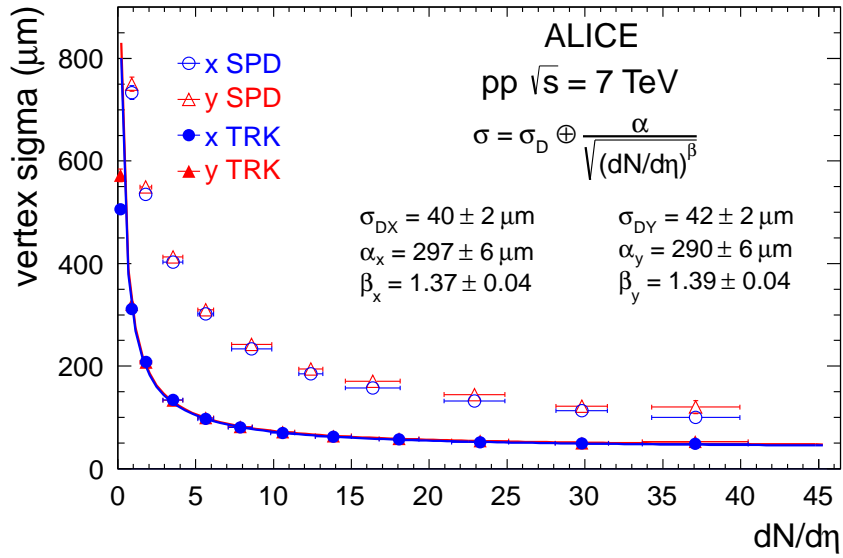}
\caption{\label{fig:vtxres} 
Transverse width of the final vertex distribution (solid points), decomposed 
into the finite size of the luminous region $\sigma_{D}$ and 
the vertex resolution $\alpha / \sqrt{(\dndeta)^\beta}$. 
For comparison, the widths of the preliminary (SPD) 
interaction vertices are shown as open points.
}
\end{figure}

Figure~\ref{fig:vtxdist} shows the $x$ (left) and $z$ (right) profiles 
of the luminous region obtained from reconstructed vertices in \pp and 
\pbpb collisions.
The transverse resolution of the preliminary interaction vertices
found with SPD (Section~\ref{sect:tracking-spd-vertex}) and of the final 
ones, found with global tracks, are shown in Fig.~\ref{fig:vtxres}. 
As expected, both resolutions scale with the square root of the number 
of contributing tracks.

\subsection{Secondary vertices}
\label{sect:secondary}

\begin{figure}[b!]
\centering
\includegraphics[scale=0.65, clip, trim = -0cm -5mm -3cm -0cm]{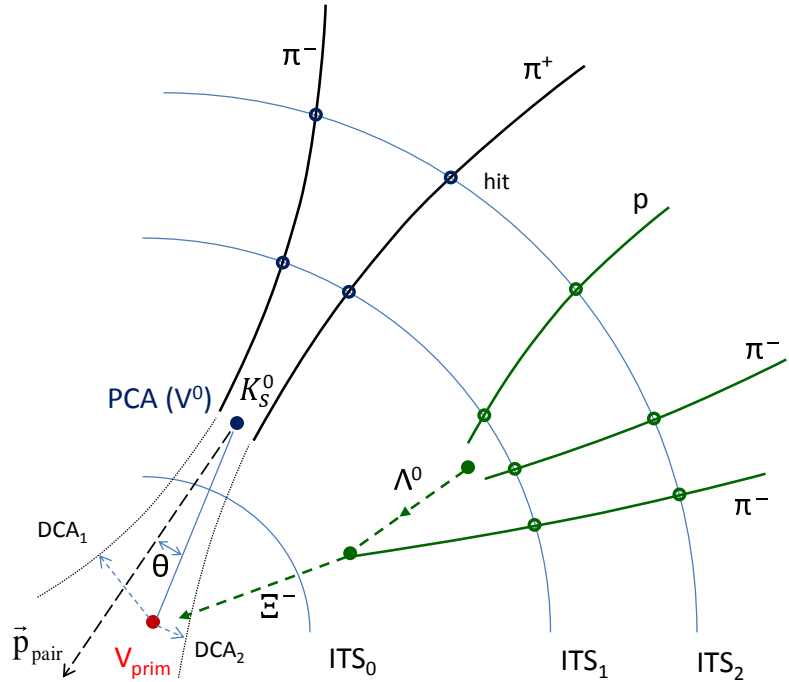}
\caption{
\label{fig:v0sketch}
Secondary vertex reconstruction principle, with \kzeros and $\Xi^{-}$ decays 
shown as an example. For clarity, the decay points were placed between the first 
two ITS layers (radii are not to scale). 
The solid lines represent the reconstructed charged particle tracks, extrapolated 
to the secondary vertex candidates. Extrapolations to the primary vertex and
auxiliary vectors are shown with dashed lines. 
}
\end{figure}

Once the tracks and the interaction vertex have been found in the course of
event reconstruction, a search for
photon conversions and secondary vertices from particle decays
is performed as shown in Fig.~\ref{fig:v0sketch}. 
Tracks with a distance of closest approach to the interaction 
vertex exceeding a certain minimum value (0.5 mm in \pp and 1 mm in \pbpb) 
are selected. 
For each unlike-sign pair of such tracks (called V$^0$ candidate) the 
point of closest approach between the two tracks is calculated. 
The V$^0$ candidates are then subjected to further cuts: 
(i) the distance between the two tracks at their PCA is requested to be 
less than 1.5~cm; 
(ii) the PCA is requested to be closer to the interaction vertex than the
innermost hit of either of the two tracks; 
(iii) the cosine of the angle~$\theta$ between the total momentum vector of the
pair $\vec p_{\rm pair}$ and the straight line connecting the primary (interaction) 
and secondary vertices must exceed 0.9.
For V$^0$ candidates with a momentum below 1.5\gevc, the latter cut is relaxed. 
This facilitates the subsequent search for cascade decays.

\begin{figure}[h]
\centering
\includegraphics[width=\textwidth]{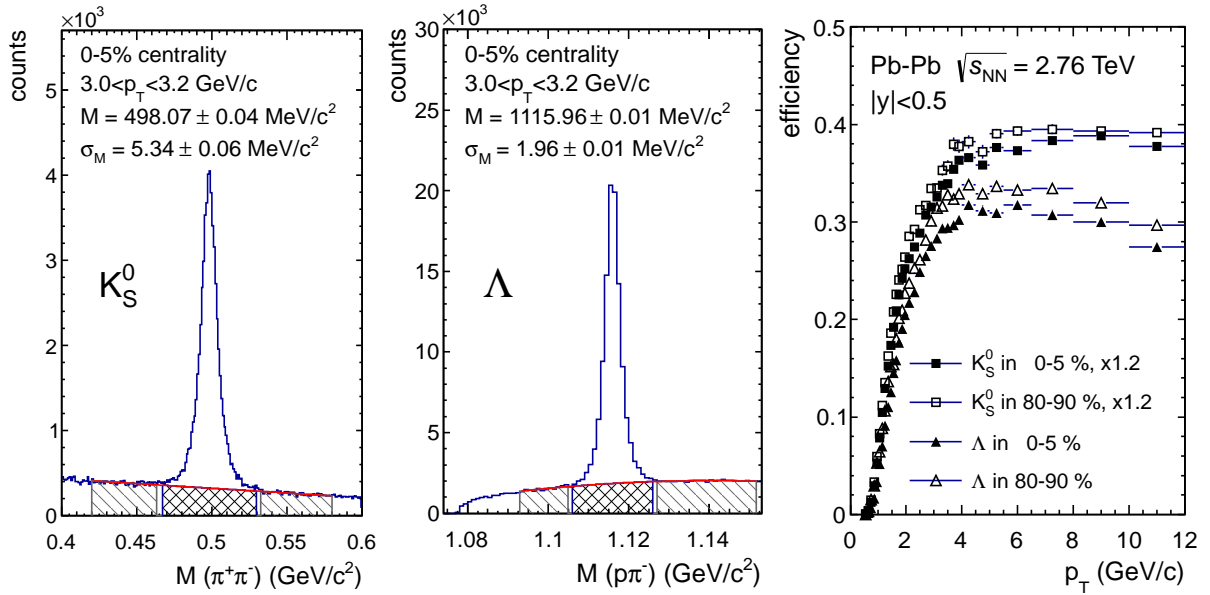}
\caption{\label{fig:LKmass} 
Invariant mass distributions of \pip\!\!\pim (left panel) and p\pim (middle panel) 
pairs in central \pbpb collisions at \mbox{\sqrtsnn = 2.76 TeV}. The hatched areas 
show the regions of the \kzeros and $\Lambda$ peaks and of the combinatorial background. 
The right-hand panel shows the reconstruction efficiencies (including the candidate 
selection cuts) as a function of transverse momentum for central (0--5$\%$) and 
peripheral (80--90$\%$) collisions.}
\end{figure}

\begin{figure}[t]
\centering
\includegraphics[width=0.75\textwidth]{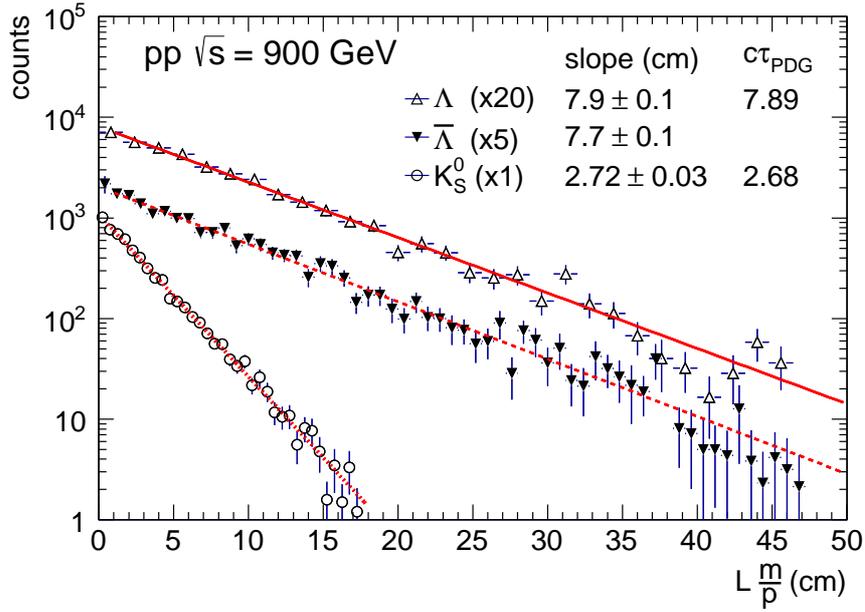}
\caption{\label{fig:ctau} 
Distance of the $\Lambda$, $\bar{\Lambda}$, and \kzeros decay
vertex from the interaction vertex, scaled by $p/m$. The slopes of the distributions
are consistent with the known lifetimes.
}
\end{figure}

Figure~\ref{fig:LKmass} shows \kzeros (left) and $\Lambda$ (center) peaks 
obtained in central \pbpb collisions. 
Proton daughters of $\Lambda$ with $\pt<1.5\gevc$ were identified by 
their energy loss in the TPC gas (see Section~\ref{sect:hadrons}).
The right plot shows the \kzeros and $\Lambda$ reconstruction efficiencies 
in central and peripheral collisions as a function of their \pt. 
The drop in $\Lambda$ reconstruction efficiency at high \pt is due to the smaller probability of decay
in the fiducial volume ($r<100$~cm) of the V$^0$ search at higher momenta.
The distributions of decay point distances from the interaction vertex agree, 
after correcting for the acceptance and efficiency, with the expectations based 
on the known lifetime of the hyperons and neutral kaons (Fig.~\ref{fig:ctau}).

After finding V$^0$ candidates, the search for the cascade ($\Xi^-$) decays is
performed as shown schematically in Fig.~\ref{fig:v0sketch}. 
V$^0$ candidates with an invariant mass in the vicinity of the $\Lambda$ 
are matched with a secondary track by cutting 
on their mutual distance at the PCA and requesting
that the latter is outside of a cylindrical volume around the interaction 
vertex ($r>0.2$~cm).

\begin{figure}[b!]
\centering
\includegraphics[width=0.6\textwidth]{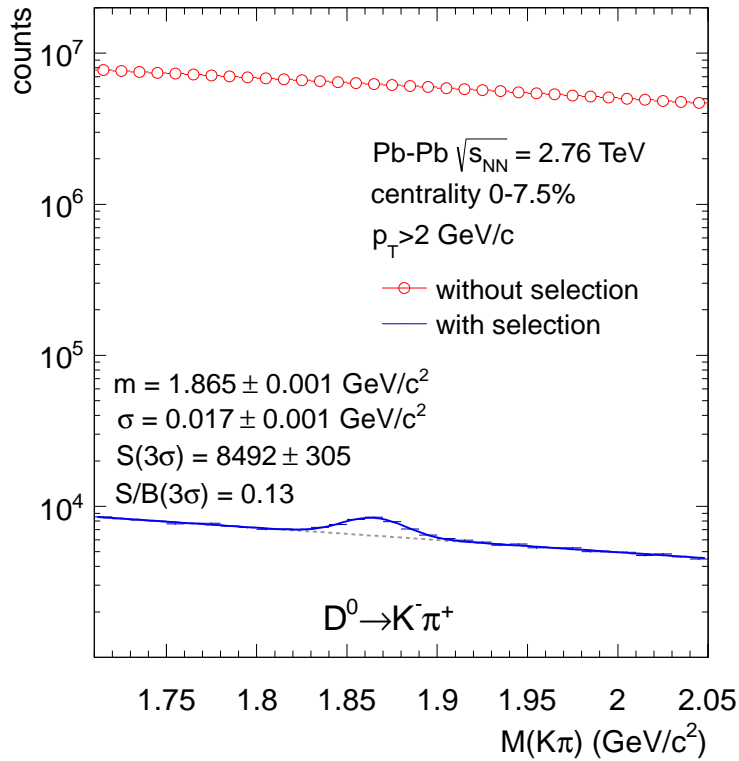}
\caption{\label{fig:d0cteff} 
Invariant mass distribution of \kam$\!\!$\pip pairs before (symbols) and after (line) 
selection cuts on the relation between the secondary (D$^{0}$ decay) and primary vertices. 
The extracted D$^{0}$ mass and its resolution as well as the significance are shown after 
selection.}
\end{figure}

The reconstruction of more complex secondary vertices is performed later, at 
the analysis stage. 
For the study of heavy-flavor decays
close to the interaction point, the secondary vertex is searched for 
by considering all unlike-sign track pairs
and selecting those passing a set of topological cuts~\cite{Abelev:2012dme}. 
In particular, the strongest improvement of the signal-to-background ratio is achieved
by cuts on the significance of the projection of the decay length in the transverse plane
$L_{xy}/\sigma_{L_{xy}}>$~7 and on the transverse pointing angle
$\cos(\theta_{xy})>$~0.998. 
$L_{xy}$ is defined as 
$\left(\vec{u}^{T} S^{-1} \vec{r}\right) / \left(\vec{u}^{T} S^{-1}
\vec{u}\right)$, where $\vec{r}$ is the vector connecting the decay
and primary vertices, $\vec{u}$ is the unit vector in direction of the 
decaying particle, and $S^{-1}$ is the inverse of the sum of the covariance
matrices of the primary and secondary vertices.
The effect of the described cuts is illustrated in Fig.~\ref{fig:d0cteff} 
which shows the resulting suppression of the combinatorial
background in the analysis of \mbox{D$^0\rightarrow$\kam\pip}.

\begin{figure}[ht]
\centering
\includegraphics[width=\textwidth]{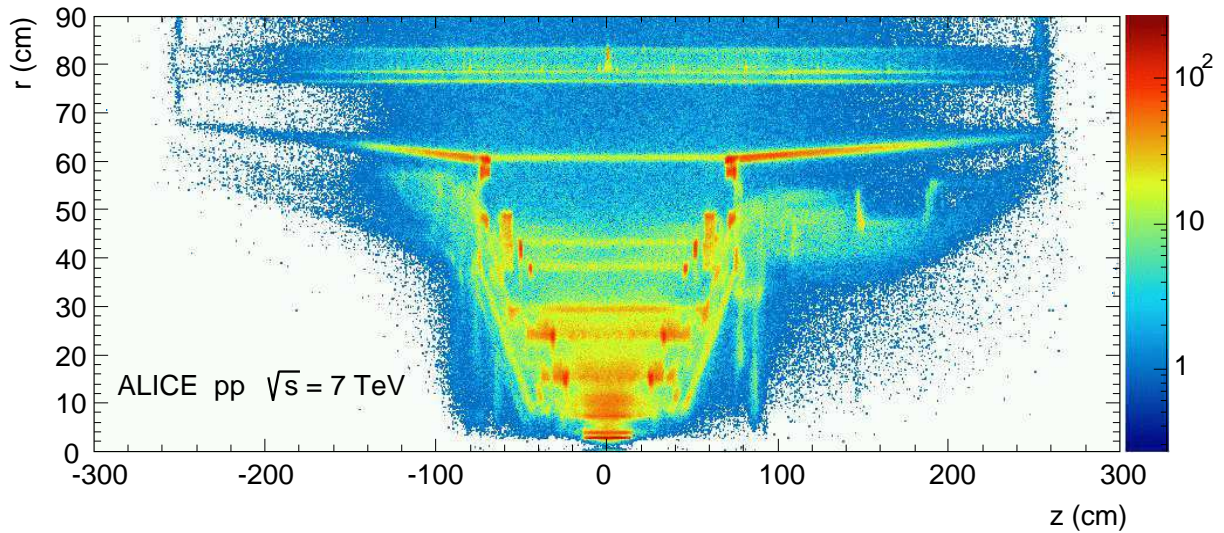}
\caption{\label{fig:tomography} 
Distribution of secondary vertices from hadronic interactions in the
ALICE material. 
The ITS layers ($r<50$~cm), the inner TPC containment vessel (60~cm~$<r<$~70~cm), 
and the inner TPC field cage ($r\sim80$~cm) are visible. 
}
\end{figure}

The implementation of the geometry and material distribution of the detectors 
in the simulation and reconstruction software is verified by comparing the 
distributions of reconstructed hadronic interaction vertices to simulations.
The hadronic interaction vertices are found at the analysis level by 
identifying groups of two or more tracks originating from a common secondary vertex.
For these, none of the track pairs should have an invariant mass of $\gamma$, \kzeros, or $\Lambda$. 
Figure~\ref{fig:tomography} shows the $r$--$z$ distribution of such vertices 
representing the material of the apparatus.

\clearpage\newpage\section{Hadron identification}
\label{sect:hadrons}

The ALICE detector has a number of different subsystems for identifying
charged hadrons and electrons. 
The following subsystems are used for hadron identification:
\begin{itemize}
\item {\it ITS:} The outer four layers of the Inner Tracking System
  have an analog readout to measure the deposited charge, thereby providing
  a \dedx measurement. This is mainly useful for low-\pt tracks
  ($\pt \lesssim 0.7$\gevc), specifically at very low \pt, where the
  ITS is used for standalone tracking.
\item {\it TPC:} The Time Projection Chamber measures the charge
  deposited on up to 159 padrows. A truncated mean \dedx (40\% 
  highest-charge clusters discarded) is calculated and used
  for a wide range of momenta. The largest separation is achieved at low \pt 
  ($\pt \lesssim 0.7$\gevc) but a good separation is also present in the relativistic 
  rise region ($\pt \gtrsim 2$\gevc) up to $\sim$20\gevc. 
\item {\it TOF:} The Time-Of-Flight detector is a dedicated detector
  for particle identification that measures the arrival time of
  particles with a resolution of $\sim$80 ps. This provides a good
  separation of kaons and protons up to $\pt \simeq 4$\gevc.
\item {\it HMPID:} The High Momentum Particle Identification
  Detector is a ring-imaging Cherenkov detector that 
  covers $|\eta|<0.6$ in pseudorapidity and 57.6$\degr$ 
  in azimuth, corresponding to 5\% acceptance of the central barrel, and provides proton/kaon
  separation up to $\pt \simeq 5$\gevc.
\end{itemize}
The measurements in the different particle identification detector
systems are then combined to further improve the separation between particle species.
This is discussed in Sections~\ref{sec:pidovw} and ~\ref{sec:pidphysan}.

The particle identification (PID) capabilities of these detectors are used 
for a wide range of physics analyses, including transverse momentum 
spectra for pions, kaons, and protons~\cite{Aamodt:2011zj,Abelev:2012wca},~\cite{Abelev:2013sm}; 
heavy-flavor decays~\cite{Abelev:2012dme}; Bose-Einstein correlations for 
pions~\cite{Aamodt:2010jj,Aamodt:2011kd,Aamodt:2011mr} and 
kaons~\cite{Abelev:2012ms,Abelev:2013fmk}; and resonance studies~\cite{Abelev:2012res}. 
The hadron identification systems is also used to identify
electrons. In addition, the calorimeters (PHOS and EMCal) and the
Transition Radiation Detector (TRD) provide dedicated electron
identification, which will be discussed in Section~\ref{sect:electrons}.

\subsection{Particle identification in the ITS}

\begin{figure}[htb]
\centering
\includegraphics[width=0.7\textwidth]{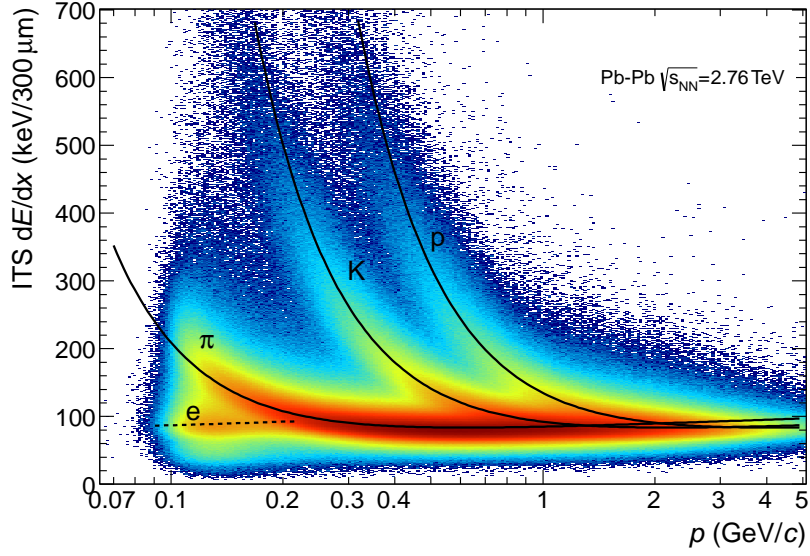}
\caption{\label{fig:its_nsig} Distribution of the energy-loss signal in
 the ITS as a function of momentum. Both the energy loss and momentum were measured by the ITS alone.
}
\end{figure}

The inner tracking system (ITS) of ALICE consists of six layers of
silicon detectors. The outer four layers provide a measurement of the
ionization energy loss of particles as they pass through the
detector. The measured cluster charge is normalized to the path
length, which is calculated from the reconstructed track parameters to
obtain a \dedx value for each layer.  For each track, the \dedx is
calculated using a truncated mean: the average of the lowest two
points if four points are measured, or a weighted sum of the lowest
(weight 1) and the second-lowest points (weight 1/2), if only three
points are measured.
An example distribution of measured truncated mean energy loss 
values as a function of momentum in the ITS is shown in Fig.~\ref{fig:its_nsig}.

\subsection{Particle identification in the TPC}
\begin{figure}[b]
\centering
\includegraphics[width=0.7\textwidth]{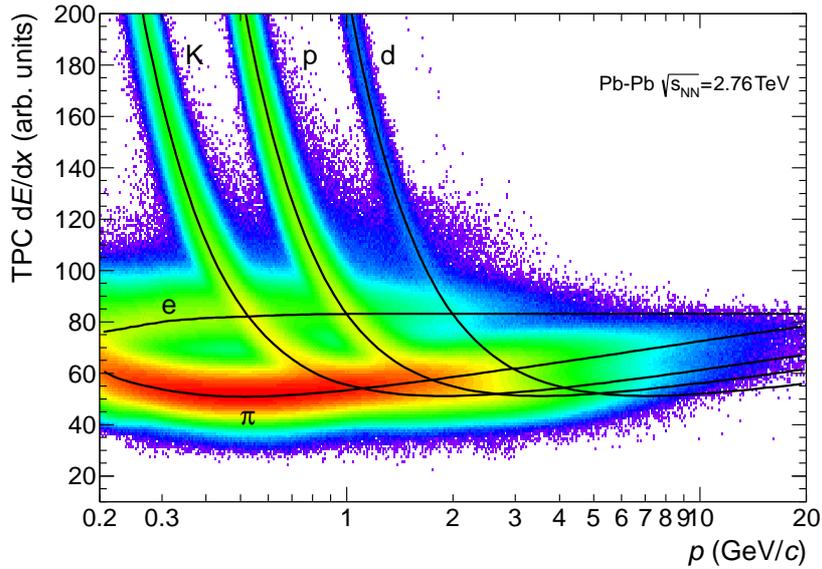}
\caption{Specific energy loss (\dedx) in the TPC vs. particle momentum 
in \pbpb collisions at \mbox{\sqrtsnn = 2.76 TeV}. The lines show the parametrizations 
of the expected mean energy loss.}
\label{fig:TPC_dEdx}
\end{figure}

The TPC~\cite{Alme:2010ke} is the main tracking detector in ALICE. 
In addition it provides information for particle identification over 
a wide momentum range. Particle identification is performed by 
simultaneously measuring the specific energy loss (\dedx), charge, 
and momentum of each particle traversing the detector gas.
The energy loss, described by the Bethe-Bloch formula, is parametrized 
by a function originally proposed by the ALEPH collaboration~\cite{blumrolandi},
\begin{equation}
f(\beta \gamma) = {P_{1} \over \beta^{P_{4}}} \Bigl(P_{2} - \beta^{P_{4}} - \ln(P_{3} + {1 \over (\beta \gamma)^{P_{5}}})\Bigr)   \;,
\end{equation}
where $\beta$ is the particle velocity, $\gamma$ is the Lorentz factor, 
and $P_{1-5}$ are fit parameters. Figure~\ref{fig:TPC_dEdx} shows the measured \dedx 
vs. particle momentum in the TPC, demonstrating the clear separation between the 
different particle species. The lines correspond to the parametrization.
While at low momenta (\mbox{$p \lesssim 1$\gevc}) particles can be identified on 
a track-by-track basis, at higher momenta particles can still be separated 
on a statistical basis via multi-Gaussian fits. 
Indeed, with long tracks ($\gtrsim 130$ samples) and with the truncated-mean method 
the resulting \dedx peak shape is Gaussian down to at least 3 orders of magnitude.

In the relativistic rise region, the \dedx exhibits a nearly constant separation
for the different particle species over a wide momentum range. 
Due to a \dedx resolution of about 5.2\,\% in \pp collisions
and 6.5\,\% in the 0--5\% most central \pbpb collisions\footnote{The 
deterioration of the energy-loss resolution in high-multiplicity events is 
caused by clusters overlapping in $z$ and/or sitting on top of a signal tail from 
an earlier cluster.}, particle ratios
can be measured at a \pt of up to 20\gevc~\cite{OrtizVelasquez:2012te}. The main
limitation at the moment is statistical precision, so it is expected
that the measurement can be extended up to $\sim 50$\gevc in the future.

\begin{figure}[h]
\centering
\includegraphics[width=0.46\textwidth]{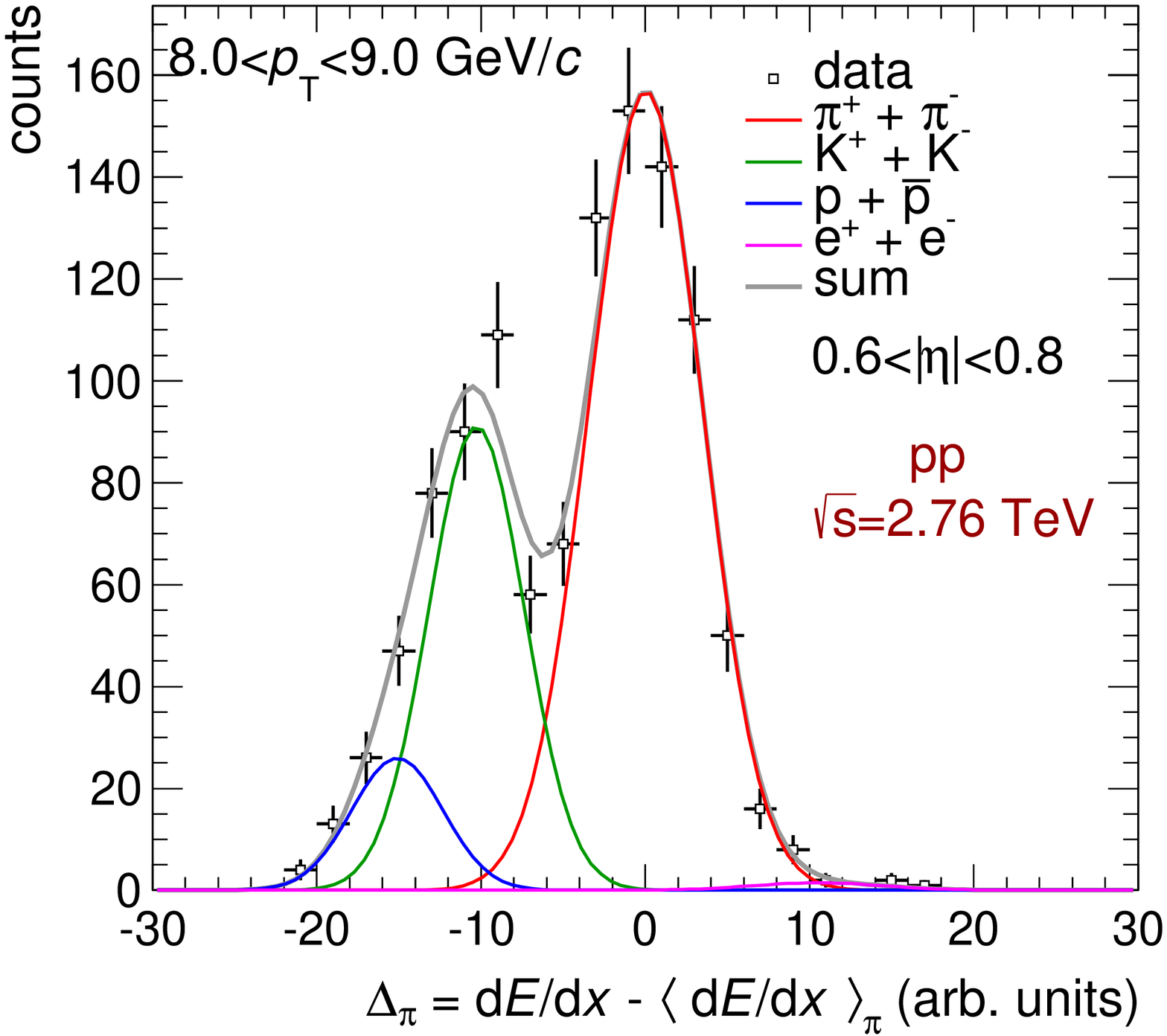}\hspace{6mm}
\includegraphics[width=0.46\textwidth]{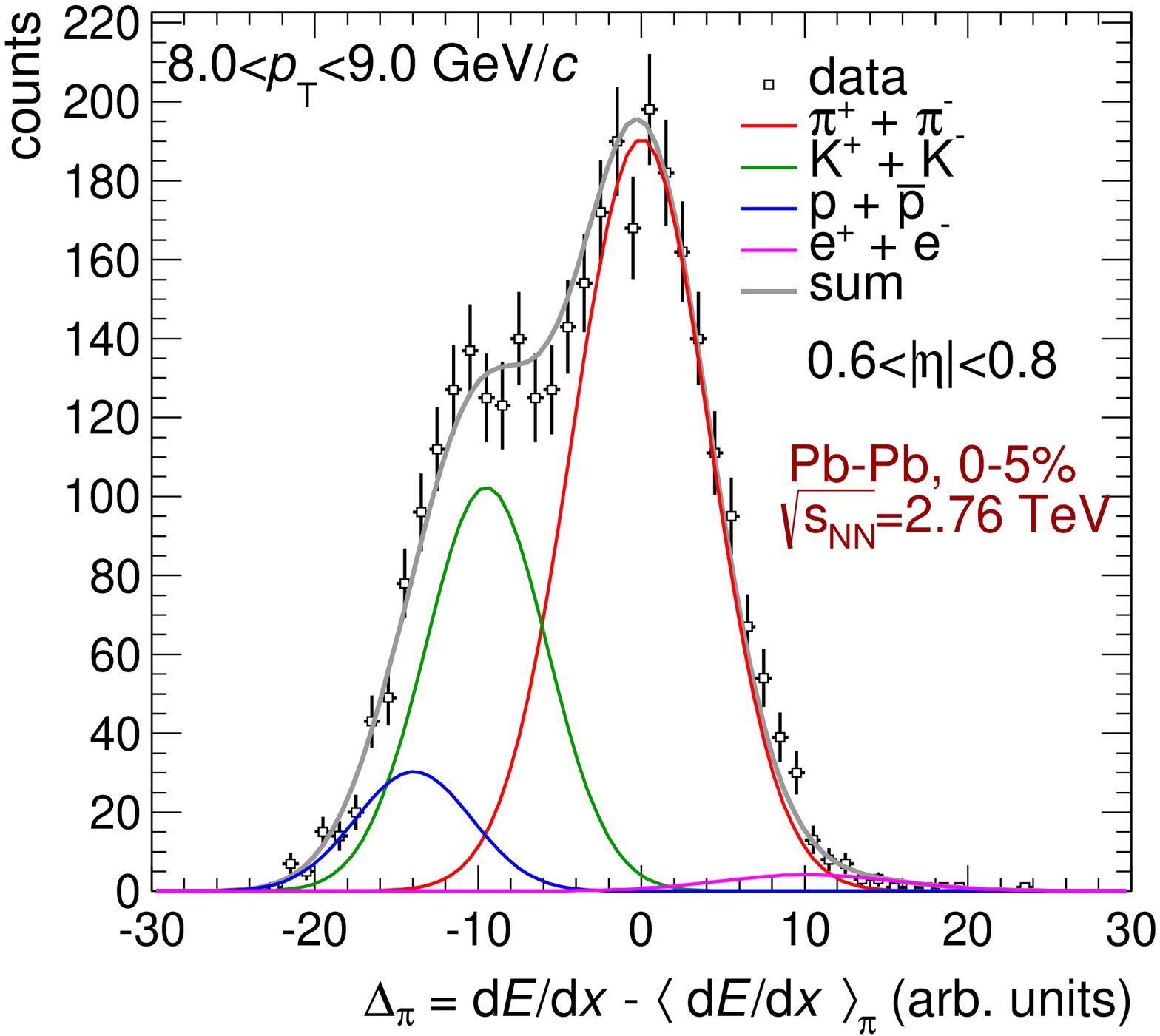}
\caption{Ionization energy loss (\dedx) distributions in the TPC in \pp (left) 
and \pbpb collisions (right) at \mbox{\sqrtsnn = 2.76 TeV}. The lines represent Gaussian 
fits as described in the main text.} 
\label{fig:TPC_multigauss}
\end{figure}

As an example, \dedx distributions for charged particles with $\pt\approx 10$\gevc are 
shown in Fig.~\ref{fig:TPC_multigauss} for \pp and the 0--5\% most central \pbpb collisions. 
Note that, for this analysis, a specific $\eta$ range was selected in order to 
achieve the best possible \dedx resolution.
The curves show Gaussian fits where the mean and width were fixed to the values obtained 
using clean samples of identified pions and protons from, respectively, \kzeros and $\Lambda$ 
decays, and assuming that the \dedx response at high \pt depends only on $\beta\gamma$. 

\subsection{Particle identification in TOF}

The Time-Of-Flight (TOF) detector~\cite{TOFdetector} of ALICE is a large
area array of Multigap Resistive Plate Chambers (MRPC), positioned at
370--399 cm from the beam axis and covering the full azimuth and the
pseudorapidity range $|\eta|<0.9$.  In \pbpb collisions, in the
centrality range 0--70\% the overall TOF resolution is 80~ps
for pions with a momentum around 1\gevc. This value includes the
intrinsic detector resolution, the contribution from electronics and
calibration, the uncertainty on the start time of the event, and the
tracking and momentum resolution~\cite{TOFperf2013}. 
TOF provides PID in the intermediate momentum range, up to 2.5\gevc 
for pions and kaons, and up to 4\gevc for protons.

\begin{figure}[hbt]
  \centering
  \includegraphics[width=0.46\textwidth]{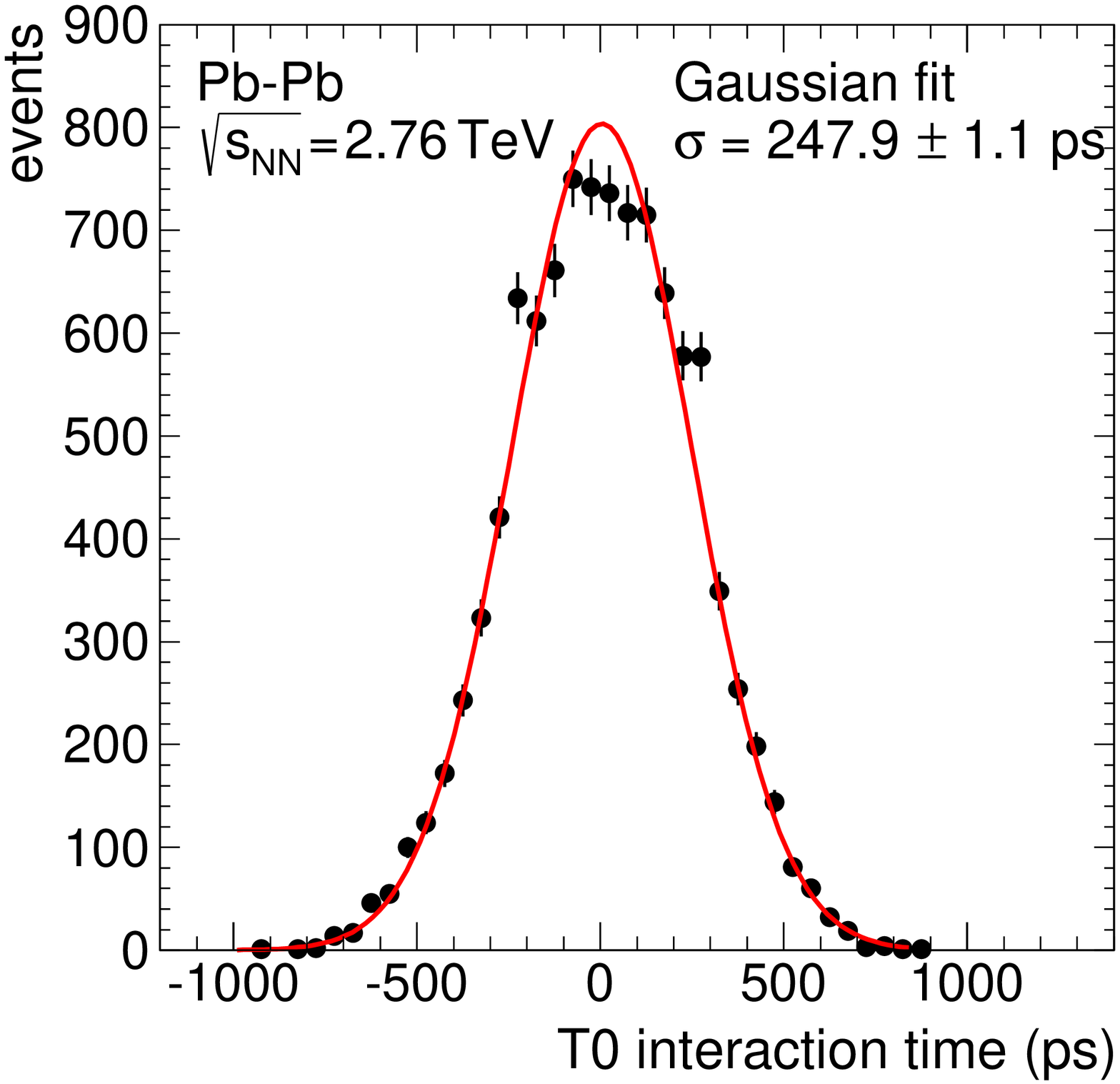}\hspace{4mm}
  \includegraphics[width=0.46\textwidth]{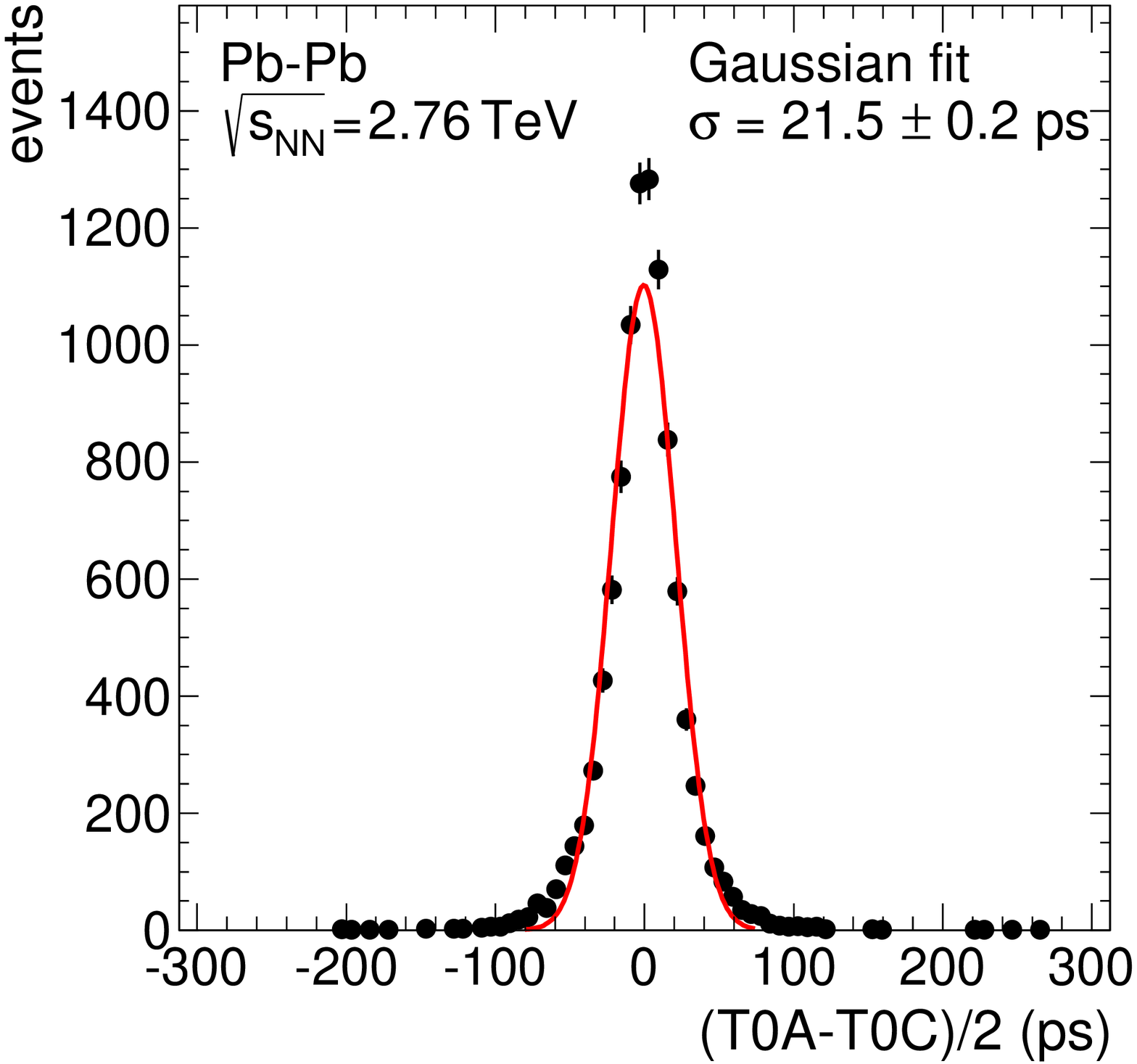}
  \caption{Interaction time of the collision with respect to the LHC clock 
measured by the T0 detector (left) and the resolution of the system 
obtained as the time difference between T0A and T0C (right). 
The time difference is corrected for the longitudinal event-vertex 
position as measured by the SPD. 
}
\label{fig:T0sumres}
\end{figure}

The start time for the TOF measurement is provided by the T0 detector, 
which consists of two arrays of Cherenkov counters T0C and
T0A, positioned at opposite sides of the interaction point
(IP) at $-3.28 < \eta < -2.97$ and $4.61 < \eta < 4.92$, respectively. 
Each array has 12 cylindrical counters equipped with a quartz
radiator and photomultiplier tube~\cite{Bondila:2005xy}.
Figure~\ref{fig:T0sumres} (left panel) shows the distribution of 
the start time (interaction time of the collision) as measured 
by the sum of the time signals from the T0A and T0C detectors in 
\pbpb collisions at \mbox{\sqrtsnn = 2.76 TeV} with respect to the nominal LHC clock 
value. The width of the distribution is indicative of how much the 
collision time can jitter with respect to its nominal value (the LHC clock edge). 
This is due to the finite size of the bunches and the clock-phase shift during 
a fill. The time resolution of the detector, 
estimated by the time difference registered in T0A and T0C, is 20--25 ps 
in \pbpb collisions (Fig.~\ref{fig:T0sumres}, right panel) and $\sim$40 ps 
in \pp collisions.
The efficiency of T0 is 100\% for the 60\% most central \pbpb collisions 
at \mbox{\sqrtsnn = 2.76 TeV}, dropping to about 50\% for events with centrality 
around 90\%. For \pp collisions at \mbox{\sqrts = 7 TeV}, the efficiency is 
about 50\% for a T0 coincidence signal (T0A-AND-T0C) and 70\% if only one 
of the T0 detectors is requested (T0A-OR-T0C).

The start time of the event $t_{\rm ev}$ is also estimated using the particle arrival 
times at the TOF detector. A combinatorial algorithm based on a $\chi^2$ 
minimization between all the possible mass hypotheses is used in the latter 
case. It can be invoked when at least three particles reach the TOF detector, 
to provide increased resolution and efficiency at larger multiplicity. With 30 tracks, 
the resolution on $t_{\rm ev}$ reaches 30~ps~\cite{TOFperf2013}.
This method is particularly useful for events in which the T0 signal is not present. 
If neither of these two methods is available, an average TOF start time for 
the run is used instead.

\begin{figure}[b!]
\centering
\includegraphics[width=0.68\textwidth]{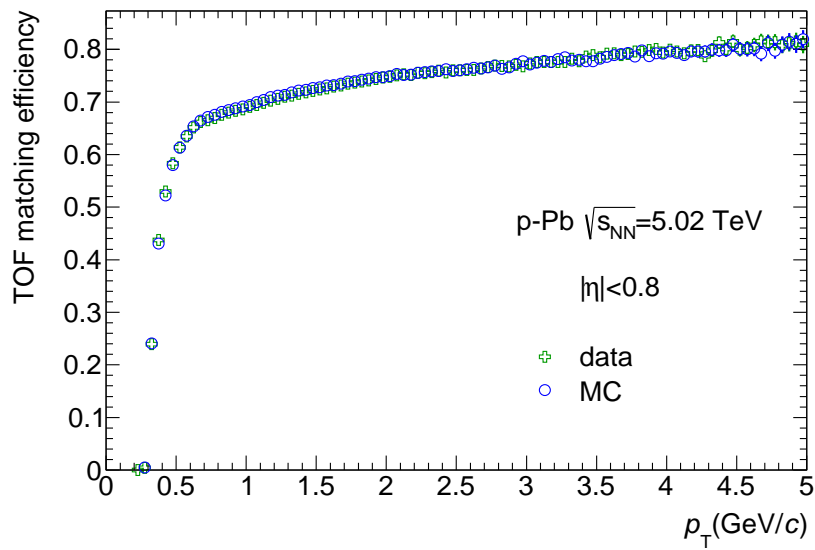}
\caption{\label{fig:TOFmatchingEff} Matching efficiency (including the geometric 
acceptance factor) at TOF for tracks reconstructed in the TPC in \ppb collisions 
at \mbox{\sqrtsnn = 5.02 TeV}, compared to Monte Carlo simulation.}
\end{figure}

The efficiency of matching TPC tracks to TOF in the 2013 \ppb run is compared
with Monte Carlo simulation in Fig.~\ref{fig:TOFmatchingEff}. 
At $\pt<0.7\gevc$, 
the matching efficiency is dominated by energy loss and the rigidity cutoff 
generated by the magnetic field. At higher transverse momenta it reflects the 
geometrical acceptance (dead space between sectors), the inactive modules, 
and the finite efficiency of the MRPCs (98.5\% on average). 

\begin{figure}[t]
\centering
\includegraphics[width=0.7\textwidth]{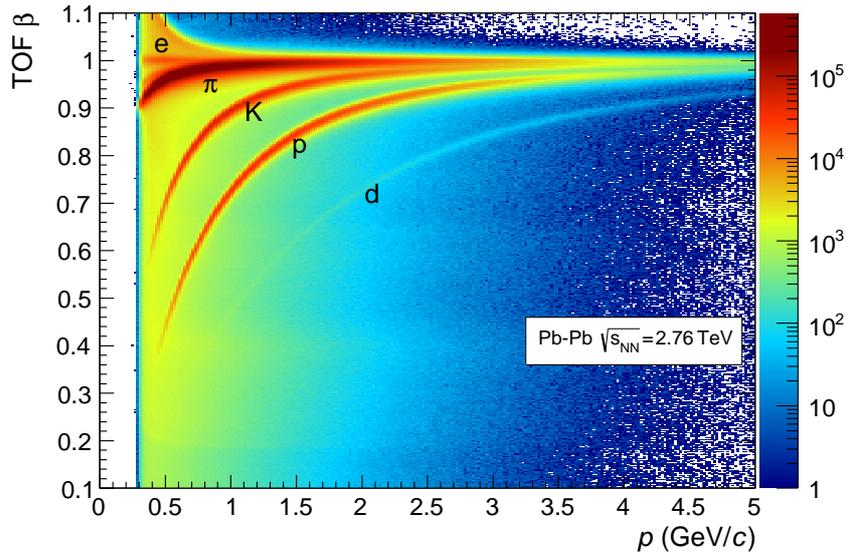}
\caption{\label{fig:tof_beta} Distribution of $\beta$ as measured by the 
TOF detector as a function of momentum for particles reaching the TOF in 
\pbpb interactions.}
\end{figure}

\begin{figure}[b]
\centering
\includegraphics[width=0.7\textwidth]{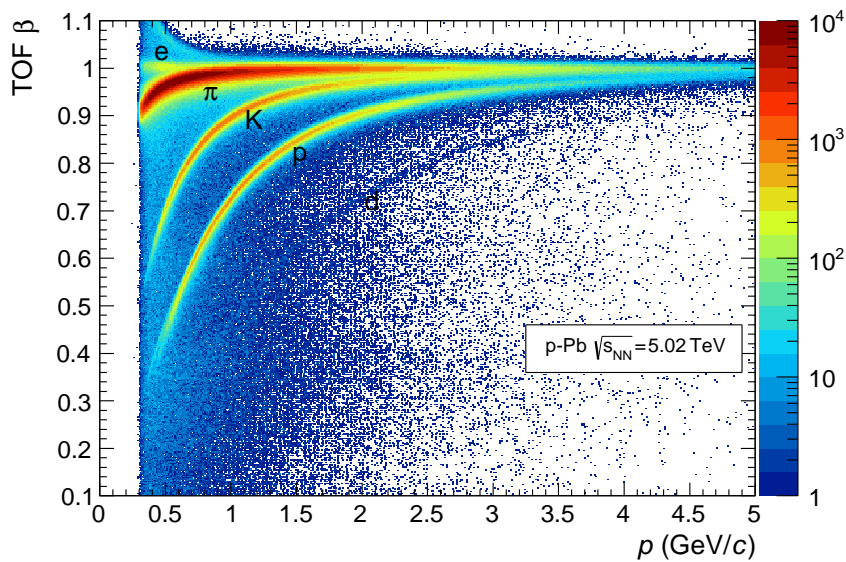}
\caption{\label{fig:tof_beta_pPb} Distribution of $\beta$ as measured by the 
TOF detector as a function of momentum for particles reaching TOF in \ppb 
interactions. The background of mismatched tracks is lower than in \pbpb.}
\end{figure}

Figure~\ref{fig:tof_beta} illustrates the performance of the TOF detector by 
showing the measured velocity $\beta$ distribution as a function of momentum 
(measured by the TPC). The background is due to tracks that are incorrectly 
matched to TOF hits in high-multiplicity \pbpb collisions. 
The distribution is cleaner in \ppb collisions (Fig.~\ref{fig:tof_beta_pPb}), 
showing that the background is not related to the resolution of the TOF 
detector, but is rather an effect of track density and the fraction of mismatched 
tracks. 
\begin{figure}[btp]
\centering
\includegraphics[width=0.72\textwidth]{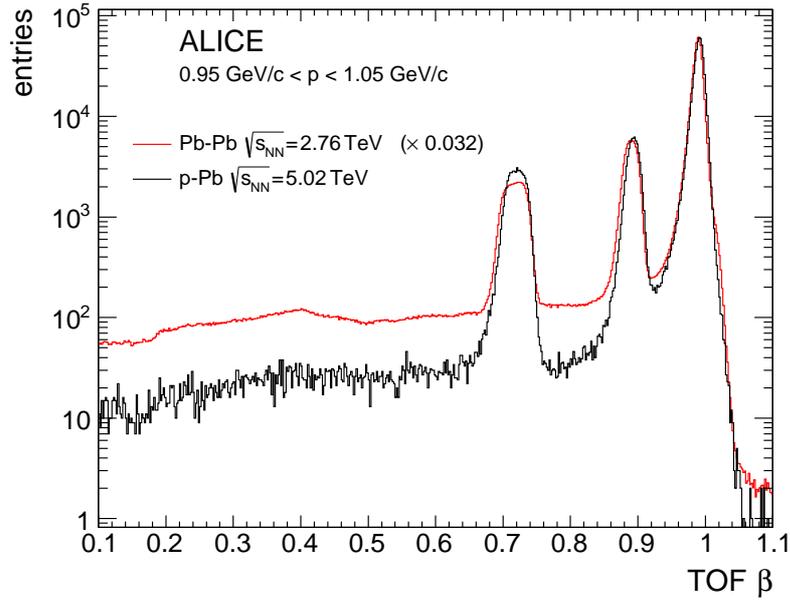}
\caption{\label{fig:TOF_mismatch} TOF $\beta$ distribution for 
tracks with momentum $0.95\gevc<p<1.05\gevc$. The \pbpb histogram is normalized to 
the \ppb one at the pion peak ($\beta=0.99$). While the resolution (width of the mass 
peaks) is the same, the background of mismatched tracks increases in the 
high-multiplicity environment of \pbpb collisions. Both samples are minimum bias.}
\end{figure}
This is also visible in Fig.~\ref{fig:TOF_mismatch} where the $\beta$ 
distribution is shown for a narrow momentum band. The pion, kaon, and proton peaks 
are nearly unchanged but the level of background due to mismatched tracks is 
higher in \pbpb. The fraction of mismatched tracks above 1\gevc in \pbpb{} events 
is closely related to the TOF occupancy. With $10^4$ hits at TOF (corresponding 
to a very central \pbpb event) the TOF pad occupancy is 6.7\% and the fraction 
of mismatched hits is around 6.5\%. 

\begin{figure}[hbt]
\centering
\includegraphics[width=0.72\textwidth]{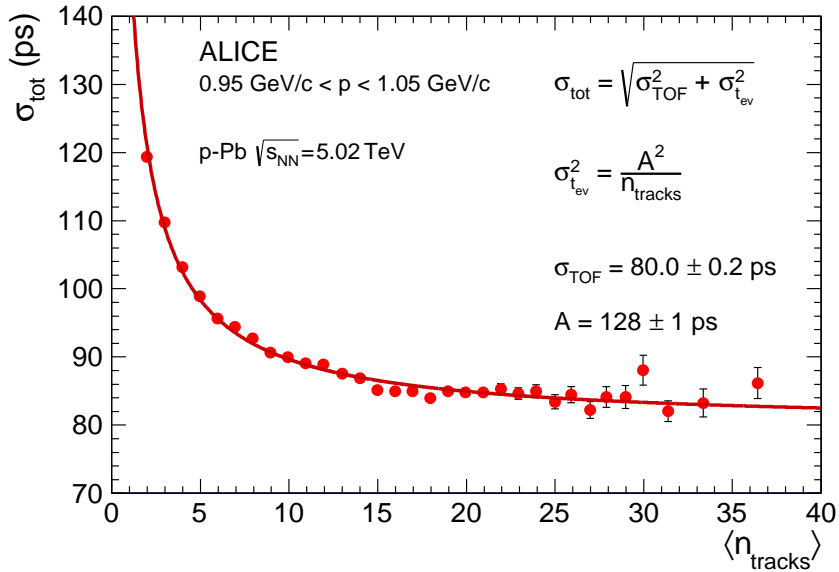}
\caption{\label{fig:tof_res}
Time resolution of pion tracks with $0.95<{\it p}<1.05$\gevc 
as a function of the number of tracks used to define the start time of the collision 
$t_{\rm ev}$\protect~\cite{TOFperf2013}. The data are from \ppb collisions.}
\end{figure}

The resolution can be studied in a given narrow momentum interval by computing 
the difference between the time of flight measured by TOF and the pion time 
expectation. The distribution is fitted with a Gaussian whose width is the 
convolution of the intrinsic time resolution of the TOF detector and the 
resolution of the event time. 
In the limit of high track multiplicity the width becomes equal to the intrinsic 
resolution of the TOF detector and has a value of 80~ps (Fig.~\ref{fig:tof_res}). 

\begin{figure}[hbt]
\centering
\includegraphics[width=0.8\textwidth]{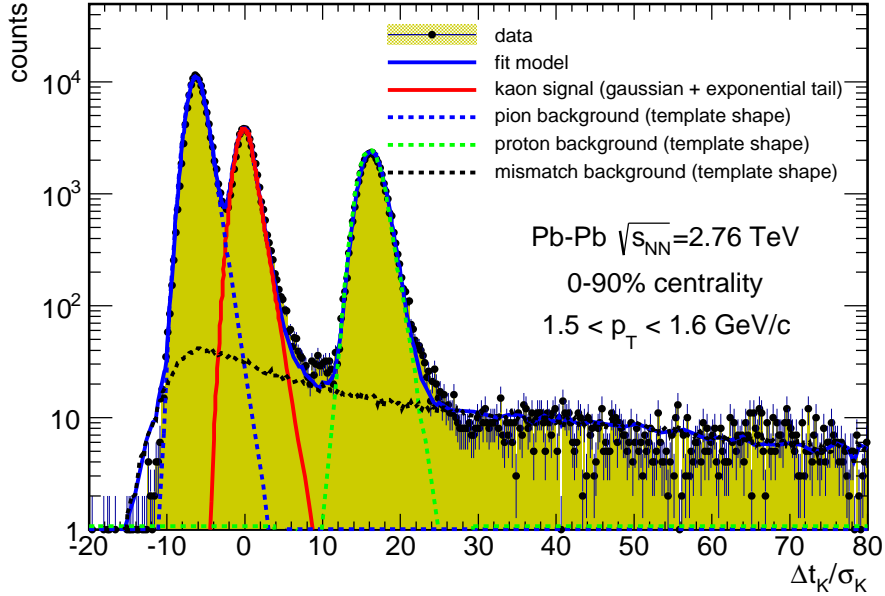}
\caption{\label{fig:tof_spectra}TOF measured in \pbpb collisions at 
\mbox{\sqrtsnn = 2.76 TeV}. The expected time of flight for kaons is subtracted 
and the result is divided by the expected resolution. 
}
\end{figure}

At those transverse momenta where the TOF resolution does not permit 
track-by-track identification, a fit of multiple Gaussian peaks is used to 
determine the particle yields. To illustrate this,
Fig.~\ref{fig:tof_spectra} shows, for tracks with $1.5 < \pt < 1.6$\gevc, the 
difference between the measured time of flight and the expectation for kaons, 
together with a template fit to the pion, kaon, and proton peaks and the 
combinatorial background from mismatched tracks.

\subsection{Particle identification in the HMPID}

\begin{figure}[hbt]
  \centering
  \includegraphics[width=0.72\textwidth]{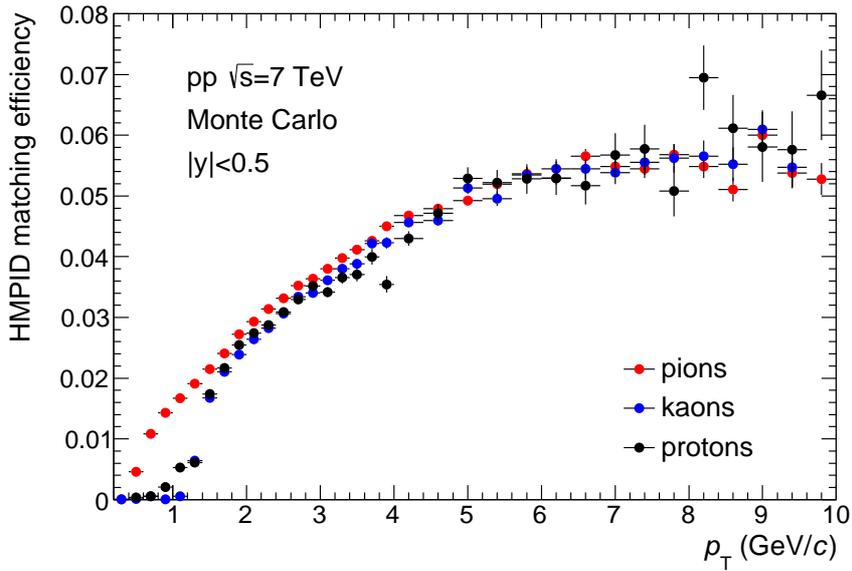}
  \caption{Matching efficiency (including the geometric acceptance factor) 
           at HMPID for tracks reconstructed in the TPC.}
  \label{fig:HMPIDmatchingEff}
\end{figure}

The High Momentum Particle Identification Detector consists of 7
identical RICH (ring-imaging Cher\-en\-kov) modules in proximity focusing
configuration, exploiting a liquid C$_6$F$_{14}$ radiator coupled to
MWPC (multiwire proportional chamber)-based photon detectors with CsI 
photocathodes and covering 11~m$^2$ ($\approx$ 5\% of TPC acceptance). 
On average 14 photoelectrons per ring are detected at saturation ($\beta\approx1$). 
The HMPID detector extends track-by-track charged hadron identification in ALICE to 
higher \pt. The identification is based on the Cherenkov angle of the ring produced by 
charged tracks. The Cherenkov angle is given by:
\begin{equation}
\cos\theta = \frac{1}{n\beta},
\label{eq:cherenkov_angle}
\end{equation}
where $n$ is the refractive index of the radiator ($n \approx$ 1.289 at
175 nm). 
The matching efficiency of tracks reconstructed in the TPC with the HMPID is 
shown in Fig.~\ref{fig:HMPIDmatchingEff} for \pp data and positive particles.
The value at large transverse momentum is dominated by the geometrical 
acceptance of the detector. At low \pt, the matching efficiency is shaped by energy
loss, a lower momentum cut due to the magnetic field, and the mass-dependent 
momentum threshold of the Cherenkov effect. 
Negative particles (not shown) have similar behavior. Antiprotons have a slightly
lower efficiency due to differing absorption behavior in the material between 
TPC and HMPID.

\begin{figure}[hbt]
  \centering
  \includegraphics[width=0.72\textwidth]{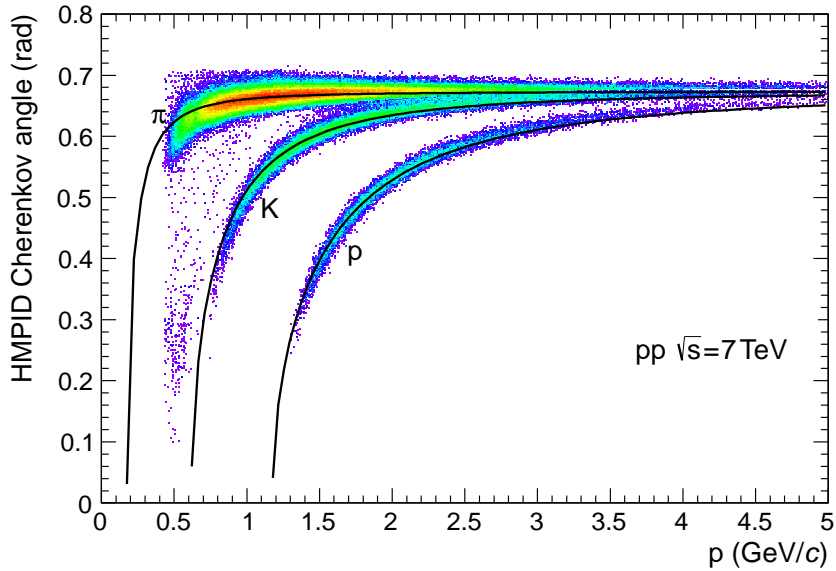}
  \caption{Mean Cherenkov angle measured by HMPID in \pp collisions at 7~TeV 
  as a function of track momentum. The lines represent parametrizations 
  of Eq.~(\ref{eq:cherenkov_angle}) for each species. }
  \label{ThetavsMom}
\end{figure}

Figure~\ref{ThetavsMom} shows the measured mean Cherenkov angle
as a function of track momentum for pions, kaons, and protons in \pp 
collisions at 7~TeV. The lines represent parametrizations of 
Eq.~(\ref{eq:cherenkov_angle}) for each species.
The separation of kaons from other charged particles, 
determined by fitting the Cherenkov angle distribution with three
Gaussians for each transverse momentum bin (the background is
negligible), is  3$\sigma$ for \pt $<$ 3\gevc for pions, and \pt $<$ 5\gevc for protons.

\begin{figure}[bt]
  \centering
  \includegraphics[width=0.8\textwidth]{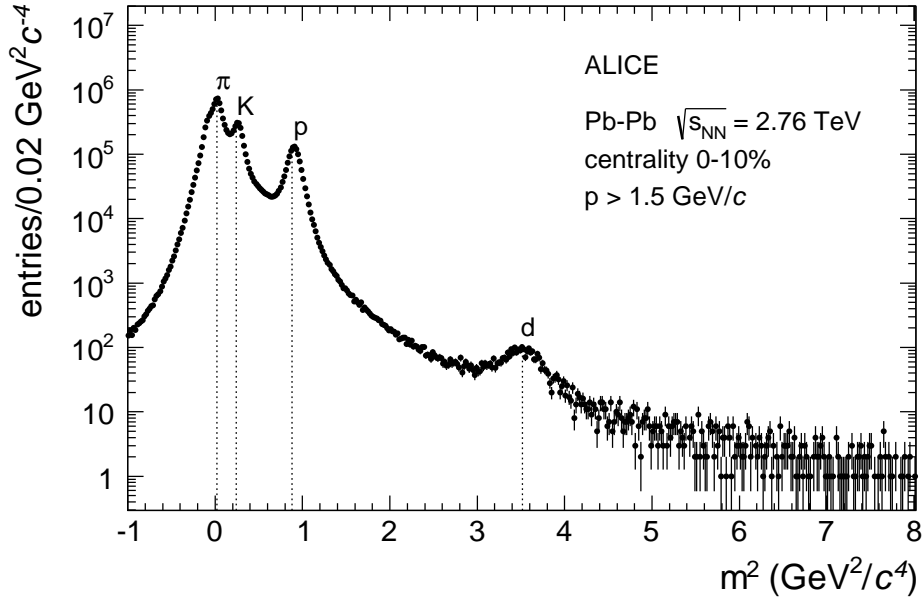}
  \caption{Squared particle masses calculated from the momentum and velocity determined  
  with ITS-TPC and HMPID, respectively, in central \pbpb collisions at \mbox{\sqrtsnn = 2.76 TeV}. 
  The velocity is calculated from the Cherenkov angle measured in the HMPID. 
  Dotted lines indicate the PDG mass values. The pion tail on the left-hand side is 
  suppressed by an upper cut on the Cherenkov angle. The deuteron peak is clearly 
  visible. }
  \label{fig:HMPID_Mass2}
\end{figure}

Figure~\ref{fig:HMPID_Mass2} shows the mass distribution of particles identified 
in the HMPID in central \pbpb collisions. The mass is calculated from the Cherenkov angle 
measured in the HMPID and the momentum determined by the central-barrel tracking detectors. 
For tracks with $p>1.5$\gevc and with 5--15 clusters per ring, the deuteron peak 
becomes clearly visible. 
This, and the fact that all of the particle peaks are at their nominal mass values, shows the 
good performance of the pattern recognition in the high-multiplicity environment of 
central \pbpb collisions.

\subsection{Overview of separation powers and combined PID}
\label{sec:pidovw}

\begin{figure}[hbt]
\centering
\includegraphics[width=\textwidth]{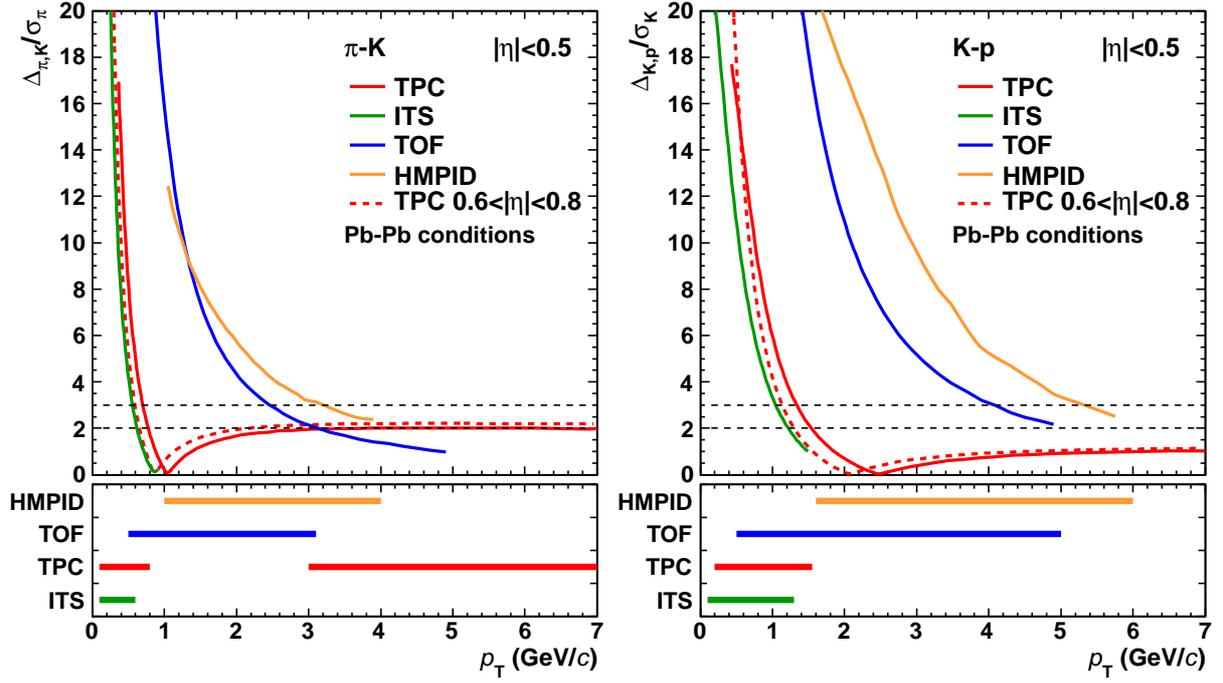}
\caption{\label{fig:hadr_separation}Separation power of hadron identification 
in the ITS, TPC, TOF, and HMPID as a function of \pt at midrapidity. 
The left (right) panel shows the separation of pions and kaons (kaons and protons), 
expressed as the distance between the peaks divided by the 
resolution for the pion and the kaon, respectively, averaged over $|\eta|<0.5$. 
For the TPC, an additional curve is shown in a narrower $\eta$ region. The lower 
panels show the range over which the different ALICE detector systems have a 
separation power of more than $2\sigma$. }
\end{figure}

Figure~\ref{fig:hadr_separation} shows the pion-kaon (left panel) 
and kaon-proton (right panel) separation power of the ITS, TPC, TOF, and 
HMPID as a function of \pt. 
The separation is calculated as the distance $\Delta$ between the peaks divided 
by the Gaussian width $\sigma$ of the pion and the kaon response, 
respectively. Note that the detector response for the individual
detectors in Figs.~\ref{fig:its_nsig}, \ref{fig:TPC_dEdx}, \ref{fig:tof_beta}, 
\ref{fig:tof_beta_pPb}, and \ref{ThetavsMom} is naturally a function 
of total momentum $p$. 
However, since most physics results are analyzed in transverse momentum bins, 
in Fig.~\ref{fig:hadr_separation} we present the separation power in 
\pt bins, averaging the momentum-dependent response over the range 
$|\eta|<0.5$. For the TPC, a forward pseudorapidity slice, relevant 
for high-\pt PID analysis, is shown as well. This also demonstrates 
the effect of averaging over a larger $\eta$ range, which mixes different 
momentum slices.

The plots demonstrate the complementarity of the different detector systems. 
At low $\pt<500$\mevc, the TPC and ITS provide the main separation,
because TOF and HMPID are not efficient. At intermediate \pt, up to 3 (4)\gevc 
for pions/kaons and 5 (6) \gevc for protons, TOF(HMPID) provides 
more than $3\sigma$ separation power. TOF has full azimuthal
coverage and it reaches lower \pt, while HMPID only covers 5\% of
the full acceptance. At higher \pt, the TPC can be used to separate
pions from protons and kaons with $\sim2\sigma$ separation,
exploiting the relativistic rise of the energy loss. Protons and kaons
can be separated statistically with a multi-Gaussian fit to the
collected signal (see Fig.~\ref{fig:TPC_multigauss}).

\begin{figure}[b!]
\centering
\includegraphics[width=0.55\textwidth]{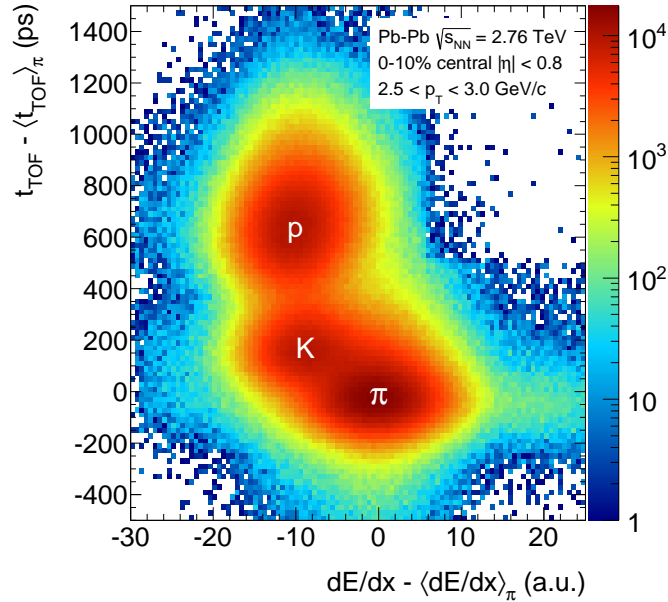}
\caption{
\label{fig:tpctofcomb} Combined pion identification with TOF and with \dedx in the TPC.}
\end{figure}

The separation of hadron species can be further improved by combining
information from multiple detectors, thus allowing a further extension of
the momentum range for identified particle measurements. An example of
this approach is shown in Fig.~\ref{fig:tpctofcomb}, where at intermediate \pt
the difference between the measured and expected PID signals for TPC and TOF are
represented. It is evident that cuts or fits using a combination
of the variables provide a better separation than just considering their
projections. This technique was used to measure the p/$\pi$ ratio in di-hadron 
correlations~\cite{Veldhoen:2012dhc} and permits, using fits of the 
bidimensional distribution, to extend the kaon/pion separation up to a 
transverse momentum of 5\gevc in \pbpb. 

A Bayesian approach to combined PID, making use of the known relative 
yields of different particle species, is under development.

\subsection{Particle identification using weak decay topology}

In addition to the direct identification of the more stable hadrons
($\pi$, K, p) using mass-dependent signals such as \dedx, TOF and
Cherenkov radiation, ALICE also identifies hadrons through their weak
decay topology. This technique is used for strange hadrons, such as
\kzeros, $\Lambda$, and the multi-strange baryons $\Xi$ and
$\Omega$, as well as for charmed hadrons. In all of these cases a full
kinematical reconstruction of the decay into charged hadrons is used, 
as described in Section~\ref{sect:secondary}. 

\begin{figure}[hbt]
\centering
\includegraphics[width=0.78\textwidth]{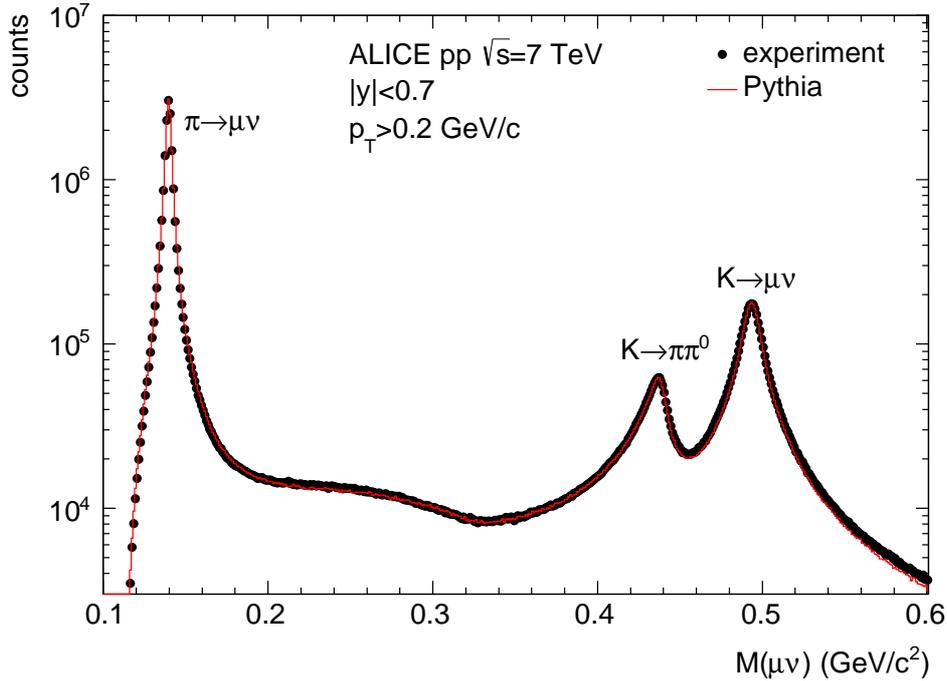}
\caption{Invariant mass of reconstructed charged particles (pions and kaons) decaying 
inside the TPC volume and producing a secondary vertex (kink). The mass is calculated 
assuming that the track segment after the kink represents a muon and that the neutral 
decay daughter is a neutrino. The neutrino momentum is taken from the difference 
between the momenta of the track segments before and after the kink. }
\label{fig:kinkMass}
\end{figure}

In addition to these, charged kaons can be identified by a distinct kink 
in the track owing to the decay into a muon and a neutrino with a branching 
ratio (BR) of 63.5\%. Figure~\ref{fig:kinkMass} 
shows an invariant mass distribution of kink-decay daughters, assumed to be a muon 
and a neutrino. 
The muon momentum is taken from the track segment after the kink. For the neutrino 
momentum, the difference between the momenta of the track segments before and after 
the kink is used. 
The distribution shows two peaks representing the muonic decays of pions and kaons, 
as well as \mbox{$\rm{K^\pm}\rightarrow\pi^\pm +\pizero$} (BR = 20.7\%) 
reconstructed with an incorrect mass assumption. 
The broad structure outside the pion mass region mainly originates from three-body 
decays of kaons. 
The efficiency for reconstruction and identification of charged kaons is $\sim$60\%  
at \pt around 1.0\gevc and decreases gradually 
at higher transverse momenta, as the angle between mother and daughter tracks
becomes smaller. 
The structures in the invariant mass distribution are well reproduced in simulation. 
The simulation also provides an estimate of the contamination. For kaons with 
transverse momenta up to 8\gevc, the contamination is below 3\%. Most of the contamination 
arises due to single charged-particle tracks with a small-angle kink caused by scattering 
rather than a decay.

\subsection{Particle identification in physics analysis}
\label{sec:pidphysan}

The use and performance of particle identification can best be
illustrated using examples of specific physics analyses. 
Transverse momentum spectra of $\pi$, K, and p, identified using ITS, TPC, 
TOF, and decay topology, were published for \pp~\cite{Aamodt:2011zj} 
and \pbpb~\cite{Abelev:2012wca,Abelev:2013sm} collisions. 
Applications of PID techniques to analyses of $\phi$, D, and light nuclei 
are briefly discussed below. 

\subsubsection{$\phi$ meson}
\begin{figure}[hbtp]
\centering
\includegraphics[width=0.95\linewidth]{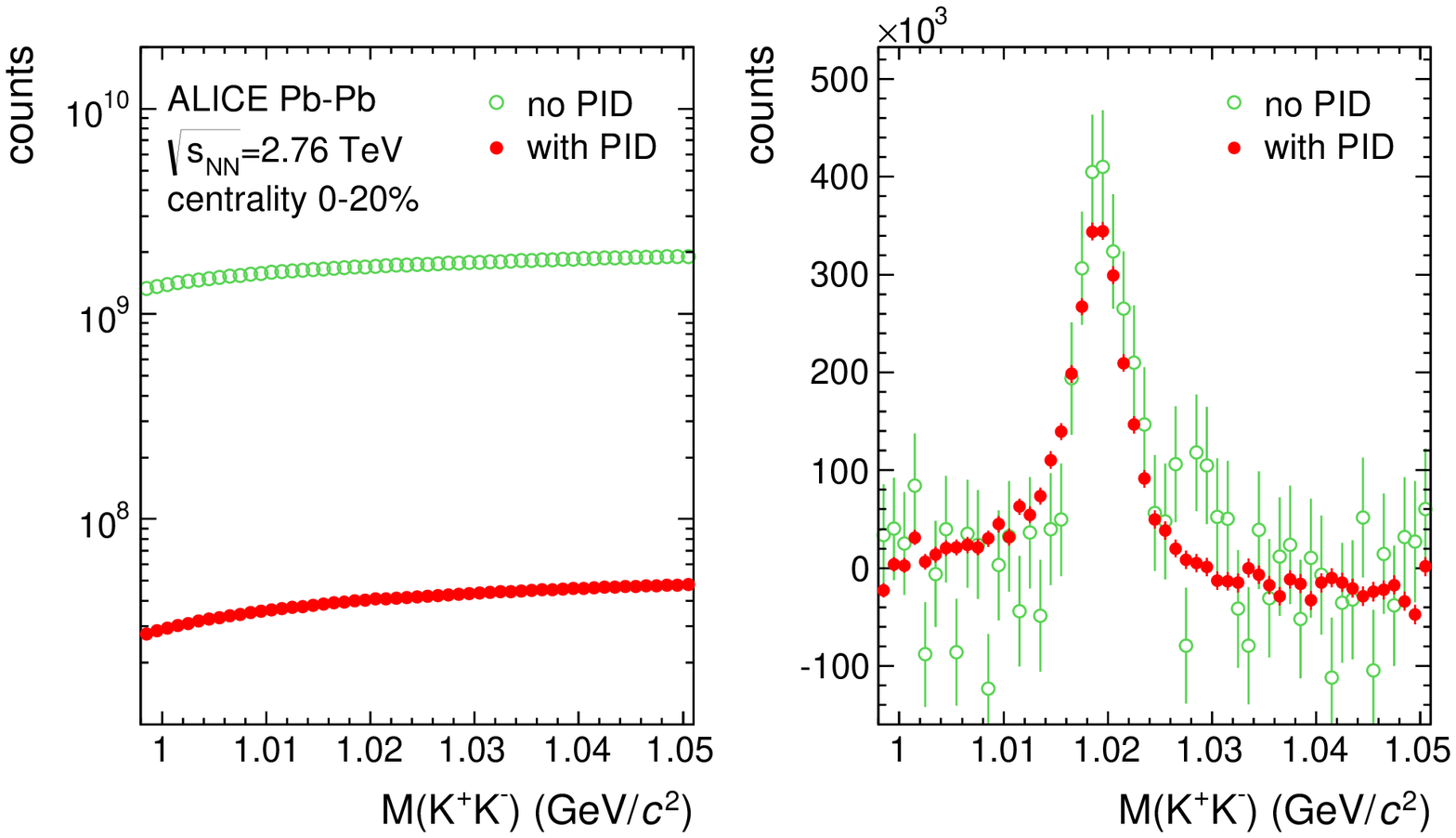}
\caption{\label{fig:phikk} Invariant mass distribution of \kap$\!\!$\kam candidate 
pairs for reconstruction of the $\phi \rightarrow$~KK decay, with and without
particle identification, before (left panel) and after (right panel) background 
subtraction. }
\end{figure}

The $\phi$ meson predominantly decays into two charged kaons $\phi \rightarrow
{\rm K}^{+}{\rm K}^{-}$. 
Since this is a strong decay, it is not possible to topologically reconstruct 
the decay. Identification of the decay products, however, dramatically improves 
the signal-to-background ratio. This is demonstrated in Fig.~\ref{fig:phikk}, 
which shows the $\phi$ signal in 3 million central \pbpb events without 
particle identification (green circles) and with particle identification 
using a 2$\sigma$ cut on the TPC \dedx (red dots). 
The signal-to-background ratio at the $\phi$ peak for \mbox{$1<\pt<1.5\gevc$} 
(\mbox{$\pt<24\gevc$}) improves from $0.3\times10^{-3}$ ($0.1\times10^{-3}$) 
to $5\times10^{-3}$ ($4\times10^{-3}$) when the PID cut is applied. 
In terms of the peak significance, the improvement is from 14 to 45 
(from 15 to 75). 

\subsubsection{D meson}
\begin{figure}[t]
\centering
\includegraphics[width=0.95\linewidth]{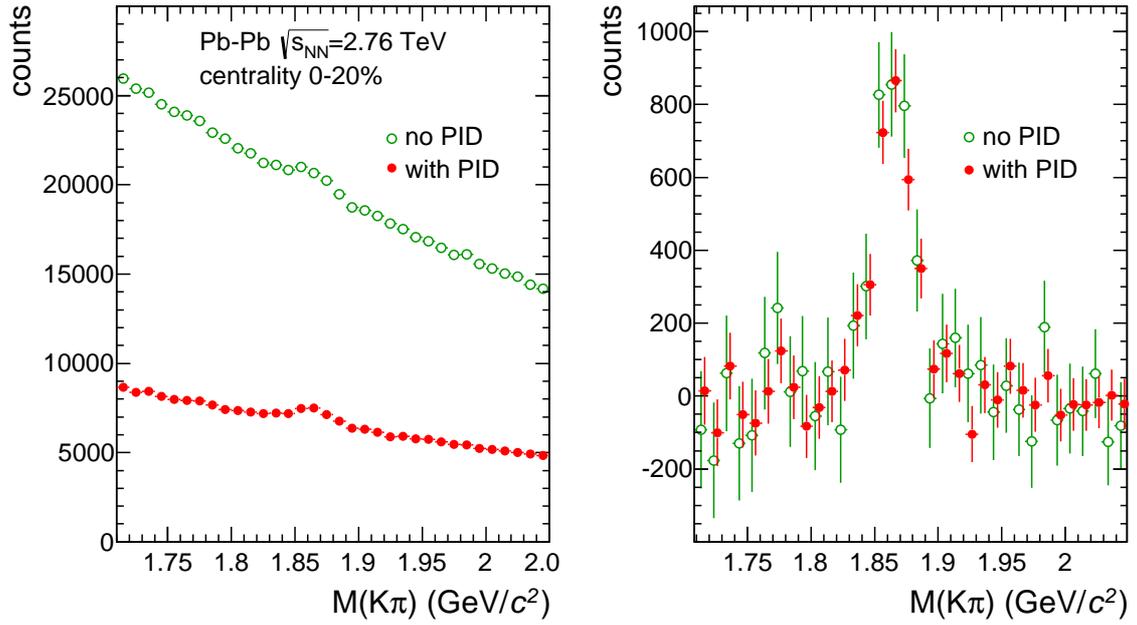}
\caption{\label{fig:Dkpi} Invariant mass distribution of K$\pi$ candidate pairs for
reconstruction of the D$^0 \rightarrow $K$\pi$ decay, with and without particle identification, 
before (left panel) and after (right panel) background subtraction.}
\end{figure}

Charm production measurements in ALICE are performed, among others, using 
hadronic decays of the charmed mesons D$^0$, D$^{\pm}$, and 
D$^{*\pm}$~\cite{ALICE:2011aa,Abelev:2012dme,Abelev:2012vra}. 
For these analyses, the identification of the kaons
greatly enhances the signal significance. As an example,
Fig.~\ref{fig:Dkpi} shows the invariant mass distribution of K$\pi$
candidate pairs with and without particle identification. The pairs
were preselected using cuts on \pt, impact parameter, and
various requirements on the decay topology. In this case, loose
particle identification cuts are used to ensure a high efficiency
in the selection. A clear reduction of the combinatorial background by
a factor of $\sim 3$ can be seen in Fig.~\ref{fig:Dkpi}, with negligible
(a few percent) loss of signal.

\subsubsection{Light nuclei}

\begin{figure}[h]
\centering
\includegraphics[width=0.66\textwidth]{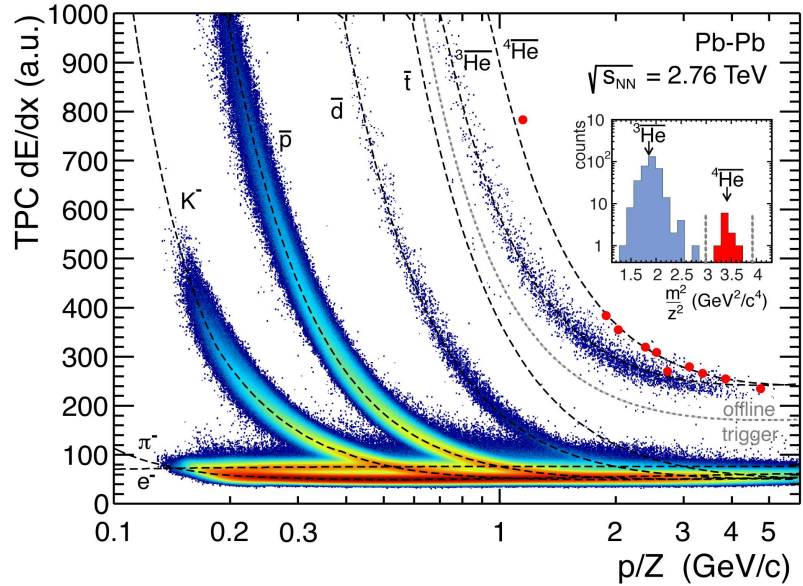}
\caption{\label{fig:antiHe} Measured \dedx signal in the ALICE TPC versus magnetic 
rigidity, together with the expected curves for negatively-charged particles. The inset 
panel shows the TOF mass measurement which provides additional separation between 
$^3\overline{\rm He}$ and $^4\overline{\rm He}$ for tracks with $p/Z > 2.3$\gevc.}
\end{figure}

In \pbpb collisions light nuclei were identified via 
the \dedx signal in the TPC and time-of-flight measurements with the
TOF detector. Figure~\ref{fig:antiHe} illustrates the separation between 
$^3\overline{\rm He}$ and $^4\overline{\rm He}$ 
in TPC and TOF.  This identification technique was used to study the 
formation of antinuclei and hyperons in \pbpb collisions.

\clearpage\newpage\section{Electron identification}
\label{sect:electrons}
The detector systems for hadron identification that are described in
the previous section are also used to identify electrons. In addition,
the following systems have dedicated electron identification
capabilities:
\begin{itemize}
\item {\it TRD:} The Transition Radiation Detector identifies electrons
  based on their specific energy loss and transition radiation (TR) 
  and covers the full central barrel\footnote{As of 2013, 5 out of 18 TRD 
  supermodules are yet to be installed. See Table~\ref{tab:installation} for details.}. 
\item {\it EMCal:} The Electromagnetic Calorimeter identifies electrons
  by measuring their energy deposition and comparing it to the measured track
  momentum ($E/p$ method). The EMCal has a partial coverage $|\eta| < 0.7$ 
  and 107\degr in $\phi$.
\item {\it PHOS:} The Photon Spectrometer is a high-granularity electromagnetic 
  calorimeter that can also identify electrons using the $E/p$ method. 
  PHOS covers $|\eta|<0.12$ with up to 5 modules, 20\degr in azimuth each.
  Three modules were installed in 2009--2013. 
\end{itemize}

The PHOS, EMCal, and TRD also have capabilities to trigger on high-momentum 
electrons, charged particles, and photons (PHOS and EMCal only). 

These detector systems provide complementary capabilities for electron
measurements: the TRD with its large acceptance and triggering
capabilities at intermediate \pt~=~2--5\gevc is particularly suited for
dilepton measurements, including quarkonia, while the trigger
capabilities of EMCal (and PHOS) make it possible to sample the full
luminosity for high-\pt electron measurements (from heavy-flavor
decays). To obtain a pure electron sample for physics analysis,
signals from multiple detectors are used (see Section~\ref{sect:electron_pid_analysis} 
for some examples).

\subsection{Electron identification in the EMCal}
\label{sect:EMCalelectronPID}

\begin{figure}[t]
\centering
\includegraphics[width=0.58\textwidth]{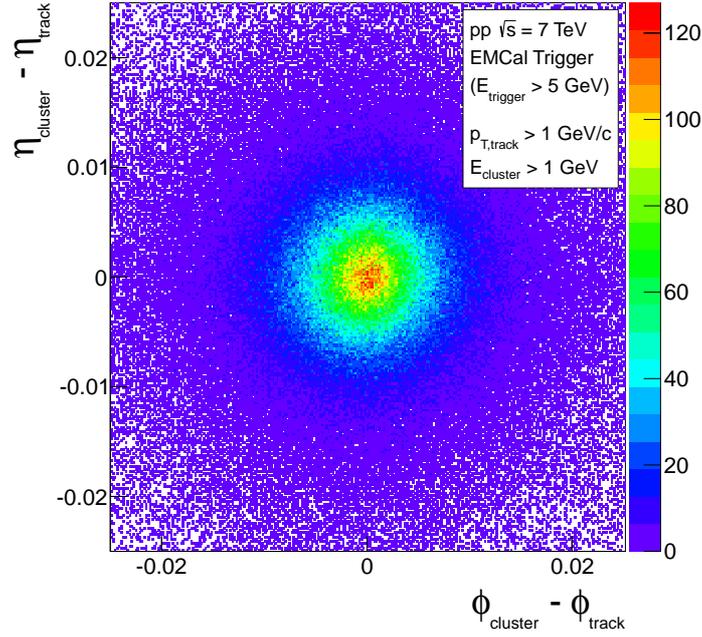}
\caption{\label{fig:EMCal_matching}Distribution of the residuals for 
the EMCal clusters to track matching in pseudorapidity 
($\eta_{\rm cluster}-\eta_{\rm track}$) vs. azimuth ($\phi_{\rm cluster}-\phi_{\rm track}$) 
in \pp collisions at \mbox{\sqrts = 7 TeV} triggered by EMCal. Only clusters 
with an energy $E_{\rm cluster}>1$~GeV and tracks with a transverse 
momentum $p_{\rm T, track}>1$\gevc are used.}
\end{figure}

Electrons 
deposit their entire energy in the calorimeter 
while hadrons typically only lose a small fraction. The ratio $E/p$
of the energy $E$ of EMCal clusters (for cluster finding see 
Section~\ref{sect:EMCalClustering}) and the momentum $p$ of
reconstructed tracks that point to the cluster is therefore used to
separate electrons and hadrons. An EMCal cluster is considered to be
matched to a track when the maximum distance between the extrapolated
track position as shown in Fig.~\ref{fig:EMCal_matching} is less than 
a predetermined cutoff value (for a minimum hadron contamination one 
uses $\Delta \eta < 0.0025$ and $\Delta \phi < 0.005$). 
The electron-hadron separation can be further enhanced by
taking into account the different electromagnetic shower shapes for
electrons and hadrons.
\begin{figure}[hb!]
\centering
\includegraphics[width=0.85\linewidth]{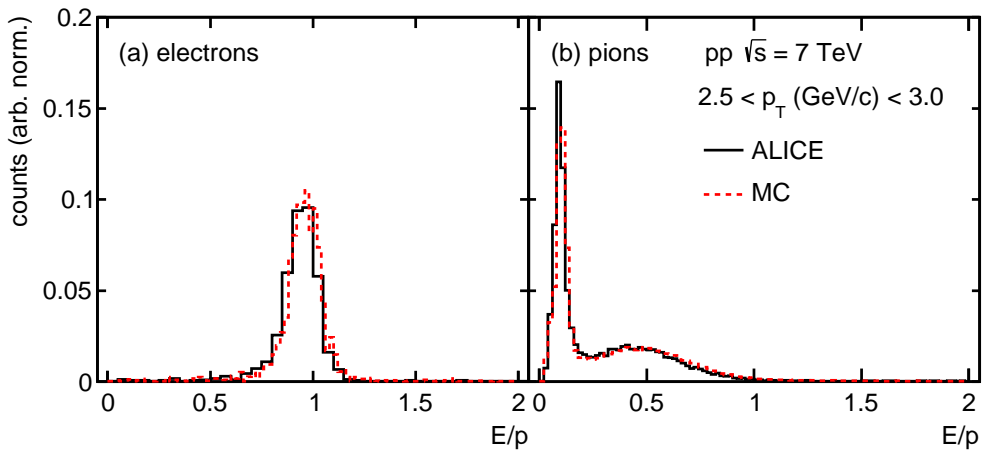}
\caption{$E/p$ distributions for (a) electrons and (b) pions in \pp
collisions at \mbox{\sqrts = 7 TeV}, measured in the experiment (red
dotted line), and compared to simulation (black full line). 
The samples of identified electrons and pions were obtained from reconstructed 
$\gamma$ conversions and $\Lambda$/\kzeros decays, respectively. 
The simulation is a Pythia simulation with realistic detector configuration and full 
reconstruction.}
\label{fig:EMCalelectronPID_1}
\end{figure}

In order to determine the $E/p$ distribution, clean electron and
hadron samples were obtained from experimental data using the 
charged tracks originating from decays of neutral particles. 
Protons and pions are identified
from the decays of $\Lambda$ and \kzeros particles and a clean
electron sample was obtained from photon conversions in the detector
material.

In Fig.~\ref{fig:EMCalelectronPID_1} the $E/p$ distributions for
electrons and pions are shown for experimental and MC data in a
transverse momentum interval $2.5\gevc<\pt<3.0\gevc$. The
normalization of both distributions is arbitrary and does not reflect
the yield ratio between the two particle species. Electrons exhibit a
clear peak at $E/pc\sim1$, with a tail at lower values due to
bremsstrahlung in the detector material in front of the EMCal.
Pions, on the other hand, are mostly minimum-ionizing 
particles, with a typical $E/pc\sim0.1$ and a shoulder at
higher values due to additional hadronic interactions in the
calorimeter.

\begin{figure}[h]
\centering
\includegraphics[width=0.53\linewidth]{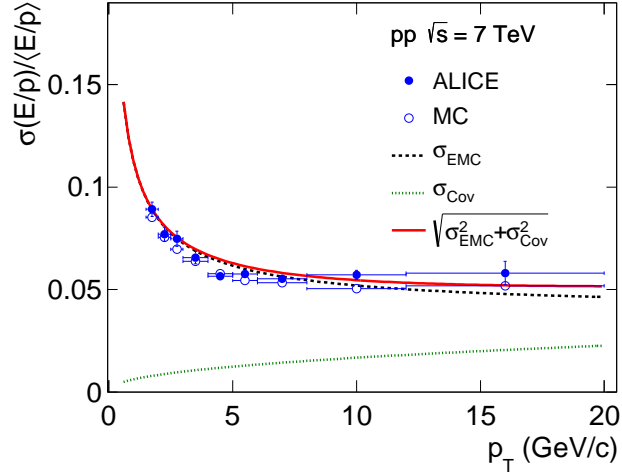}
\caption{Relative resolution of $E/p$ vs. transverse momentum \pt for 
electrons in experimental data (full dots) and from a fully reconstructed 
MC (open circles) in \pp collisions at \mbox{\sqrts = 7 TeV}. The EMCal energy 
resolution deduced from the width of the \pizero and $\eta$ invariant mass 
peaks (black dotted line), added in quadrature to the TPC \pt resolution 
(green dash-dotted line), describes  the 
measurement reasonably well (red solid line). 
}
\label{fig:EMCalelectronPID_2}
\end{figure}

The $E/p$ distribution for electrons can be characterized using a
Gaussian fit, which then can be cut on for electron
identification or used to calculate probability densities and a Bayesian
probability.  For the latter, a parametrization of the hadron
distributions is determined as well. Figure~\ref{fig:EMCalelectronPID_2} 
shows the relative resolution of $E/p$ as a function of
transverse momentum as measured in the experiment, compared to detector 
simulations of full events from Pythia. The experimental data are compatible 
with the simulation within uncertainties. Shown in the same figure are  
the EMCal energy resolution, deduced from the width of the \pizero and $\eta$ 
peaks in the invariant mass distribution of photon pairs, and the momentum 
resolution of electrons from tracking (relevant at high momentum).
The two contributions added in quadrature describe the measured $E/p$ 
resolution reasonably well. 

\subsection{Electron identification in the TRD}

\begin{figure}[b]
\centering
\includegraphics[width=0.72\textwidth]{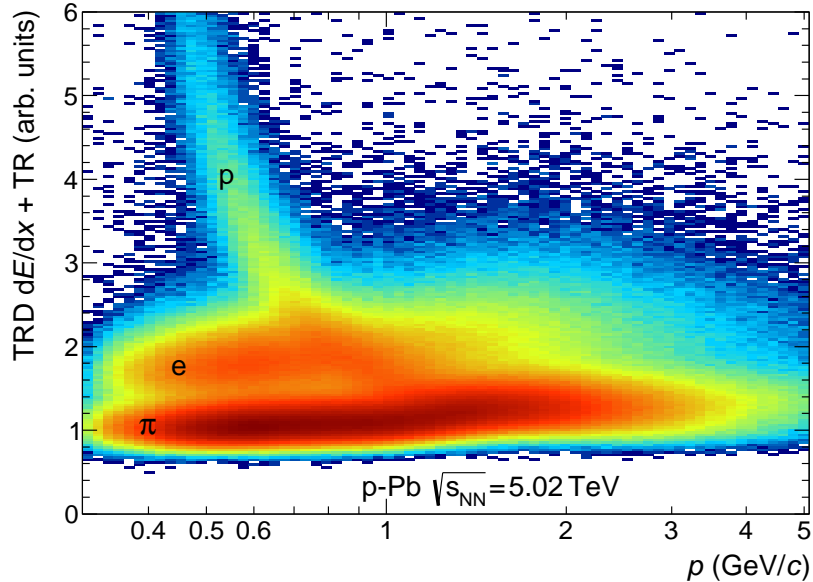}
\caption{\label{fig:TRD2dplot}Sum of the TRD signal (ionization energy 
loss plus transition radiation) as a function of momentum for protons from 
$\Lambda$ decays, charged pions from \kzeros{} decays, and electrons from 
$\gamma$ conversions in \ppb collisions.
}
\end{figure}

\begin{figure}[t]
   \centering
\includegraphics[width=0.9\textwidth]{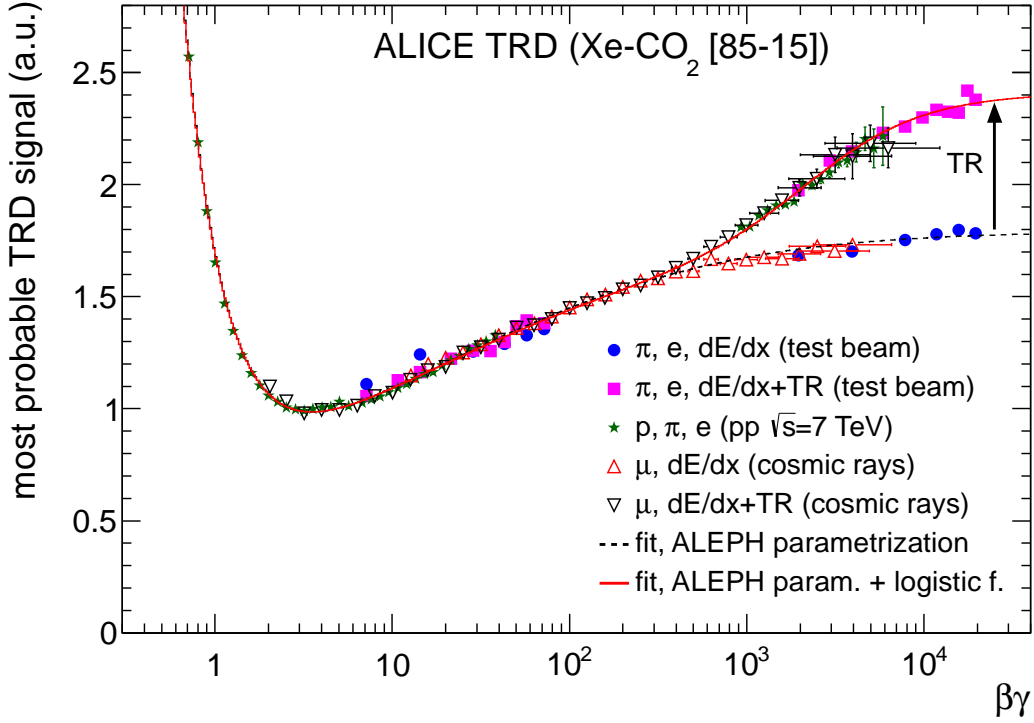} 
  \caption[]{The most probable TRD signal as a function of $\beta\gamma$. 
  Measurements performed in test beam runs, \pp collisions at 
  \mbox{\sqrts = 7 TeV}, and cosmic rays are compared.}
   \label{TRD_MPV}
\end{figure}

The Transition Radiation Detector provides
electron identification in the central barrel ($|\eta|<$0.9)~\cite{ALICE:2011ae} 
and can also be used to trigger (L1 hardware trigger, as discussed in 
Section~\ref{sect:datataking:trigger}) on electrons with high transverse momenta 
and on jets~\cite{Klein:2011hs}. The electron identification is based on the specific
energy loss and transition radiation. The TRD is composed of six layers consisting 
of a radiator followed by a drift chamber. Transition radiation is
produced when a relativistic charged particle ($\gamma \rm \gtrsim 10^3$) 
traverses many interfaces of two media of different dielectric
constants~\cite{Andronic:2011an} composing the radiator.  On average,
for each electron with a momentum above 1\gevc, one TR photon \mbox{(energy
range: 1--30 keV)} is absorbed and converted in the
high-$Z$ gas mixture (Xe-CO$_2$ [85-15]) in each layer of the detector.
Figure~\ref{fig:TRD2dplot} shows the combined TRD signal (\dedx and TR) as a 
function of momentum for \ppb collisions. 
The dependence of the most probable TRD signal on $\beta\gamma$  is shown in 
Fig.~\ref{TRD_MPV}. 
The data are from measurements with pions and electrons in test beam runs at 
CERN PS, performed with and without the radiator~\cite{Bailhache:2006hs}; protons, 
pions, and electrons in \pp collisions at \mbox{\sqrts = 7 TeV}~\cite{MFasel}; 
and cosmic muons triggered by subdetectors of the ALICE setup~\cite{Collaboration:2012zsa}. 
With cosmic muons, the selection of 
the flight direction allows one to measure only the specific energy loss (\dedx) 
or the summed signal (\dedx + TR). The onset of TR production is visible for 
$\beta\gamma \gtrsim 800$, both for electrons and high-energy (TeV scale) 
cosmic muons. Also note that the muon signal is consistent with that from 
electrons at the same $\beta\gamma$.

\begin{figure}[t]
   \centering
\includegraphics[width=0.5\textwidth]{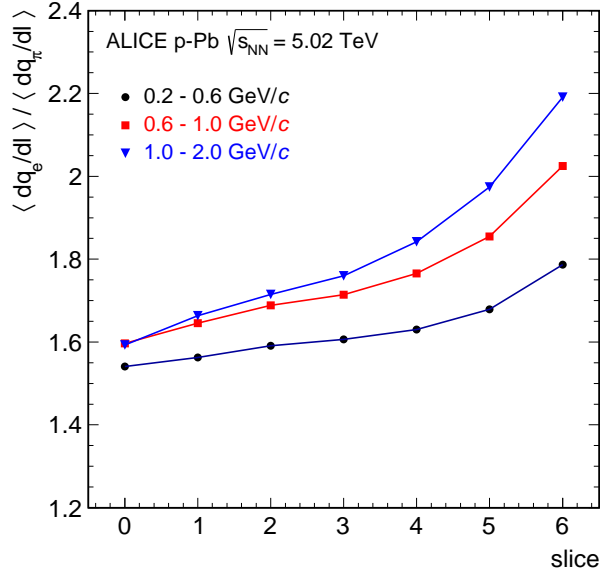} 
  \caption[]{The ratio of the average signal of electrons to that of pions 
  as a function of the depth in the detector (slice number; the lowest slice 
  number is farthest away from the radiator).}
  \label{TRD_AvCharge}
\end{figure}

For particle identification, the signal of each chamber is divided into seven 
slices (starting the numbering at the read-out end farthest away from the radiator), 
each integrating the sampled signal in about \mbox{5 mm} of detector thickness.
Figure~\ref{TRD_AvCharge} 
shows the ratio of the average signal for electrons 
to that of pions as a function of slice number. The TR contribution is visible at
large slice numbers (corresponding to long drift times) because the TR is predominantly 
absorbed at the entrance of the detector.

\begin{figure}[t]
\begin{tabular}{lr} \hspace{-5mm}\begin{minipage}{.5\textwidth}
\includegraphics[width=\textwidth]{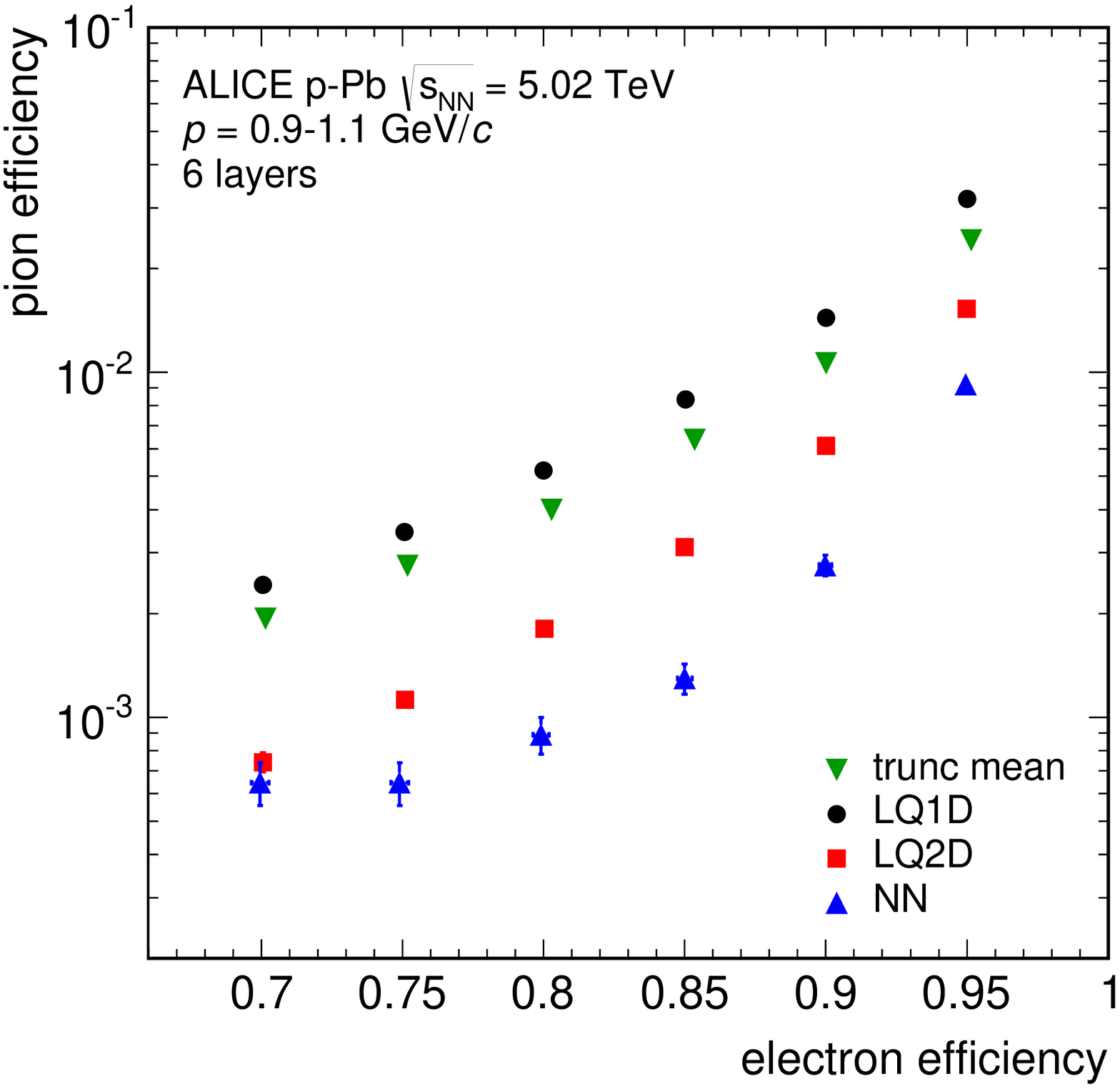} 
\end{minipage} &\begin{minipage}{.5\textwidth}
\includegraphics[width=1.\textwidth]{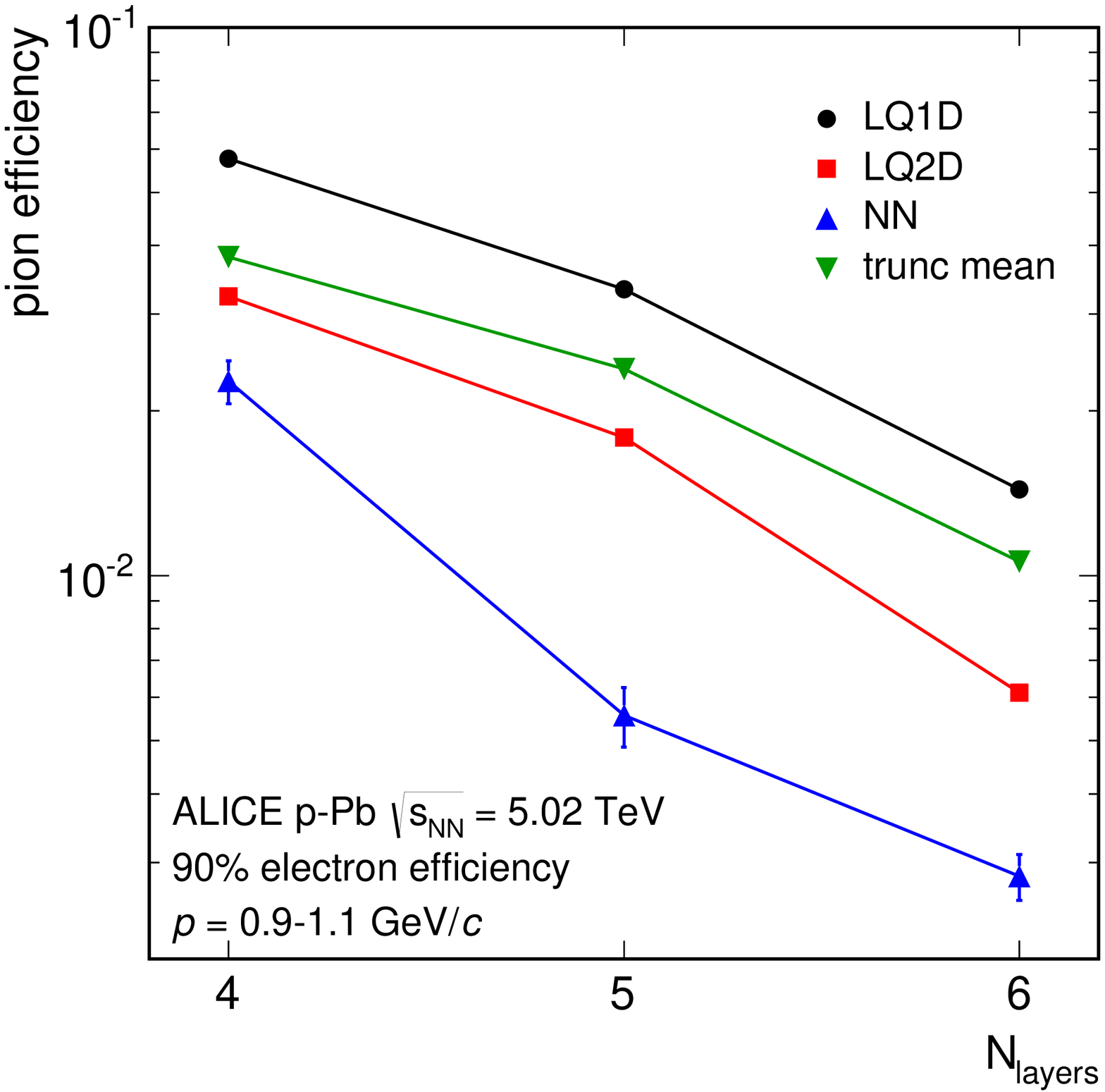} 
\end{minipage}  \end{tabular}
  \caption[]{Pion efficiency as a function of electron efficiency (left panel, 
for 6 layers) and as a function of the number of layers (right panel, for 90\% 
electron efficiency) for the momentum range 0.9--1.1\gevc. 
The results are compared for the truncated mean, LQ1D, LQ2D, and NN methods. }
   \label{TRD_PID}
\end{figure}

The above plot was produced using data collected in the recent \ppb run at 
\mbox{\sqrtsnn = 5.02 TeV}. The same data are used to quantify the TRD identification 
performance. Clean samples of electrons from $\gamma$ conversions and pions from \kzeros 
decays~\cite{MFasel} are selected using topological cuts and TPC and TOF particle identification. 
The performance of the detector is expressed in terms of the pion efficiency, 
which is the fraction of pions that are incorrectly identified as electrons. 
The pion rejection factor is the inverse of the pion efficiency.
We employ the following methods:
(i) truncated mean~\cite{blumrolandi,xlu}; (ii) one-dimensional likelihood on the 
total integrated charge (LQ1D)~\cite{MFasel}; (iii) two-dimensional likelihood 
on integrated charge (LQ2D)~\cite{dlohner}; and (iv) neural networks (NN)~\cite{Adler:2005rn}. 
The results are compared in Fig.~\ref{TRD_PID}, where the pion efficiency 
is shown as a function of the electron efficiency and as a function of the 
number of layers providing signals.
The truncated mean and the LQ1D are simple and robust methods which provide 
reasonable pion rejection. The LQ2D and NN methods also make use of the 
temporal distribution of the signal, which provides about a factor of two 
improvement of the pion rejection compared to the truncated mean and LQ1D methods.
The present pion rejection factors obtained from collision data confirm the 
design value found in test beams with prototypes~\cite{Bailhache:2006hs}.

\begin{figure}[ht]
   \centering
\includegraphics[width=.53\textwidth]{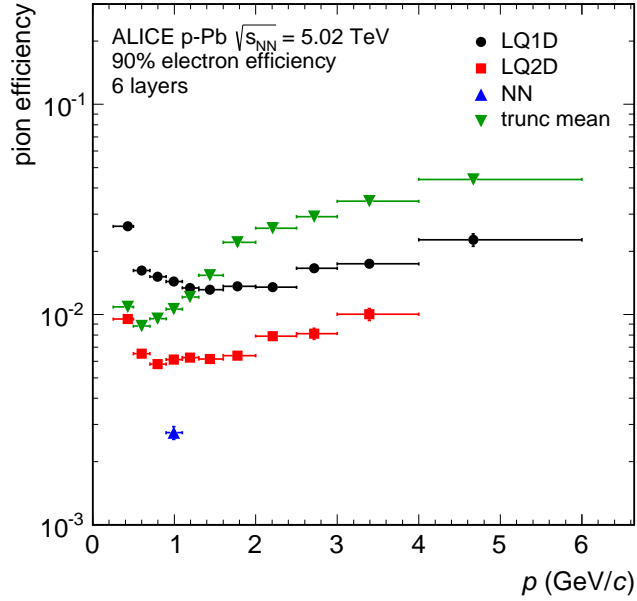} 
  \caption[]{Momentum dependence of the pion efficiency for the truncated mean, LQ1D, LQ2D, 
  and NN methods. The results are for 90\% electron efficiency and for tracks 
  with signals in six layers.}
   \label{TRD_PIDvsmom}
\end{figure}

The momentum dependence of the pion efficiency is shown in Fig.~\ref{TRD_PIDvsmom}. 
The pion rejection with the LQ1D and LQ2D methods first improves with increasing 
momentum because of the onset of the transition radiation. Starting from 1--2\gevc, 
the saturation of the TR production and the relativistic rise of the specific energy 
loss of pions lead to a gradual reduction of the electron--pion separation power. 
The LQ2D method lacks necessary references for momenta above 4\gevc. Studies 
with parametrizations of the respective charge-deposit distributions are ongoing 
and the first results look promising. 
The truncated-mean method shows very good pion rejection at low momenta where the 
energy loss dominates the signal. At higher momenta, the rejection power decreases 
because the TR contribution, yielding higher charge deposits, is likely to be removed 
in the truncation.

In addition to the identification efficiency, there is a finite matching efficiency 
between TPC tracks and TRD clusters, which is $\geq 85 \%$  for $\pt>0.8$ in the 
azimuthal area covered by the TRD. Losses are mostly due to chamber boundaries.

\subsection{Electron identification in physics analysis}
\label{sect:electron_pid_analysis}
One of the important uses of electron identification in physics
analysis is the measurement of the electron
spectrum from semileptonic decays of heavy-flavor hadrons. For this
measurement, a very pure electron sample is selected, using a
combination of various detectors, such as ITS+TPC+TOF+TRD, or
EMCal+TPC. 

\begin{figure}[t]
\centering
\includegraphics[width=0.7\textwidth]{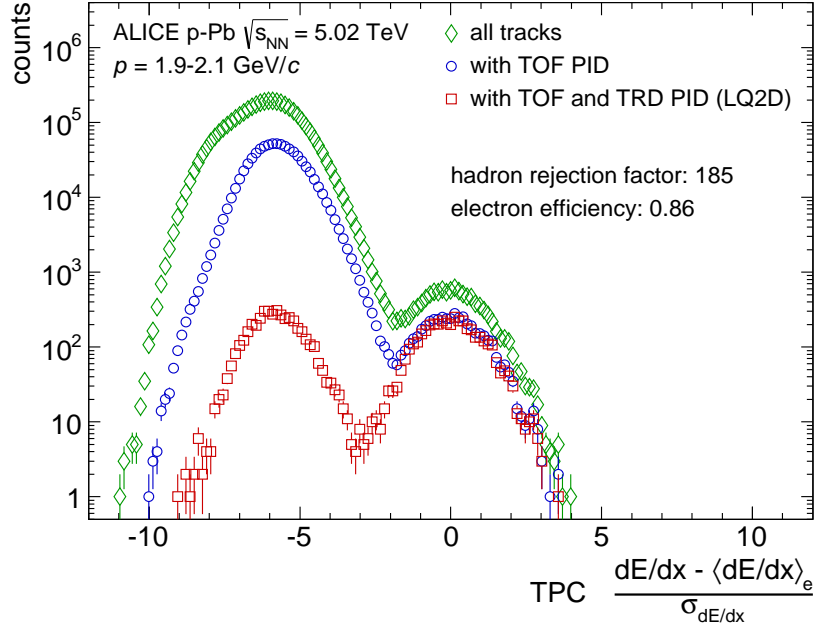}
\caption{\label{fig:combPIDelectrons}\dedx distribution of electron candidate 
tracks, with TOF and TRD selections (using 6 tracklets in the TRD) in \pp 
collisions. Only tracks with six TRD tracklets are included.}
\end{figure}

To illustrate the strength of combined PID for electrons, we show in
Fig.~\ref{fig:combPIDelectrons} 
the TPC \dedx distribution of tracks 
with $p=2\gevc$ and compare with track samples where cuts are applied on TOF 
and TRD to select electrons. It can be seen in the figure that the TOF and
TRD cuts reduce the hadron contamination in the track sample, allowing
the selection of a very pure electron sample when combined with TPC \dedx.
For details, we refer to the corresponding publication~\cite{Abelev:2012xe}. 

Another illustrative case for the application of electron identification is
the reconstruction of the decay of the \jpsi meson into an electron
and a positron. In this case, rather loose selection cuts are applied 
on electrons, since the hadronic contamination only enters in the
combinatorial background in the invariant mass distribution.

\begin{figure}[t!]
\centering
\includegraphics[width=0.72\linewidth]{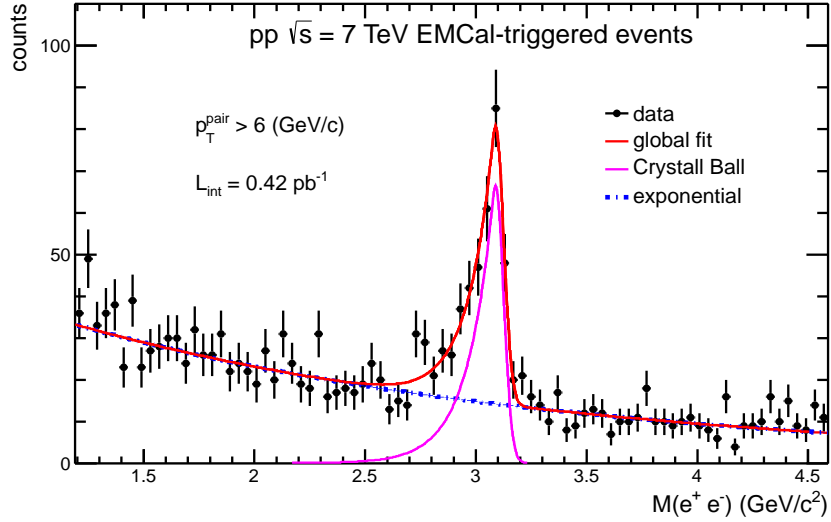}
\caption{Invariant mass distribution for \jpsi candidates from
EMCal-triggered events in \pp collisions at \mbox{\sqrts = 7 TeV}
($\mathcal{L} \approx 0.4$~pb$^{-1}$, 8M events). Electrons are
identified by their energy loss in the TPC ($\dedx > 70$) and the $E/p$ 
ratio in the EMCal ($0.9<E/p<1.1$) for both legs. A fit to the signal
(Crystal Ball\protect~\cite{Gaiser:1982yw}) and the background (exponential) 
is shown in addition.}
\label{fig:EMCalelectronPID_3} \end{figure}

Figure~\ref{fig:EMCalelectronPID_3} shows the invariant mass
distribution of \jpsi candidates decaying into \ee,
identified using the EMCal. In this analysis, electrons from EMCal-triggered
events are identified by a combination of TPC energy loss
and the $E/p$ ratio. This allows the extension of the \pt
interval and leads to a better $S/B$ ratio. More analysis details and results 
can be found in Refs.~\refcite{Abelev:2012kr,Abelev:2012gx}.

\begin{figure}[t]
   \centering
\includegraphics[width=.62\textwidth]{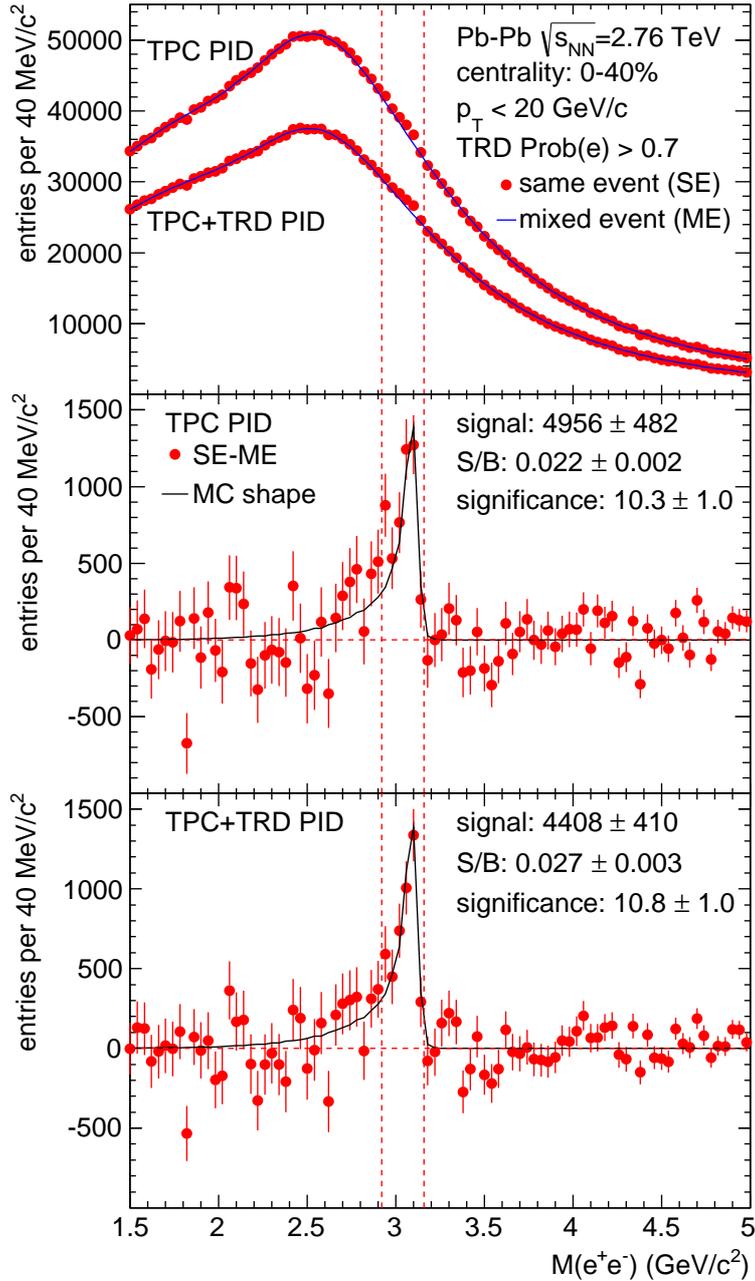} 
  \caption[]{\ee invariant-mass distribution with TPC-only as well as TPC and 
TRD particle identification in 0--40\% centrality in \pbpb collisions at 
\mbox{\sqrtsnn = 2.76 TeV}.}
   \label{TRD_JPsi}
\end{figure}

In Fig.~\ref{TRD_JPsi} we show the effect of the TRD electron
identification for the \jpsi measurement in the 40\% most
central \pbpb collisions at \mbox{\sqrtsnn = 2.76 TeV}.  
In both the TPC-only and the 
TPC+TRD combined analysis, electrons were identified through their specific 
energy loss in the TPC, applying a $(-1.5\sigma,+3\sigma)$ inclusion cut.  
Pions and protons are rejected via $\pm3.5\sigma$ and $\pm4\sigma$ exclusion 
cuts, respectively. For the TPC-only analysis, the total number of candidates 
after background subtraction is 4956$\pm$482 \jpsi in the invariant mass region 
2.92--3.16\gevcc, with a signal-to-background ratio of $0.022 \pm 0.002$ and 
a significance of $\sim$10. 

In the TPC+TRD combined analysis, the LQ2D method was applied, requiring an 
electron likelihood of at least 0.7. 
For the data shown here, collected in year 2011, the TRD had only 
partial coverage (10 out of the 18 TRD supermodules were installed).
Thus the TRD particle identification was used whenever a candidate \jpsi leg 
had a signal in at least 4 TRD layers.  Despite reduced coverage, the signal 
to background ratio improved by roughly 20\% compared to the TPC-only analysis. 
The impact of TRD on the significance of the \jpsi yield is small but
will increase once all 18 TRD supermodules have been installed. 

To significantly enrich the quarkonium sample, the TRD detector was 
used to select events with electrons at the trigger level 1 
(see Section~\ref{sect:datataking}). 
For this, track segments (tracklets) were reconstructed locally 
in the front-end electronics mounted on each chamber.
The tracklets were calculated as a straight line fit through the positions 
of the clusters, determined taking into account the pad response function. 
The tracklets from different TRD layers are combined using again a straight 
line fit and the transverse momentum was determined for tracks which were 
detected in at least four TRD layers. 
The \pt resolution was better than 20\% over the target \pt range of 2--8\gevc. 
For the particle identification, the total charge of each tracklet was 
translated into an electron probability by a look-up table based on 
reference data with clean electron and pion samples. 
Pad-by-pad gain variations were corrected for in the front-end electronics, 
based on Kr calibration. 
To ensure stable drift velocity and gas amplification, a feedback system 
was implemented to compensate for environmental changes (mostly of the 
pressure) by high voltage adjustments. 
A global electron probability was calculated by averaging over the contributing 
tracklets. For an electron efficiency of 40\%, a pion rejection factor of 200 
was achieved in \pp collisions. 
The dominant background was from (low-\pt) photons, which convert into \ee at large 
radii and thus produce electrons with small apparent deflection. 
For an overview of the TRD trigger see Ref.~\refcite{Klein:2011hs}. 

\clearpage\newpage
\newif\ifcomment

\section{Photons}
\label{sect:photons}
Photon identification at midrapidity in ALICE is performed 
either by reconstructing the electromagnetic shower developed in the PHOS and EMCal calorimeters, 
or by reconstructing electron-positron pairs originating from photons converted in the material of the inner 
detector (``conversion electrons'') with the ITS and TPC using the Photon Conversion Method~(PCM).

\subsection{Photon reconstruction with calorimeters}
The central barrel of the ALICE setup contains two calorimeters for photon detection: 
the Photon Spectrometer (PHOS)~\cite{Dellacasa:1999kd,Zhou:2007zzh} 
and the Electromagnetic Calorimeter (EMCal)~\cite{Abeysekara:2010ze}. 
Both calorimeters have cellular structure with square cells with a transverse size of 
$2.2\times2.2$ cm in PHOS (``crystals'') and $6\times6$ cm in EMCal (``towers''), 
which is roughly equal to (or slightly larger than) the Moli{\`e}re radius. 
With this choice of cell size, the electromagnetic showers produced by photons and electrons cover 
groups of adjacent cells (clusters). The material budget of the cells along the particle
path is $20X_0$ which is sufficient for photons, electrons, and positrons with about 100 \gevc 
to deposit their full energy. For hadronic interactions, the thickness of the cells is about one nuclear 
radiation length, i.e.\ the calorimeters are rather transparent for hadrons. The energy deposited 
by hadrons is small compared with their full energy (see Fig.~\ref{fig:EMCalelectronPID_1}). 

The cells of the calorimeters are packed into rectangular matrices called modules in PHOS and 
supermodules in EMCal. As of 2012, the PHOS detector consists of 3 modules of 
$64 \times 56$ cells each~($|\eta|<0.12$, $260\degr<\phi<320\degr$), and the EMCal contains 
10 supermodules of $48 \times 24$ cells and 2 supermodules of $48 \times 8$ 
cells~($|\eta|<0.7$, $80\degr<\phi<187\degr$).

Below, we briefly discuss the cluster finding methods and the photon reconstruction 
performance of \mbox{EMCal} and \mbox{PHOS}. The electron identification capabilities of the two 
calorimeters are described in Section~\ref{sect:electrons}. 

\subsubsection{Cluster finder in PHOS}
In PHOS, the cluster finding algorithm starts from any cell with a measured amplitude above some 
threshold, referred to as the seed energy, $E_{\rm seed}$~\cite{Alessandro:2006yt}. The choice of this 
seed energy depends on the event environment. In \pp collisions the occupancy of the PHOS detector is low, 
and thus the probability of showers overlapping is small. The seed energy is set to 
$E_{\rm seed}=0.2$~GeV, slightly below the MIP threshold. 
In the high-multiplicity environment of \mbox{\pbpb} collisions, the overlap 
probability becomes significant. In order to suppress the hadronic background the seed energy is set to 
$E_{\rm seed}=0.4$~GeV. Cells with an energy above the noise level, which share a common edge with the seed 
cell, are added to the cluster. Subsequently, further cells above the noise level are added if they are 
adjacent to cells that have already been added.

Clusters can be produced either by a single electromagnetic or hadronic shower, or by 
several overlapping showers. In the latter case, the cluster may have distinct local maxima, i.e.\
cells with large energy separated by at least one cell with smaller energy. The presence of such 
local maxima in a cluster initiates cluster unfolding, which is a procedure that separates the 
cells of the primary cluster from several clusters corresponding to individual particles. 
The cluster unfolding algorithm is based on the knowledge of the transverse 
profile of electromagnetic showers.

\subsubsection{Cluster finder in EMCal}
\label{sect:EMCalClustering}
Due to the larger cell size in EMCal compared to PHOS, 
the cluster finding algorithm in EMCal varies depending on the event environment~\cite{Alessandro:2006yt}. 
The default algorithm is the same as that implemented in PHOS, used with a seed energy of 
$E_{\rm seed}=0.3$~GeV, slightly above the MIP threshold. 
At pion transverse momenta $\pt>6$\gevc, showers from decay photons of \pizero start to overlap, thus 
reducing the performance of the \pizero reconstruction. For such overlapping clusters, a slightly modified 
version of the cluster finding algorithm stops adding cells at the first local minimum to avoid shower merging 
from the two decay photons. An alternative algorithm, originally developed for heavy-ion collisions where the 
cell occupancy of the EMCal detectors is high, uses a fixed shape of $3 \times 3$ cells centered around the 
seed cell. 

\subsubsection{Cluster parameters}
Clusters found in the calorimeters are characterized by several parameters. Since photons and electrons are 
expected to deposit their full energy in the PHOS and EMCal, the sum of cell energies $e_i$ is 
used as the estimator of the photon or electron energy~$E = \sum_{i=1}^{N} e_i$.
The photon coordinate $\bar{x}$ in the reference system of the module can be determined as the first 
moment of the coordinates $x_i$ of the cells contributing to the cluster, weighted by the logarithms of 
the cell energies
$w_i = \max\left[0,w_0 + \log(e_i/E)\right]$ with $w_0=4.5$. 
For inclined photons, the center of gravity 
of the shower is displaced towards the inclination direction. As the actual incidence angle 
of photons is not known, one assumes that all detected photons are produced in the primary 
vertex, meaning that the incidence angle is determined geometrically from the photon hit coordinate.
The shape of showers which develop in the calorimeters can be characterized by the eigenvalues 
$\lambda_0$, $\lambda_1$ of the covariance matrix built from the cell coordinates and weights 
$w_i$~\cite{Alessandro:2006yt}, and may be used to differentiate between different incident particle 
species.
A cluster can be further characterized by the time of flight of a particle from the 
interaction point to the calorimeter, which is selected as the shortest time among the digits 
making up the cluster.

For PHOS, another cluster parameter defined for high-multiplicity environments using the 
cluster cell content is the core energy.
The core energy is given by the sum of cell energies within a circle of radius $R=3.5$~cm around the 
cluster coordinate, where $R$ is defined such that 98\% of the electromagnetic shower energy is deposited 
within this circle. 

\subsubsection{Photon identification in calorimeters}
\begin{figure}[hbt]
\includegraphics[width=0.4\hsize]{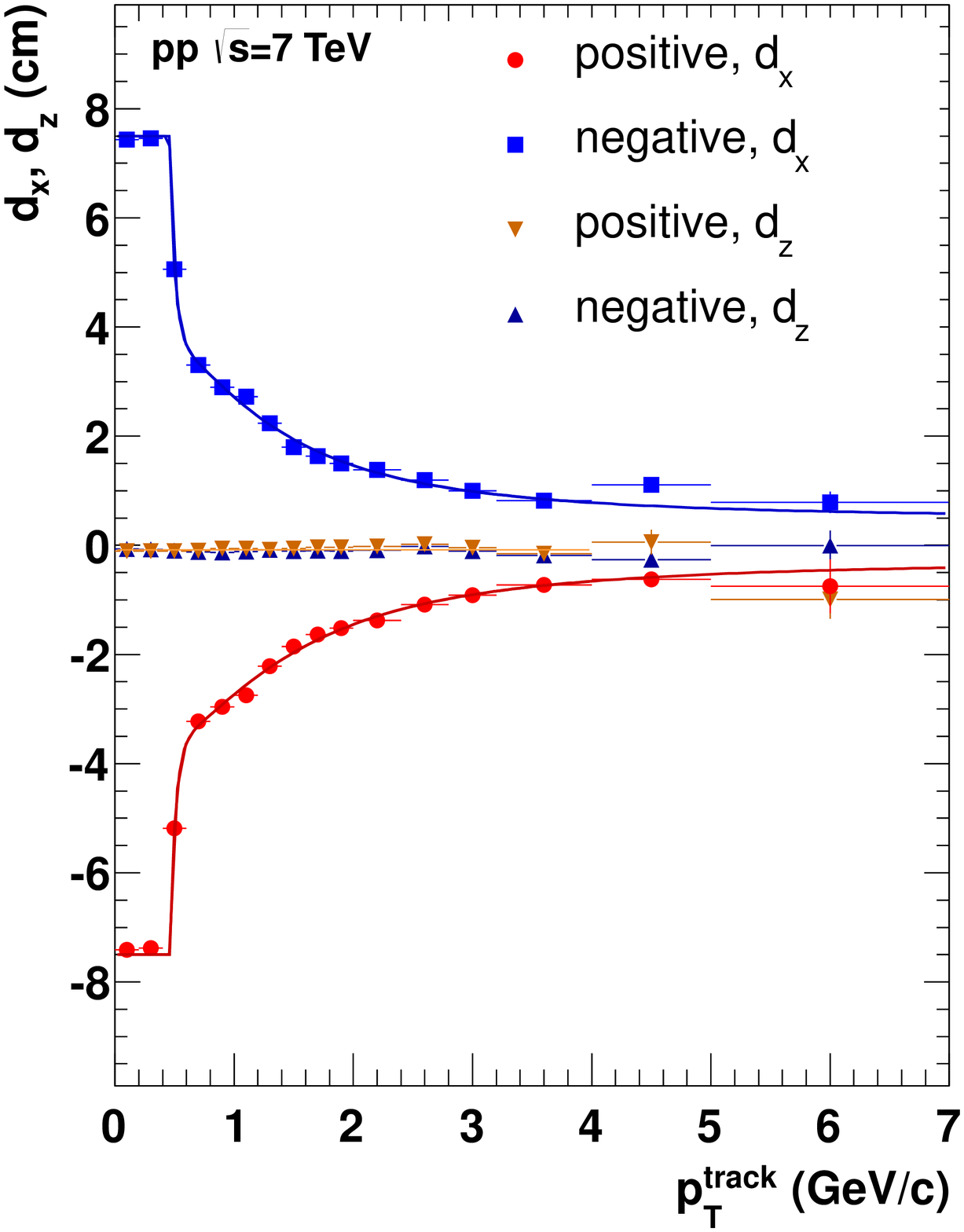}
\hfil
\includegraphics[width=0.4\hsize]{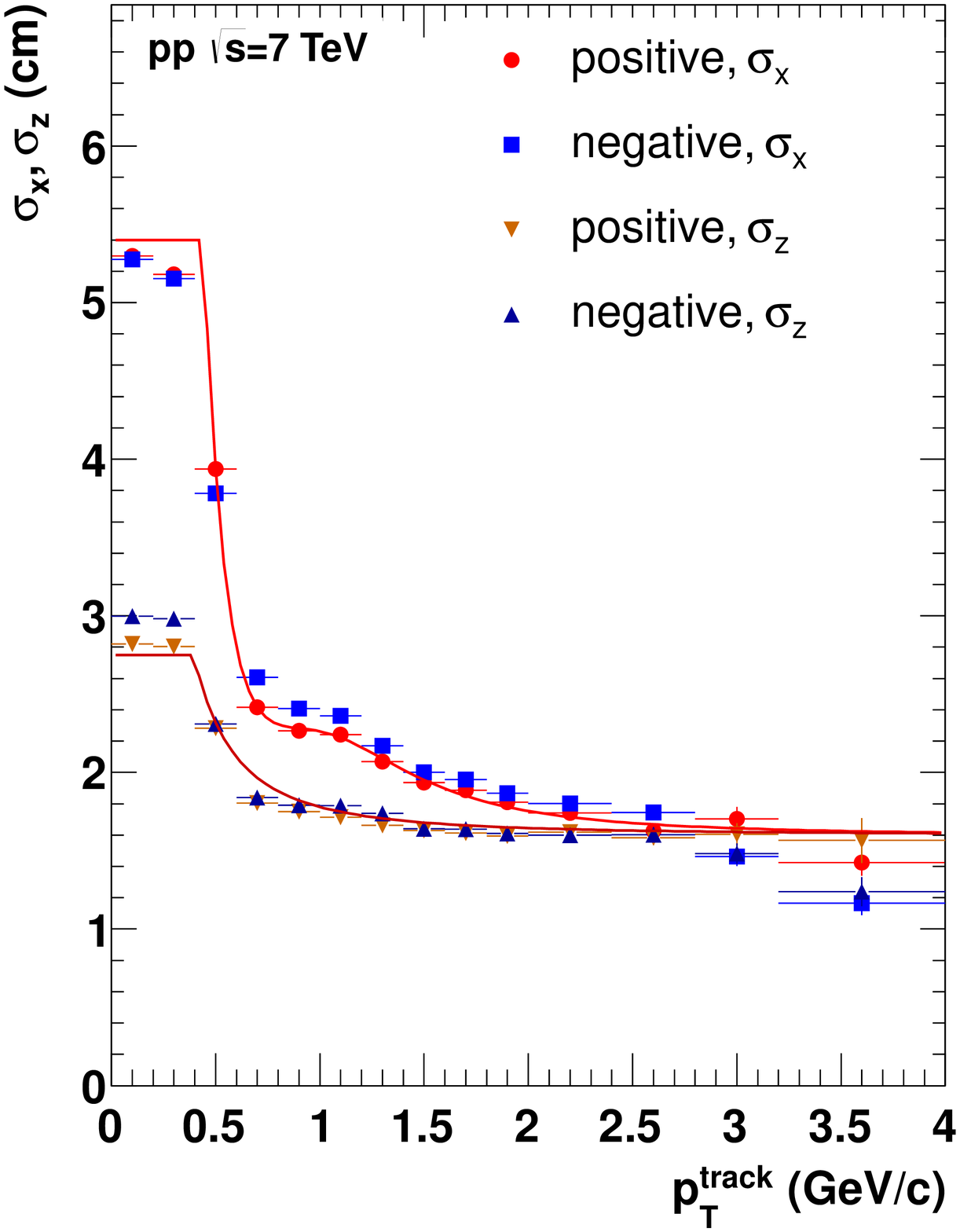}
\caption{Mean track matching distance (left) and RMS of the track matching distance 
distribution (right) for PHOS. The lines are fits to phenomenological parameterizations.}
\label{fig:sec11:PIDCPV}
\end{figure}

Photon identification in the calorimeters is based on three complementary criteria:
\begin{enumerate}
\item Since photons cannot be traced by the tracking system, a cluster with no reconstructed 
tracks in the vicinity (as propagated to the calorimeter surface) is considered as a neutral 
particle candidate.
\item Showers produced in the active calorimeter medium by photons and hadrons differ by the 
transverse profile. Shower shape parameters $\lambda_0$, $\lambda_1$, $E_{\rm core}$ are used 
to discriminate electromagnetic showers from hadronic ones.
\item The time-of-flight information of the cluster can be used to identify fast particles and suppress 
clusters produced by nucleons.
\end{enumerate}
Neutral particle identification is based on the distance between the cluster 
center and the nearest charged particle track at the face of the calorimeter. 
As the calorimeter signal for charged hadrons is generated at a finite depth, 
the centroids of the cluster--track matching distributions are systematically 
shifted as shown in the left panel of Fig.~\ref{fig:sec11:PIDCPV} for PHOS. 
Knowing the positions and widths (right panel of Fig.~\ref{fig:sec11:PIDCPV}) 
of these distributions, one can recognize and suppress clusters produced by 
charged hadrons. The selection parameters for PHOS and EMCal depend on the 
cluster energy and the purity of the photon sample required for particular 
analyses. Typical values for the selection are 0.005 in the azimuthal 
and 0.003 in pseudorapidity direction.

The shower shape helps in distinguishing between showers produced by single photons, hadrons, 
and photons from the decay of high-momentum \pizero. The latter is more relevant 
for EMCal, in which photons from the \pizero decay start overlapping from $\pt>6$\gevc. 
Single photons tend to have spherically shaped showers, while the clusters with merged showers 
from high-\pt \pizero decays are elongated. The elongation is quantified by the parameter 
$\lambda_0^2$, which is the weighted RMS of the shower energy along the major ellipse
axis. For photons the typical value of this shower shape parameter $\lambda_0^2$ is around 0.25 
independent of the cluster energy, while for \pizero 
it has a value of $\lambda_0^2\approx 2.0$ for $\pt\sim$6\gevc and decreases to
$\lambda_0^2\approx 0.4$ at $\pt\sim$30\gevc, allowing for good discrimination between 
these two kinds of clusters. This feature is especially interesting for the identification of high-momentum 
\pizero{s} because the invariant mass method (see Section~\ref{subsec11:pi0}) has low efficiency 
above $\pt>20$\gevc for EMCal and $\pt>60$\gevc for PHOS.

\begin{figure}[b]
\centering 
\includegraphics[width=0.7\textwidth]{{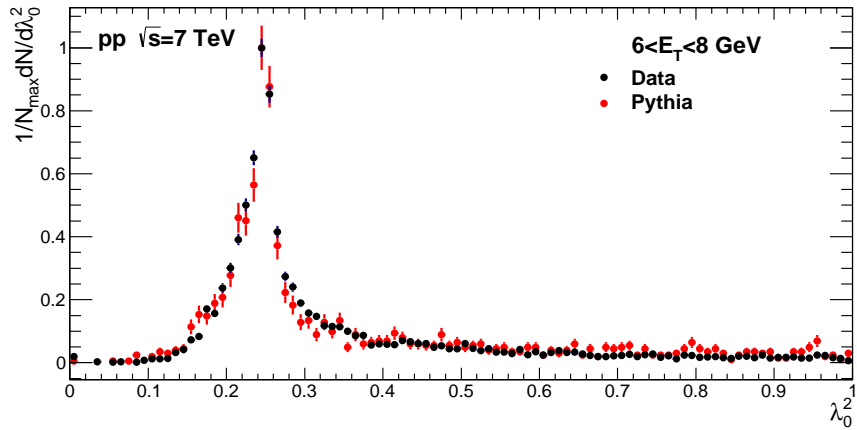}}
\caption{$\lambda_0^2$ distribution of photon clusters in the EMCal with transverse energy of 
$6\gevc<E_{\rm T}<8\gevc$ originating from ``semi-converted'' \pizero{s} in \pp collisions at 
7 TeV compared to Monte Carlo simulation.}
\label{fig:sec11:M02}
\end{figure}

To test the quality of the photon identification with the EMCal, \pizero{s} with one 
of the decay photons converting in the inner material of the experiment (see Sect.~\ref{subsec11:PCM}) and 
the other decay photon reaching the EMCal~(semi-converted \pizero) are used to select a photon-enriched 
sample of clusters. This is achieved by
reconstructing the invariant mass of the cluster-conversion pairs and selecting those clusters whose 
pair masses lie in the \pizero mass range. The $\lambda_0^2$ distribution of these clusters 
is compared to Monte Carlo simulations in Fig.~\ref{fig:sec11:M02}. 
In this simulation, Pythia~\cite{Sjostrand:2006za} events are fully reconstructed in the ALICE
experiment and subject to standard analysis cuts. The two distributions show satisfactory agreement.
The application of these criteria depends on the specific physics analysis being undertaken. 
For processes with a high signal-to-background ratio, one of the criteria may be sufficient to reach 
an adequate purity, while in other cases it may become necessary to combine all three photon 
identification methods.

\subsection{Photon Conversion Method}
\label{subsec11:PCM}
At energies above 5 MeV, the interaction of photons with detector material is dominated by 
the creation of positron--electron (\ee) pairs~\cite{Alessandro:2006yt}.
The converted photon and its conversion point can be reliably measured by reconstructing the 
electron and positron with the ITS and TPC for conversions within $180$~cm 
from the beam axis. Within the fiducial acceptance ($|\eta|<0.9$) the main sources for 
conversions are the beam pipe, 
the $6$ layers of the ITS, the TPC vessels, and part of the TPC drift gas. Outside the fiducial 
acceptance, the ITS services and the ITS and TPC support structures lead to additional 
contributions. The photon conversion probability is very sensitive to the amount, 
geometry, and chemical composition of the traversed material. Therefore, it is vital to have 
accurate knowledge of the material budget before photon production can be assessed quantitatively. 

\begin{figure}[h]
\centering
\includegraphics[width=0.6\textwidth]{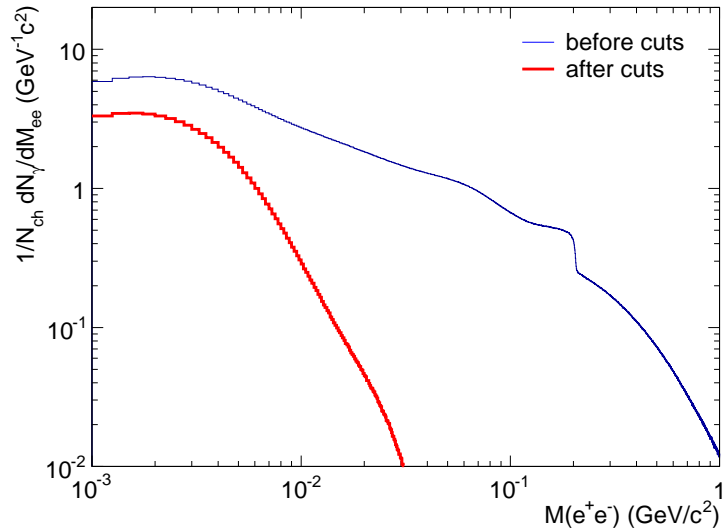}
\caption{Invariant mass distribution of all reconstructed secondaries (blue) and of the 
selected photon candidates (red) after all cuts were applied.}
\label{fig:sec11:v0invmass}
\end{figure}

The converted photons are obtained by employing a secondary vertex algorithm (V$^0$ finder), as 
explained in Section~\ref{sect:secondary}. The same algorithm is used to reconstruct 
$\kzeros, \Lambda, \bar{\Lambda}$, and $\gamma$ conversions from reconstructed tracks. 
In order to obtain a clean photon sample, the PID capabilities of the TPC and TOF are exploited 
as described in Section~\ref{sect:electrons}. 
Electron and positron track candidates are selected by requiring the specific energy loss 
d$E$/d$x$ in the TPC and the time of flight in TOF to be within 
$(-4\sigma_{\dedx}, +5\sigma_{\dedx})$ and $(-2\sigma_{\rm TOF}, +3\sigma_{\rm TOF})$, 
respectively, from the values expected for electrons. 
Tracks close to the pion line in Fig.~\ref{fig:TPC_dEdx} -- within $(-0.5\sigma_{\dedx}, +0.5\sigma_{\dedx})$ 
and $(-\infty, +0.5\sigma_{\dedx})$ for momenta below and above 0.3\gevc, respectively -- 
are rejected.
The precision of the photon conversion point estimate can be improved with respect to the one obtained 
from the V$^0$ algorithm by requiring that the momentum vectors of the \ee pair are 
almost parallel at the conversion point. 
The final photons are selected by a cut of the $\chi^2(\gamma)$/ndf after applying constraints 
on the photon candidate mass and on the opening angle between the reconstructed photon momentum 
and the vector joining the collision vertex and the conversion point. 
The invariant mass distributions of all V$^0$s calculated with the electron mass hypothesis 
before and after all selection criteria are shown in Fig.~\ref{fig:sec11:v0invmass}.

\begin{figure}[b]
\centering
\includegraphics[width=0.6\textwidth]{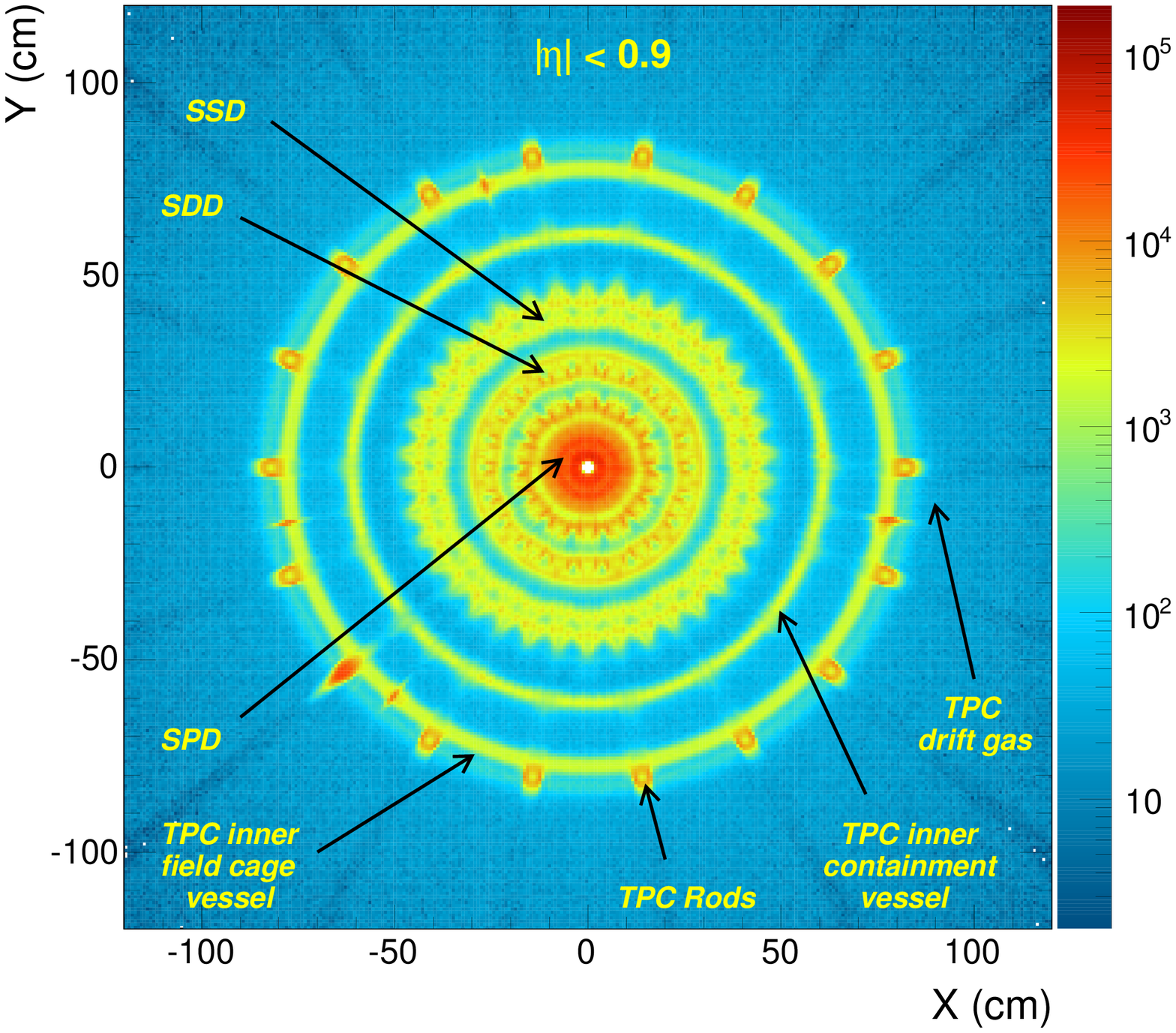}
\caption{Transverse distribution of the reconstructed photon conversion points for $|\eta|<0.9$.}
\label{fig:sec11:XYdist}
\centering
\includegraphics[width=0.62\textwidth]{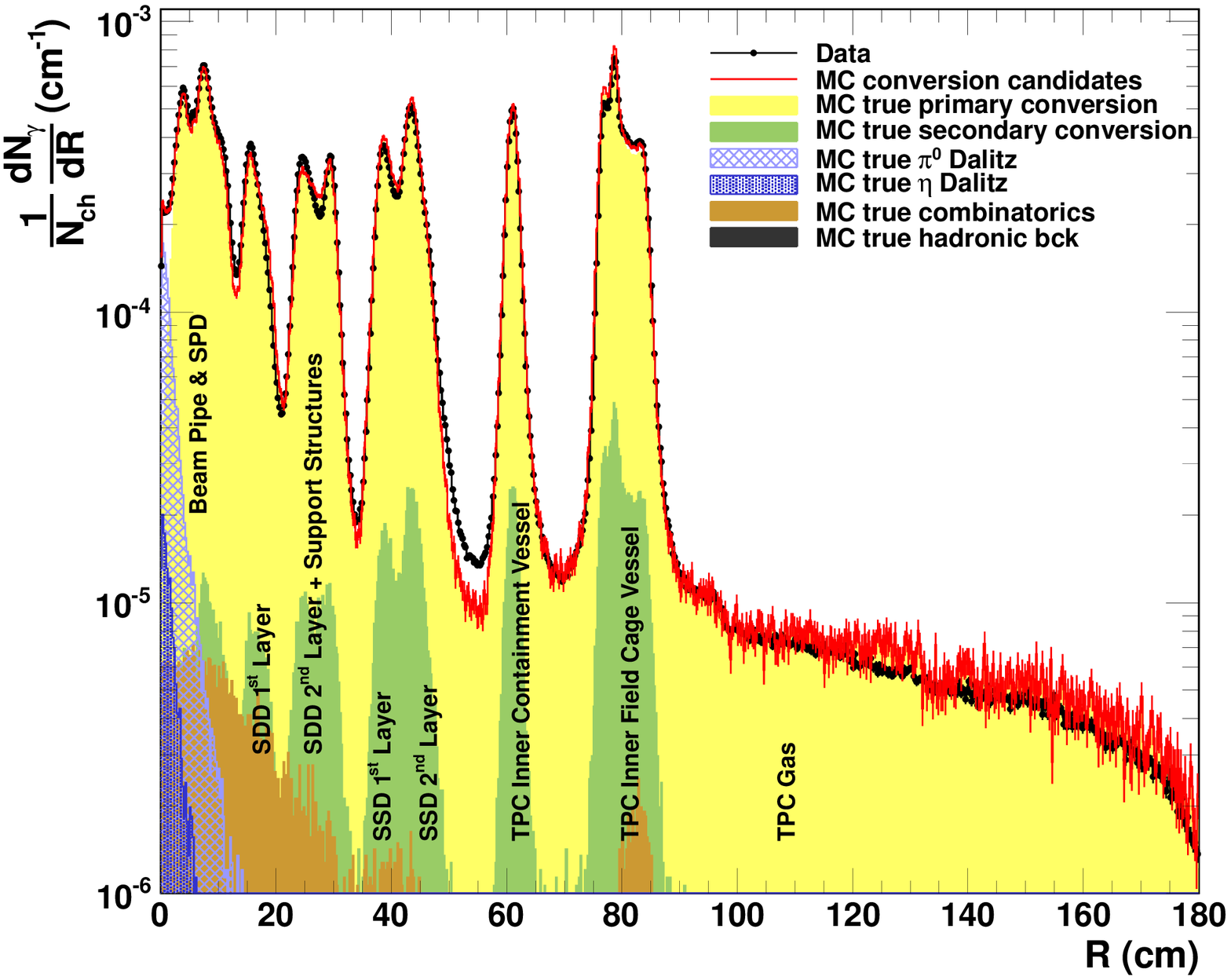}
\caption{Radial distribution of the reconstructed photon conversion points for $|\eta|<0.9$ (black) 
compared to MC simulations performed with PHOJET (red). Distributions for true converted photons 
are shown in yellow. Physics contamination from true \pizero and $\eta$ Dalitz decays, where 
the primary \ee are reconstructed as photon conversions, are shown as dashed blue 
histograms. Random combinatorics and true hadronic background are also shown.}
\label{fig:sec11:radialdist}
\end{figure}

The distribution of the reconstructed photon conversion points, shown in 
Figs.~\ref{fig:sec11:XYdist} and \ref{fig:sec11:radialdist} 
for $|\eta|<0.9$, represents a precise 
$\gamma$-ray tomography of the ALICE inner barrel detectors. Different layers of the ITS and 
the TPC are clearly separated. The radial distribution is compared to Monte Carlo (MC) simulations generated with 
PHOJET~\cite{Engel:1995yda}. The integrated detector material for $R<180$~cm and $|\eta|<0.9$ amounts 
to a radiation thickness of $11.4\pm0.5$\% $X_0$, and results in a conversion probability of about $8.5$\%. 
The differences between the measured and simulated distributions (apparent mainly at $R=50$~cm) 
are taken into account when estimating systematic uncertainties in the analyses that 
rely on the knowledge of the material.
Further details relating to the analysis of the ALICE material distribution, the photon conversion 
probability and reconstruction efficiency in the inner parts of the detector are discussed in 
Ref.~\refcite{nimpaperpcm}.

\subsection{\pizero and $\eta$ reconstruction}
\label{subsec11:pi0}
\begin{figure}[b]
\includegraphics[width=0.31\hsize]{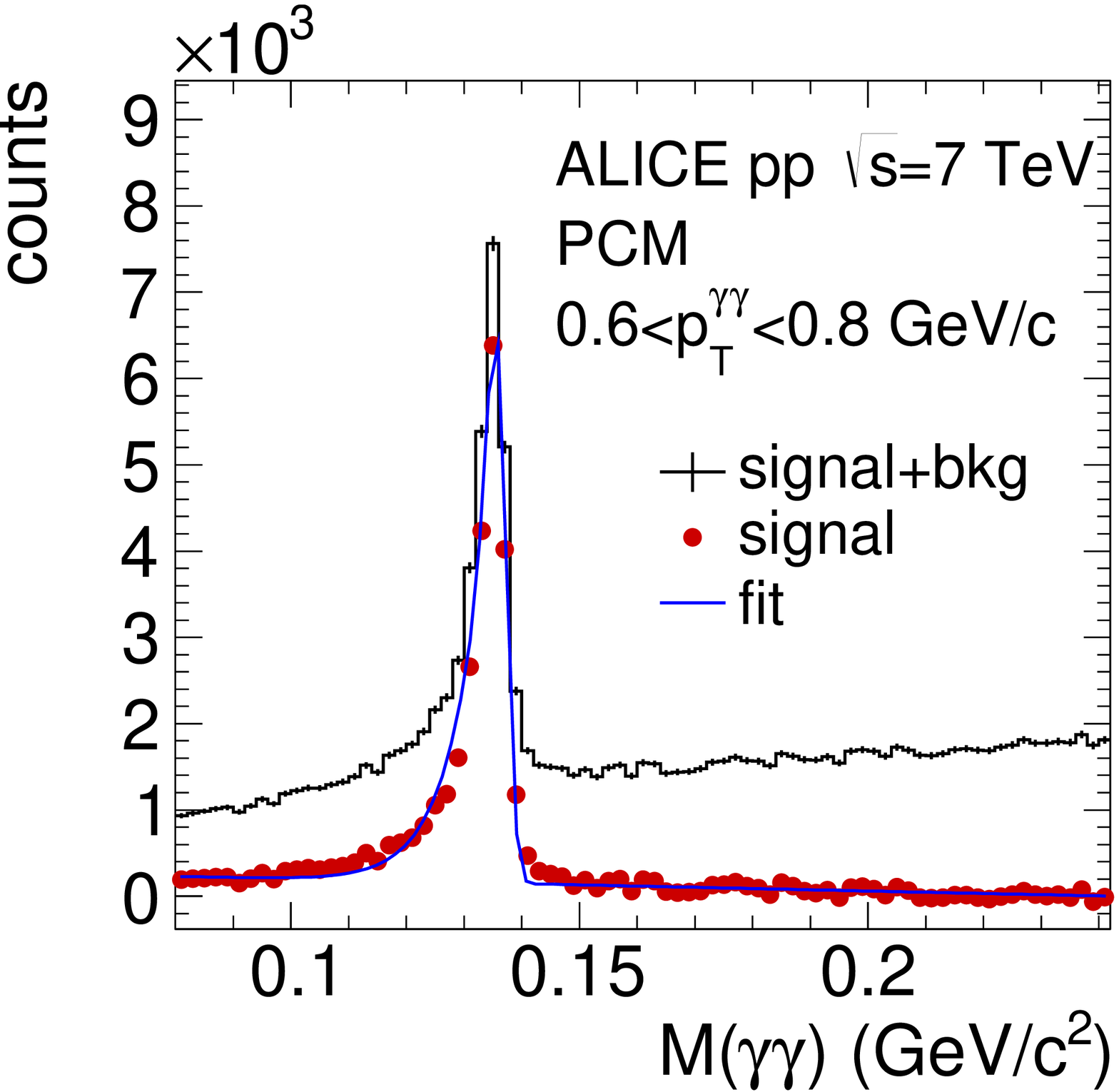}
\hfil
\includegraphics[width=0.31\hsize]{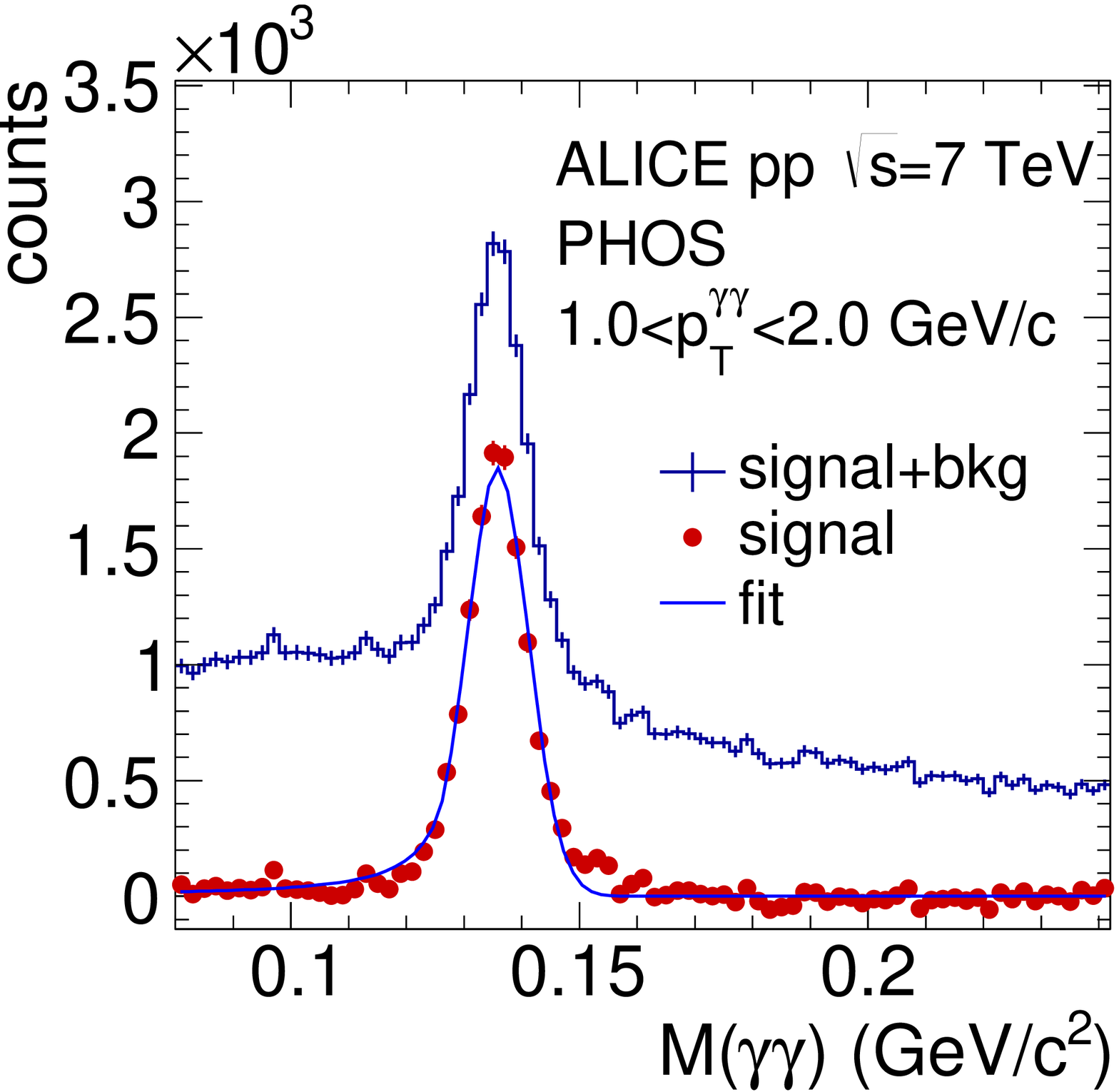}
\hfil
\includegraphics[width=0.31\hsize]{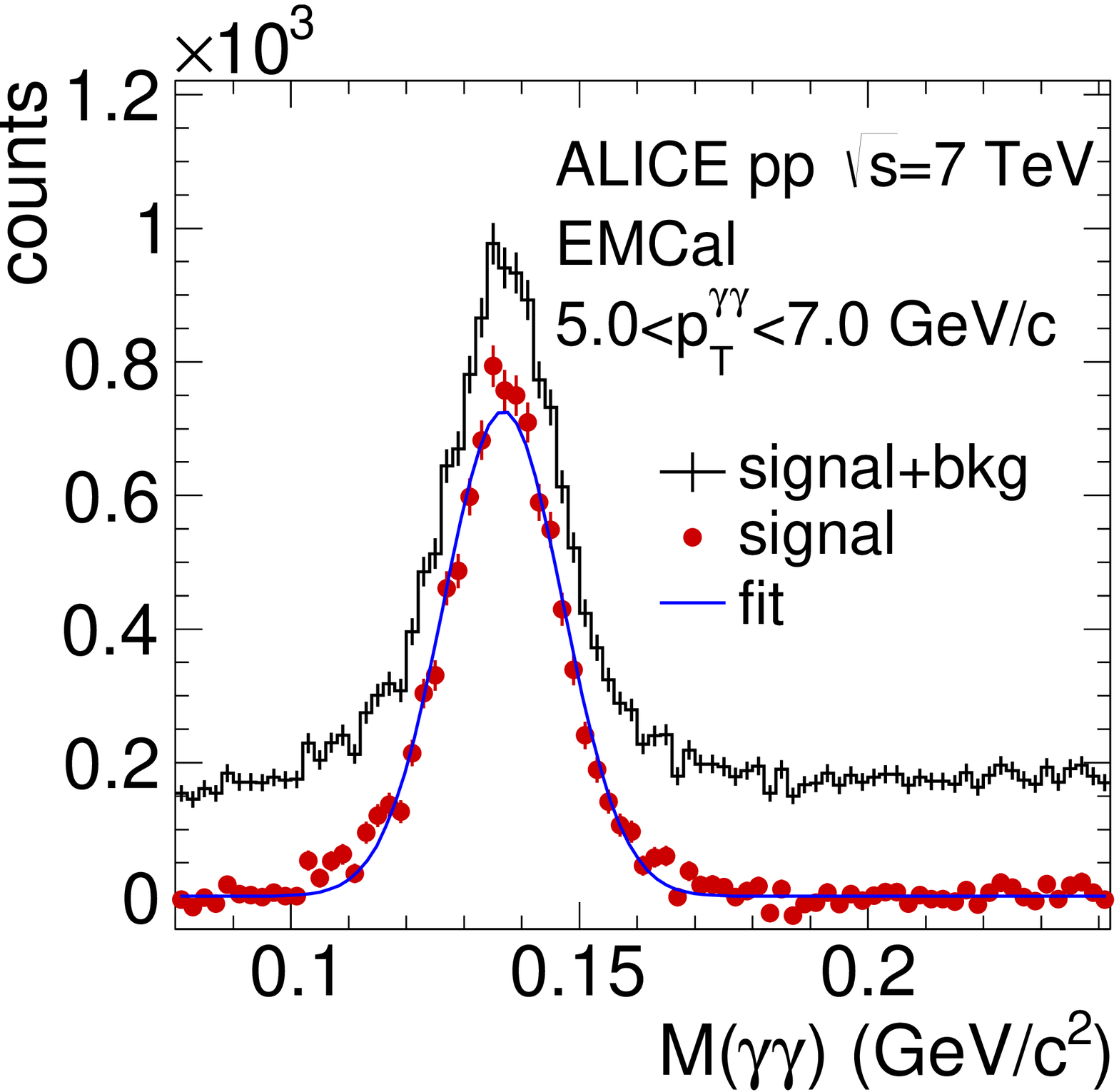}
\caption{Invariant mass spectra of photon candidate pairs for \pp collisions at 7 TeV by PCM, PHOS and EMCal.}
\label{fig:sec11:2gamInvMass}
\includegraphics[width=0.31\hsize]{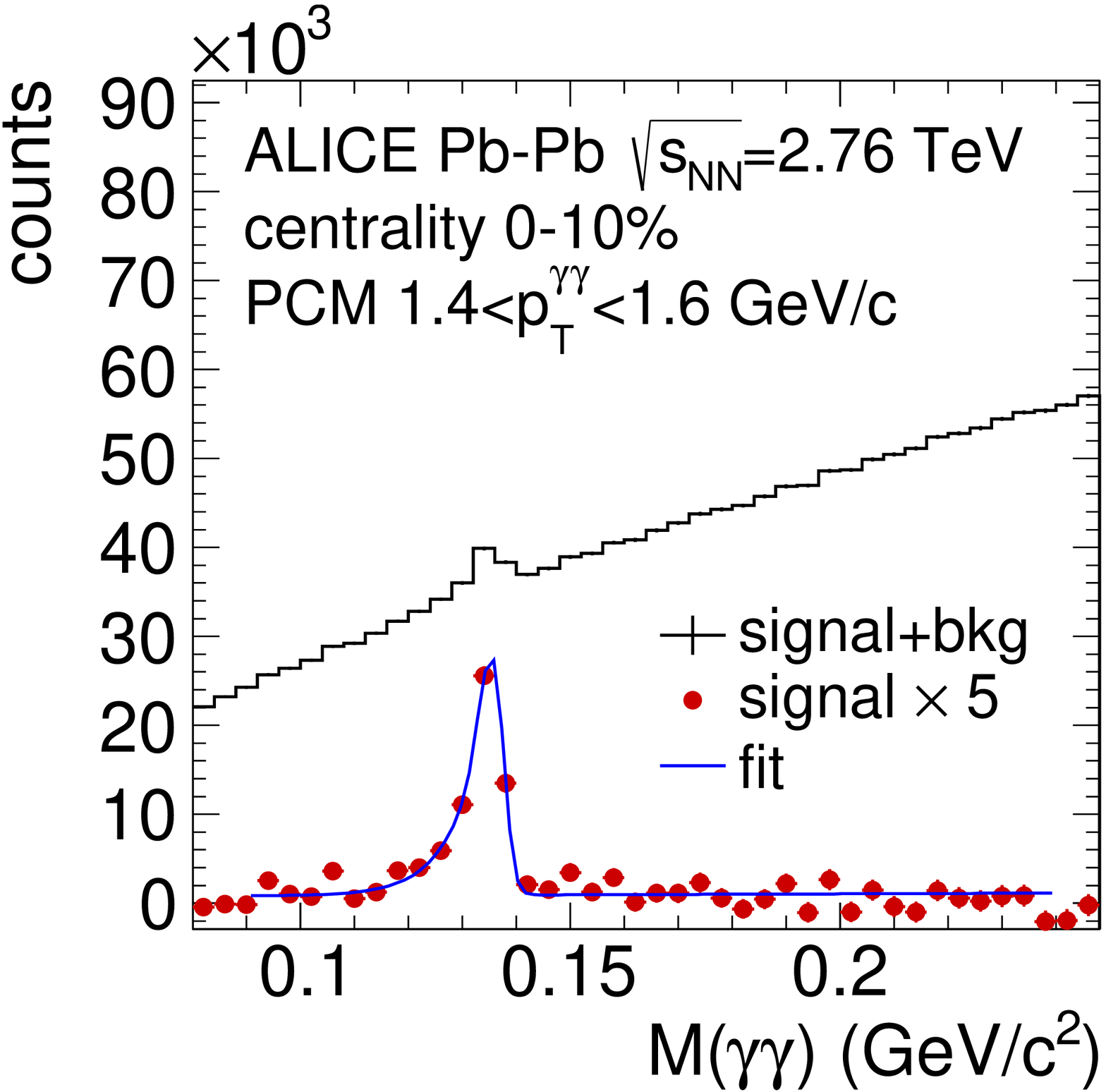}
\hfil
\includegraphics[width=0.31\hsize]{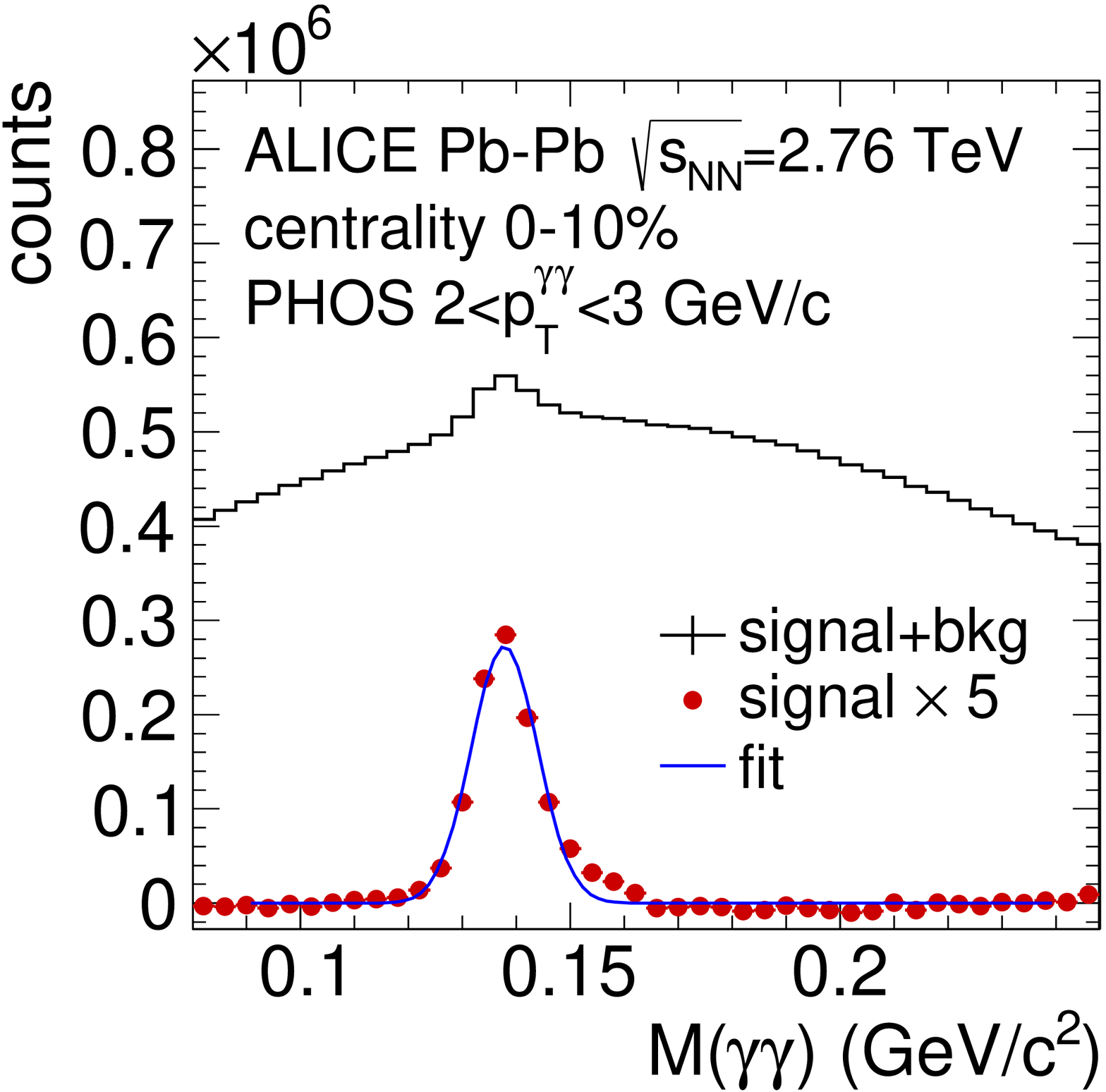}
\hfil
\includegraphics[width=0.31\hsize]{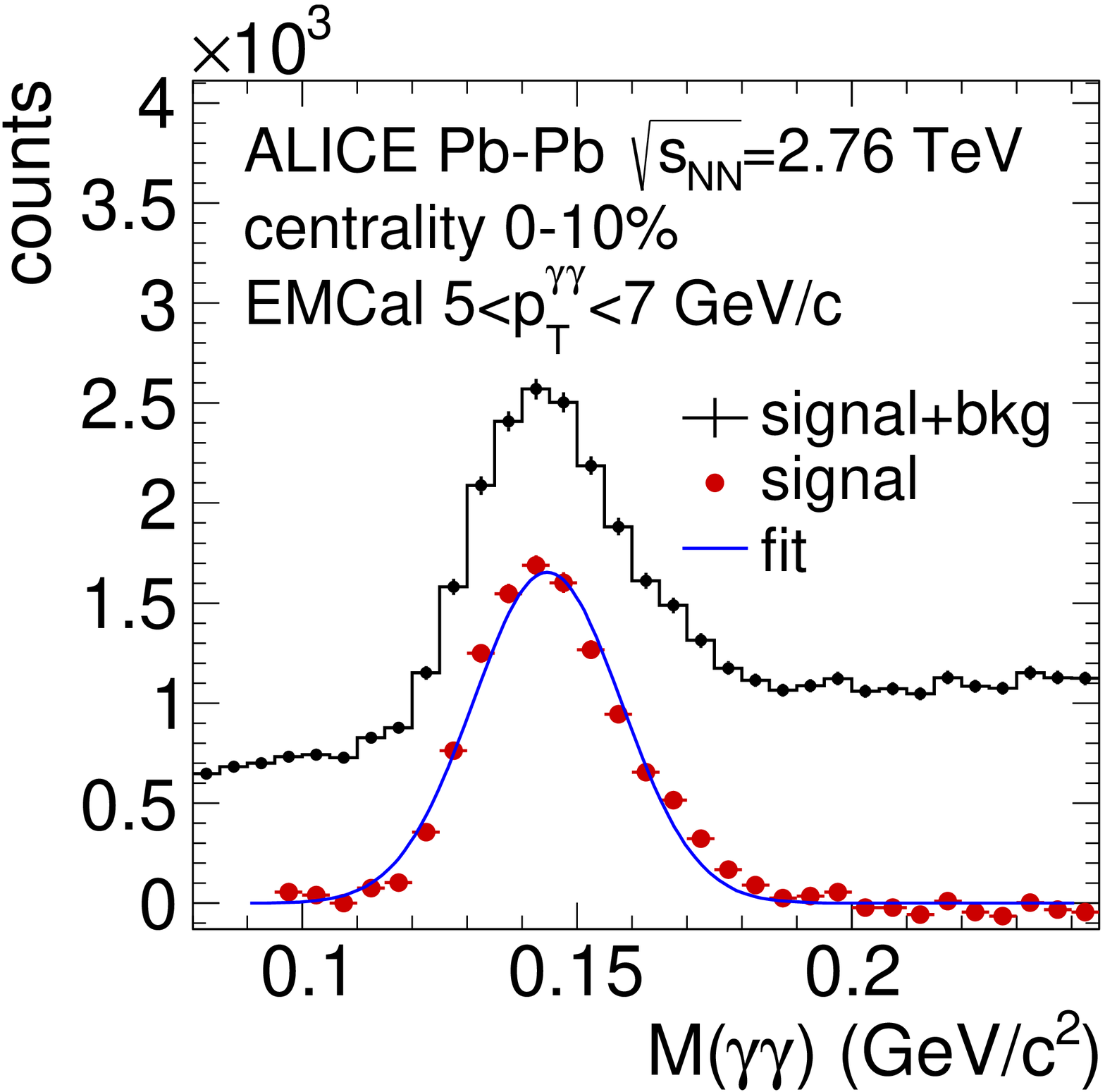}
\caption{Invariant mass spectra of photon candidate pairs for 0--10\% central \pbpb collisions at \mbox{\sqrtsnn = 2.76 TeV}
by PCM, PHOS and EMCal.} 
\label{fig:sec11:2gamInvMassPbPb}
\end{figure}

\begin{figure}[p]
\centering
\includegraphics[width=0.63\hsize]{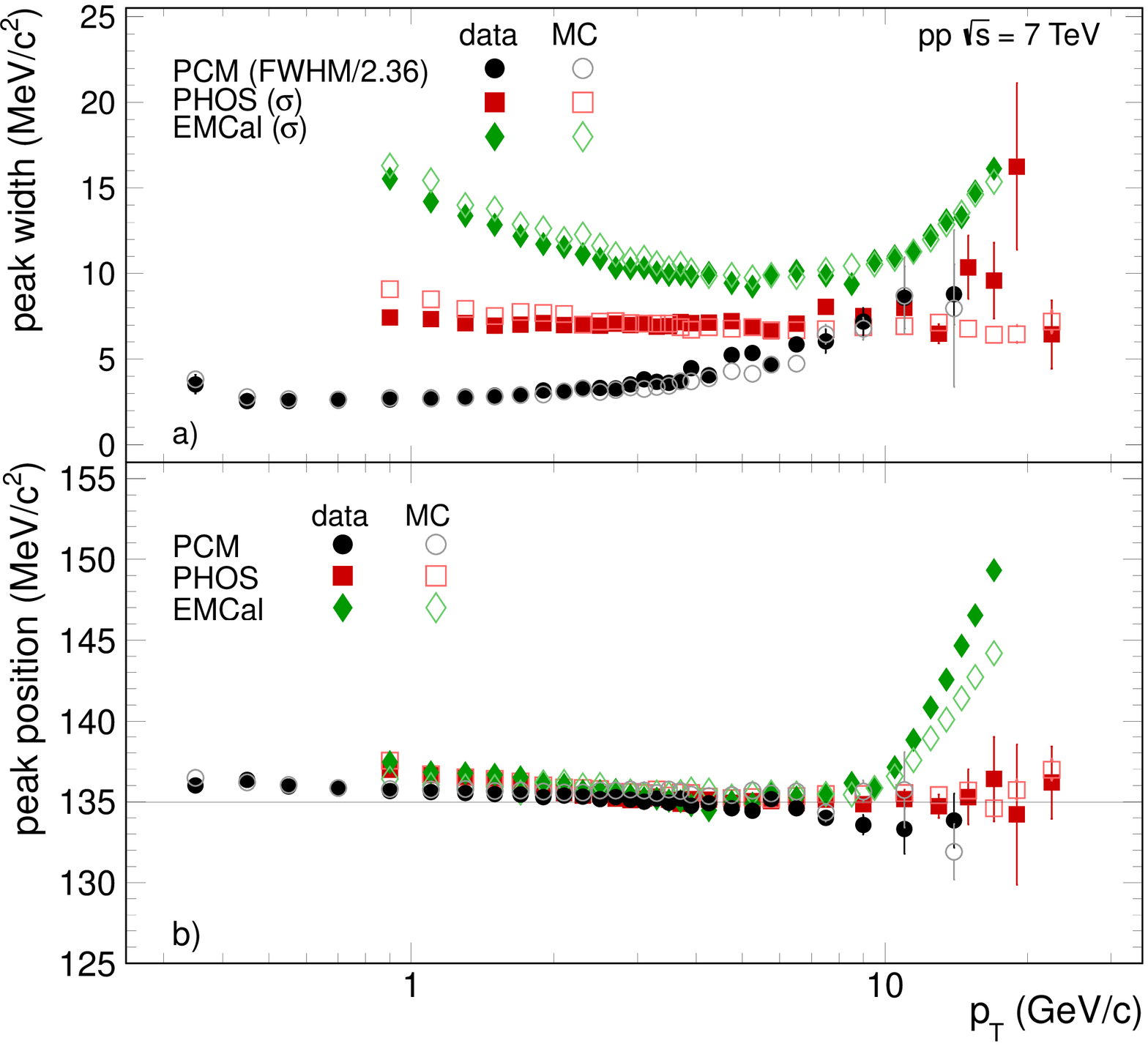}
\caption{Reconstructed \pizero peak width~(a) and position~(b) in \pp collisions at \mbox{\sqrts = 7 TeV} 
for PCM, PHOS, and EMCal compared to Monte Carlo simulations~(Pythia for PCM and PHOS, 
and embedding of clusters from single \pizero in data for EMCal).}
\label{fig:sec11:mcdatatuning:pp}
\centering
\includegraphics[width=0.63\hsize]{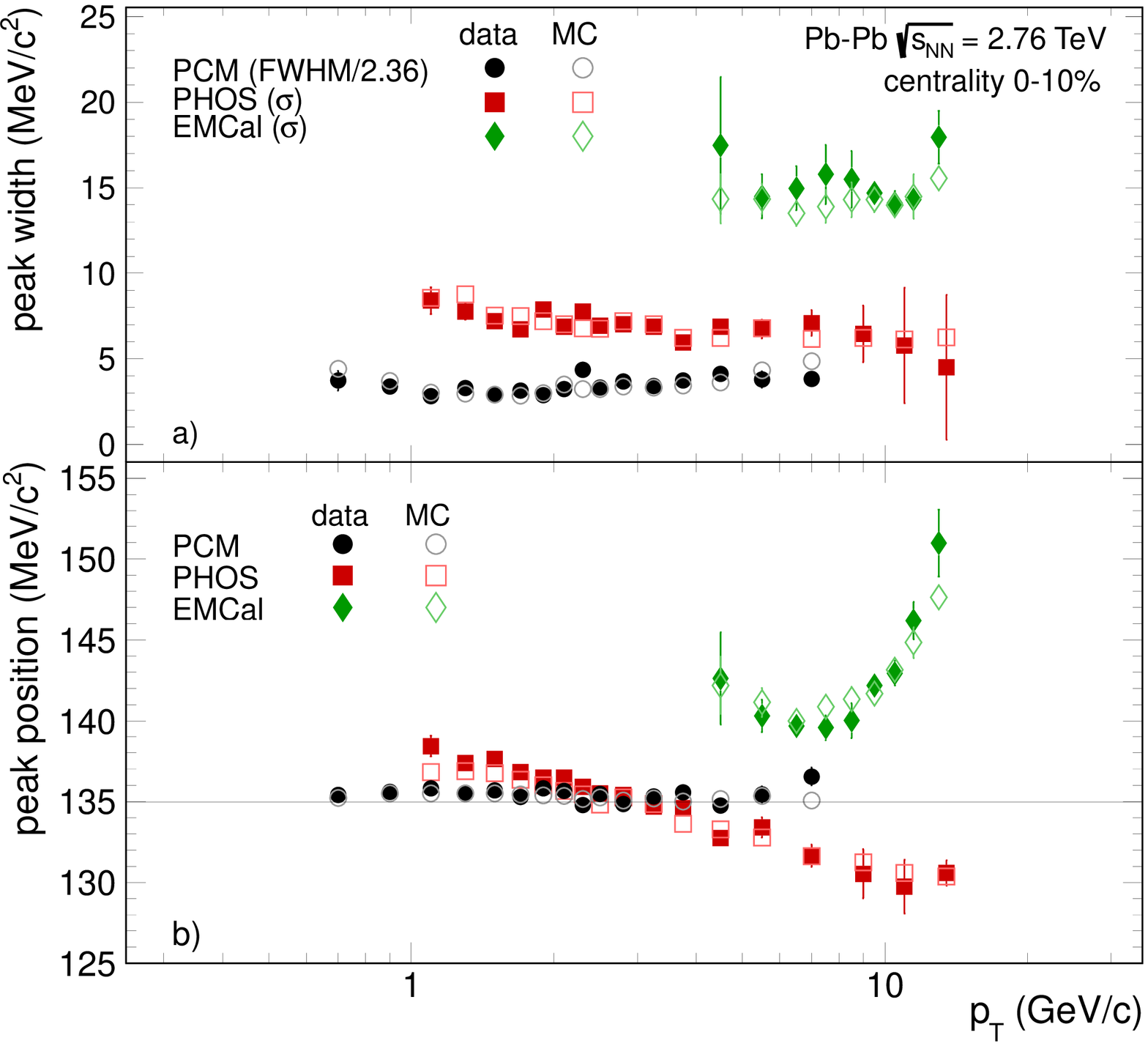}
\caption{Reconstructed \pizero peak width~(a) and position~(b) in 0--10\% central \pbpb collisions at 
\mbox{\sqrtsnn = 2.76 TeV} for PCM, PHOS, and EMCal compared to Monte Carlo simulations~(Hijing for PCM, and
embedding of clusters from single \pizero in data for PHOS and EMCal).}
\label{fig:sec11:mcdatatuning:pbpb}
\end{figure}

The detection of light neutral mesons like \pizero and $\eta$ is a benchmark for photon detectors. 
The mesons are identified via the invariant mass of photon candidate pairs~\cite{Abelev:2012cn}. 
For the calorimeters, rather loose photon identification criteria are sufficient to extract the 
\pizero peak from invariant-mass spectra in \pp collisions.
In particular, all clusters with an energy $E>0.3$~GeV (and with 3 or more cells in PHOS)
are considered as photon candidates for \pizero measurement. Figure~\ref{fig:sec11:2gamInvMass} shows the 
invariant mass spectra of photon pairs in the mass range around the \pizero peak measured in \pp collisions 
at \mbox{\sqrts = 7 TeV} for $0.6<\pt^{\gamma\gamma}<0.8$, $1.0<\pt^{\gamma\gamma}<2.0$, and $5<\pt^{\gamma\gamma}<7$\gevc 
by PCM, PHOS, and EMCal, respectively.
The invariant mass distributions
are fitted using a Gaussian distribution, leading to a mass position of $135.8$ and $136.8$~MeV/$c^2$ 
with a width of $5.3$ and $10.3$~MeV/$c^2$ for PHOS and EMCal, respectively. In the case of
PCM, the peak is asymmetric, but nevertheless is fitted by a pure Gaussian to the right of the mass
peak, leading to a mass position of $135.8$ with a width of $1.5$~MeV/$c^2$. 
The background is estimated using first-order polynomials after the uncorrelated contribution
estimated using the event mixing technique has been subtracted.
To contrast the low occupancy environment present in \pp collisions, Fig.~\ref{fig:sec11:2gamInvMassPbPb} 
shows similar invariant mass distributions in the 0--10\% most central \pbpb collisions at 
\mbox{\sqrtsnn = 2.76 TeV} for $1.4<\pt^{\gamma\gamma}<1.6$, $2.0<\pt^{\gamma\gamma}<3.0$, and 
$5<\pt^{\gamma\gamma}<7$\gevc by PCM, PHOS, and EMCal. 
For the PHOS and PCM, we show a low \pt range illustrating how the S/B worsens in the high-multiplicity 
environment of central \pbpb collisions, while for the EMCal the focus is on higher \pt values. 
To cope with the large occupancy in the calorimeters, the cluster energy is approximated with 
the core energy $E_{\rm core}$ for PHOS, while for EMCal the minimum cluster energy is increased to $E>2$~GeV and
a mild cut on the shower shape of $\lambda_0^2<0.5$ is required.
The mass position and width obtained from the Gaussian fits are $135.6$, $137.8$, and $144.6$~MeV/$c^2$ for the position, 
and $1.9$, $6.1$, and $13.4$~MeV/$c^2$ for the width in PCM, PHOS, and EMCal, respectively.
The dependence of the pion mass position and width on the transverse momentum shown in 
Figs.~\ref{fig:sec11:mcdatatuning:pp} and \ref{fig:sec11:mcdatatuning:pbpb} 
is used for tuning the Monte Carlo simulations.

\begin{figure}[h]
\includegraphics[width=0.47\hsize]{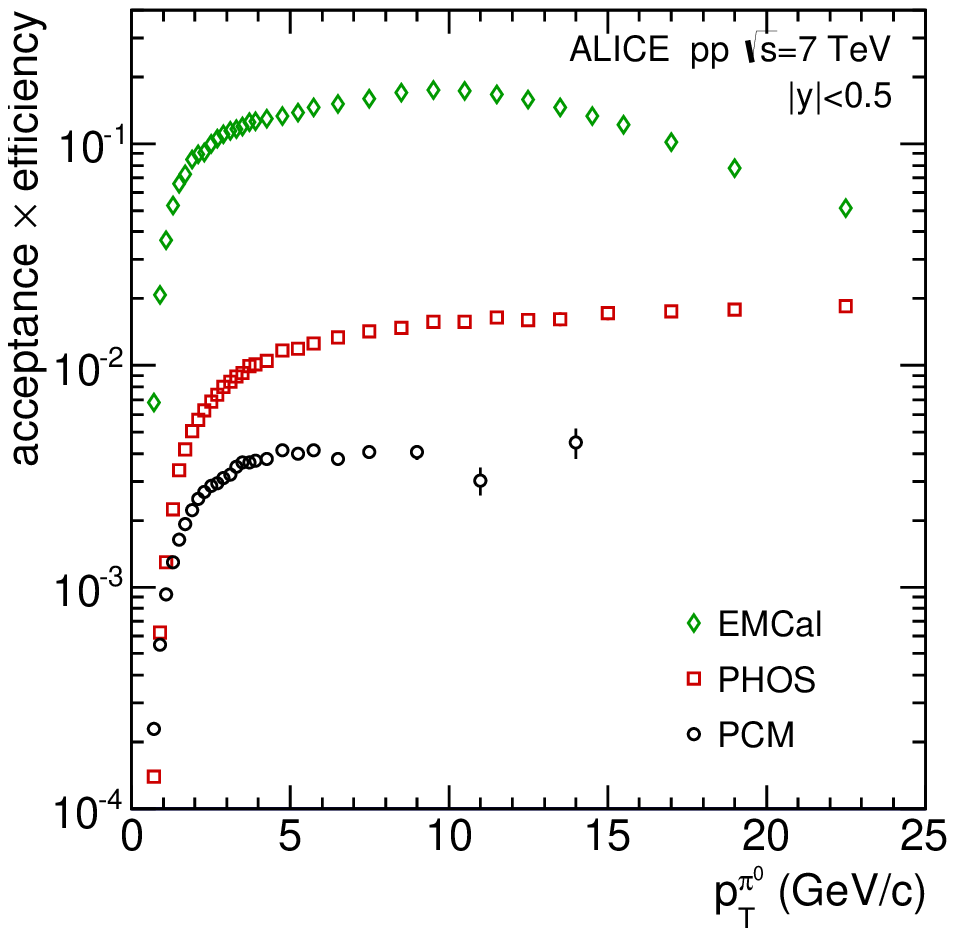}
\hfil
\includegraphics[width=0.47\hsize]{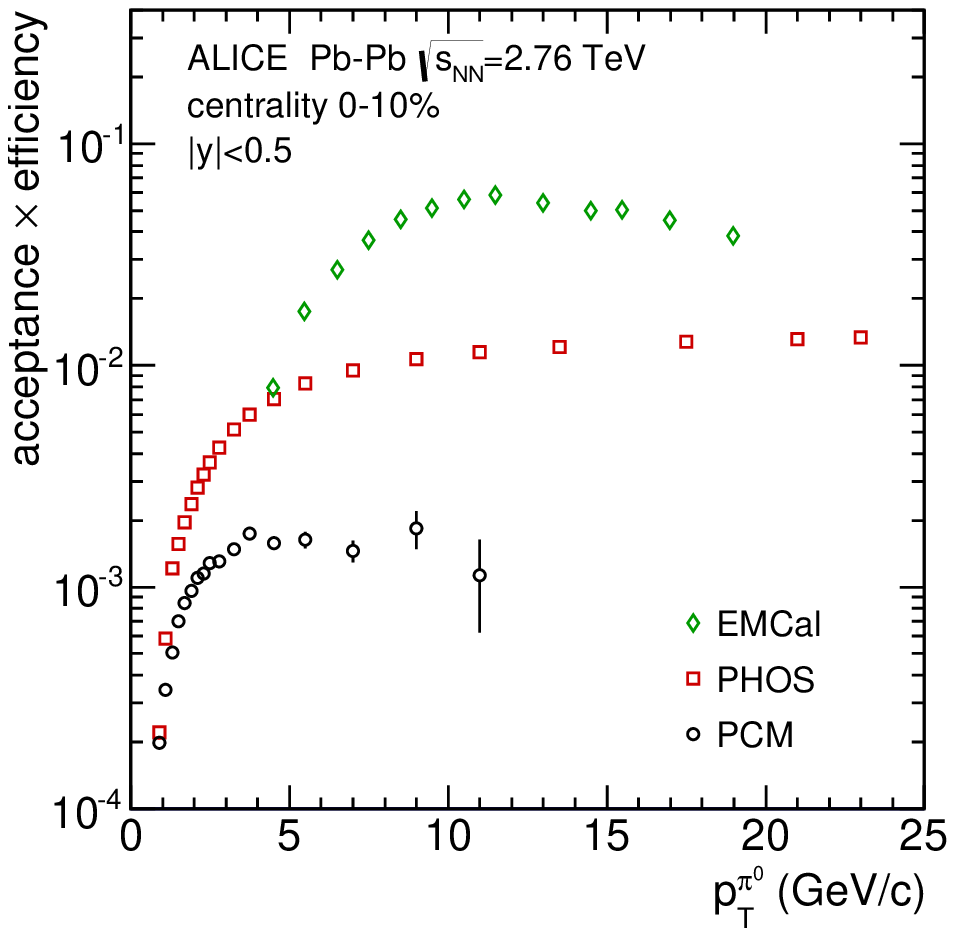}
\caption{Total correction~(efficiency and acceptance) for $|y|<0.5$ for \pizero reconstruction via
two-photon invariant mass determination in \pp collisions at \mbox{\sqrts = 7 TeV}~(left panel) and 
in 0--10\% central \pbpb collisions at \mbox{\sqrtsnn = 2.76 TeV}~(right panel) for PCM, PHOS, and EMCal.}
\label{fig:sec11:totalcorrection}
\end{figure}

The increasing difference in the mass position between the data and simulation, which gets apparent for the EMCal 
at momenta above 10\gevc in \pp collisions, may be improved with a cluster unfolding algorithm based on a 
model of the transverse profile of the shower in the EMCal.
Compared to the calorimeters, the PCM method can be used to measure the \pizero down to very low momentum, but with 
a rather small efficiency due to the small probability of about 0.7\% for both photons to convert. 
Compared with PHOS, the EMCal 
has a worse \pizero resolution, but a $\sim\!$10 times larger acceptance. 
This is illustrated in Fig.~\ref{fig:sec11:totalcorrection}, which compares the total correction~(product of efficiency 
and acceptance) for $|y|<0.5$ for PCM, PHOS, and EMCal 
in \pp collisions at \mbox{\sqrts = 7 TeV}~(left panel) and in 0--10\% central \pbpb collisions at 2.76~TeV~(right panel).
The \pizero reconstruction efficiency for the EMCal decreases at around 10\gevc 
due to the fact that the showers from the two decay photons start to overlap significantly. 
For PHOS, the \pizero reconstruction efficiency is affected by the shower merging only above 25\gevc (not shown).

\clearpage\newpage\section{Jets}
\label{sect:jets}

Jet measurements in relativistic nuclear collisions are of particular interest
due to the phenomenon of ``jet quenching'' (Ref.~\refcite{Majumder:2010qh} and
references therein), in which an energetic parton interacts with the
color-charged, hot and dense matter prior to its fragmentation into
hadrons. This interaction modifies the hadronic structure and transverse momentum 
of jets generated in the medium relative to those in vacuum, producing a variety
of phenomena that are observable experimentally and can be calculated
theoretically~\cite{Majumder:2010qh}. Measurements of jet quenching
thus provide unique information on the properties of hot QCD matter.

Operationally, a jet is specified in terms of a reconstruction
algorithm~\cite{Cacciari:2011ma} that clusters hadrons within a
specified distance $R$ in angular space,
i.e. $\sqrt{(\Delta\eta)^2+(\Delta\phi)^2}<R$. The algorithm should be
applicable in comparable fashion to both experimental data and
theoretical calculations based on perturbative QCD, dictating that
it be both infrared safe (jet measurement stable against additional
soft radiation) and colinear-safe (independent of the details of
fragmentation of the parton shower into final-state hadrons)~\cite{Cacciari:2011ma}.

Jet reconstruction in nuclear collisions is especially challenging,
owing to the large and inhomogeneous background in such events. The accurate
measurement of jets in heavy-ion collisions requires careful 
accounting of both the overall level of underlying event background,
and the influence of its region-to-region 
fluctuations~\cite{Abelev:2012ej,Cacciari:2010te,Cacciari:2011tm}.

Jets are measured within ALICE in the central detector, utilizing charged
particle tracking in ITS and TPC (see Section~\ref{sect:intro}) for the 
charged hadronic energy and electromagnetic (EM) calorimetry to measure the 
neutral hadronic energy carried by photons (\pizero, $\eta$, ...)~\cite{Abeysekara:2010ze}. 
This approach is closely related to ``Particle Flow'' 
methods~\cite{CMS:2010eua} and enables detailed control of the constituent
particles used in the jet reconstruction. This is of especial
importance in the complex heavy-ion collision environment. The
inclusive jet cross section, measured using this technique in \pp
collisions at \mbox{\sqrts = 2.76 TeV}, has been reported by
ALICE~\cite{Abelev:2013fn}.  Jet measurements using a similar approach
have also been reported for \pp collisions at
RHIC~\cite{Abelev:2006uq,Abelev:2007vt,Adamczyk:2012qj}.

In this section we present the current performance of ALICE jet
reconstruction. The emphasis is on the recently completed measurement
of the inclusive jet cross section in \pp collisions at
\mbox{\sqrts = 2.76 TeV}~\cite{Abelev:2013fn}, together with considerations
for ongoing heavy-ion jet analyses.

\subsection{EMCal jet trigger}
\label{subsect:JetTrigger}

The ALICE EMCal~\cite{Abeysekara:2010ze}, a lead-scintillator electromagnetic
calorimeter covering 107 degrees in azimuth and $|\eta|<0.7$, is used
to trigger on jets. The jet trigger in Ref.~\refcite{Abelev:2013fn} is
based on the EMCal single shower (SSh) trigger, labeled E0
in Table~\ref{tab:triggers}, which utilizes the
fast hardware sum of transverse energy (\Et) in groups of $4\times4$
adjacent EMCal towers, implemented as a sliding window. An SSh trigger
accept is issued if the threshold is exceeded by at least one EMCal
tower group. The nominal threshold was 3.0~GeV for the data recorded
in \pp collisions at \mbox{\sqrts = 2.76 TeV}. An event is accepted 
if it also passes the minimum bias (MB) trigger requirements.

The EMCal Jet Patch (JP) trigger (EJE and EJE2 in
Table~\ref{tab:triggers}) sums tower energies within a sliding
window of 32 by 32 adjacent EMCal towers, corresponding to
$\Delta\eta\times\Delta\phi\approx0.46 \times 0.46$. For
heavy-ion running, the JP integrated energy is corrected for the
underlying event in the collision prior to comparison to the trigger
threshold. This correction is based on the analog charge sum in the
V0 detectors at forward rapidity (see Table~\ref{tab:dets}), which
is observed to be highly correlated with the transverse energy measured in
the EMCal acceptance. The V0 signal provides a centrality estimator
that is used by the programmable logic of the EMCal Summable Trigger
Unit to adjust the JP trigger threshold on an event-wise 
basis~\cite{Abeysekara:2010ze}.

\begin{figure}[t]
\centering
\includegraphics[width=0.95\textwidth]{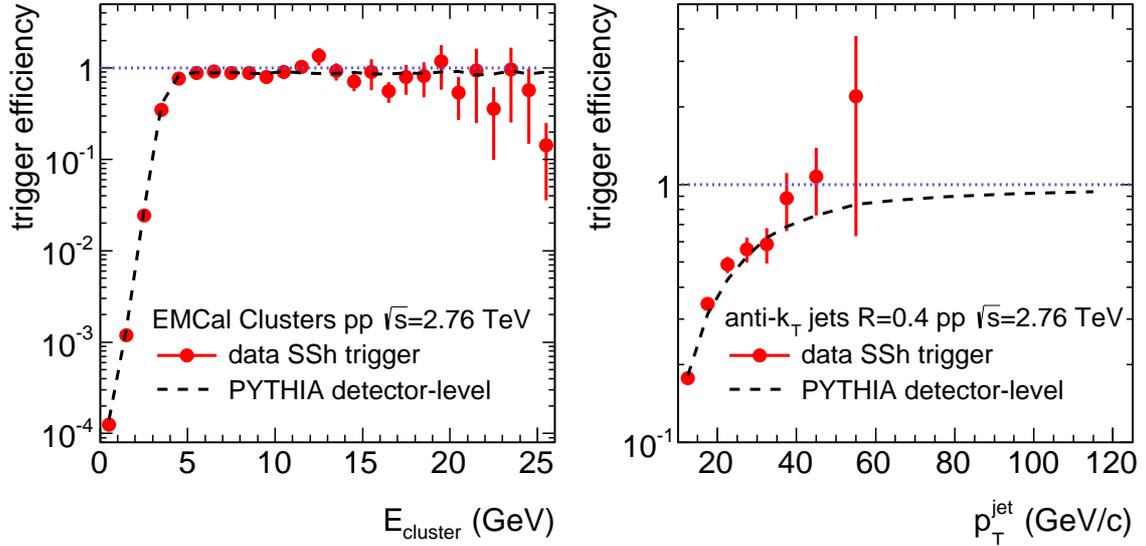}
\caption{SSh trigger efficiency in \pp collisions at \mbox{\sqrts = 2.76 TeV}. 
  Efficiency for single EM clusters (left panel) and reconstructed
  jets (\mbox{anti-kT}, $R=0.4$, right panel) for data (red points) is well 
  reproduced in simulation (black dashed line). See text for details.}
\label{fig:JetTrig}
\end{figure}

Figure~\ref{fig:JetTrig}, left panel, represents the SSh trigger efficiency
for single EM clusters in \pp collisions at \mbox{\sqrts = 2.76 TeV},
measured by comparing to MB data. Also shown is a
calculation of the SSh trigger efficiency from a detailed,
detector-level simulation based on the PYTHIA event generator (Perugia
2010 tune) and GEANT3. The distribution of data is normalized to the
simulated distribution in the region \mbox{$\pt>5\gevc$}. Good agreement is
observed between measurement and simulation in the turn-on
region of the trigger.

Figure~\ref{fig:JetTrig}, right panel, shows the efficiency of the SSh
trigger for jets in \pp collisions at \mbox{\sqrts = 2.76 TeV}. 
Jets are reconstructed offline using the \mbox{anti-kT} 
algorithm~\cite{Cacciari:2008gp}, $R=0.4$. The red points show the 
trigger efficiency measured in data as the ratio of jet yields
in SSh-triggered and MB data. Since the kinematic reach
of the MB dataset is limited, we also assess the jet trigger bias by a
data-driven simulation, shown by the black dashed line. This calculation
utilizes the measured EM cluster trigger efficiency (left panel, red points),
together with the detailed detector-level simulation (PYTHIA6 +
GEANT3) to model the jet response. The simulation and data differ in
the trigger turn-on region by $\sim18\%$ in yield, corresponding to a
shift in Jet Energy Scale of $\sim$1--2~GeV. This shift is within the
precision of the simulation, and is accounted for in the systematic
uncertainties of the corresponding cross section measurement~\cite{Abelev:2013fn}.

\subsection{Jets in \pp collisions}
\label{subsect:ppJets}

Instrumental corrections and systematic uncertainties of jet
measurements depend on the jet observable under consideration. In this
section we discuss the main instrumental corrections for measurement
of the inclusive jet cross section in \pp collisions at
\mbox{\sqrts = 2.76 TeV}, with more detail found in Ref.~\refcite{Abelev:2013fn}.

\subsubsection{Undetected hadronic energy} 

Long-lived neutral hadrons (principally, neutrons and \kzerol), will
not be detected by the tracking system and will most often deposit
only a small fraction of their energy in the EMCal. Correction for
this unobserved component of jet energy is based on
simulations. PYTHIA predictions for high-\pt identified particle
production have been compared with ALICE inclusive measurements of
high-\pt protons and charged kaons in 2.76~TeV \pp collisions, with
good agreement observed. The systematic uncertainty in the jet energy
correction arising from this comparison of simulations and measurement
is negligible~\cite{Abelev:2013fn}.

\begin{figure}[h]
\centering
\includegraphics[width=\textwidth]{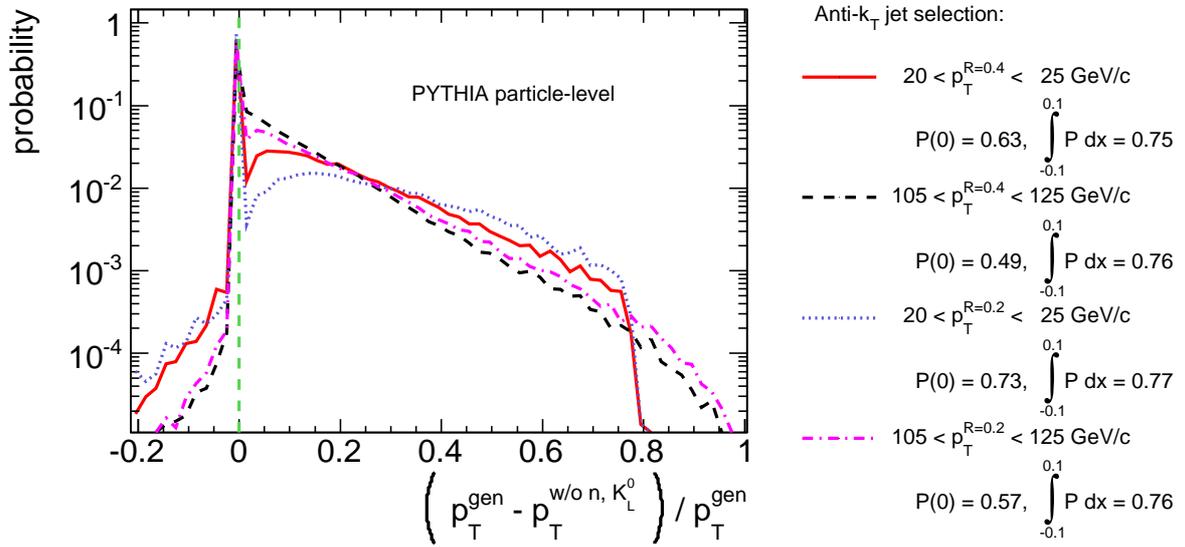}
\caption{PYTHIA particle-level simulation of jet-by-jet energy shift
  due to unobserved contributions from neutrons and \kzerol.}
\label{fig:Missing NK}
\end{figure}

Figure~\ref{fig:Missing NK} shows a PYTHIA particle-level simulation
of the shift in jet energy due to unobserved neutral hadronic energy,
calculated on a jet-by-jet basis. Jet reconstruction (\mbox{anti-kT}, $R=0.2$
and 0.4) was carried out twice on each simulated event: first
including all stable particles except neutrinos, and then excluding
the neutron and \kzerol component. The distribution of the relative
difference in reconstructed jet energy is shown for various intervals
in jet \pt, where the difference is normalized by the jet energy
calculated without contribution from neutrons and \kzerol.  The
calculation exhibits no shift in jet energy for between 50\% and 70\%
of the jet population, corresponding to the probability for jets not
to contain an energetic neutron or \kzerol among its fragments. A tail to 
positive momentum shift $\Delta\pt$ is observed, corresponding to energy 
lost due to the unobserved energy. A small tail to negative $\Delta\pt$ is
also observed, corresponding to rare cases in which the exclusion of a
neutron or \kzerol shifts the jet centroid significantly, causing the
jet reconstruction algorithm to include additional hadrons from the
event.  For jets reconstructed with \mbox{anti-kT}, $R=0.4$, the Jet Energy
Scale correction and systematic uncertainty due to this effect is
$(4\pm0.2)\%$ for jet $\pt=20$\gevc, and $(6\pm0.5)\%$ at 
100\gevc~\cite{Abelev:2013fn}.

\subsubsection{Charged particle energy deposition in EMCal}
\label{sect:chapaendeinem}

Charged hadrons and electrons shower in the EMCal, and are also
measured by the ALICE tracking system. Their contribution to EMCal
cluster energy must be accounted for, in order not to double-count a
fraction of their energy in the measured jet energy. The correction
procedure minimizes dependence on the simulation of 
hadronic and EM showers. 

Charged-particle trajectories are propagated to a depth of 10$X_{0}$
in the EMCal, with each track then matched to the nearest EMCal
cluster falling within $\Delta\eta = 0.015$ and
$\Delta\phi=0.03$. Multiple charged tracks can be matched to a
single cluster, though the probability for multiple matches is less
than 0.5\% for \pp collisions. We then define \sump\ to be the sum of
the 3-momentum magnitude of all matched tracks.  For measured cluster
energy \Eclust, the corrected cluster energy \Ecorr\ is set to zero if
$\Eclust<\fsub\cdot\sump\>c$; otherwise, $\Ecorr=\Eclust-\fsub\cdot\sump\>c$, 
where $\fsub=1$ for the primary analysis and is varied for systematic
checks. The correction to the cluster energy, $\DEcorr=\Eclust-\Ecorr$,
takes the following values:

\begin{equation}
\DEcorr = \left\{ \begin{array}{rcl}
\Eclust & \textrm{for} & \Eclust < \fsub \cdot \sump \>c\\
 & & \\
\fsub \cdot \sump & \textrm{for} & \Eclust > \fsub \cdot \sump \>c.
\end{array}\right.
\end{equation}

\noindent
To examine the distribution of \DEcorr, we specify $\fsub=1$ and consider
the following ratio, which is calculated on a cluster-by-cluster
basis:

\begin{equation}
\Rcorr = \frac{\DEcorr}{\sump\>c}.
\label{eq:Rcorr}
\end{equation}

\begin{figure}[h]
\centering
\includegraphics[width=\textwidth]{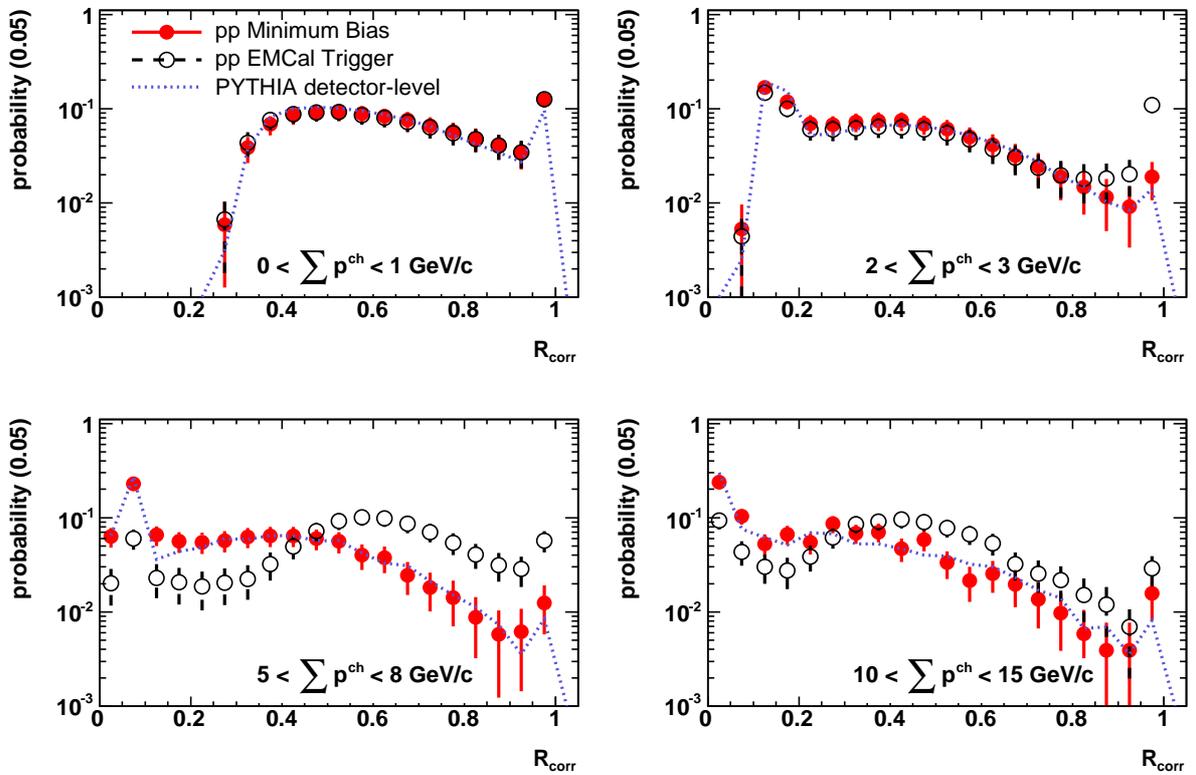}
\caption{Probability distribution of \Rcorr\ (Eq.~(\ref{eq:Rcorr})) for
  various intervals of \sump, measured in MB and EMCal-triggered \pp
  collisions, compared to detector-level simulations based on PYTHIA.}
\label{fig:JetHadronicCorrection}
\end{figure}

\noindent
Figure~\ref{fig:JetHadronicCorrection} shows the normalized
probability distribution of \Rcorr measured in four different bins of
\sump for MB and EMCal-triggered \pp collisions, each compared to a
detector-level simulation (PYTHIA6). For a cluster whose energy
arises solely from matched charged tracks, i.e. which does not contain
photons or untracked charged particles, the
ratio $\Rcorr=E/pc$, where $E$ is the EMCal shower energy and $p$ is
the momentum of the charged tracks contributing to the shower. The
probability per cluster for pileup from photons or untracked charged
particles in \pp collisions is less than 0.5\%, so that
Fig.~\ref{fig:JetHadronicCorrection} represents, to good accuracy, the
in-situ measurement of $E/p$ for the EMCal.

The peak at unity in Fig.~\ref{fig:JetHadronicCorrection} corresponds
to 100\% of the matched track momenta being subtracted from the
cluster energy. Full containment of a hadronic shower in the EMCal is
unlikely, and the peak at unity originates in part from
over-subtraction from pileup due to neutral particles and unmeasured
charged particles.

The figure shows that the distribution of \Rcorr\ for the MB trigger
is modeled well by a PYTHIA-based detector-level calculation. The
variation in the distribution for the EMCal-triggered data is due to
the trigger bias: the EMCal trigger at threshold favors highly
abundant low \pt charged hadrons that deposit above-average energy in
the EMCal.

Detector-level simulations show that the above procedure corrects the
Jet Energy Scale to within 1--2\% in \pp collisions, for choices of
\fsub\ between 0.7 and 1.0. The contribution of this correction to the
Jet Energy Resolution is about 5\% at $\ptjet=40\gevc$, and 8\% at
$\ptjet=100\gevc$.

\subsubsection{Other corrections} 

Other significant corrections to the inclusive jet cross section
measurement are due to the tracking efficiency and track momentum
resolution. A brief discussion of these effects is found below; for
further details see Ref.~\refcite{Abelev:2013fn}.

Jets in \pp\ collisions are made up of a limited number of particles,
with large jet-to-jet fluctuations in both the \pt\ distribution of
the constituents and the relative fraction of jet energy carried by
neutral or charged particles. The effect of tracking efficiency on
measured jet \pt\ is therefore not modelled well by a Gaussian
distribution, but has a more complex form. This distribution has been
studied using PYTHIA-based simulations, which show that for 74\% 
of jets with particle-level \pt\ in the range 105--125\gevc 
(\mbox{anti-kT}, $R$=0.4) the \pt shift due to tracking efficiency is 
below 10\%. For 30\% of the population, the shift is negligible.
For \pp collisions at \mbox{\sqrts = 2.76 TeV}, 
tracking efficiency generates a Jet Energy Scale uncertainty of
2.4\% and a multiplicative correction to the inclusive jet cross 
section of a factor of $1.37\pm0.12$~\cite{Abelev:2013fn}.

The \pt\ resolution of tracking and the energy resolution of the EMCal
contribute an uncertainty in Jet Energy Scale of 1--2\%, generating a
systematic uncertainty in the inclusive jet cross section that is
small compared to other contributions~\cite{Abelev:2013fn}. This
arises because jets are multi-hadron objects whose energy is carried
to a significant extent by a number of relatively low \pt\ constituents, with
average constituent \pt\ increasing only gradually with jet \pt.

\subsubsection{Jet structure} 

\begin{figure}[h]
\centering
\includegraphics[width=\textwidth]{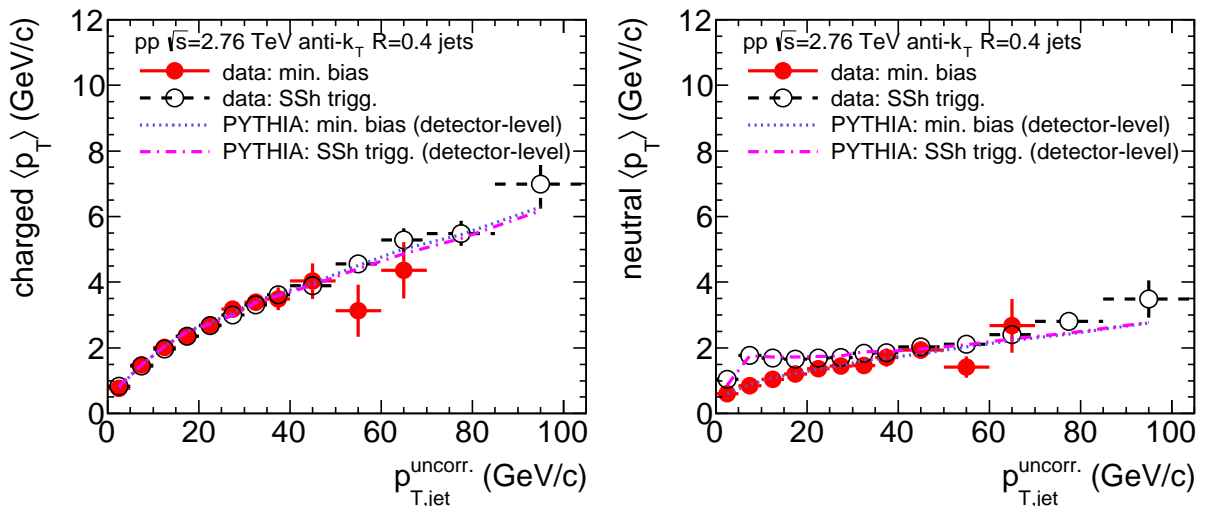}
\caption{Mean transverse momentum, $\left<\pt\right>$, of constituents
  measured in reconstructed jets in 2.76~TeV \pp collisions (\mbox{anti-kT},
  $R=0.4$) vs. jet \pt. Left: charged tracks; Right: neutral
  clusters. Data are shown for MB and SSh triggers,
  and are compared to detector-level simulations.}
\label{fig:ppJetMeanpt}
\end{figure}

We next compare specific features of reconstructed jet structure in
data and PYTHIA-based detector-level
simulations. Figure~\ref{fig:ppJetMeanpt} shows the jet \pt dependence
of the mean hadron \pt within the jet, $\left<\pt\right>$, for charged
tracks (left) and neutral clusters (right), for both MB and
SSh-triggered event populations. The value of $\left<\pt\right>$ rises
slowly with jet \pt, and is well described by the detector-level
PYTHIA simulation over the full measured range.

\begin{figure}[h]
\centering
\includegraphics[width=\textwidth]{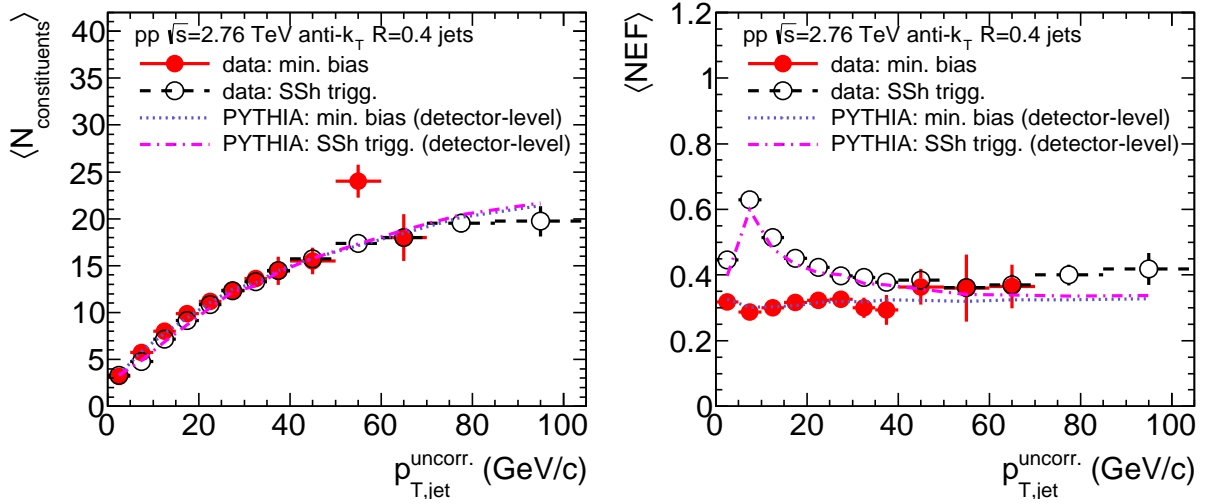}
\caption{Mean total number of constituents (left) and mean neutral
  energy fraction (right) measured in reconstructed jets in 2.76~TeV \pp
  collisions (\mbox{anti-kT}, $R=0.4$), vs. jet \pt. Data are shown for MB and SSh 
  triggers, and are compared to detector-level simulations.}
\label{fig:ppJetConstituents}
\end{figure}

Figure~\ref{fig:ppJetConstituents}, left panel, shows the mean number
of jet constituents (total number of charged tracks and neutral
clusters), while the right panel shows the mean Neutral Energy
Fraction (NEF). Both distributions are presented as a function of jet
\pt. PYTHIA detector-level simulations describe both distributions
accurately, for both the MB and SSh-triggered datasets. The NEF
distributions are discussed in more detail in
Ref.~\refcite{Abelev:2013fn}.

\subsubsection{Jet energy resolution}

Jet Energy Resolution is calculated using simulations, with all
significant components of the simulation validated against data
(e.g. Figs.~\ref{fig:ppJetMeanpt} and \ref{fig:ppJetConstituents}; see
further discussion in Ref.~\refcite{Abelev:2013fn}). Jet reconstruction is
carried out on each generated event at both particle and
detector level. Reconstructed jets whose centroids lie close in
$(\eta,\phi)$ at the particle and detector level are identified, and
their relative difference in reconstructed jet energy is calculated
according to:

\begin{equation}
\Delta\pt = \frac{\pt^{\rm det} - \pt^{\rm part}}{\pt^{\rm part}}
\label{eq:DpT}
\end{equation}

\begin{figure}[t]
\centering
\includegraphics[width=\textwidth]{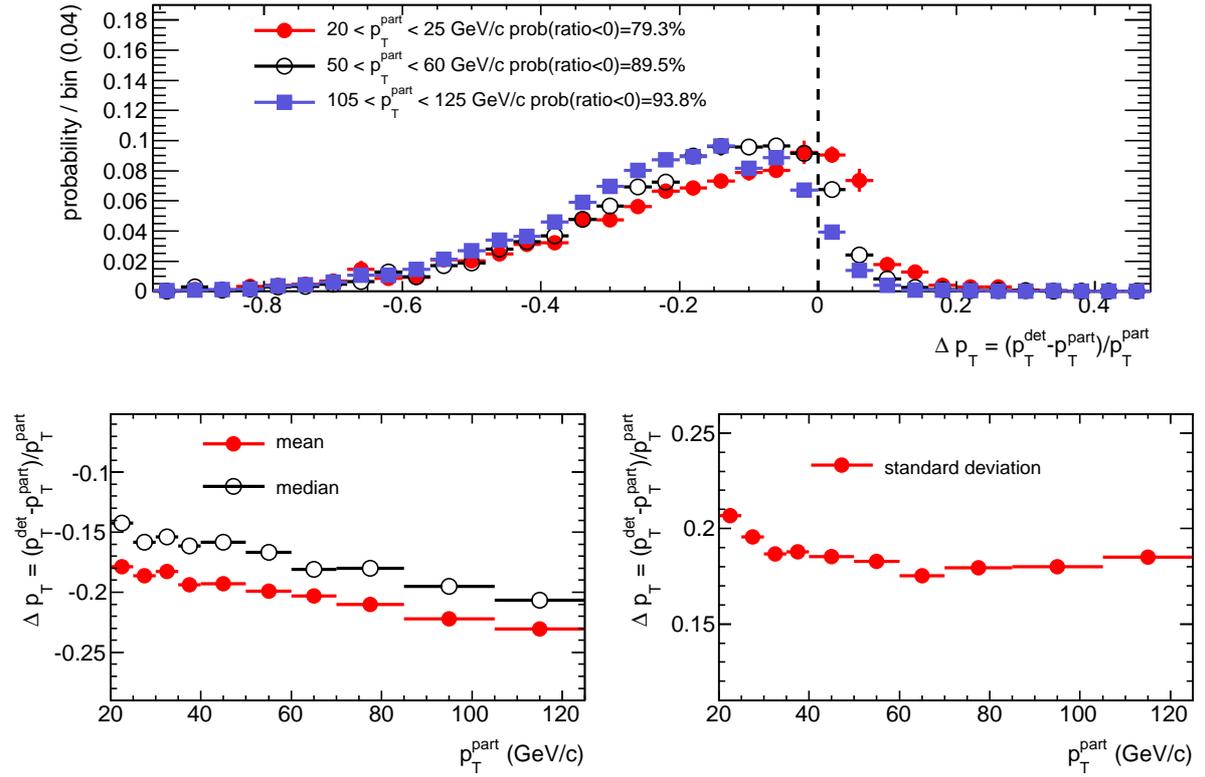}
\caption{Instrumental effects on jet energy measurement
  (Eq.~(\ref{eq:DpT})). Upper panel: jet-by-jet distribution for
  various intervals in jet \pt. Lower panels: Mean and median (left)
  and standard deviation (right) of these distributions.}
\label{fig:JetResolution}
\end{figure}

Figure~\ref{fig:JetResolution}, upper panel, shows the distribution of
$\Delta\pt$ for three ranges of jet \pt. The distributions are
weighted towards negative values, corresponding to lower energy at the
detector level. The lower panels show the median and mean (left) and
RMS (right) of the upper distributions, as a function of
particle-level \pt. The mean relative energy shift (Jet Energy
Scale, or JES, correction) is seen to be \pt-dependent, ranging
between 17\% and 22\%. The RMS, corresponding to the Jet Energy
Resolution (JER), is seen to be a weak function of jet \pt in the
range 40--100~GeV, varying between 18\% and 20\%.

\subsection{Jets in heavy-ion collisions}
\label{subsect:JetsinHIC}

\begin{figure}[t]
\centering
\includegraphics[width=0.8\textwidth]{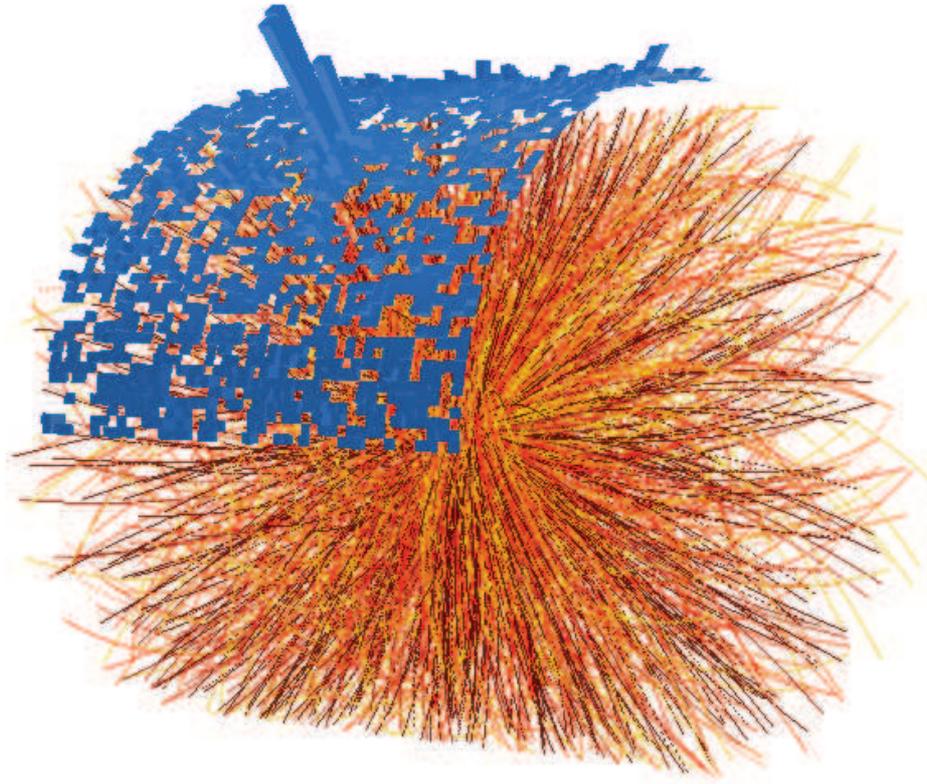}
\caption{Event display of a central \pbpb collision containing a high \pt\ jet in the EMCal
  acceptance. The event was triggered using the EMCal SSh trigger.}
\label{fig:JetEvtDisplay}
\end{figure}

Full jet reconstruction in heavy-ion collisions offers the possibility
to measure jet quenching effects at the partonic level, without the
biases intrinsic to measurements based on high \pt single hadrons,
which suppress direct observation of the structure of quenched
jets. While hard jets are clearly visible in event displays of single
heavy-ion collisions (see Fig.~\ref{fig:JetEvtDisplay}), accurate
measurement of the energy of such jets on an event-by-event basis is
challenging, due to the large and inhomogeneous underlying
background. The mean background energy in a cone of radius $R=0.4$ is
about 60~GeV in a central \pbpb collision, though the distribution
of this quantity has a large tail to much higher values. It is not
possible to discriminate the hadronic component of a hard jet from
that of the background on a rigorous basis, and any jet reconstruction
algorithm applied to such events will therefore incorporate hadrons
arising from multiple incoherent sources (hard jets, mini-jets, soft
production) into the same jet. This results in a significant distortion
(``smearing'') of the hard jet energy distribution, together with
generation of a large population of ``combinatorial'' jets comprising
solely hadrons generated in soft processes. The latter population has
no distinct physical origin, and is experimental noise.

Since jet quenching is generically expected both to soften and to
broaden the fragmentation pattern of jets in medium relative to jets
in vacuum, care must be taken in the choice of instrumentation and
algorithm to preserve the soft component of jets in heavy-ion
measurements. ALICE's unique capabilities to measure hadrons
efficiently down to very low \pt raise the possibility of jet
reconstruction with very low infrared cutoff ($\sim0.2$\gevc), even
in heavy-ion collisions. Techniques to remove the combinatorial
component from the measured jet population and to correct the
remaining hard-jet distribution for the effects of background, while
preserving the low infrared cutoff, are outlined in
Refs.~\refcite{Abelev:2012ej,deBarros:2012ws,Verweij:2012bk}. These
techniques have recently been applied to ALICE data to measure the
inclusive jet cross section~\cite{Verweij:2012bk,Reed:2012sy} and
hadron-jet coincidences~\cite{Cunqueiro:2012vga} in \pbpb collisions.
Full analyses of jets in heavy-ion collisions will be reported in
forthcoming ALICE publications. Correction for
background depends upon the physics observable under consideration,
and we do not consider it further here. 

The remainder of this section discusses instrumental corrections for
heavy-ion jet measurements, which are similar to those applied in \pp
collisions (see Ref.~\refcite{Abelev:2013fn} and discussion above). The main
difference arises in the correction for charged particle energy
deposition in the EMCal, due to the greater pileup contribution of
photons and untracked charged particle energy to EMCal clusters,
arising from the high multiplicity in heavy-ion events. For \pp
collisions, the cluster pileup probability is less than 0.5\%, whereas
in central \pbpb collisions the probability of having two or more
particles contributing above noise threshold to the cluster energy is
about 5\%.

\begin{figure}[hbt]
\centering
\includegraphics[width=0.99\textwidth]{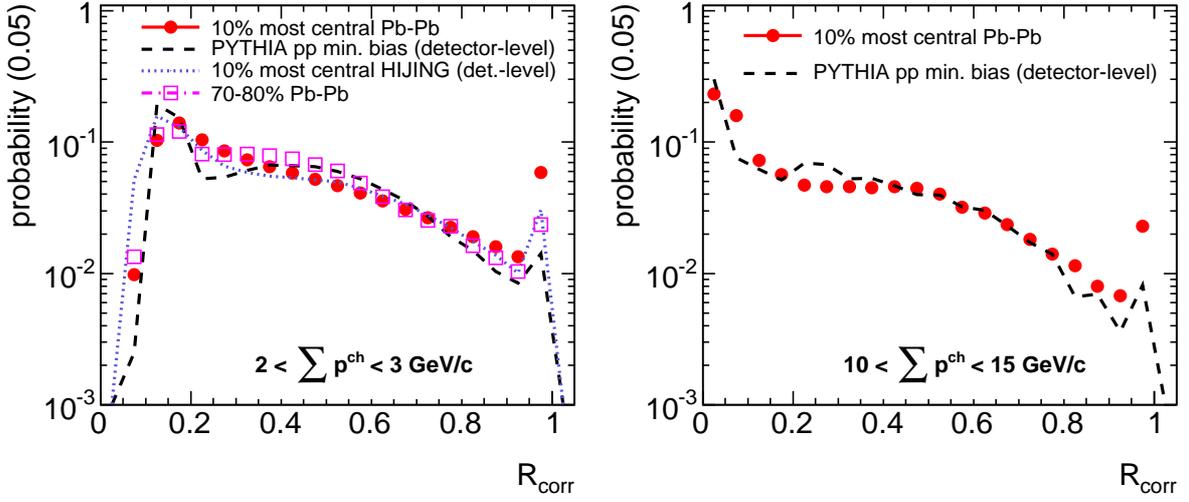}
\caption{Probability distribution of \Rcorr\ (Eq.~(\ref{eq:Rcorr})) in
  two different intervals of \sump, measured in central (0--10\%) and
  peripheral (70\%--80\%, left panel only) \pbpb collisions. Also
  shown are detector-level simulations for MB \pp  collisions based 
  on PYTHIA (same distributions as Fig.~\ref{fig:JetHadronicCorrection}), 
  and for central \pbpb collisions based on HIJING (left panel only).} 
\label{fig:PbPbJetHadronicCorrection}
\end{figure}

We utilize the probability distribution of \Rcorr\
(Eq.~(\ref{eq:Rcorr})), which corresponds to the EMCal $E/p$
distribution in the absence of cluster pileup, to assess the effects
of pileup in the heavy-ion environment. The \Rcorr\ probability
distribution is shown in Fig.~\ref{fig:JetHadronicCorrection} for \pp
collisions, and in Fig.~\ref{fig:PbPbJetHadronicCorrection} for
central (0--10\%) and peripheral (70--80\%) \pbpb collisions, in
two different intervals of
\sump. Figure~\ref{fig:PbPbJetHadronicCorrection} also shows two
different detector-level simulations: the PYTHIA distribution is the
same as that shown in Fig.~\ref{fig:JetHadronicCorrection}, which
accurately describes the \Rcorr\ distribution for MB \pp collisions,
while Hijing is used to model the \Rcorr\ probability distribution for
0--10\% central \pbpb collisions.

All data and simulated distributions in
Fig.~\ref{fig:PbPbJetHadronicCorrection} are qualitatively similar:
the most probable value of \Rcorr\ ($\approx 0.15$) matches within
10\% and the medians are compatible within 4\%. In the left panel, for 
\mbox{$2 < \sump < 3$~GeV}, the peripheral \pbpb distribution does not match 
that for \pp in detail at (and slightly above) the minimum ionizing 
particle (MIP) peak. The probability at the saturation peak, $\Rcorr=1$, is
largest for central \pbpb, with lower probability for peripheral
\pbpb, and even lower for \pp. This is due to a larger contribution
from cluster pileup, which increases the probability for large cluster
energy. However, the increase in probability for the saturation peak
from peripheral to central collisions is seen to be only 3\%. Since
the probability is normalized to unity, this difference between the
systems at $\Rcorr=1$ must be accompanied by differences for $\Rcorr<1$,
which are visible but are of moderate magnitude. The Hijing simulation
models the \Rcorr distribution for central collisions reasonably
well, though its estimate of the probability for $\Rcorr=1$ is lower
than seen in data, and it undershoots the data slightly in the region
just above the MIP peak.

The right panel in Fig.~\ref{fig:PbPbJetHadronicCorrection}, for $10 <
\sump < 15$~GeV (and correspondingly for more energetic EMCal
clusters), also exhibits minor differences between \Rcorr\
distributions in central \pbpb and \pp. Since the magnitude of cluster
pileup energy is independent of the true cluster energy, its relative effect
on the \Rcorr\ probability distribution is expected to be smaller for
larger cluster energy.

The above observations indicate that the magnitude of cluster pileup
effects in central \pbpb collisions due to neutral particles and
unmeasured charged particles is modest.  While the pileup contribution
cannot be measured explicitly on a cluster-by-cluster basis, its
average magnitude can be estimated, based on the distributions in
Fig.~\ref{fig:PbPbJetHadronicCorrection}, to correspond to about 50~MeV 
of additional energy per EMCal tower for central \pbpb relative to
\pp collisions. However, subtraction of this average value from each
tower in a cluster does not improve the overall agreement of the
distributions in Fig.~\ref{fig:PbPbJetHadronicCorrection}, and such a
correction is not applied in the physics analysis of jets. Rather, the
difference between the distributions is incorporated into the
systematic uncertainty of the measurement.

\clearpage\newpage\section{Muons}
\label{sect:muons}

\begin{figure}[hbt]
\begin{center}
\includegraphics[width=0.9\textwidth]{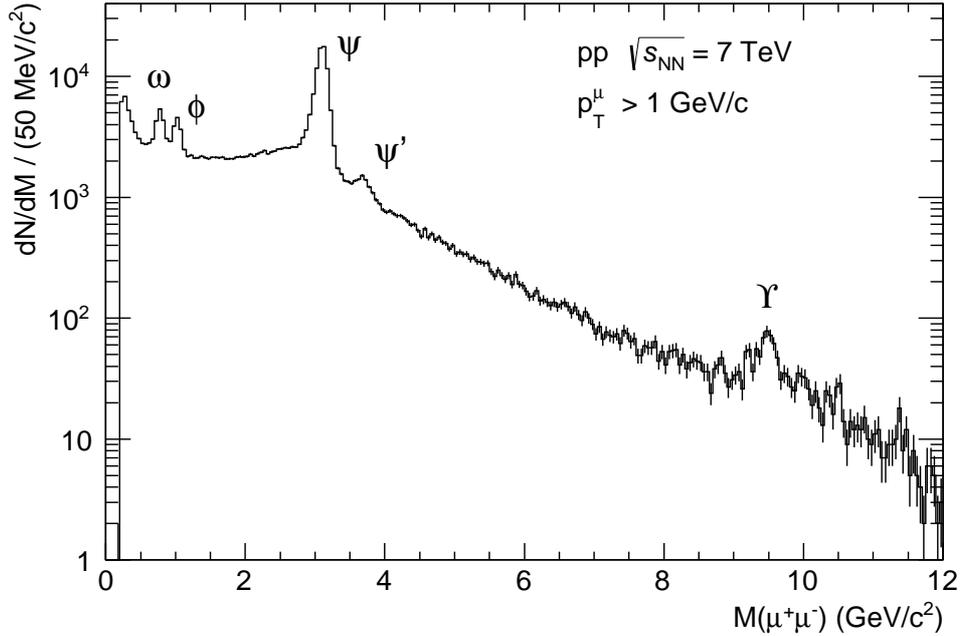}
\caption{\label{fig:fullminvrange}Invariant mass distribution of $\mu^+ \mu^-$ 
pairs measured by ALICE for \pp collisions at \mbox{\sqrts = 7 TeV} 
(\lumi~=~1.35~pb$^{-1}$, corresponding to the full 2011 dimuon-triggered data sample).}
\end{center}
\end{figure}

Light ($\omega$ and $\phi$) and heavy (\jpsi and $\Upsilon$ families) vector 
mesons are measured in ALICE in their $\mu^+\mu^-$ decay channel using the 
muon spectrometer. The invariant mass reach with the statistics collected in one 
year of running with \pp collisions is illustrated in Fig.~\ref{fig:fullminvrange}. 
The spectrometer is also used to measure the production of single muons from 
decays of heavy-flavor hadrons~\cite{Abelev:2012qh} and W$^\pm$ bosons. 
Below we discuss the performance of the spectrometer, with an emphasis on the \jpsi 
measurement. 

The muon spectrometer covers $-4.0 < \eta < -2.5$ and consists of the 
following components: a passive front absorber (4.13~m, 
$\sim\!\!10 \ \lambda_{\rm int}, \sim\!\!60 \ X_{0}$) suppressing charged hadrons and 
muons from $\pi$/K decays; a high-granularity tracking system of ten detection planes 
(five stations, two Cathode Pad Chambers each); a large dipole magnet ($\int B \ \mathrm{d} z=3~\mathrm{Tm}$, 
bending tracks vertically); a passive muon-filter wall (1.2 m thick, 
$\sim\!\!7.2 \ \lambda_{\rm int}$) followed by four planes of Resistive Plate Chambers 
for triggering; and inner beam shielding to protect the detection 
chambers from the primary and secondary particles produced at large rapidities.  

The key features of the muon spectrometer are good \jpsi acceptance 
down to $\pt=0$ and high 
readout granularity resulting in an occupancy of 2\% in central \pbpb 
collisions. The combined effect of the front absorber (which stops primary 
hadrons) and of the muon-filter wall (which suppresses the low-momentum 
muons from pion and kaon decays) leads to a detection threshold of $p\gtrsim 4$\gevc 
for tracks matching the trigger.

During the heavy-ion run in 2011, about 20\% of the electronic channels in the 
tracking chambers had to be discarded because of faulty electronics or high voltage 
instabilities. In a similar way, the noisy strips in the trigger chambers 
(0.3\%)~\cite{Bossu:2012jt} have also been excluded from data taking. 

The clusters of charge deposited by the particles crossing the muon tracking 
chambers are unfolded using the Maximum Likelihood Expectation Maximization (MLEM) 
algorithm~\cite{Zinchenko:2003yg} and fitted with a 2D Mathieson~\cite{Mathieson:1988rn} 
function to determine their spatial location. 
A tracking algorithm based on the Kalman filter reconstructs 
the trajectory of the particles across the five tracking stations. These tracks 
are then extrapolated to the vertex position measured by the ITS (SPD only in 
most cases) and their kinematic parameters are further corrected for multiple 
scattering and energy loss of muons in the front absorber~\cite{offline:2009}.

While the actual detector occupancy measured in real \pbpb collisions, 2\%,  is 
well below the design value (5\%), it was still important to fine tune the 
reconstruction parameters to keep the fraction of fake tracks as low as 
possible. The size of the roads (defined in the tracking algorithm that searches  
for new clusters to be attached to the track candidates) is limited by the 
intrinsic cluster resolution and the precision of the alignment of the 
apparatus. 

Since the background in \pbpb collisions is large, tight selection 
criteria have to be imposed on single muon tracks in order to preserve the 
purity of the muon sample. Tracks 
reconstructed in the tracking chambers are required to match a trigger track, they 
must lie within the pseudorapidity range $-4 < \eta < -2.5$, and 
their transverse radius coordinate at the end of the front absorber must be in the 
range $17.6~{\rm cm} < R_{\rm abs} < 89$~cm. An additional cut on $p \times $DCA, 
the product of the track momentum and the distance between the vertex 
and the track extrapolated to the vertex transverse plane, may also be 
applied to further reduce residual contamination. With such cuts, a large 
fraction of the remaining fake tracks are removed. 

\subsection{Reconstruction efficiency}

\begin{figure}[b]
\begin{center}
\includegraphics[width=0.75\textwidth]{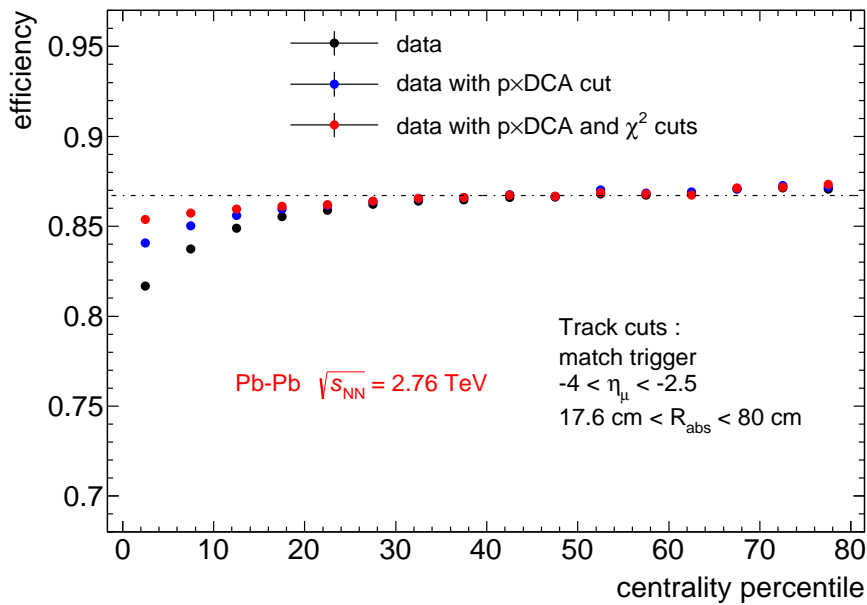}
\caption{\label{fig:muon-tracking-efficiency-vs-centrality}
Measured muon track reconstruction efficiency in \pbpb collisions as a function 
of the collision centrality.}
\end{center}
\end{figure}

The track reconstruction efficiency (Fig.~\ref{fig:muon-tracking-efficiency-vs-centrality})
is determined with experimental data using a method that takes advantage of 
the redundancy of the detector, i.e. the fact that a subset of all chambers 
is sufficient for a track to be reconstructed. The tracking algorithm requires 
at least one cluster in each of the first three stations and at least three 
clusters in three different chambers in the last two stations in order to 
validate a track. As a result, the efficiency of a given chamber can be determined 
by the ratio of the number of reconstructed tracks detected in that chamber over 
the total number of reconstructed tracks. In order to avoid any bias that may be 
introduced by the reconstruction criteria themselves, only tracks that still satisfy 
these criteria when that chamber is not taken into account must be considered when 
computing the ratio. For instance, in the first station, the efficiency of one of 
the two chambers is determined by dividing the number of tracks detected in both 
chambers by the number of tracks detected by the other chamber. By combining the 
individual chamber efficiencies according to the reconstruction criteria, one can 
determine the overall reconstruction efficiency. 

The resulting efficiency (black points in Fig.~\ref{fig:muon-tracking-efficiency-vs-centrality}) 
exhibits a drop for central \pbpb collisions. This drop can, however, be 
largely ascribed to the remaining fake tracks, which inherently contain less clusters 
than the others. To cure this problem, the $p \times $DCA cut is applied first, 
strongly reducing this contamination (blue points in 
Fig.~\ref{fig:muon-tracking-efficiency-vs-centrality}).
Then a second cut on the normalized $\chi^2$ of the tracks ($\chi^2 < 3.5$) is 
added to further cut the remaining contamination at very low \pt 
($<$1--2\gevc), where the $p \times $DCA cut is not 100\% efficient (red 
points on the figure). After all these cuts have been applied, the relative loss of 
efficiency as a function of centrality is very low (of the order of 1.5\% in the centrality 
bin 0--10\%). 

\begin{figure}[h]
\begin{center}
\includegraphics[width=0.9\textwidth]{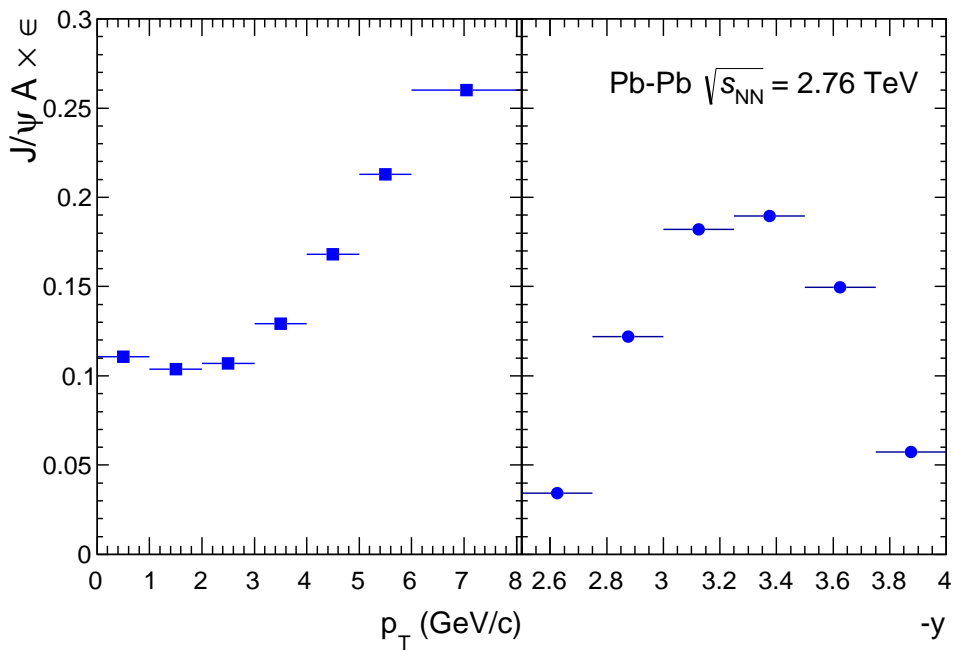}
\caption{\label{fig:jpsiacceptance}Muon spectrometer acceptance times efficiency 
for \jpsi within $-4.0<y<-2.5$ during the \pbpb 2011 campaign, as a function of 
the \jpsi transverse momentum (left) or rapidity (right).}
\end{center}
\end{figure}

The product of acceptance $A$ and efficiency $\trueepsilon$ for measuring \jpsi mesons 
emitted within $-4.0<y<-2.5$, obtained from Monte Carlo (MC) simulations of pure \jpsi 
signal with input $y$ and \pt distributions tuned to the measured ones, is 
sizable down to $\pt=0$. 
The transverse momentum dependence (for \jpsi within $-4.0<y<-2.5$) and the 
rapidity dependence (for a realistic \pt distribution) of this quantity are 
shown in Fig.~\ref{fig:jpsiacceptance}.

\subsection{Trigger efficiency}

While it has been verified with data that the efficiency of 
the trigger chambers themselves does not vary with the centrality of the 
collision, the overall reconstruction efficiency of the trigger tracks can do so.
The reason is that the trigger algorithm can only produce one trigger track per 
local board, and the detector is divided into 234 local boards. 
So even if the occupancy in the trigger system is small, the probability that 
two tracks are close enough to interfere in the trigger response can be sizable.
The response of the algorithm, taking  this effect into account, is nevertheless 
well reproduced in simulations using the embedding technique (see below). 
In these simulations we observe a relative loss of trigger track 
reconstruction efficiency of 3.5\% in the most central collisions.

\begin{figure}[h]
\begin{center}
\includegraphics[width=0.72\textwidth]{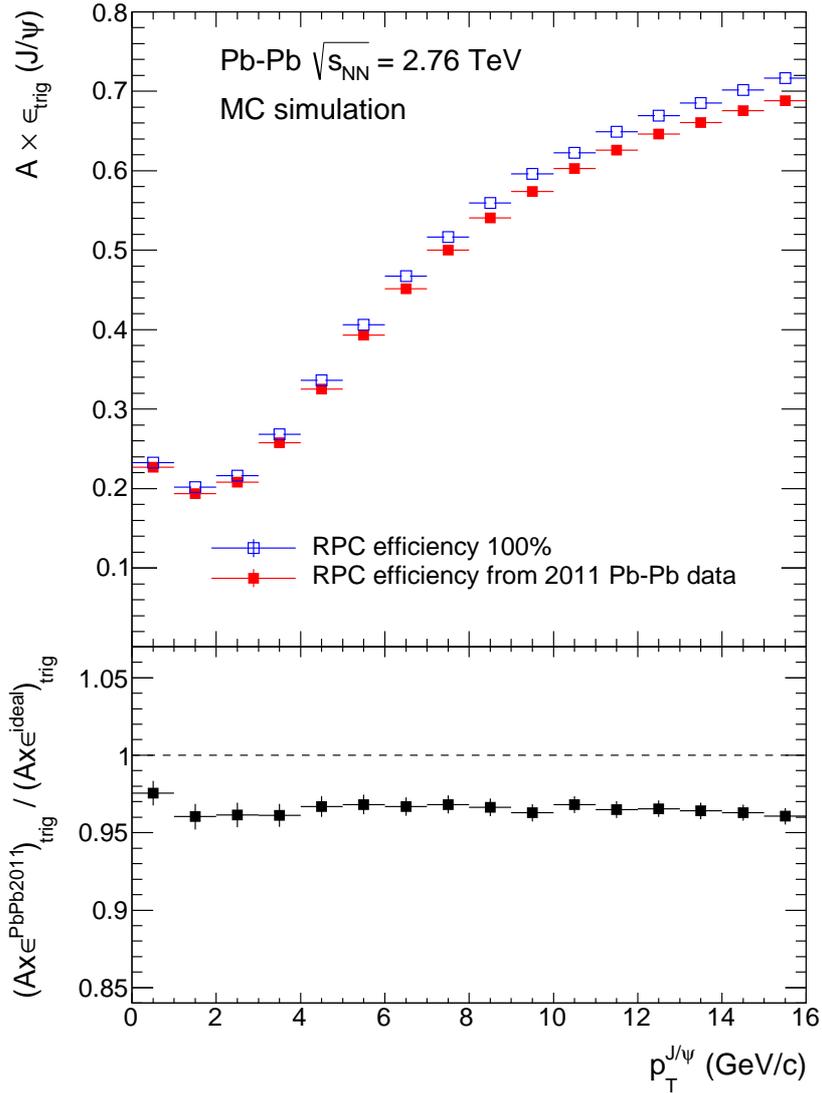}
\caption{\label{fig:trigger efficiency}Unlike-sign dimuon trigger efficiency 
for \jpsi, calculated using a realistic (filled squares) and ideal (open squares) 
chamber efficiency. The ratio of the two curves is shown in the bottom panel.}
\end{center}
\end{figure}

The trigger used for \jpsi measurements~\cite{Abelev:2013ila} in the 2011 
\pbpb run was an unlike-sign dimuon trigger (MUL) with a \pt threshold of 1\gevc 
for each muon. The  centrality-integrated efficiency of this trigger for \jpsi 
is shown in Fig.~\ref{fig:trigger efficiency} as a function of the \jpsi 
transverse momentum. The trigger efficiency is evaluated via a MC simulation 
having as input the trigger chamber efficiency, determined from experimental 
data~\cite{Bossu:2012jt}. In order to separate the detector efficiency from 
acceptance effects, the simulation was also run assuming a chamber efficiency 
of 100\%. The effect of the trigger chamber inefficiencies is smaller than 5\%, 
with weak (if any) \pt dependence.

\subsection{Invariant-mass resolution}

The momentum resolution of the muon spectrometer crucially depends 
on the detector alignment. 
Each of the 156 detection elements of the muon spectrometer's tracking 
chambers has six spatial degrees of freedom, three translations 
and three rotations. In addition, since the detection elements are mounted in 
independent support structures, six further degrees of freedom per half-chamber 
need to be considered. The initial position of the (half-)chambers was measured 
by the CERN survey group with about 1~mm resolution in three directions. 
The displacements of the (half-)chambers relative to a reference chamber has been 
monitored by the Geometry Monitoring System (GMS)~\cite{Grigorian:2005sia} 
with about 40~$\mu$m resolution in three directions.
The optimal method for aligning the tracking detectors is to use 
reconstructed tracks taken with and without magnetic field and perform a least-square 
minimization of the cluster-to-track residuals with respect to the alignment and 
the track parameters simultaneously. 
A special computation-efficient implementation~\cite{Castillo:1047110} allowed 
the minimization to be performed on a sample of 500000 tracks, which corresponded 
to a few hours of data taking. The resulting alignment resolution was~$\sim$100~$\mu$m. 

\begin{figure}[hbt]
\begin{center}
\includegraphics[width=0.75\textwidth]{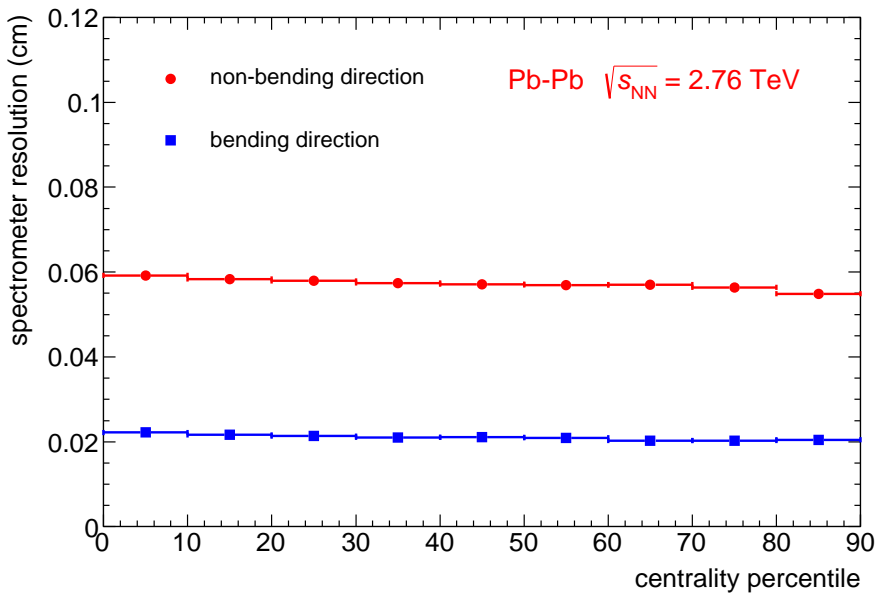}
\caption{\label{fig:cluster-resolution-vs-centrality}
Muon spectrometer resolution measured as a function of the centrality 
of the collision. The main contributions come from the cluster 
resolution and the residual misalignment of the tracking chambers.}
\end{center}
\end{figure}

The overall detector 
resolution, including the cluster resolution and the residual misalignment, can 
be measured using the distance between the position of the clusters and the 
position of the reconstructed tracks they belong to. Within chambers it ranges 
between 450 and 800~$\mu$m in the non-bending direction, and between 100 and 
400~$\mu$m in the bending direction. The degradation in resolution due to the 
large occupancy in central heavy ion collisions is less than 5\% 
(Fig.~\ref{fig:cluster-resolution-vs-centrality}).

\begin{figure}[p]
\begin{center}
\includegraphics[width=0.75\textwidth]{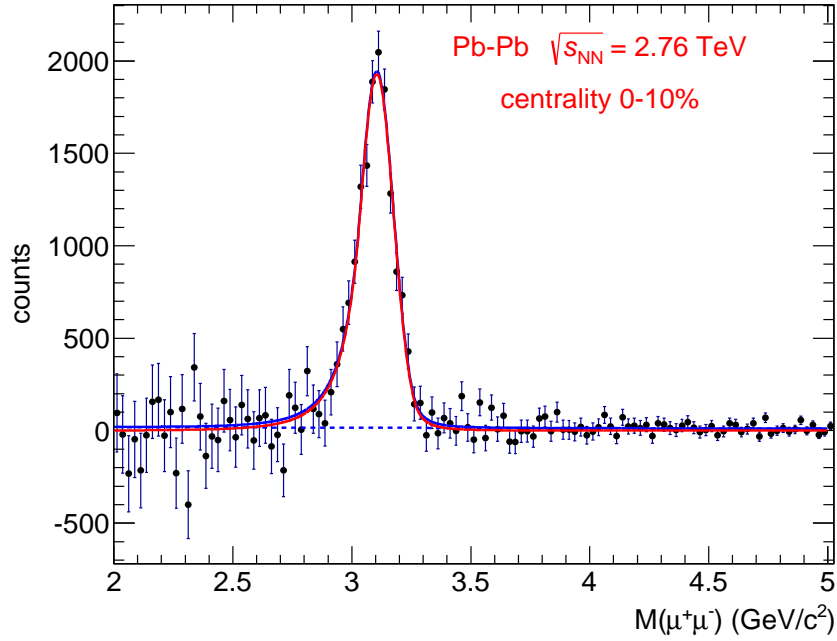}
\caption{\label{fig:jpsi-invariant-mass-0-10}
Invariant-mass distribution of $\mu^+ \mu^-$ pairs in 0--10\% most central 
\pbpb collisions at \mbox{\sqrtsnn = 2.76 TeV} with the \jpsi peak fitted
by an extended Crystal Ball function. The combinatorial background was 
determined by the event mixing method and subtracted.}
\end{center}
\end{figure}

\begin{figure}[p]
\begin{center}
\includegraphics[width=0.75\textwidth]{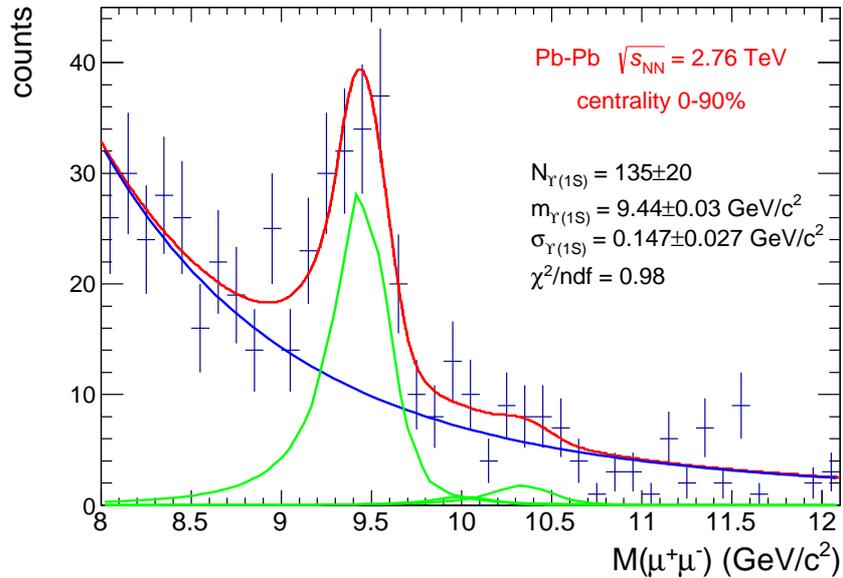}
\caption{\label{fig:upsilon-invariant-mass}
Invariant-mass distribution of $\mu^+ \mu^-$ pairs in \pbpb collisions at 
\mbox{\sqrtsnn = 2.76 TeV} with the $\Upsilon(1S)$,  $\Upsilon(2S)$,  and 
$\Upsilon(3S)$ peaks fitted by the sum of three extended Crystal Ball 
functions with identical relative widths and identical relative 
displacements from the PDG mass values. The tail shape is fixed by the 
embedding-MC simulation and the combinatorial background is parametrized 
by an exponential.}
\end{center}
\end{figure}

To extract the invariant mass distributions of muon pairs in \pbpb collisions, 
the standard track cuts previously described (trigger matching, $R_{\rm abs}$ and 
pseudorapidity cuts) are applied to both muon tracks. The \jpsi peak in the 
$\mu^+\mu^-$ invariant mass spectra can be fitted by an extended Crystal Ball 
function~\cite{Gaiser:1982yw} (Fig.~\ref{fig:jpsi-invariant-mass-0-10}). 
The mass resolution at the \jpsi peak in central \pbpb collisions, 
$\sim$73\mevcc, is in agreement with the design value. 
An analogous fit of the $\Upsilon$ peak in minimum-bias \pbpb collisions yields 
a mass resolution of $147\pm27$\mevcc. This is shown in Fig.~\ref{fig:upsilon-invariant-mass}, 
representing the full statistics of the 2011 run. 
The mass resolution, in general, is determined by multiple scattering and 
energy loss in the front absorber, intrinsic spatial resolution of the 
chambers, and alignment. At the \jpsi and $\Upsilon$ peaks the resolution is 
dominated by multiple scattering in the front absorber and the overall detector 
resolution, respectively. 

\begin{figure}[t]
\begin{center}
\includegraphics[width=0.75\textwidth]{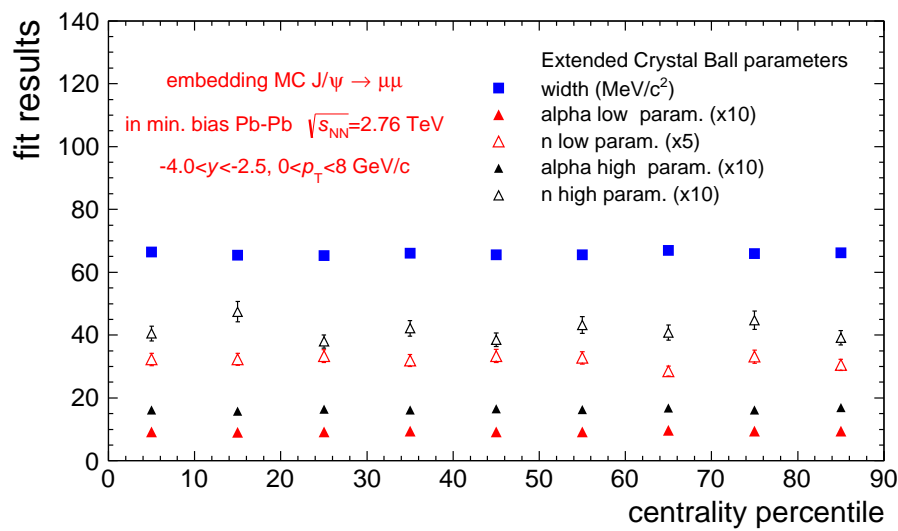}
\caption{\label{fig:jpsi-mass-shape-vs-centrality}
Centrality dependence of \jpsi invariant mass fit parameters.}
\end{center}
\end{figure}

The aforementioned increase of the detector occupancy with the centrality of 
the collision could alter the shape of the \jpsi mass peak. 
This effect has been studied using a Monte Carlo embedding procedure, in which 
a simulated signal particle (a \jpsi in our case) is embedded into a real 
raw-data event. The embedded event is then reconstructed as if it were a  
real event. This technique has the advantage of providing the most realistic 
background conditions. With such a technique it was shown 
(Fig.~\ref{fig:jpsi-mass-shape-vs-centrality}) that the \jpsi signal fit 
parameters do not depend on centrality. 
The peak widths obtained from the simulation agree within errors 
(from 3\% for central collisions to 10\% for the most peripheral ones) with those observed in experimental data.
The same embedding technique has also been used to confirm the small drop in the 
track reconstruction efficiency for the most central collisions mentioned above.

\clearpage\newpage\section{Conclusion and outlook}
\label{sect:conclusion}

\renewcommand{\thefootnote}{\alph{footnote}}

ALICE recorded data for all collision systems and energies offered by the Large 
Hadron Collider in its first running period from 2009 to 2013. 
The performance of the experiment was in good agreement with expectations. 
This is shown in Table~\ref{tab:summary} where, for selected parameters, the 
achieved performance is compared to the expectations contained in the ALICE 
Physics Performance Report from 2006~\cite{Alessandro:2006yt}~\footnotetext[1]{without track quality cuts}.
\begin{table}[b]
\ifcern
  \renewcommand{\arraystretch}{1.2}
\fi
\ifijmp
  \renewcommand{\arraystretch}{1.05}
\fi
\centering
{\begin{tabular*}{\textwidth}{@{\extracolsep{\fill}}lrr}
\toprule
Parameter & Expected & Achieved \\
\colrule
\multicolumn{3}{l}{\it Event vertex resolution with ITS--TPC tracks} \\
vertex resolution at $\dndeta=5$, transverse         & 85 \mum                      & 97 \mum         \\
vertex resolution at $\dndeta=25$, transverse        & 35 \mum                      & 32 \mum         \\
\colrule
\multicolumn{3}{l}{\it DCA resolution of ITS--TPC tracks in central \pbpb collisions} \\
transverse DCA resolution at $\pt=0.3\gevc$         & 200\mum                       & 200\mum         \\
transverse DCA resolution at $\pt=3\gevc$           &  30\mum                       &  30\mum         \\
transverse DCA resolution at $\pt=20\gevc$          &  15\mum                       &  15\mum         \\
\colrule
\multicolumn{3}{l}{\it DCA resolution of ITS--TPC tracks in \pp collisions (including vertex resolution)} \\
transverse DCA resolution at $\pt=0.2\gevc$         & 300\mum                       & 300\mum         \\
transverse DCA resolution at $\pt=3\gevc$           &  50\mum                       &  45\mum         \\
transverse DCA resolution at $\pt=30\gevc$          &  25\mum                       &  20\mum         \\
\colrule
\multicolumn{3}{l}{\it Barrel tracking efficiency in central \pbpb collisions} \\
TPC track finding efficiency at $\pt>0.2\gevc$      &       $>78\%$\footnotemark[1] &         $>70\%$ \\
TPC track finding efficiency at $\pt>1.0\gevc$      &       $>90\%$\footnotemark[1] &         $>78\%$ \\
ITS matching efficiency at $\pt>0.2\gevc$           &       $>95\%$                 &         $>92\%$ \\
\colrule
\multicolumn{3}{l}{\it Barrel \pt resolution} \\
$\Delta\pt/\pt$ of TPC tracks at $\pt=10\gevc$      &           4--6\%              &        6\%      \\
$\Delta\pt/\pt$ of TPC tracks at $\pt=30\gevc$      &           10--15\%            &       18\%      \\
$\Delta\pt/\pt$ of ITS--TPC tracks at $\pt=10\gevc$ &           1--2\%              &       1.5\%     \\
$\Delta\pt/\pt$ of ITS--TPC tracks at $\pt=30\gevc$ &           2--3\%              &       2.5\%     \\
\colrule
\multicolumn{3}{l}{\it Barrel particle identification} \\
TPC \dedx resolution in \pp                         &            5.4\%              &         5.2\%   \\
TPC \dedx resolution in central \pbpb               &            6.8\%              &         6.5\%   \\
TOF resolution                                      &           60-110 ps           &         80 ps   \\
T0 resolution                                       &           15-50 ps            &         21 ps   \\
\colrule
\multicolumn{3}{l}{\it Muon spectrometer} \\
MUON track finding efficiency                       &            95\%               &         85-87\% \\
invariant mass resolution at \jpsi peak in central \pbpb      & 70-74\mevcc         &        73\mevcc \\
invariant mass resolution at $\Upsilon$ peak in central \pbpb & 99--115\mevcc       &   147(27)\mevcc \\
\botrule
\end{tabular*} 
\tbl{Selection of parameters characterizing the performance of the ALICE experiment 
in Run~1 of the LHC. The expectations published in 2006 in the ALICE PPR\protect~\cite{Alessandro:2006yt} 
(column~2) and the achieved performance (column~3) are compared. For the vertex resolution, 
the approximation $\dndy\equiv\dndeta$ is used. \hfill{}}
\label{tab:summary}}
\end{table}

ALICE measurements during the full-energy LHC Run~2 (2015--2017) will, on one hand, 
focus on low-\pt observables where triggering is not possible. The goal here is to 
increase the statistics to $\sim$500 million minimum bias \pbpb events. Concerning 
rare probes, it is planned to inspect 1~nb$^{-1}$ \pbpb interactions in the rare-trigger 
running mode. 
This requires increasing the collision rates to 10--20~kHz, for which consolidation 
work is ongoing. The TPC electronics will be upgraded and the maximum readout rate of this 
detector will be doubled. The completion of TRD and PHOS, and extension 
of EMCal by adding calorimeter modules on the opposite side (Di-Jet Calorimeter, 
DCal)~\cite{Allen:1272952} are other important ingredients of the preparation for Run~2. 

In Run~3 (after 2018), the LHC will provide \pbpb collisions at a rate of 50~kHz. 
With the planned continuous readout of the ALICE TPC, the statistics available for 
data analysis could be increased compared to Run~2 by two orders of magnitude. To achieve 
this, the ALICE Collaboration has presented a plan to upgrade its detector systems. 
The current ITS will be replaced and the overall rate capabilities of the experiment 
will be enhanced. The goal is to have sampled, by the mid-2020s, an integrated 
luminosity of 10~nb$^{-1}$. In addition, three new detectors have been proposed. 
For more information, the reader is referred to the upgrade documents cited in 
Table~\ref{tab:upgrades}.  
\begin{table}[h]
\centering
\renewcommand{\arraystretch}{1.2}
{\begin{tabular*}{\textwidth}{@{\extracolsep{\fill}}lp{5.8cm}l}
\toprule
System & Upgrade & Documents\\
\colrule
ITS     & reduced material, improved resolution, topological trigger at L2
        & CDR~\cite{Musa:1431539}, LoI~\cite{Musa:1475243}, TDR~\cite{tdr-its}\\
TPC     & faster gas, GEM readout chambers, new readout electronics, continuous readout   
        & LoI~\cite{Musa:1475243}, TDR~\cite{Appelshaeuser:1622286} \\
Trigger / Readout & fast readout of ITS, TPC, TRD, TOF, EMCal, PHOS, MTR, MCH, and ZDC; replacing T0/V0/FMT with a new detector FIT; new trigger system
        & LoI~\cite{Musa:1475243}, TDR~\cite{ALICE:1603472} \\
O$^2$   & new combined DAQ, HLT, and offline computing system for high-rate and continuous readout & LoI~\cite{Musa:1475243} \\
MFT     & Muon Forward Tracker, pixel Si before absorber, $-4<\eta<-2.5$, better resolution and S/B for heavy flavors 
        & Addendum to LoI~\cite{CERN-LHCC-2013-014}\\
VHMPID  & Very High Momentum PID, gas Cherenkov, $\pi$/K/p separation in $5<p<25\gevc$ 
        & Ref.~\refcite{Acconcia:2013ptg} \\
FoCal   & Forward EM Calorimeter, W+Si $2.5<\eta<4.5$, $\gamma/\pi$ discrimination
        & Ref.~\refcite{Peitzmann:1572890}\\
\botrule
\end{tabular*} 
\tbl{ALICE upgrades considered for the time after Run 2.}
\label{tab:upgrades}}
\end{table}

\newenvironment{acknowledgement}{\relax}{\relax}
\begin{acknowledgement}
\section*{Acknowledgements}
The ALICE Collaboration would like to thank all its engineers and technicians for their invaluable contributions to the construction of the experiment and the CERN accelerator teams for the outstanding performance of the LHC complex.
\\
The ALICE Collaboration gratefully acknowledges the resources and support provided by all Grid centres and the Worldwide LHC Computing Grid (WLCG) collaboration.
\\
The ALICE Collaboration acknowledges the following funding agencies for their support in building and
running the ALICE detector:
State Committee of Science,  World Federation of Scientists (WFS)
and Swiss Fonds Kidagan, Armenia,
Conselho Nacional de Desenvolvimento Cient\'{\i}fico e Tecnol\'{o}gico (CNPq), Financiadora de Estudos e Projetos (FINEP),
Funda\c{c}\~{a}o de Amparo \`{a} Pesquisa do Estado de S\~{a}o Paulo (FAPESP);
National Natural Science Foundation of China (NSFC), the Chinese Ministry of Education (CMOE)
and the Ministry of Science and Technology of China (MSTC);
Ministry of Education and Youth of the Czech Republic;
Danish Natural Science Research Council, the Carlsberg Foundation and the Danish National Research Foundation;
The European Research Council under the European Community's Seventh Framework Programme;
Helsinki Institute of Physics and the Academy of Finland;
French CNRS-IN2P3, the `Region Pays de Loire', `Region Alsace', `Region Auvergne' and CEA, France;
German BMBF and the Helmholtz Association;
General Secretariat for Research and Technology, Ministry of
Development, Greece;
Hungarian OTKA and National Office for Research and Technology (NKTH);
Department of Atomic Energy and Department of Science and Technology of the Government of India;
Istituto Nazionale di Fisica Nucleare (INFN) and Centro Fermi -
Museo Storico della Fisica e Centro Studi e Ricerche "Enrico
Fermi", Italy;
MEXT Grant-in-Aid for Specially Promoted Research, Ja\-pan;
Joint Institute for Nuclear Research, Dubna;
National Research Foundation of Korea (NRF);
CONACYT, DGAPA, M\'{e}xico, ALFA-EC and the EPLANET Program
(European Particle Physics Latin American Network)
Stichting voor Fundamenteel Onderzoek der Materie (FOM) and the Nederlandse Organisatie voor Wetenschappelijk Onderzoek (NWO), Netherlands;
Research Council of Norway (NFR);
Polish Ministry of Science and Higher Education;
National Science Centre, Poland;
Ministry of National Education/Institute for Atomic Physics and CNCS-UEFISCDI - Romania;
Ministry of Education and Science of Russian Federation, Russian
Academy of Sciences, Russian Federal Agency of Atomic Energy,
Russian Federal Agency for Science and Innovations and The Russian
Foundation for Basic Research;
Ministry of Education of Slovakia;
Department of Science and Technology, South Africa;
CIEMAT, EELA, Ministerio de Econom\'{i}a y Competitividad (MINECO) of Spain, Xunta de Galicia (Conseller\'{\i}a de Educaci\'{o}n),
CEA\-DEN, Cubaenerg\'{\i}a, Cuba, and IAEA (International Atomic Energy Agency);
Swedish Research Council (VR) and Knut $\&$ Alice Wallenberg
Foundation (KAW);
Ukraine Ministry of Education and Science;
United Kingdom Science and Technology Facilities Council (STFC);
The United States Department of Energy, the United States National
Science Foundation, the State of Texas, and the State of Ohio.

\end{acknowledgement}

\ifcern
\fi
\ifijmp
\fi

\clearpage
\appendix
\section{The ALICE Collaboration}
\label{app:collab}



\begingroup
\small
\begin{flushleft}
B.~Abelev\Irefn{org71}\And
A.~Abramyan\Irefn{org1}\And
J.~Adam\Irefn{org37}\And
D.~Adamov\'{a}\Irefn{org79}\And
M.M.~Aggarwal\Irefn{org83}\And
M.~Agnello\Irefn{org90}\textsuperscript{,}\Irefn{org107}\And
A.~Agostinelli\Irefn{org26}\And
N.~Agrawal\Irefn{org44}\And
Z.~Ahammed\Irefn{org126}\And
N.~Ahmad\Irefn{org18}
% ...
\renewcommand\labelenumi{\textsuperscript{\theenumi}~}

\section*{Affiliation notes}
\renewcommand\theenumi{\roman{enumi}}
\begin{Authlist}
\item \Adef{0}Deceased
\item \Adef{idp1109504}{Also at: St. Petersburg State Polytechnical University}
\item \Adef{idp2988992}{Also at: Department of Applied Physics, Aligarh Muslim University, Aligarh, India}
\item \Adef{idp3689696}{Also at: M.V. Lomonosov Moscow State University, D.V. Skobeltsyn Institute of Nuclear Physics, Moscow, Russia}
\item \Adef{idp3941072}{Also at: University of Belgrade, Faculty of Physics and "Vin\v{c}a" Institute of Nuclear Sciences, Belgrade, Serbia}
\item \Adef{idp4222720}{Permanent Address: Permanent Address: Konkuk University, Seoul, Korea}
\item \Adef{idp4767104}{Also at: Institute of Theoretical Physics, University of Wroclaw, Wroclaw, Poland}
\item \Adef{idp5683104}{Also at: University of Kansas, Lawrence, KS, United States}
\end{Authlist}

\section*{Collaboration Institutes}
\renewcommand\theenumi{\arabic{enumi}~}
\begin{Authlist}

\item \Idef{org1}A.I. Alikhanyan National Science Laboratory (Yerevan Physics Institute) Foundation, Yerevan, Armenia
\item \Idef{org2}Benem\'{e}rita Universidad Aut\'{o}noma de Puebla, Puebla, Mexico
\item \Idef{org3}Bogolyubov Institute for Theoretical Physics, Kiev, Ukraine
\item \Idef{org4}Bose Institute, Department of Physics and Centre for Astroparticle Physics and Space Science (CAPSS), Kolkata, India
\item \Idef{org5}Budker Institute for Nuclear Physics, Novosibirsk, Russia
\item \Idef{org6}California Polytechnic State University, San Luis Obispo, CA, United States
\item \Idef{org7}Central China Normal University, Wuhan, China
\item \Idef{org8}Centre de Calcul de l'IN2P3, Villeurbanne, France
\item \Idef{org9}Centro de Aplicaciones Tecnol\'{o}gicas y Desarrollo Nuclear (CEADEN), Havana, Cuba
\item \Idef{org10}Centro de Investigaciones Energ\'{e}ticas Medioambientales y Tecnol\'{o}gicas (CIEMAT), Madrid, Spain
\item \Idef{org11}Centro de Investigaci\'{o}n y de Estudios Avanzados (CINVESTAV), Mexico City and M\'{e}rida, Mexico
\item \Idef{org12}Centro Fermi - Museo Storico della Fisica e Centro Studi e Ricerche ``Enrico Fermi'', Rome, Italy
\item \Idef{org13}Chicago State University, Chicago, USA
\item \Idef{org14}Commissariat \`{a} l'Energie Atomique, IRFU, Saclay, France
\item \Idef{org15}COMSATS Institute of Information Technology (CIIT), Islamabad, Pakistan
\item \Idef{org16}Departamento de F\'{\i}sica de Part\'{\i}culas and IGFAE, Universidad de Santiago de Compostela, Santiago de Compostela, Spain
\item \Idef{org17}Department of Physics and Technology, University of Bergen, Bergen, Norway
\item \Idef{org18}Department of Physics, Aligarh Muslim University, Aligarh, India
\item \Idef{org19}Department of Physics, Ohio State University, Columbus, OH, United States
\item \Idef{org20}Department of Physics, Sejong University, Seoul, South Korea
\item \Idef{org21}Department of Physics, University of Oslo, Oslo, Norway
\item \Idef{org22}Dipartimento di Fisica dell'Universit\`{a} 'La Sapienza' and Sezione INFN Rome, Italy
\item \Idef{org23}Dipartimento di Fisica dell'Universit\`{a} and Sezione INFN, Cagliari, Italy
\item \Idef{org24}Dipartimento di Fisica dell'Universit\`{a} and Sezione INFN, Trieste, Italy
\item \Idef{org25}Dipartimento di Fisica dell'Universit\`{a} and Sezione INFN, Turin, Italy
\item \Idef{org26}Dipartimento di Fisica e Astronomia dell'Universit\`{a} and Sezione INFN, Bologna, Italy
\item \Idef{org27}Dipartimento di Fisica e Astronomia dell'Universit\`{a} and Sezione INFN, Catania, Italy
\item \Idef{org28}Dipartimento di Fisica e Astronomia dell'Universit\`{a} and Sezione INFN, Padova, Italy
\item \Idef{org29}Dipartimento di Fisica `E.R.~Caianiello' dell'Universit\`{a} and Gruppo Collegato INFN, Salerno, Italy
\item \Idef{org30}Dipartimento di Scienze e Innovazione Tecnologica dell'Universit\`{a} del  Piemonte Orientale and Gruppo Collegato INFN, Alessandria, Italy
\item \Idef{org31}Dipartimento Interateneo di Fisica `M.~Merlin' and Sezione INFN, Bari, Italy
\item \Idef{org32}Division of Experimental High Energy Physics, University of Lund, Lund, Sweden
\item \Idef{org33}Eberhard Karls Universit\"{a}t T\"{u}bingen, T\"{u}bingen, Germany
\item \Idef{org34}European Organization for Nuclear Research (CERN), Geneva, Switzerland
\item \Idef{org35}Faculty of Engineering, Bergen University College, Bergen, Norway
\item \Idef{org36}Faculty of Mathematics, Physics and Informatics, Comenius University, Bratislava, Slovakia
\item \Idef{org37}Faculty of Nuclear Sciences and Physical Engineering, Czech Technical University in Prague, Prague, Czech Republic
\item \Idef{org38}Faculty of Science, P.J.~\v{S}af\'{a}rik University, Ko\v{s}ice, Slovakia
\item \Idef{org39}Frankfurt Institute for Advanced Studies, Johann Wolfgang Goethe-Universit\"{a}t Frankfurt, Frankfurt, Germany
\item \Idef{org40}Gangneung-Wonju National University, Gangneung, South Korea
\item \Idef{org41}Gauhati University, Department of Physics, Guwahati, India
\item \Idef{org42}Helsinki Institute of Physics (HIP), Helsinki, Finland
\item \Idef{org43}Hiroshima University, Hiroshima, Japan
\item \Idef{org44}Indian Institute of Technology Bombay (IIT), Mumbai, India
\item \Idef{org45}Indian Institute of Technology Indore, Indore (IITI), India
\item \Idef{org46}Inha University, Incheon, South Korea
\item \Idef{org47}Institut de Physique Nucl\'eaire d'Orsay (IPNO), Universit\'e Paris-Sud, CNRS-IN2P3, Orsay, France
\item \Idef{org48}Institut f\"{u}r Informatik, Johann Wolfgang Goethe-Universit\"{a}t Frankfurt, Frankfurt, Germany
\item \Idef{org49}Institut f\"{u}r Kernphysik, Johann Wolfgang Goethe-Universit\"{a}t Frankfurt, Frankfurt, Germany
\item \Idef{org50}Institut f\"{u}r Kernphysik, Westf\"{a}lische Wilhelms-Universit\"{a}t M\"{u}nster, M\"{u}nster, Germany
\item \Idef{org51}Institut Pluridisciplinaire Hubert Curien (IPHC), Universit\'{e} de Strasbourg, CNRS-IN2P3, Strasbourg, France
\item \Idef{org52}Institute for Nuclear Research, Academy of Sciences, Moscow, Russia
\item \Idef{org53}Institute for Subatomic Physics of Utrecht University, Utrecht, Netherlands
\item \Idef{org54}Institute for Theoretical and Experimental Physics, Moscow, Russia
\item \Idef{org55}Institute of Experimental Physics, Slovak Academy of Sciences, Ko\v{s}ice, Slovakia
\item \Idef{org56}Institute of Physics, Academy of Sciences of the Czech Republic, Prague, Czech Republic
\item \Idef{org57}Institute of Physics, Bhubaneswar, India
\item \Idef{org58}Institute of Space Science (ISS), Bucharest, Romania
\item \Idef{org59}Instituto de Ciencias Nucleares, Universidad Nacional Aut\'{o}noma de M\'{e}xico, Mexico City, Mexico
\item \Idef{org60}Instituto de F\'{\i}sica, Universidad Nacional Aut\'{o}noma de M\'{e}xico, Mexico City, Mexico
\item \Idef{org61}iThemba LABS, National Research Foundation, Somerset West, South Africa
\item \Idef{org62}Joint Institute for Nuclear Research (JINR), Dubna, Russia
\item \Idef{org63}Konkuk University, Seoul, South Korea
\item \Idef{org64}Korea Institute of Science and Technology Information, Daejeon, South Korea
\item \Idef{org65}KTO Karatay University, Konya, Turkey
\item \Idef{org66}Laboratoire de Physique Corpusculaire (LPC), Clermont Universit\'{e}, Universit\'{e} Blaise Pascal, CNRS--IN2P3, Clermont-Ferrand, France
\item \Idef{org67}Laboratoire de Physique Subatomique et de Cosmologie (LPSC), Universit\'{e} Joseph Fourier, CNRS-IN2P3, Institut Polytechnique de Grenoble, Grenoble, France
\item \Idef{org68}Laboratori Nazionali di Frascati, INFN, Frascati, Italy
\item \Idef{org69}Laboratori Nazionali di Legnaro, INFN, Legnaro, Italy
\item \Idef{org70}Lawrence Berkeley National Laboratory, Berkeley, CA, United States
\item \Idef{org71}Lawrence Livermore National Laboratory, Livermore, CA, United States
\item \Idef{org72}Moscow Engineering Physics Institute, Moscow, Russia
\item \Idef{org73}National Centre for Nuclear Studies, Warsaw, Poland
\item \Idef{org74}National Institute for Physics and Nuclear Engineering, Bucharest, Romania
\item \Idef{org75}National Institute of Science Education and Research, Bhubaneswar, India
\item \Idef{org76}Niels Bohr Institute, University of Copenhagen, Copenhagen, Denmark
\item \Idef{org77}Nikhef, National Institute for Subatomic Physics, Amsterdam, Netherlands
\item \Idef{org78}Nuclear Physics Group, STFC Daresbury Laboratory, Daresbury, United Kingdom
\item \Idef{org79}Nuclear Physics Institute, Academy of Sciences of the Czech Republic, \v{R}e\v{z} u Prahy, Czech Republic
\item \Idef{org80}Oak Ridge National Laboratory, Oak Ridge, TN, United States
\item \Idef{org81}Petersburg Nuclear Physics Institute, Gatchina, Russia
\item \Idef{org82}Physics Department, Creighton University, Omaha, NE, United States
\item \Idef{org83}Physics Department, Panjab University, Chandigarh, India
\item \Idef{org84}Physics Department, University of Athens, Athens, Greece
\item \Idef{org85}Physics Department, University of Cape Town, Cape Town, South Africa
\item \Idef{org86}Physics Department, University of Jammu, Jammu, India
\item \Idef{org87}Physics Department, University of Rajasthan, Jaipur, India
\item \Idef{org88}Physik Department, Technische Universit\"{a}t M\"{u}nchen, Munich, Germany
\item \Idef{org89}Physikalisches Institut, Ruprecht-Karls-Universit\"{a}t Heidelberg, Heidelberg, Germany
\item \Idef{org90}Politecnico di Torino, Turin, Italy
\item \Idef{org91}Purdue University, West Lafayette, IN, United States
\item \Idef{org92}Pusan National University, Pusan, South Korea
\item \Idef{org93}Research Division and ExtreMe Matter Institute EMMI, GSI Helmholtzzentrum f\"ur Schwerionenforschung, Darmstadt, Germany
\item \Idef{org94}Rudjer Bo\v{s}kovi\'{c} Institute, Zagreb, Croatia
\item \Idef{org95}Russian Federal Nuclear Center (VNIIEF), Sarov, Russia
\item \Idef{org96}Russian Research Centre Kurchatov Institute, Moscow, Russia
\item \Idef{org97}Saha Institute of Nuclear Physics, Kolkata, India
\item \Idef{org98}School of Physics and Astronomy, University of Birmingham, Birmingham, United Kingdom
\item \Idef{org99}Secci\'{o}n F\'{\i}sica, Departamento de Ciencias, Pontificia Universidad Cat\'{o}lica del Per\'{u}, Lima, Peru
\item \Idef{org100}Sezione INFN, Bari, Italy
\item \Idef{org101}Sezione INFN, Bologna, Italy
\item \Idef{org102}Sezione INFN, Cagliari, Italy
\item \Idef{org103}Sezione INFN, Catania, Italy
\item \Idef{org104}Sezione INFN, Padova, Italy
\item \Idef{org105}Sezione INFN, Rome, Italy
\item \Idef{org106}Sezione INFN, Trieste, Italy
\item \Idef{org107}Sezione INFN, Turin, Italy
\item \Idef{org108}SSC IHEP of NRC "Kurchatov institute" , Protvino, Russia
\item \Idef{org109}SUBATECH, Ecole des Mines de Nantes, Universit\'{e} de Nantes, CNRS-IN2P3, Nantes, France
\item \Idef{org110}Suranaree University of Technology, Nakhon Ratchasima, Thailand
\item \Idef{org111}Technical University of Split FESB, Split, Croatia
\item \Idef{org112}The Henryk Niewodniczanski Institute of Nuclear Physics, Polish Academy of Sciences, Cracow, Poland
\item \Idef{org113}The University of Texas at Austin, Physics Department, Austin, TX, USA
\item \Idef{org114}Universidad Aut\'{o}noma de Sinaloa, Culiac\'{a}n, Mexico
\item \Idef{org115}Universidade de S\~{a}o Paulo (USP), S\~{a}o Paulo, Brazil
\item \Idef{org116}Universidade Estadual de Campinas (UNICAMP), Campinas, Brazil
\item \Idef{org117}University of Houston, Houston, TX, United States
\item \Idef{org118}University of Jyv\"{a}skyl\"{a}, Jyv\"{a}skyl\"{a}, Finland
\item \Idef{org119}University of Liverpool, Liverpool, United Kingdom
\item \Idef{org120}University of Tennessee, Knoxville, TN, United States
\item \Idef{org121}University of Tokyo, Tokyo, Japan
\item \Idef{org122}University of Tsukuba, Tsukuba, Japan
\item \Idef{org123}University of Zagreb, Zagreb, Croatia
\item \Idef{org124}Universit\'{e} de Lyon, Universit\'{e} Lyon 1, CNRS/IN2P3, IPN-Lyon, Villeurbanne, France
\item \Idef{org125}V.~Fock Institute for Physics, St. Petersburg State University, St. Petersburg, Russia
\item \Idef{org126}Variable Energy Cyclotron Centre, Kolkata, India
\item \Idef{org127}Vestfold University College, Tonsberg, Norway
\item \Idef{org128}Warsaw University of Technology, Warsaw, Poland
\item \Idef{org129}Wayne State University, Detroit, MI, United States
\item \Idef{org130}Wigner Research Centre for Physics, Hungarian Academy of Sciences, Budapest, Hungary
\item \Idef{org131}Yale University, New Haven, CT, United States
\item \Idef{org132}Yonsei University, Seoul, South Korea
\item \Idef{org133}Zentrum f\"{u}r Technologietransfer und Telekommunikation (ZTT), Fachhochschule Worms, Worms, Germany
\end{Authlist}
\endgroup

\end{document}